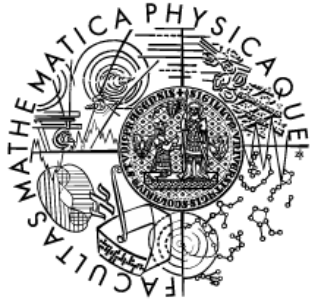

**FACULTY OF MATHEMATICS AND PHYSICS**
Charles University

# Dissertation

MSc Georg Stettinger

## SELECTED TOPICS IN STRING FIELD THEORY

Institute of Theoretical Physics

| | |
|---|---|
| Supervisor: | Prof. Doc. Martin Schnabl |
| Study programme: | Physics |
| Study branch: | Theoretical Physics, Astronomy and Astrophysics |

Prague 2025

I hereby declare that I have written this dissertation independently, using only the mentioned and duly cited sources and literature, and that the work has not been used in another university study programme or to obtain the same or another academic title.

In Prague on 15.05.2025,                  MSc Georg Stettinger

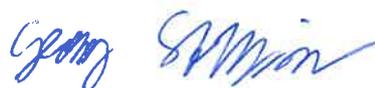

Title: Selected topics in string field theory

Author: Georg Stettinger

Institute: Institute of Theoretical Physics

Supervisor: Mgr. Martin Schnabl, Ph.D., CEICO, Institute of Physics, Czech Academy of Sciences

Abstract: This thesis deals with two topics in string field theory (SFT) which are of interest in current research: The role of stubs in open bosonic SFT and its variational principle in presence of a target space boundary.

The first part is a pedagogical introduction into the field of bosonic SFT. After discussing some general properties we expect, Witten's open SFT and Zwiebach's closed SFT are presented in detail. This means we set up the action, explain the algebraical and geometrical structure and mention physical applications. In the open case we review the most important analytic solutions including the necessary operator technology whileas in the closed case we focus on quantization and the more elaborate construction of the vertices.

The second part contains the papers I have completed within my PhD [71, 72, 79, 78], where the first three deal with stubs in OSFT and the last one with boundary terms.

Stubs in string field theory are interesting for various reasons, including building a bridge between the algebraic structures of open and closed SFT or acting as a regulator for otherwise ill-defined solutions. We have shown how stubs can be implemented into Witten theory using a generalized version of the homological perturbation lemma. The resulting theory is based on an $A_\infty$-action and hence algebraically similar to CSFT. We also constructed maps (cohomomorphisms) with different properties, which map solutions of Witten theory to solutions of the stubbed theory.

Since the most important analytic solutions are formulated in the sliver frame, it was necessary to carry over the whole stub construction to the sliver frame as well. Here we encountered some interesting subtleties related to sliver frame perturbation theory. By computing the 4-point amplitude, we could show why it is necessary to include higher vertices from a purely geometrical viewpoint. In the end, using the maps mentioned above, we were able to write down explicit solutions to the $A_\infty$-type equations of motion of the stubbed theory.

In the third paper we generalize some results of [13] and [50] as well: In [13] it has been shown that stubs can be introduced via an auxiliary field whileas in [50] everything is lifted up to the full quantum level. Both results can be shown to hold for sliver frame stubs as well, as a possible application we discuss regularization of identity-like solutions.

The last paper deals with the question about what happens to open string field theory if the target space has a boundary. We show that for the kinetic term one needs to add a Gibbons-Hawking-like additional term to the action. The cubic term in this context is problematic due to its non-local nature, hence we provide a purely qualitative treatment.

**Acknowledgements**

First of all I want to thank my supervisor Martin Schnabl for guiding me through this PhD and suggesting many interesting projects to work on. I really enjoyed the inspiring discussions and the always nice and cheerful atmosphere. Second, my thank goes to all my fellow researchers that I met in this last couple of years who I could exchange ideas and knowledge with. In a random order I mention Vinicius Bernardes, Paolo Rossi, Anna Morozova, Matěj Kudrna, Tomáš Procházka, Carlo Maccaferri, Barton Zwiebach, Raphaela Wutte, Ted Erler, Ashoke Sen, Yuji Okawa, Jojiro Yoshinaka, Alberto Ruffino, Riccardo Poletti, Atakan Fırat, Harold Erbin, Christoph Chiaffrino, Jaroslav Scheinpflug, Branislav Jurčo, Igor Kavkhine, Jakub Vošmera, Franz Stettinger and Ruggero Noris. Third, I would like to thank Radka Švecová and Dominika Vyštejnová for the kind and patient help with all the administrative challenges. I also want to thank Martin Schnabl and Vinicius Bernardes for helpful comments on the draft.

Last but not least I thank all the other people present in my life within the last years without whom this work would have never been completed.

My work has been funded by the Grant Agency of Czech Republic under the grants EXPRO 20-25775X and co-funded by the European Union and supported by the Czech Ministry of Education, Youth and Sports (Project No. FORTE – CZ.02.01.01/00/22_008/0004632).

# Contents

# Part I

# Introduction

From a broad perspective, the aim of theoretical high-energy physics is to understand and mathematically describe the fundamental particles and interactions that we observe in our universe. So far, four different basic forces have been identified: The strong nuclear force, the weak nuclear force, the electromagnetic force and gravity. While the first three are combined in the *Standard Model of Particle Physics* within the framework of *quantum field theory*, the last one is governed by Albert Einstein's theory of *General Relativity* within the framework of *differential geometry*. Although both of these theories were highly successfull and are experimentally verified to a large extent, they are not totally complete yet: The Standard Model cannot yet explain the origin of the *neutrino mass* and also suffers from fine-tuning problems such as the *hierarchy problem* or the *strong-CP problem*. On the other hand, GR does not give any explanation yet for the very small but non-zero value of the *cosmological constant*. There also exist the famous issues with *dark matter* and *inflation*, which could be caused by either some yet unknown type of particles which are missing in the Standard Model or some modification of Einstein gravity.

However, there is one unsolved problem which is even more fundamental and is therefore considered the most important one in theoretical high-energy physics: General Relativity and quantum field theory are incompatible. At length scales comparable to the Planck scale, gravity must be quantum, but until now it is not clear how to consistently quantize gravity. If one takes Einstein gravity and naively follows the standard quantization procedure after expanding around some background, one finds that the theory is *non-renormalizable*. This means that certain infinities occur, which cannot be absorbed by redefinitions of the parameters of the theory. Therefore, quantum general relativity on its own is inconsistent and not predictive on all scales.

There have been numerous attempts to come up with an alternative, consistent theory of quantum gravity, including *loop quantum gravity, causal dynamical triangulation, non-commutative geometry* or *Penrose's twistor theory*. The framework that got by far the most attention and is right now probably the most promising candidate is called *string theory*. The main idea is very simple: If one goes back to the non-renormalizability of Einstein gravity, one sees that the troublesome infinities basically occur because the gravitons are pointlike objects and interact in a spacetime portion of zero volume. It is therefore a natural idea to consider extended objects to model particles. The simplest extended object is a one-dimensional object, called *string*. Strings can be either *open*, with two endpoints, or *closed* circles, both of them turn out to be necessary in a realistic model of the universe. If one sets up a theory of interacting strings moving through spacetime via a bottom-up approach, one already encounters an extremely rich theory with many intriguing features. Every excitation mode of the string now represents a particle, giving rise to an infinite tower of particles of all different spins and arbitrarily high masses. Maybe the most striking feature is that Einstein gravity emerges very naturally in the framework of string theory: The closed string spectrum contains a massless spin-2 particle that represents the graviton and, moreover, a gravitational background where strings can move consistently has to obey the Einstein equations![1] This strongly suggests that string theory is the desired theory of quantum gravity. Of course, the positive features do not come without a price: String theory requires two more profound assumptions that have not been observed, namely *supersymmetry* and *extra dimensions*. Supersymmetry predicts a partner particle with the same mass but different spin for every particle in the Standard Model. Despite quite a huge experimental effort, none of those partner particles have been observed yet. The problem of extra dimensions refers to the fact that bosonic (supersymmetric) string theory is only consistent in 26 (10) spacetime dimensions. Since we observe only four dimensions, the remaining ones must be compactified on some small, unobservable scale. Those two issues have raised criticism against string theory, however, they also have opened up very interesting research areas with remarkable physical and mathematical results. Examples are the relation between string theory and *supergravity* during the second string revolution and the whole topic of *flux compactifications* with its connection to algebraic geometry. As a result, string theory is nowadays the dominating candidate amongst all attempts of unifying gravity and quantum field theory.

String theory in its standard form is a *first-quantized* theory, this means one is always working with only one worldsheet. This is perfectly fine for computing amplitudes at tree-level or loop-level, deduce consistency conditions on different backgrounds and many more things. However, suppose one is interested in phenomena involving a large, unknown number of worldsheets. Many strings are propagating through spacetime and giving rise to a collective phenomenon, such as a laser beam, a quark-gluon plasma or a black hole. Then first-quantized string theory is not sufficient anymore, one needs a more general and more comprehensive theory called *string field theory*. Despite its undeniable necessity, string field theory is still a quite small research area compared to other directions in string theory. After the first seminal paper by Edward Witten in 1986 [85] the field was mainly developed by Barton Zwiebach and Ashoke Sen in the 90s. The formalism, however, was still quite untractable for actual calculations and many results relied on numerical analysis. A breakthrough was the first analytic solution of open string field theory representing the tachyon vacuum in 2005 [70], which was followed by several other exact solutions [27, 1, 24, 25] during the following years. In closed string field theory, no analytic solutions exist to this date, mostly due to the much more complicated algebraic and geometric structure. A main obstacle is to find a consistent set of vertices, which are simple enough to allow for concrete computations. Some progress has been made recently mainly through the work of Barton Zwiebach and Atakan Fırat, see [6, 32, 29, 30]. It is to hope that we are now entering an era where it will be possible to attain more explicit results in closed string field theory.

The thesis is organized as follows: In part two we give a general introduction into bosonic string field theory. After discussing some basic properties we expect from any string field theory in section 1, we develop open and closed string field theory in more detail. It is written in a pedagogical way and should be readable with basic knowledge of first-quantized string theory (e. g. chapter 1-5 of Polchinski [65]). In some cases, more involved calculations are omitted and an adequate reference is given. Part three is the collection of research papers I have completed within my PhD; three of them are about how to deform open string field theory using stubs while the last one deals with

---
[1]At least to first order in the $\alpha'$-expansion.

boundary terms in open string field theory:

*Open string field theory with stubs,* Martin Schnabl, Georg Stettinger; JHEP 07 (2023) 032

*More on stubs in open string field theory,* Martin Schnabl, Georg Stettinger; JHEP 02 (2025) 017

*Sliver frame stubs in OSFT via auxiliary fields,* Georg Stettinger; Phys.Rev.D 111 (2025) 10, 106019

*A boundary term for open string field theory,* Georg Stettinger; JHEP 05 (2025) 226

**Note added**

During the defence process it turned out that there are some overlaps between some of the results of this thesis and prior existing literature, see [37, 54, 59]. Other important references in this context include [39, 38, 40, 57]. The author wants to thank Ivo Sachs for pointing this out.

# Part II

# String field theory

## 1    General aspects

*String field theory (SFT)* is a second-quantized version of string theory, describing the collective behaviour of a large number of interacting strings. Phenomena that have been studied within string field theory include tachyon condensation, instantons, UV- and IR-divergences, D-branes and multi-branes.

To gain an understanding of the importance and the expected properties of SFT a comparison with Einstein's theory of general relativity is useful: Imagine you are given all rules of graviton scattering, such that you know how to compute any graviton amplitude. How would you calculate the metric of a black hole? Finding, for instance, the Schwarzschild metric directly from the amplitudes is at least highly non-trivial, if not impossible. Of course, we are missing an essential concept here, which is the Einstein-Hilbert action. Given the action, we can proceed in both directions: If we expand the metric around some background, $g_{\mu\nu} = \eta_{\mu\nu} + h_{\mu\nu}$, where $\eta_{\mu\nu}$ is in the simplest case just the flat Minkowski metric, we can read off the propagator and the vertices for the graviton field $h_{\mu\nu}$. The Feynman rules then yield all possible graviton amplitudes. On the other hand, by solving the equations of motion under certain symmetry assumptions, we can find all known black hole solutions.

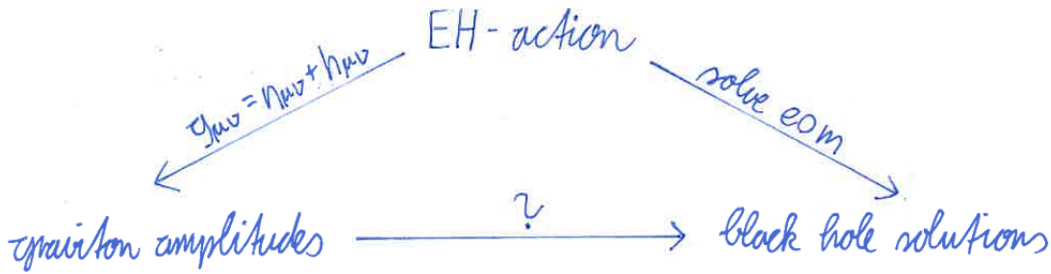

For string theory the story is similar: First-quantized string theory gives us in principle all of the amplitudes (up to computational difficulties). We can also deduce conditions on backgrounds in which strings can consistently propagate. As mentioned in the introduction, this is how the Einstein equations show up in string theory. But to find solutions where infinitely many excitation modes contribute, such as the tachyon vacuum, is impossible in pure first-quantized string theory. Again, there is a field theory defined via an action principle missing and this is precisely SFT. Analogously to the gravity case, SFT should give rise to the well-known string amplitudes by applying the Feynman rules. The classical solutions to its equations of motion then represent collective off-shell configurations of a large number of strings.

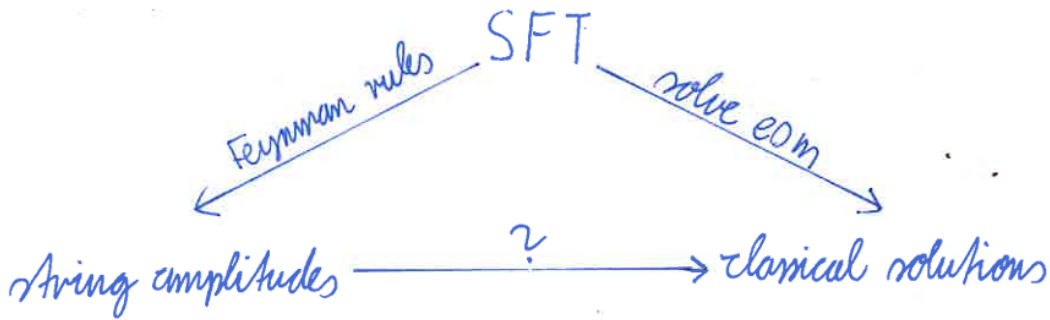

Let us collect a few important properties and features that we expect from an SFT:

- **The string field.** The first question we need to ask is, what should be the fundamental field in our theory? Following the route of second quantization, it should be an element of the Hilbert space of the first quantized theory, treated as a classical field. This would mean that the string field lives in the Hilbert space of the worldsheet CFT. In fact, the question is very subtle and until now there is no precise definition of the Hilbert space of string fields available. We know that the Fock space of the CFT is dense in the string field space, however, there exist important string fields which are not Fock states, see section 2.6.

- **Gauge invariance.** Since SFT should contain gauge theories like electrodynamics, we expect the whole theory to have some kind of gauge invariance. Most SFTs are based on BRST-quantization[2], where a state $\Psi$ is equivalent to $\Psi + Q\Lambda$ for some $\Lambda$. This redundancy in the description has to turn into a gauge symmetry in the full SFT. Gauge invariance typically requires the theory to have a very special algebraic structure [9].

- **Geometric interpretation.** In contrast to ordinary quantum field theory, where Feynman diagrams are just graphs, a string Feynman diagram is a two-dimensional surface. We expect the propagators and the vertices to respect this geometric structure, such that every Feynman diagram corresponds to a smooth Riemann surface. Moreover, it follows from unitarity that within an amplitude, the Feynman diagrams have to cover the full moduli space of Riemann surfaces with the given number of insertions and boundaries. We demand that this property is fulfilled in every SFT, it will become particularly important when stubs are introduced, see section 5.

- **Off-shell states.** Being a quantum theory, every interaction vertex has to be well-defined on off-shell states. It means, we have to specify local coordinates around the punctures because the insertion will not be independent of coordinate changes anymore. This will be a major complication in the construction of closed SFT, see section 3.5.

From these properties we see that a string field theory essentially consists of three pieces: A *CFT*, which defines our space of string fields, a *special algebraic structure* that gives rise to gauge invariance and a *set of vertices*, defined as subspaces of the respective moduli space of Riemann surfaces, with a well-defined action on off-shell string fields.

---

[2]Light-cone string field theory is an exception to that.

The SFTs which have been constructed so far split naturally into bosonic and supersymmetric as well as into open and closed SFTs. In this thesis we will only discuss bosonic SFTs, namely the two most well-known ones: *Witten's open string field theory (OSFT)* and *Zwiebach's closed string field theory (CSFT)*. In section four we give a short overview of other known SFTs.

# 2 Open string field theory

## 2.1 The open string background

We will now start our systematic discussion of Witten's OSFT by describing the boundary CFT which defines the space of string fields. Typically, this CFT is called "background", although this terminology might seem strange at first sight; one would maybe be more entitled to call it the "core" or the "essence". However, the CFT carries all the information about spacetime and the D-brane configuration in which the strings are moving, hence the name background is justified.

In fact, string field theory is consistent for a very large class of backgrounds, it does not even have to have a clear spacetime interpretation.[3] Here we want to focus on the simplest case of 26 scalar fields with Neumann boundary conditions representing a spacetime filling D-brane. Different structures such as lower dimensional branes or non-trivial background fields can be incorporated in a straightforward manner.

The *mode expansion* for the scalar fields reads

$$X^\mu(z,\overline{z}) = x^\mu - i\alpha' p^\mu \ln|z|^2 + i\sqrt{\frac{\alpha'}{2}} \sum_{n\neq 0} \frac{\alpha_n^\mu}{n}\left(\frac{1}{z^n} + \frac{1}{\overline{z}^n}\right) \quad (2.1)$$

where $z, \overline{z}$ are the complex worldsheet coordinates. Its (anti-)holomorphic derivative is a conformal primary of dimension (1,0) ((0,1))

$$\partial X^\mu(z) = -i\sqrt{\frac{\alpha'}{2}} \sum_n \frac{\alpha_n^\mu}{z^{n+1}}, \qquad \overline{\partial} X^\mu(z) = -i\sqrt{\frac{\alpha'}{2}} \sum_n \frac{\alpha_n^\mu}{\overline{z}^{n+1}}, \quad (2.2)$$

where

$$\alpha_0^\mu = \sqrt{2\alpha'} p^\mu. \quad (2.3)$$

In principle our worldsheet is represented as the upper half plane only but for some calculations it is useful to use the *doubling trick* and define

$$\partial X^\mu(z) = \overline{\partial} X^\mu(\overline{z}) \quad \text{for Im}(z) < 0. \quad (2.4)$$

---

[3]For example, in [68] OSFT was used as a tool to explore conformal perturbation theory, without any interpretation of propagating strings.

Hence, instead of working with a holomorphic and an antiholomorphic field in the upper half plane, we just consider the holomorphic field extended to the whole plane. The same story also holds for the reparametrization ghosts

$$b(z) = \sum_n \frac{b_n}{z^{n+2}}, \qquad c(z) = \sum_n \frac{c_n}{z^{n-1}} \tag{2.5}$$

being primary fields of holomorphic dimension plus two and minus one, respectively. In the following, we will freely pass around between the upper half plane and the doubling trick description, depending on the purposes.

The mode expansion formulas can be inverted via

$$\alpha_n^\mu = i\sqrt{\frac{2}{\alpha'}} \oint \frac{dz}{2\pi i} z^n \partial X^\mu(z), \qquad b_n = \oint \frac{dz}{2\pi i} z^{n+1} b(z), \qquad c_n = \oint \frac{dz}{2\pi i} z^{n-2} c(z). \tag{2.6}$$

The basic *operator product expansions* of the fields are

$$\partial X^\mu(z) \partial X^\nu(w) = -\frac{\alpha'}{2} \eta^{\mu\nu} \frac{1}{(z-w)^2} + \text{reg}, \qquad b(z) c(w) = \frac{1}{z-w} + \text{reg} \tag{2.7}$$

with all other combinations being non-singular. In terms of the modes they translate into the commutation relations

$$[\alpha_n^\mu, \alpha_m^\nu] = n \eta^{\mu\nu} \delta_{n+m}, \qquad \{b_n, c_m\} = \delta_{n+m}. \tag{2.8}$$

If one denotes the (holomorphic) conformal dimension of a primary field $\phi$ by $h$, then all modes with $n > -h$ annihilate the $SL(2,\mathbb{R})$-invariant vacuum. All the other modes are creation operators and can be used to construct the *Fock space* of the CFT. Explicitly, an arbitrary Fock state has the form

$$\Psi = \alpha_{-n_1}^{\mu_1} ... \alpha_{-n_r}^{\mu_r} b_{-m_1} ... b_{-m_s} c_{-p_1} ... c_{-p_t} |0, k\rangle, \qquad n_i \geq 1, \quad m_i \geq 2, \quad p_i \geq -1 \tag{2.9}$$

with the momentum dependent vacuum defined as

$$|0, k\rangle =: e^{ik_\mu X^\mu(0,0)} : |0\rangle = e^{ik_\mu x^\mu} |0\rangle. \tag{2.10}$$

The Fock states form a basis of our Hilbert space $\mathcal{H}_{BCFT}$ if one allows for infinite linear combinations of non-zero elements. To every state we can associate a local operator which creates the state in the infinite past. By a slight abuse of notation we will denote it with the same letter as the state itself and write

$$\Psi \equiv \Psi(0) |0\rangle. \tag{2.11}$$

This relation is called the *state-operator-correspondence*. For instance, a tachyon field would be given by

$$t(x) = \int \frac{d^{26}k}{(2\pi)^{26}} T(k) \left(c e^{ik_\mu X^\mu}\right)(0) |0\rangle = \int \frac{d^{26}k}{(2\pi)^{26}} T(k) e^{ik_\mu x^\mu} c_1 |0\rangle \tag{2.12}$$

where $T(k)$ is the spacetime field in the momentum representation. All higher-level fields take a similar form.

In SFT, an alternative representation of states called the *Schrödinger representation* is useful as well: Consider a surface conformally equivalent to the unit half disc, prepared with some boundary operator insertions (not necessarily at the origin). After specifying some boundary conditions on the upper semicircle, one can evaluate the path integral over the surface to obtain a complex number. The prepared semidisc therefore provides a functional from the space of field configurations at unit time to the complex numbers, which is the definition of a quantum state. The Schrödinger representation will be particularly useful for the geometric interpretation of gluing surfaces.

The description we gave here is suitable for "in-states", created in the infinite past. What about "out-states", created in the infinite future, i. e. on the point $\infty$ on the world-sheet? First, we should compactify the worldsheet by adding a point at infinity and turn it into the Riemann semisphere. In the Schrödinger representation, such a state should be prepared in the complement of the unit half disk within the upper half sphere. For Fock states, we can use the map $I(z) = -\frac{1}{z}$ which interchanges those two regions and define the *BPZ-conjugate* of a state $\Psi$ as

$$\Psi^* = \langle 0 | I \circ \Psi(0). \tag{2.13}$$

The notation $f \circ \phi(z)$ for a conformal transformation $f(z)$ and a holomorphic primary $\phi(z)$ just encodes the transformation rule

$$f \circ \phi(z) = \left(\frac{df(z)}{dz}\right)^h \phi(f(z)). \tag{2.14}$$

If we apply that to the generic Fock state (2.9) we get

$$\Psi^* = (-1)^{\Sigma n_i + \Sigma m_j + \Sigma p_k + r + t} \langle 0, k | \alpha_{n_1}^{\mu_1} ... \alpha_{n_r}^{\mu_r} b_{m_1} ... b_{m_s} c_{p_1} ... c_{p_t}. \tag{2.15}$$

The BPZ-conjugate induces a very natural bilinear form on the space of string fields which corresponds to the pairing of an in-state with an out-state.[4] In the Schrödinger representation one just glues the two surfaces together and integrates over the boundary conditions on the unit semicircle. In the Fock state picture, this boils down to a two-point function of a state with a BPZ-dual state and is therefore given by

$$\langle \Psi_1, \Psi_2 \rangle \equiv \langle 0 | I \circ \Psi_1(0) \Psi_2(0) | 0 \rangle. \tag{2.16}$$

One can straightforwardly show the relation

$$\langle \Psi_1, \Psi_2 \rangle = (-1)^{gh(\Psi_1)gh(\Psi_2)} \langle \Psi_2, \Psi_1 \rangle. \tag{2.17}$$

Here, the *ghost number* $gh(\Psi)$ of a Fock state $\Psi$ (2.9) provides a natural grading on the space of string fields and is given by the difference of *c*- and *b*-ghost insertions,

$$gh(\Psi) = t - s. \tag{2.18}$$

---
[4]One should mention that additionally to BPZ-conjugation we also have the standard *hermitian conjugation*. It can be used to define a *reality condition* on the string field, see for instance [64, 33].

The BPZ-product of two momentum-dependent vacua will be normalized as

$$\langle 0, k | c_{-1} c_0 c_1 | 0, k' \rangle = (2\pi)^{26} \delta^{(26)}(k + k').  \tag{2.19}$$

One should note at this point that a correlation function is only non-vanishing if the total ghost number is three. We normalize the basic ghost three-point function to

$$\langle c(z_1) c(z_2) c(z_3) \rangle = (z_1 - z_2)(z_1 - z_3)(z_2 - z_3).  \tag{2.20}$$

A special field which is present in any CFT is the *energy-momentum tensor*, given by[5]

$$T(z) = T^{(m)}(z) + T^{(gh)}(z) = -\frac{1}{\alpha'} : \partial X^\mu \partial X_\mu : (z) + :(\partial b) c : (z) - 2\partial (:bc:)(z).  \tag{2.21}$$

It generates infinitesimal conformal transformations $z \to z + \epsilon(z)$ according to

$$\delta_\epsilon \phi(z) = \oint_{C(z)} \frac{dw}{2\pi i} \epsilon(w) T(w) \phi(z).  \tag{2.22}$$

For any holomorphic primary field $\phi(z)$ of dimension $h$ the OPE takes the form

$$T(z) \phi(w) = \frac{h \phi(w)}{(z - w)^2} + \frac{\partial \phi(w)}{z - w} + \text{reg}.  \tag{2.23}$$

In contrast, we have

$$T(z) T(w) = \frac{c}{2(z-w)^4} + \frac{2T(w)}{(z-w)^2} + \frac{\partial T(w)}{z - w} + \text{reg}.  \tag{2.24}$$

with $c$ being the *central charge* of the theory, indicating that $T(z)$ is in general not primary. Its modes $L_n$ given by

$$L_n = \oint \frac{dz}{2\pi i} z^{n+1} T(z), \qquad T(z) = \sum_n \frac{L_n}{z^{n+2}}  \tag{2.25}$$

obey the *Virasoro algebra*

$$[L_n, L_m] = (n - m) L_{n+m} + \frac{c}{12}(n^3 - n) \delta_{n+m}.  \tag{2.26}$$

The central charge $c$ is given as the sum of its matter and ghost contribution, $c = c_{matter} + c_{ghost}$. $c_{matter}$ is equal to the spacetime dimension $D$ whileas $c_{ghost} = -26$, hence since we work in the critical dimension $D = 26$, the total central charge vanishes. For a primary field $\phi(z)$ the relations

$$[L_0, \phi(z)] = h\phi(z), \qquad [L_{-1}, \phi(z)] = \partial \phi(z).  \tag{2.27}$$

can be deduced from (2.23).

---

[5]In the presence of a boundary in target space, $T^{(m)}$ is not holomorphic anymore, see section 8.4

The *BRST-operator* $Q$ is defined via the BRST current $j_B$ as a contour integral around the origin,

$$Q\Psi = \oint_{C(0)} \frac{dz}{2\pi i} j_B(z) \Psi(0) |0\rangle \qquad (2.28)$$

with

$$j_B(z) = cT^{(m)}(z) + :bc\partial c:(z) + \frac{3}{2}\partial^2 c(z). \qquad (2.29)$$

It obeys the important relations

$$Q^2 = 0, \qquad [Q, b(z)] = T(z), \qquad [Q, c(z)] = c\partial c(z). \qquad (2.30)$$

By pulling the integration contour over the whole Riemann sphere until it encircles the point at infinity we can also show

$$\langle \Psi_1, Q\Psi_2 \rangle = -(-1)^{gh(\Psi_1)} \langle Q\Psi_1, \Psi_2 \rangle, \qquad (2.31)$$

i. e. $Q$ is BPZ-odd.[6] The space of physical, or on-shell, states is defined to be the *cohomology* of $Q$ at ghost number one:

$$Q\Psi = 0, \qquad \Psi \simeq \Psi + Q\Lambda \qquad (2.32)$$

## 2.2 The Witten action

Having summarized the main properties of our background CFT, it is time now to set up the action of open string field theory.

Let us start with the *kinetic term*: In any QFT, the kinetic term gives rise to the linearized equations of motion which describe free on-shell particles. In our case, on-shell string fields obey $Q\Psi = 0$ together with the equivalence relation (2.32). A natural choice would hence be [85]

$$S_{kin} = \frac{1}{2} \langle \Psi, Q\Psi \rangle. \qquad (2.33)$$

This expression fulfills the ghost number constraint and using (2.31) we can deduce the desired linearized equations of motion

$$Q\Psi = 0. \qquad (2.34)$$

Taking a different representative of the cohomology results in

$$\frac{1}{2} \langle \Psi + Q\Lambda, Q(\Psi + Q\Lambda) \rangle = \frac{1}{2} \langle \Psi, Q\Psi \rangle, \qquad (2.35)$$

so the kinetic term stays invariant. As anticipated already, the redundancy in the description of the first-quantized theory turns into a gauge symmetry of the second-quantized theory. This is a necessary mechanism, without this gauge symmetry we would introduce new degrees of freedom not present in the original string theory.

---

[6]Again, this relation is modified in the presence of a target space boundary, see 8.3

The more non-trivial part of the theory is the interaction part and we will find that actually only one *cubic term* is already enough to make it consistent. This is a very special property of Witten theory and not the case for many other SFTs. If the cubic term should have a similar form involving the BPZ-product, then what we need is a product of ghost number zero, usually denoted by $*$, which takes two string fields as inputs:

$$S_{cub} = \frac{1}{3} \langle \Psi, \Psi * \Psi \rangle. \tag{2.36}$$

The *star product* constructed by Witten in [85] has the following interpretation: Consider two incoming strings and separate both into a left half and a right half. The whole space of string states can then be seen as the tensor product of the spaces of the left- and right half degrees of freedom. Now glue the right half of the first string to the left half of the second string and integrate over those degrees of freedom. What is left is a new string state made out of the left half of the first string and the right half of the second string:

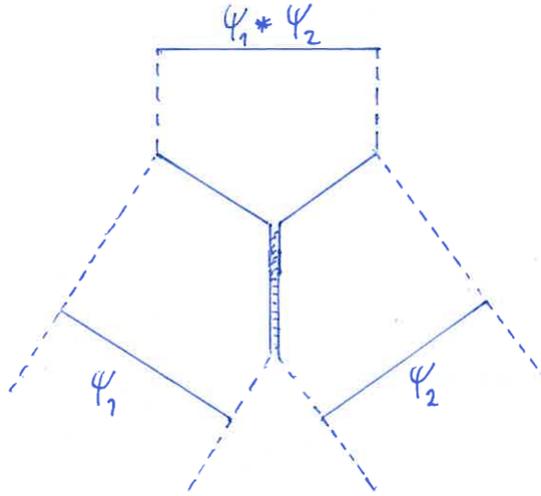

It is easy to see from this graphical definition that $*$ is associative but not commutative. Moreover, if we look at the full vertex (2.36) we observe the following: From the Schrödinger representation we learn that the BPZ-product effectively just glues the two strings together and integrates over their degrees of freedom. That means that within the 3-vertex, the right half of the outgoing string (with reversed orientation) is glued to the left half of the product string and vice versa. This suggests that the vertex is actually fully cyclically symmetric:

$$\langle \Psi_1, \Psi_2 * \Psi_3 \rangle = \langle \Psi_2, \Psi_3 * \Psi_1 \rangle. \tag{2.37}$$

(A mathematical proof will be given in section 2.4.) Now consider the BRST-operator in the form (2.28) acting on the two input strings of the product. Since the contour integral is conformally invariant, we can imagine it running around the strings $\Psi_1$ and $\Psi_2$ in the figure above. At the part where the strings are glued together, the contour integrations cancel out and what remains just encircles the product. This means that $Q$ acts as a derivative with respect to the star product.

Before we give a more explicit mathematical definition of the three-vertex in section 2.4, let us now write down the full action:

$$S(\Psi) = -\frac{1}{\alpha'^3 g_o^2} \left( \frac{1}{2} \langle \Psi, Q\Psi \rangle + \frac{1}{3} \langle \Psi, \Psi * \Psi \rangle \right) \tag{2.38}$$

The prefactor is there for dimensional reasons, $g_o$ is the open string coupling constant. The action is of Chern-Simons type, with its equations of motion given by

$$Q\Psi + \Psi * \Psi = 0. \tag{2.39}$$

The algebraic relations are summarized as follows: First we have the nilpotency of $Q$, the Leibniz rule and the associativity of the star product

$$Q^2 = 0, \tag{2.40}$$
$$Q(\Psi_1 * \Psi_2) = Q\Psi_1 * \Psi_2 + (-1)^{gh(\Psi_1)} \Psi_1 * Q\Psi_2,$$
$$(\Psi_1 * \Psi_2) * \Psi_3 = \Psi_1 * (\Psi_2 * \Psi_3)$$

and second the compatibility relations with the BPZ-product called *cyclicity conditions*

$$\langle \Psi_1, Q\Psi_2 \rangle = -(-1)^{gh(\Psi_1)} \langle Q\Psi_1, \Psi_2 \rangle, \tag{2.41}$$
$$\langle \Psi_1, \Psi_2 * \Psi_3 \rangle = \langle \Psi_2, \Psi_3 * \Psi_1 \rangle.$$

The first set tells us $Q$ and $*$ form a differential graded algebra whereas the latter imply that this algebra is cyclic with respect to the BPZ-product. Together they guarantee that the gauge invariance of the kinetic term extends to the non-linear level: The variation

$$\delta\Psi = Q\Lambda + \Psi * \Lambda - \Lambda * \Psi, \tag{2.42}$$

where $\Lambda$ is some ghost number zero gauge parameter, leaves the action invariant. Checking that explicitly is straightforward but tedious, in the next section we will give a more elegant proof in the coalgebra formalism.

## 2.3 The coalgebra formalism

Witten theory is undoubtedly the simplest SFT with a well-known algebraic structure. However, other SFTs, for instance the modification using stubs discussed in section 5, some superstring field theories as well as CSFT require more elaborate structures. In this section we will introduce a powerful formalism which encompasses those in an elegant way.

Let us start by defining the *tensor coalgebra $TV$* associated to a (graded) vector space $V$ as

$$TV = V^{\otimes 0} + V^{\otimes 1} + V^{\otimes 2} + ... = \sum_{n=0}^{\infty} V^{\otimes n} \tag{2.43}$$

We can project on the *n*-th tensor power using the projection operator $\pi_n$, i. e.

$$\pi_n TV = V^{\otimes n}. \tag{2.44}$$

Suppose we are given some multilinear map $m_n$ on $V$, $m_n : V^{\otimes n} \to V$. We can extend the domain of $m_n$ to the whole $TV$ be defining the *coderivation*[7] $\mathbf{m_n}$ as

$$\mathbf{m_n} = \sum_{j=n}^{\infty} \sum_{k=0}^{j-n} 1^{\otimes k} \otimes m_n \otimes 1^{\otimes j-k-n}. \tag{2.45}$$

On tensor power $V^{\otimes k}$ with $k < n$ the action of $\mathbf{m_n}$ shall give zero. Note that the formula also makes sense for $n = 0$, then it just corresponds to the insertion of some special vector. A general coderivation $\mathbf{m}$ can be defined as a linear combination of expressions of the form of (2.45) for different $n$. The individual products can then be recovered as

$$m_n = \pi_1 \mathbf{m} \pi_n. \tag{2.46}$$

The product of two coderivations is in general not a coderivation, but their commutator is, as can be checked straightforwardly.

Now we want to apply those concepts to string field theory: First, our vector space $V$ shall be the space of string fields $\mathcal{H}_{BCFT}$. It is convenient to introduce a new degree for the string fields defined by

$$|\Psi| = gh(\Psi) + 1. \tag{2.47}$$

The degree of a map is then just defined as the difference of the degree of the output and the sum of the input degrees. We start by defining the degree-odd maps

$$m_1(\Psi) \equiv Q\Psi, \qquad m_2(\Psi_1, \Psi_2) \equiv (-1)^{|\Psi_1|} \Psi_1 * \Psi_2. \tag{2.48}$$

The coalgebra formalism provides a method to add those two maps into a single quantity: After defining

$$\mathbf{m} \equiv \mathbf{m_1} + \mathbf{m_2} \tag{2.49}$$

one can straightforwardly show that the single equation

$$\mathbf{m}^2 = 0 \tag{2.50}$$

encodes all three relations in (2.40). Moreover, if we use the short-cut notation

$$\frac{1}{1-\Psi} \equiv \sum_{n=0}^{\infty} \Psi^{\otimes k} \tag{2.51}$$

---

[7]This section is just a very short introduction to the topic; for a more detailed review, where it is also explained why the name coderivation is justified, see [84] and also [17, 18].

we can write the equations of motion (2.39) as

$$\mathbf{m}\frac{1}{1-\Psi} = 0. \tag{2.52}$$

These simplifications already reflect the utility of the formalism. Suppose now we want to generalize our theory and allow for higher vertices as well. From the above discussion, the obvious generalization is to take an odd coderivation that squares to zero but do not restrict the number of inputs of the individual maps. Such a structure is called an $A_\infty$-*algebra*, a collection of multi-linear maps with arbitrarily many inputs that obey quadratic relations stemming from (2.50). For example, the nilpotency of $m_1$ and the Leibniz rule stay the same, but the associativity relation changes into

$$\begin{aligned} & m_2(m_2(\Psi_1, \Psi_2), \Psi_3) + (-1)^{|\Psi_1|} m_2(\Psi_1, m_2(\Psi_2, \Psi_3)) + \\ & m_1(m_3(\Psi_1, \Psi_2, \Psi_3)) + m_3(m_1(\Psi_1), \Psi_2, \Psi_3) + \\ & (-1)^{|\Psi_1|} m_3(\Psi_1, m_1(\Psi_2), \Psi_3) + (-1)^{|\Psi_1|+|\Psi_2|} m_3(\Psi_1, \Psi_2, m_1(\Psi_3)) = 0. \end{aligned} \tag{2.53}$$

We see that in an $A_\infty$-algebra the two-product need not be associative, with the price of introducing higher products. The equations of motion of an $A_\infty$-theory, called Maurer-Cartan equation, read

$$\pi_1 \mathbf{m} \frac{1}{1-\Psi} = \sum_{n=1}^\infty m_n\left(\Psi^{\otimes n}\right) = 0. \tag{2.54}$$

To write down the corresponding action, we need to incorporate the BPZ-product into the formalism: Define the *symplectic form*

$$\omega(\Psi_1, \Psi_2) = -(-1)^{|\Psi_1|} \langle \Psi_1, \Psi_2 \rangle, \tag{2.55}$$

then the $A_\infty$-infinity action is

$$S(\Psi) = \omega\left(\Psi, \sum_{n=1}^\infty \frac{1}{n+1} m_n\left(\Psi^{\otimes n}\right)\right). \tag{2.56}$$

To correctly produce the equations of motion (2.54), the maps $m_n$ must be cyclic, i. e. obey

$$\omega(\Psi_1, m_n(\Psi_2...\Psi_{n+1})) = -(-1)^{|\Psi_1|} \omega(m_n(\Psi_1...\Psi_n), \Psi_{n+1}). \tag{2.57}$$

For $n = 1, 2$ these conditions are identical to (2.41).

We will now give a proof that every $A_\infty$-infinity action of the form (2.56) has gauge invariance. First, let us get rid of the $\frac{1}{n+1}$-factor by the following trick: Define an interpolating string field $\Psi(t)$ with $\Psi(0) = 0$ and $\Psi(1) = \Psi$ (the details of the interpolation are not important). The action (2.56) can then be written equivalently as

$$S(\Psi) = \int_0^1 dt\, \omega\left(\pi_1 \partial_t \frac{1}{1-\Psi(t)}, \pi_1 \mathbf{m} \frac{1}{1-\Psi(t)}\right), \tag{2.58}$$

where $\boldsymbol{\partial_t}$ is the coderivation corresponding to $\partial_t$. In fact, the above expression is just an integral over a total derivative where only the surface term contributes[8],

$$S(\Psi) = \int_0^1 dt\, \boldsymbol{\partial_t} \sum_{n=1}^\infty \frac{1}{n+1}\omega\left(\pi_1 \frac{1}{1-\Psi(t)}, \pi_1 \mathbf{m_n} \frac{1}{1-\Psi(t)}\right). \tag{2.59}$$

Let us now define an infinitesimal gauge transformation as

$$\delta\Psi(t) = \pi_1 \{\boldsymbol{\Lambda}(t), \mathbf{m}\} \frac{1}{1-\Psi(t)} \tag{2.60}$$

where $\Lambda(t)$ is an odd-degree gauge parameter, interpolated from $\Lambda(0) = 0$ to $\Lambda(1) = \Lambda$. $\boldsymbol{\Lambda}(t)$ is its associated coderivation, which means according to formula (2.45), a tensor factor of $\Lambda(t)$ is inserted at all possible positions, without taking any inputs. Plugging into the action yields

$$\begin{aligned}
\delta S(\Psi) &= \int_0^1 dt\, \omega\left(\pi_1 \boldsymbol{\partial_t}\{\boldsymbol{\Lambda}(t), \mathbf{m}\}\frac{1}{1-\Psi(t)}, \pi_1 \mathbf{m}\frac{1}{1-\Psi(t)}\right) \\
&\quad + \int_0^1 dt\, \omega\left(\pi_1 \boldsymbol{\partial_t}\frac{1}{1-\Psi(t)}, \pi_1 \mathbf{m}\{\boldsymbol{\Lambda}(t), \mathbf{m}\}\frac{1}{1-\Psi(t)}\right) \\
&= \int_0^1 dt\, \omega\left(\pi_1 \boldsymbol{\partial_t}\{\boldsymbol{\Lambda}(t), \mathbf{m}\}\frac{1}{1-\Psi(t)}, \pi_1 \mathbf{m}\frac{1}{1-\Psi(t)}\right) \\
&\quad + \int_0^1 dt\, \omega\left(\pi_1 \{\boldsymbol{\Lambda}(t), \mathbf{m}\}\frac{1}{1-\Psi(t)}, \pi_1 \boldsymbol{\partial_t}\mathbf{m}\frac{1}{1-\Psi(t)}\right) \\
&= \omega\left(\pi_1 \{\boldsymbol{\Lambda}, \mathbf{m}\}\frac{1}{1-\Psi}, \pi_1 \mathbf{m}\frac{1}{1-\Psi}\right) = \omega\left(\pi_1 \mathbf{m}\boldsymbol{\Lambda}\frac{1}{1-\Psi}, \pi_1 \mathbf{m}\frac{1}{1-\Psi}\right) \\
&= \omega\left(\pi_1 \boldsymbol{\Lambda}\frac{1}{1-\Psi}, \pi_1 \mathbf{m}^2 \frac{1}{1-\Psi}\right) = 0. \tag{2.61}
\end{aligned}$$

In the first step we used cyclicity of $\mathbf{m}$ (2.57) and the symmetry of the BPZ-product (2.17), in the second step we evaluated the integral and in the third step we noticed that after the application of $\boldsymbol{\Lambda}$, the result has at least tensor power two and gets killed by $\pi_1$. As desired, we could show that every $A_\infty$-action based on an odd, nilpotent coderivation has gauge invariance given by (2.60). The proof also holds for Witten theory as a special case.

## 2.4 The three-vertex

After discussing the CFT background and the algebraic structure of Witten theory, it is now time to look at geometric properties of the three-vertex. The definition given in section (2.2) was very heuristic and we are still in need for a precise CFT description in terms of correlation functions. First, it will be useful to introduce a new coordinate of the worldsheet given by

$$h(z) = \frac{1+iz}{1-iz}. \tag{2.62}$$

---

[8]Observe that the Leibniz rule holds for tensor products.

It maps the upper half plane to the unit disk, with the original unit half disk mapped to the right half disk. The exterior of the unit half disk is mapped to the left half disk, such that a BPZ-product becomes a correlation function of insertions at plus one and minus one. The gluing of the string states at time $\tau = 0$ is done on the segment from $-i$ to $i$ on the imaginary axis.

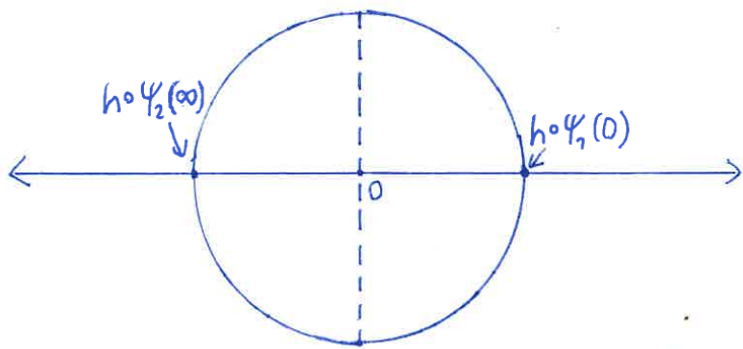

For the Witten vertex we would like to glue together three half disks in a similar way, however, then we would end up with a conical excess of $\pi$ at the origin. To remedy this, we "shrink" the unit half disks to wedges of opening angle $\frac{2\pi}{3}$. Since we know already that the vertex has to be cyclically symmetric, it makes sense to shrink all three surface states by the same factor. The shrinking can be implemented by taking the map $h(z)$ to the power $\frac{2}{3}$. To define maps for all three string states into the unit disk, we must also include a rotation factor, which yields

$$g_1(z) = e^{-\frac{2\pi i}{3}} h^{\frac{2}{3}}(z), \qquad g_2(z) = h^{\frac{2}{3}}(z), \qquad g_3(z) = e^{\frac{2\pi i}{3}} h^{\frac{2}{3}}(z). \tag{2.63}$$

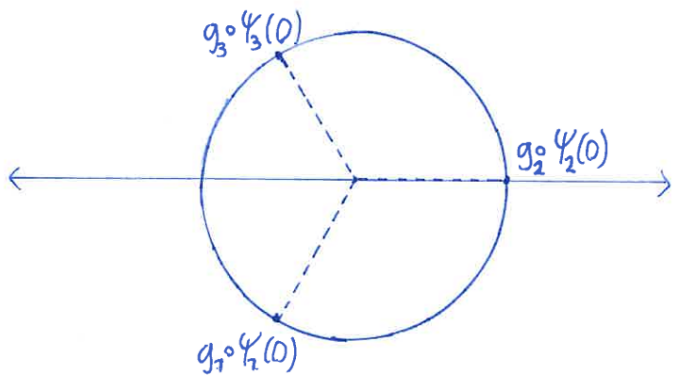

If we want a description of the three-vertex on the original UHP, we need to transform back with the map $h^{-1}(z) = i\frac{1-z}{1+z}$. Now one can show that a remarkable simplification of the coordinate

maps occurs: We define

$$f_1(z) = h^{-1}\left(e^{-\frac{2\pi i}{3}} h^{\frac{2}{3}}(z)\right) = \tan\left(\frac{2}{3}\left(\arctan z - \frac{\pi}{2}\right)\right),$$

$$f_2(z) = h^{-1}\left(h^{\frac{2}{3}}(z)\right) = \tan\left(\frac{2}{3}(\arctan z)\right),$$

$$f_3(z) = h^{-1}\left(e^{\frac{2\pi i}{3}} h^{\frac{2}{3}}(z)\right) = \tan\left(\frac{2}{3}\left(\arctan z + \frac{\pi}{2}\right)\right). \tag{2.64}$$

The three-vertex is now described as a three-point function in the upper half plane, as desired:

$$\langle \Psi_1, \Psi_2 * \Psi_3 \rangle = \langle f_1 \circ \Psi_1(0)\, f_2 \circ \Psi_2(0)\, f_3 \circ \Psi_3(0) \rangle. \tag{2.65}$$

We can also verify the algebraic properties that were argued graphically in section (2.2) in a mathematical way: For instance, for showing cyclicity we define the function

$$\tilde{f}(z) = \tan\left(\arctan z + \frac{\pi}{3}\right) \tag{2.66}$$

which is an element of $SL(2,\mathbb{R})$. From the invariance of correlation functions under global conformal transformations we find

$$\langle f_1 \circ \Psi_1(0)\, f_2 \circ \Psi_2(0)\, f_3 \circ \Psi_3(0) \rangle$$
$$= \langle \tilde{f} \circ f_1 \circ \Psi_1(0)\, \tilde{f} \circ f_2 \circ \Psi_2(0)\, \tilde{f} \circ f_3 \circ \Psi_3(0) \rangle$$
$$= \langle f_2 \circ \Psi_1(0)\, f_3 \circ \Psi_2(0)\, f_1 \circ \Psi_3(0) \rangle$$
$$= \langle \Psi_3, \Psi_1 * \Psi_2 \rangle \tag{2.67}$$

because of

$$\tilde{f} \circ f_j(z) = f_{j+1}(z) \quad \text{modulo 3}. \tag{2.68}$$

The Leibniz rule can be shown in the following way: We can consider the BPZ-product of $Q(\Psi_1 * \Psi_2)$ with an arbitrary test state which yields

$$\langle \Phi, Q(\Psi_1 * \Psi_2) \rangle = -(-1)^{gh(\Phi)} \langle Q\Phi, \Psi_1 * \Psi_2 \rangle$$
$$= -(-1)^{gh(\Phi)} \left\langle f_1 \circ \oint_{C(0)} \frac{dz}{2\pi i} j_B(z)\, \Phi(0)\, f_2 \circ \Psi_2(0)\, f_3 \circ \Psi_3(0) \right\rangle. \tag{2.69}$$

We know that the integration of the BRST-current commutes with conformal transformations. We can therefore pull it over the whole Riemann sphere such that it encircles the two other punctures. The result is

$$\left\langle f_1 \circ \Phi(0)\, f_2 \circ \oint_{C(0)} \frac{dz}{2\pi i} j_B(z)\, \Psi_2(0)\, f_3 \circ \Psi_3(0) \right\rangle$$
$$+ \left\langle f_1 \circ \Phi(0)\, f_2 \circ \Psi_2(0)\, f_3 \circ \oint_{C(0)} \frac{dz}{2\pi i} j_B(z)\, \Psi_3(0) \right\rangle$$
$$= \langle \Phi, Q\Psi_1 * \Psi_2 \rangle + (-1)^{gh(\Psi_1)} \langle \Phi, \Psi_1 * Q\Psi_2 \rangle. \tag{2.70}$$

Since the test state was arbitrary and the BPZ-product is non-degenerate, we conclude that the Leibniz rule holds.

It is instructive to explicitly compute the three-vertex for a tachyon field

$$t(x) = \int \frac{d^{26}k}{(2\pi)^{26}} T(k) \left(ce^{ik_\mu X^\mu}\right)(0)|0\rangle. \tag{2.71}$$

We want to apply formula (2.14) so we calculate

$$f_1(0) = -\sqrt{3}, \qquad f_2(0) = 0, \qquad f_3(0) = \sqrt{3} \tag{2.72}$$

$$f_1'(0) = \frac{8}{3}, \qquad f_2'(0) = \frac{2}{3}, \qquad f_3'(0) = \frac{8}{3}. \tag{2.73}$$

The conformal weight $h$ of the combined object $\left(ce^{ik_\mu X^\mu}\right)$ is $\alpha' k^2 - 1$ so we get for the whole expression

$$\langle t(x), t(x) * t(x) \rangle$$
$$= \int \frac{d^{26}k_1}{(2\pi)^{26}} T(k_1) \int \frac{d^{26}k_2}{(2\pi)^{26}} T(k_2) \int \frac{d^{26}k_3}{(2\pi)^{26}} T(k_3) \tag{2.74}$$
$$\left\langle \left(\frac{8}{3}\right)^{\alpha' k_1^2 - 1} \left(ce^{ik_1 \cdot X}\right)(-\sqrt{3}) \left(\frac{2}{3}\right)^{\alpha' k_2^2 - 1} \left(ce^{ik_2 \cdot X}\right)(0) \left(\frac{8}{3}\right)^{\alpha' k_3^2 - 1} \left(ce^{ik_3 \cdot X}\right)(\sqrt{3}) \right\rangle$$
$$= \int \frac{d^{26}k_1}{(2\pi)^{26}} \int \frac{d^{26}k_2}{(2\pi)^{26}} \int \frac{d^{26}k_3}{(2\pi)^{26}} T(k_1) T(k_2) T(k_3) \frac{27}{128} \left(\frac{8}{3}\right)^{\alpha'\left(k_1^2 + k_3^2\right)} \left(\frac{2}{3}\right)^{\alpha' k_2^2}$$
$$\left\langle e^{ik_1 \cdot X}(-\sqrt{3}) e^{ik_2 \cdot X}(0) e^{ik_3 \cdot X}(\sqrt{3}) \right\rangle \left\langle c(-\sqrt{3}) c(0) c(\sqrt{3}) \right\rangle \tag{2.75}$$

The ghost correlator can be evaluated using (2.20) and just gives a factor of $-6\sqrt{3}$. The tachyon 3-point function is derived for instance in [65] and reads

$$\left\langle e^{ik_1 \cdot X}(z_1) e^{ik_2 \cdot X}(z_2) e^{ik_3 \cdot X}(z_3) \right\rangle = (2\pi)^{26} \delta\left(\sum k_i\right) \sum_{i<j} |z_i - z_j|^{2\alpha' k_i \cdot k_j}. \tag{2.76}$$

Plugging that in yields

$$= -\int \frac{d^{26}k_1}{(2\pi)^{26}} \int \frac{d^{26}k_2}{(2\pi)^{26}} T(k_1) T(k_2) T(-k_1 - k_2) \frac{81\sqrt{3}}{64}$$
$$\left(\frac{8}{3}\right)^{\alpha'\left(k_1^2 + (k_1+k_2)^2\right)} \left(\frac{2}{3}\right)^{\alpha' k_2^2} \sqrt{3}^{2\alpha'(k_1 \cdot k_2 - k_2 \cdot (k_1+k_2))} \left(2\sqrt{3}\right)^{-2\alpha'(k_1 \cdot (k_1+k_2))} \tag{2.77}$$

where we already integrated over the $\delta$-function. The exponentials in the second line can be

simplified as

$$\exp\left(\alpha'\ln 2 \cdot \left(6k_1^2 + 3k_2^2 + 6k_1 \cdot k_2 + k_2^2 - 2k_1^2 - 2k_1 \cdot k_2\right)\right.$$
$$\left. - \alpha'\ln 3 \cdot \left(2k_1^2 - k_2^2 - 2k_1 \cdot k_2 - k_2^2 - k_2^2 - k_1^2 - k_1 \cdot k_2\right)\right)$$
$$= \exp\left(\alpha'\left(k_1^2 + k_2^2 + k_1 \cdot k_2\right)(4 \cdot \ln 2 - 3 \cdot \ln 3)\right)$$
$$= \exp\left(\alpha'\ln\left(\frac{4}{3\sqrt{3}}\right)\left(k_1^2 + k_2^2 + (-k_1 - k_2)^2\right)\right) \tag{2.78}$$

and after Fourier-transforming back the tachyon fields we have

$$-\frac{81\sqrt{3}}{64}\int \frac{d^{26}k_1}{(2\pi)^{26}} \int \frac{d^{26}k_2}{(2\pi)^{26}} \int d^{26}x_1 \int d^{26}x_2 \int d^{26}x_3$$
$$e^{\alpha'\ln\left(\frac{4}{3\sqrt{3}}\right)\left(k_1^2+k_2^2+(-k_1-k_2)^2\right)} e^{-ik_1 \cdot x_1} t(x_1) e^{-ik_2 \cdot x_2} t(x_2) e^{-i(-k_1-k_2) \cdot x_3} t(x_3)$$
$$= -\frac{81\sqrt{3}}{64}\int \frac{d^{26}k_1}{(2\pi)^{26}} \int \frac{d^{26}k_2}{(2\pi)^{26}} \int d^{26}x_1 \int d^{26}x_2 \int d^{26}x_3$$
$$e^{-\alpha'\ln\left(\frac{4}{3\sqrt{3}}\right)(\Box_1+\Box_2+\Box_3)} e^{-ik_1 \cdot x_1} t(x_1) e^{-ik_2 \cdot x_2} t(x_2) e^{-i(-k_1-k_2) \cdot x_3} t(x_3)$$
$$= -\frac{81\sqrt{3}}{64}\left(e^{-\alpha'\ln\left(\frac{4}{3\sqrt{3}}\right)\Box} t(x)\right)^3. \tag{2.79}$$

This is the final result for the three-vertex evaluated for a tachyon field. It has infinitely many derivatives acting on the function $t(x)$, hence the interaction is *non-local*. This was expected, because a string itself is a non-local object, but still, non-locality is an important feature of SFT that has profound consequences.

## 2.5 Tachyon condensation and Sen's conjectures

We turn now to a physical problem which has been addressed successfully within the framework of OSFT and provided evidence of its practicality and adequacy. Bosonic string theory contains the tachyon mode as its lowest level state, a state with negative energy. Usually this is considered as a sickness of the theory since it implies the vacuum to be unstable. However, Ashoke Sen has given a more refined interpretation for the case of open strings: The assumption is that the presence of tachyons leads to the decay of the corresponding D-brane. This means, at some point, when the whole D-brane has decayed, a true vacuum is reached, where no string excitations are present because the strings cannot end anywhere anymore. This is a collective phenomenon that involves a large number of worldsheets and can therefore not be studied in first-quantized string theory. In string field theory however, we expect the tachyon vacuum to show up as a solution of the classical field equations. More precisely, *Sen's conjectures* on tachyon condensation read the following:

1. There exists a solution to the equations of motion of Witten theory (2.39) that describes the true vacuum of the theory. It is universal, i. e., independent of the details of the BCFT and its total energy is equal to minus the tension of the D-brane that had decayed.

2. The tachyon vacuum does not support any excitation modes.

3. Lower dimensional D-branes show up as lump solutions of (2.39), again with their total energy being equal to the difference of the respective D-brane tensions.

A few comments are in order: First, "universal" means that the conjectured solution $\Psi_{TV}$ is constructed solely out of fields present in any BRST-quantized BCFT. These will turn out to be the $b$-ghost, the $c$-ghost and the energy momentum tensor. Second, since the tachyon vacuum is a time-independent solution, its total energy will be just the on-shell action (up to the volume of the time direction). The tension of the spacetime filling D-brane is given by $\frac{1}{2\pi^2 \alpha'^3 g_o^2}$ [65], hence we expect the relation

$$S\left(\Psi_{TV}\right) = -\frac{1}{2\pi^2 \alpha'^3 g_o^2}. \tag{2.80}$$

As a last point, we should clarify what the second conjecture precisely means. Let us assume now that $\Psi_{TV}$ is the true vacuum of the theory and expand all other string fields around it:

$$\Psi^* = \Psi - \Psi_{TV}. \tag{2.81}$$

If we rewrite the action in the new variable $\Psi^*$ we get

$$S\left(\Psi^*\right) = S^*\left(\Psi^*\right) + S\left(\Psi_{TV}\right). \tag{2.82}$$

$S\left(\Psi_{TV}\right)$ is constant (and given by (2.80)) whileas the new shifted action takes the form

$$S^*\left(\Psi^*\right) = -\frac{1}{\alpha'^3 g_o^2} \left(\frac{1}{2}\left\langle\Psi^*, Q^*\Psi^*\right\rangle + \frac{1}{3}\left\langle\Psi^*, \Psi^* * \Psi^*\right\rangle\right) \tag{2.83}$$

with

$$Q^* = Q + \{\Psi_{TV}, \cdot\}_* . \tag{2.84}$$

One should observe that the shifted BRST-operator $Q^*$ still squares to zero and obeys the Leibniz rule. Its cohomology defines the space of physical excitations around $\Psi_{TV}$. A sufficient condition for a vanishing cohomology is the existence of a so-called *homotopy operator* $A$. It is a string field that fulfills

$$Q^* A = \mathbb{1}, \tag{2.85}$$

where $\mathbb{1}$ is the identity string field, hence it basically inverts $Q^*$. Now every $Q^*$-closed state is also $Q^*$-exact because of

$$Q^* \Psi = 0 \quad \Rightarrow \quad \Psi = Q^* \left(A * \Psi\right). \tag{2.86}$$

The first conjecture has been proven in [70] by providing an analytic expression for $\Psi_{TV}$, a proof of the second one followed shortly after [11]. The third conjecture has been verified numerically to high precision in [55] and later proven analytically in full generality in [24].

## 2.6 Analytic solutions

In this section we want to give the explicit expressions of the most important analytic solutions of OSFT. First of all, we need to introduce some algebraic framework.

### 2.6.1 The sliver frame

The non-trivial ingredient in the Witten equations of motion (2.39) is the star product, so it is desirable to find a convenient representation of it. Looking at the formulas (2.64) defining the three-vertex, it is hard to resist trying the conformal transformation

$$w = \frac{2}{\pi} \arctan z. \tag{2.87}$$

It maps the upper half plane to a semiinfinite cylinder with the identification $w \sim w + 2$. The unit half disk gets mapped to the semiinfinite strip bounded by the vertical lines $\operatorname{Re} w = \pm\frac{1}{2}$, respectively, which represent the right and left half of the string. The string midpoint is sent to infinity. In the $w$-coordinate, the transformations (2.64) simply become

$$g_1(w) = \frac{2}{3}w - \frac{2}{3}, \qquad g_2(w) = \frac{2}{3}w, \qquad g_3(w) = \frac{2}{3}w + \frac{2}{3}. \tag{2.88}$$

and instruct us to first rescale the strip and then translate along the real axis. The halves of the strings are then glued together as expected from the graphical definition of the star product, moreover, through the cylinder identification, the two outer halves are also glued together representing the BPZ-product. Since from conformal invariance we are allowed to arbitrarily rescale the cylinder, it is convenient to undo the rescaling in (2.88) and work on a cylinder of circumference three. For Fock states, the three-vertex then just becomes a correlation function of operators inserted at zero, one and two on this cylinder denoted by $C_3$. The simplicity of the star product suggests to work in the $w$-coordinate, called the *sliver frame*, from now on.

### 2.6.2 The KBc-algebra

The full star algebra of string fields is a huge and very complicated object, so it is desirable to construct some simpler subalgebra to work with. The most fundamental state in our theory is the $SL(2, \mathbb{R})$-invariant vacuum, denoted as $\Omega$, so let us analyze expressions of the form

$$W_\alpha = \Omega * \Omega * ... * \Omega \quad (\alpha \text{ times}). \tag{2.89}$$

Those states are called *wedge states* and play an important role in the theory of analytic solutions. They obey the (obvious) algebra

$$W_\alpha * W_\beta = W_{\alpha+\beta} \tag{2.90}$$

and are represented geometrically just as strips of width $\alpha$ without any insertions. For $\alpha = 0$ we have a strip of vanishing width which can be identified as the identity $\mathbb{1}$ of the star algebra.

For $\alpha \to \infty$ we obtain a projector state called the sliver, which typically gives rise to singular expressions.

From this geometrical perspective, there is actually no reason why $\alpha$ should be an integer. An algebraic description for $W_\alpha$ for any real positive $\alpha$ can be given by the following calculation [64, 20, 76]: Consider the overlap with an arbitrary test state $\Phi$ of dimension $h$ prepared as an operator $\Phi(0)$ inserted within the vacuum strip $\Omega$. The result is a one-point function on a cylinder of width $\alpha + 1$:

$$\langle \Phi, W_\alpha \rangle = \langle \Phi(0) \rangle_{C_{\alpha+1}} \tag{2.91}$$

The cylinder can be rescaled to unit circumference by the transformation $w \to \frac{w}{\alpha+1}$:

$$\langle \Phi(0) \rangle_{C_{\alpha+1}} = \left( \frac{1}{\alpha+1} \right)^h \langle \Phi(0) \rangle_{C_1} \tag{2.92}$$

Now we can take a derivative with respect to $\alpha$ and find

$$\left\langle \Phi, \frac{d}{d\alpha} W_\alpha \right\rangle = -h \left( \frac{1}{\alpha+1} \right)^{h+1} \langle \Phi(0) \rangle_{C_1} = -\frac{h}{\alpha+1} \langle \Phi(0) \rangle_{C_{\alpha+1}}. \tag{2.93}$$

The factor of $h$ can be extracted from $\Phi$ using the energy-momentum tensor: $L_0 \Phi = h\Phi$, so using (2.25) we get

$$\left\langle \Phi, \frac{d}{d\alpha} W_\alpha \right\rangle = -\frac{1}{\alpha+1} \left\langle \oint_{C(0)} \frac{dw}{2\pi i} w T(w) \Phi(0) \right\rangle_{C_{\alpha+1}}. \tag{2.94}$$

The expression can be simplified by a contour deformation: Let us expand the contour until it runs along the outer edges of the strip which are identified: (It is useful here to apply the doubling trick such that we have a full infinite cylinder.)

$$\left\langle \Phi, \frac{d}{d\alpha} W_\alpha \right\rangle = -\frac{1}{\alpha+1} \left( \left\langle \int_{\alpha+\frac{1}{2}-i\infty}^{\alpha+\frac{1}{2}+i\infty} \frac{dw}{2\pi i} w T(w) \Phi(0) \right\rangle_{C_{\alpha+1}} - \left\langle \int_{-\frac{1}{2}-i\infty}^{-\frac{1}{2}+i\infty} \frac{dw}{2\pi i} w T(w) \Phi(0) \right\rangle_{C_{\alpha+1}} \right). \tag{2.95}$$

After a shift $w \to w + \alpha + 1$ in the second term we see that the terms linear in $w$ cancel and we are left with

$$\left\langle \Phi, \frac{d}{d\alpha} W_\alpha \right\rangle = -\left\langle \Phi(0) \int_{\alpha+\frac{1}{2}-i\infty}^{\alpha+\frac{1}{2}+i\infty} \frac{dw}{2\pi i} T(w) \right\rangle_{C_{\alpha+1}} \equiv -\langle \Phi, W_\alpha * K \rangle. \tag{2.96}$$

In the last equality we defined a new string field $K$ as an infinitesimally thin strip with an insertion of $\int \frac{dw}{2\pi i} T(w)$ along an infinite vertical line. Since $\Phi$ was an arbitrary primary we can deduce the following differential equation:

$$\frac{d}{d\alpha} W_\alpha = -W_\alpha * K. \tag{2.97}$$

When dealing with expressions in $K$ (and later also $B$ and $c$) it is customary to write the star product just as an ordinary multiplication, hence we can write the solution of (2.97) as

$$W_\alpha = e^{-\alpha K}. \tag{2.98}$$

This equation tells us that $K$ is just minus the star product logarithm of the vacuum and allows us to algebraically write down wedge states for arbitrary positive real $\alpha$.

Since $K$ is the generator of all wedge states, we conclude it should be part of our desired subalgebra. Moreover, we know that a classical open string field is of ghost number one, so we must also incorporate the $c$-ghost somehow. The simplest way to do that is to define a string field $c$ by an infinitesimal strip with an insertion of $c(0)$. For instance, the zero-momentum tachyon can then be written as

$$c(0)\left|0\right\rangle = e^{-\frac{K}{2}}ce^{-\frac{K}{2}}. \tag{2.99}$$

It is natural to include the $b$-ghost as well and due to the similarity between $b(w)$ and $T(w)$ we define it as an infinitesimal strip with an insertion of $\int_{-i\infty}^{i\infty}\frac{dw}{2\pi i}b(w)$, in analogy to $K$. We can straightforwardly deduce the commutators

$$[K,B] = 0, \qquad \{B,c\} = \mathbb{1}, \qquad c^2 = B^2 = 0. \tag{2.100}$$

The commutator between $c$ and $K$ can be evaluated using contour deformation: The insertion within the infinitesimal strip yields

$$\oint_{C(0)}\frac{dw}{2\pi i}T(w)c(0) = \partial c(0). \tag{2.101}$$

We can now define a new string field $\partial c$ as an infinitesimal strip with $\partial c(0)$ inserted. However, it is important to note that $\partial c$ is not a linear combination of $K$, $B$ and $c$, hence they do not form a Lie algebra. What we mean by the *KBc-algebra* (see [63]) is actually the *universal enveloping algebra* of $K$, $B$ and $c$ with respect to (2.100). This means, we take the set of all words formed by the letters $K$, $B$ and $c$ and mod out by the commutators (2.100). The most general state of ghost number zero is then always a linear combination of states of the form

$$f_1(K), \qquad f_2(K)cf_3(K)B \tag{2.102}$$

with $f_i$ arbitrary power series in $K$. If there is more then one $B$ present, the $B$s can be commuted through and brought together whileas the last $B$ is brought to the most right. Similarly, ghost number one states are linear combinations of

$$f_1(K)cf_2(K), \qquad f_3(K)cf_4(K)cf_5(K)B. \tag{2.103}$$

From (2.30) we can deduce that the KBc-algebra is actually closed under BRST transformations:

$$QB = K, \qquad QK = 0, \qquad Qc = c\partial c = cKc. \tag{2.104}$$

These nice properties as well as the fact that it was purely constructed out of universal elements suggests that we can find analytical solutions such as the tachyon vacuum within the KBc-algebra.

### 2.6.3 Sliver frame operators

Before coming to that, we introduce a few more sliver frame operators which will play an important role later on. We start by transforming the Hamiltonian $L_0$ of our original BCFT to the sliver frame and define

$$\mathcal{L}_0 = \oint_{C(0)} \frac{dw}{2\pi i} wT(w) = \oint_{C(0)} \frac{dz}{2\pi i} \arctan z \left(1 + z^2\right) T(z) = L_0 + \frac{2}{3} L_2 - \frac{2}{15} L_4 - \cdots. \quad (2.105)$$

$\mathcal{L}_0$ only contains positive Virasoro generators hence it obeys $\mathcal{L}_0 \Omega = 0$ and $\mathcal{L}_0 \phi = h\phi$ for conformal primaries, however, it is not BPZ-even anymore, $\mathcal{L}_0 \neq \mathcal{L}_0^*$. For later purposes we define the combinations $\hat{\mathcal{L}} \equiv \mathcal{L}_0 + \mathcal{L}_0^*$ and $\mathcal{L}^- \equiv \mathcal{L}_0 - \mathcal{L}_0^*$.

We will now examine how $\mathcal{L}_0$ and $\mathcal{L}_0^*$ act on an arbitrary wedge states with possible insertions. On the vacuum we can write

$$\begin{aligned}
\mathcal{L}_0 \Omega &= \int_{-i\infty + \frac{1}{2}}^{i\infty + \frac{1}{2}} \frac{dw}{2\pi i} wT(w) - \int_{-i\infty - \frac{1}{2}}^{i\infty - \frac{1}{2}} \frac{dw}{2\pi i} wT(w) \\
&= \int_{-i\infty}^{i\infty} \frac{dw}{2\pi i} \left(w + \frac{1}{2}\right) T\left(w + \frac{1}{2}\right) - \int_{-i\infty}^{i\infty} \frac{dw}{2\pi i} \left(w - \frac{1}{2}\right) T\left(w - \frac{1}{2}\right)
\end{aligned} \quad (2.106)$$

as well as

$$\mathcal{L}_0^* \Omega = \int_{-i\infty}^{i\infty} \frac{dw}{2\pi i} \left(-w + \frac{1}{2}\right) T\left(w + \frac{1}{2}\right) + \int_{-i\infty}^{i\infty} \frac{dw}{2\pi i} \left(w + \frac{1}{2}\right) T\left(w - \frac{1}{2}\right). \quad (2.107)$$

Similarly, on a strip of arbitrary width (which we place for convenience between $-\frac{1}{2}$ and $\alpha - \frac{1}{2}$) we get

$$\begin{aligned}
\mathcal{L}_0 \Omega^\alpha &= \int_{-i\infty}^{i\infty} \frac{dw}{2\pi i} \left(w + \frac{1}{2}\right) T\left(w + \alpha - \frac{1}{2}\right) - \int_{-i\infty}^{i\infty} \frac{dw}{2\pi i} \left(w - \frac{1}{2}\right) T\left(w - \frac{1}{2}\right) \\
&= \frac{1}{2} \left( \int_{-i\infty}^{i\infty} \frac{dw}{2\pi i} T\left(w + \alpha - \frac{1}{2}\right) + \int_{-i\infty}^{i\infty} \frac{dw}{2\pi i} T\left(w - \frac{1}{2}\right) \right) \\
&\quad + \int_{-i\infty}^{i\infty} \frac{dw}{2\pi i} wT\left(w + \alpha - \frac{1}{2}\right) - \int_{-i\infty}^{i\infty} \frac{dw}{2\pi i} wT\left(w - \frac{1}{2}\right),
\end{aligned} \quad (2.108)$$

$$\begin{aligned}
\mathcal{L}_0^* \Omega^\alpha &= \int_{-i\infty}^{i\infty} \frac{dw}{2\pi i} \left(-w + \frac{1}{2}\right) T\left(w + \alpha - \frac{1}{2}\right) + \int_{-i\infty}^{i\infty} \frac{dw}{2\pi i} \left(w + \frac{1}{2}\right) T\left(w - \frac{1}{2}\right) \\
&= \frac{1}{2} \left( \int_{-i\infty}^{i\infty} \frac{dw}{2\pi i} T\left(w + \alpha - \frac{1}{2}\right) + \int_{-i\infty}^{i\infty} \frac{dw}{2\pi i} T\left(w - \frac{1}{2}\right) \right) \\
&\quad - \int_{-i\infty}^{i\infty} \frac{dw}{2\pi i} wT\left(w + \alpha - \frac{1}{2}\right) + \int_{-i\infty}^{i\infty} \frac{dw}{2\pi i} wT\left(w - \frac{1}{2}\right).
\end{aligned} \quad (2.109)$$

Observe that we are not allowed to shrink the contour since the vertex operator generating a wedge state is non-local. The integrals have to be placed on the outer edges of the strip. We can now easily deduce

$$\hat{\mathcal{L}}\Psi = \left( \int_{-i\infty}^{i\infty} \frac{dw}{2\pi i} T\left(w + \alpha - \frac{1}{2}\right) + \int_{-i\infty}^{i\infty} \frac{dw}{2\pi i} T\left(w - \frac{1}{2}\right) \right) \Psi = K\Psi + \Psi K \qquad (2.110)$$

$$\mathcal{L}^{-}\Psi = 2 \left( \int_{-i\infty}^{i\infty} \frac{dw}{2\pi i} wT\left(w + \alpha - \frac{1}{2}\right) - \int_{-i\infty}^{i\infty} \frac{dw}{2\pi i} wT\left(w - \frac{1}{2}\right) \right) \Psi \qquad (2.111)$$

where the string field $K$ was recognized in the first line. The second line shows that $\mathcal{L}^{-}$ is actually a derivative of the star product,

$$\mathcal{L}^{-}(\Psi_1 * \Psi_2) = \mathcal{L}^{-}\Psi_1 * \Psi_2 + \Psi_1 * \mathcal{L}^{-}\Psi_2 \qquad (2.112)$$

because the two contour integrals in the middle just cancel out. We can decompose

$$\mathcal{L}_0 \Psi = \frac{1}{2}(K\Psi + \Psi K) + \frac{1}{2}\mathcal{L}^{-}\Psi \qquad (2.113)$$

which implies

$$\mathcal{L}_0\left(e^{-\frac{K}{2}} X e^{-\frac{K}{2}}\right) = e^{-\frac{K}{2}} \frac{\mathcal{L}^{-}}{2}(X) e^{-\frac{K}{2}} \qquad (2.114)$$

for any string field $X$. The factors $e^{-\frac{K}{2}}$ are called *security strips*: They are important since they ensure that the strip has non-vanishing width and does not give rise to ill-defined correlation functions, see the discussion on identity-like solutions below. The action of $\mathcal{L}^{-}$ on the $KBc$-algebra is easy to determine: Since (2.111) acting on an infinitesimal strip is just a contour integral of $wT(w)$, it gives back the conformal dimension of the operator insertion and we get

$$\frac{1}{2}\mathcal{L}^{-}K = K, \qquad \frac{1}{2}\mathcal{L}^{-}B = B, \qquad \frac{1}{2}\mathcal{L}^{-}c = -c. \qquad (2.115)$$

This result together with (2.113) enables us to calculate the action of $\mathcal{L}_0$ on any $KBc$-element straightforwardly.

The last operator we want to introduce is

$$\mathcal{B}_0 = \oint_{C(0)} \frac{dw}{2\pi i} w\, b(w) = b_0 + \frac{2}{3}b_2 - \frac{2}{15}b_4 - \cdots \qquad (2.116)$$

in the sliver frame. It can be used instead of $b_0$ to implement a gauge condition called the *sliver gauge*, which will be discussed in the next section. Analogously to the case of $\mathcal{L}_0$ we define $\hat{\mathcal{B}} \equiv \mathcal{B}_0 + \mathcal{B}_0^*$ and $\mathcal{B}^{-} \equiv \mathcal{B}_0 - \mathcal{B}_0^*$ and find a similar decomposition:

$$\mathcal{B}_0 \Psi = \frac{1}{2}\hat{\mathcal{B}}\Psi + \frac{1}{2}\mathcal{B}^{-}\Psi = \frac{1}{2}\left(B\Psi + (-1)^{gh(\Psi)}\Psi B\right) + \frac{1}{2}\mathcal{B}^{-}\Psi. \qquad (2.117)$$

Also, $\mathcal{B}^{-}$ is a derivative of the star product and its action on the $KBc$-elements is given by

$$\frac{1}{2}\mathcal{B}^{-}K = B \quad , \frac{1}{2}\mathcal{B}^{-}B = 0 \quad , \frac{1}{2}\mathcal{B}^{-}c = 0. \qquad (2.118)$$

These relations can be derived via the OPEs between $b(w)$, $T(w)$ and $c(w)$.

### 2.6.4 Explicit examples

Finally we are ready to give explicit examples of analytic solutions in OSFT. We will start with the most famous one, the tachyon vacuum in its original form introduced in [70] which reads

$$\Psi_{TV} = e^{-\frac{K}{2}} c \frac{KB}{1 - e^{-K}} c e^{-\frac{K}{2}}. \tag{2.119}$$

The original derivation is complicated and will not be presented here, a rather quick way to get the result can be found in [20]. It obeys the above mentioned gauge condition

$$\mathcal{B}_0 \Psi_{TV} = 0 \tag{2.120}$$

called the *sliver gauge*. The equations of motion can be checked straightforwardly using (2.104):

$$Q\Psi_{TV} = e^{-\frac{K}{2}} cKc \frac{KB}{1 - e^{-K}} ce^{-\frac{K}{2}} - e^{-\frac{K}{2}} c \frac{K^2}{1 - e^{-K}} ce^{-\frac{K}{2}} + e^{-\frac{K}{2}} c \frac{KB}{1 - e^{-K}} cKce^{-\frac{K}{2}} \tag{2.121}$$

$$\begin{aligned}
\Psi_{TV} * \Psi_{TV} &= e^{-\frac{K}{2}} c \frac{KB}{1 - e^{-K}} c e^{-K} c \frac{KB}{1 - e^{-K}} c e^{-\frac{K}{2}} \\
&= e^{-\frac{K}{2}} c \frac{K}{1 - e^{-K}} \left( Bce^{-K} - Bce^{-K} Bc \right) \frac{K}{1 - e^{-K}} ce^{-\frac{K}{2}} \\
&= e^{-\frac{K}{2}} c \frac{K}{1 - e^{-K}} \left[ Bc, e^{-K} \right] \frac{K}{1 - e^{-K}} ce^{-\frac{K}{2}} \\
&= -e^{-\frac{K}{2}} c \frac{K}{1 - e^{-K}} \left[ Bc, 1 - e^{-K} \right] \frac{K}{1 - e^{-K}} ce^{-\frac{K}{2}} \\
&= -e^{-\frac{K}{2}} c \frac{K}{1 - e^{-K}} BcKce^{-\frac{K}{2}} + e^{-\frac{K}{2}} cKBc \frac{K}{1 - e^{-K}} ce^{-\frac{K}{2}} \\
&= -e^{-\frac{K}{2}} c \frac{K}{1 - e^{-K}} BcKce^{-\frac{K}{2}} + e^{-\frac{K}{2}} c \frac{K^2}{1 - e^{-K}} ce^{-\frac{K}{2}} - e^{-\frac{K}{2}} cKcB \frac{K}{1 - e^{-K}} ce^{-\frac{K}{2}} \\
&= -Q\Psi_{TV}. \tag{2.122}
\end{aligned}$$

It is instructive to expand $\Psi_{TV}$ in eigenstates of $\mathcal{L}_0$, one finds

$$\begin{aligned}
\mathcal{L}_0 = -1 : &\quad e^{-\frac{K}{2}} ce^{-\frac{K}{2}} \\
\mathcal{L}_0 = 0 : &\quad -\frac{1}{2} e^{-\frac{K}{2}} cKBce^{-\frac{K}{2}} \\
\mathcal{L}_0 = 1 : &\quad \frac{1}{12} e^{-\frac{K}{2}} cK^2 Bce^{-\frac{K}{2}} \\
&\quad \vdots \\
\mathcal{L}_0 = n - 1 : &\quad \frac{(-1)^n B_n}{n!} e^{-\frac{K}{2}} cK^n Bce^{-\frac{K}{2}} \tag{2.123}
\end{aligned}$$

where the coefficients $B_n$ are the *Bernoulli numbers.* So we have a sum of wedge states of width one with some operator insertion along the imaginary axis. The first term is just the zero-momentum tachyon of eigenvalue minus one, as expected. A special role is played by the zero eigenvalue state since it is BRST-exact,

$$e^{-\frac{K}{2}}cKBce^{-\frac{K}{2}} = Q\left(e^{-\frac{K}{2}}Bce^{-\frac{K}{2}}\right). \tag{2.124}$$

This implies that $e^{-\frac{K}{2}}cKBce^{-\frac{K}{2}}$ is annihilated by $\mathcal{B}_0$, $\mathcal{L}_0$ and $Q$, which has consequences for amplitudes in sliver gauge, see section 6.3.

The computation of the on-shell action is technically involved and will not be presented here, instead we will verify Sen's second conjecture. Define the homotopy operator

$$A_{TV} = B\frac{1-e^{-K}}{K} = B\int_0^\infty dt\, e^{-tK}\left(1-e^{-K}\right) = B\int_0^1 dt\, e^{-tK} \tag{2.125}$$

where a Schwinger-type representation of $\frac{1}{K}$ was used. Now to check equation (2.85) we calculate

$$\begin{aligned}&QA_{TV} + A_{TV}\Psi_{TV} + \Psi_{TV}A_{TV}\\ &= 1 - e^{-K} + B\frac{1-e^{-K}}{K}e^{-\frac{K}{2}}c\frac{KB}{1-e^{-K}}ce^{-\frac{K}{2}} + e^{-\frac{K}{2}}c\frac{KB}{1-e^{-K}}ce^{-\frac{K}{2}}B\frac{1-e^{-K}}{K}\\ &= 1 - e^{-K} + e^{-\frac{K}{2}}Bce^{-\frac{K}{2}} + e^{-\frac{K}{2}}cBe^{-\frac{K}{2}} = 1\end{aligned} \tag{2.126}$$

so indeed, the tachyon vacuum does not support any string excitations.

In fact, $\Psi_{TV}$ belongs to a more general class of solutions given by the expression [63]

$$\Psi = f(K)c\frac{KB}{1-f^2(K)}cf(K). \tag{2.127}$$

The equations of motion can be checked along the lines of (2.122), see also [64]. Apart from $f(K) = e^{-\frac{K}{2}}$ there is also the interesting choice $f(K) = \frac{1}{\sqrt{1+K}}$ leading to

$$\begin{aligned}\Psi_{STV} &= \frac{1}{\sqrt{1+K}}c\frac{KB}{1-\frac{1}{1+K}}c\frac{1}{\sqrt{1+K}} = \frac{1}{\sqrt{1+K}}(c+cKBc)\frac{1}{\sqrt{1+K}}\\ &= \frac{1}{\pi}\int_0^\infty \frac{dt}{\sqrt{t}}e^{-t(K+1)}(c+cKBc)\int_0^\infty \frac{ds}{\sqrt{s}}e^{-s(K+1)}.\end{aligned} \tag{2.128}$$

This solution is called the *simple tachyon vacuum* and was introduced in [27]. From the last step[9] one can see that unlike $\Psi_{TV}$, it is a continuous superposition of wedge states of width ranging from zero to infinity. The reason why it is called the simple tachyon vacuum is because computations are typically easier than those involving $\Psi_{TV}$. For example, for calculating the on-shell action

$$S_{on-shell} = -\frac{1}{\alpha'^3 g_o^2}\frac{1}{6}\langle \Psi_{STV}, Q\Psi_{STV}\rangle \tag{2.129}$$

---

[9]In can be proven using the new variable $u = \sqrt{t}$.

one can omit the second term in (2.128) since it is BRST-exact, see (2.124). What remains is just an integral over correlation functions involving $c$ and $\partial c$, which can be evaluated quite easily, see [27]. The homotopy operator also takes a simple form

$$A_{STV} = B \frac{1}{1+K} \tag{2.130}$$

and obeys

$$\begin{aligned}
&QA_{STV} + A_{STV}\Psi_{STV} + \Psi_{STV}A_{STV} \\
&= \frac{K}{1+K} + \frac{1}{(1+K)^{\frac{3}{2}}} B\left(c + cKBc\right) \frac{1}{\sqrt{1+K}} + \frac{1}{\sqrt{1+K}} \left(c + cKBc\right) B \frac{1}{(1+K)^{\frac{3}{2}}} \\
&= \frac{K}{1+K} + \frac{1}{(1+K)^{\frac{3}{2}}} \left(Bc + KBc\right) \frac{1}{\sqrt{1+K}} + \frac{1}{\sqrt{1+K}} \left(cB + cBK\right) \frac{1}{(1+K)^{\frac{3}{2}}} \\
&= \frac{K}{1+K} + \frac{1}{\sqrt{1+K}} Bc \frac{1}{\sqrt{1+K}} + \frac{1}{\sqrt{1+K}} cB \frac{1}{\sqrt{1+K}} = \frac{K}{1+K} + \frac{1}{1+K} \\
&= 1.
\end{aligned} \tag{2.131}$$

Given the simplicity of the Witten equations of motion and the BRST properties (2.104), one might ask if there are no simpler solutions to find within the $KBc$-algebra. Indeed, just by playing around, it is easy to find the ghost number one solutions $-cK$, $-Kc$ and $cKBc$ [16]. All of them have the property that they consist of a strip with vanishing width with insertions and are therefore called *identity-like solutions*. They are problematic in the sense that it is typically not possible to compute the on-shell action unambiguously: One would have to evaluate a ghost correlator on a cylinder of vanishing circumference and the result is dependent on the regularization procedure. For this reason, most identity-like solutions are not considered physically meaningful. One exception is the string field

$$\Psi_{IL} = c\left(1 - K\right) \tag{2.132}$$

which can be interpreted as a representative of the tachyon vacuum. It obeys the Witten equations of motion as can be checked easily:

$$Q\Psi_{IL} + \Psi_{IL} * \Psi_{IL} = cKc\left(1-K\right) + c\left(1-K\right)c\left(1-K\right) = cKc\left(1-K\right) - cKc\left(1-K\right) = 0. \tag{2.133}$$

Although for the above-mentioned reason it is not possible to compute the action, we can find a homotopy operator

$$A_{IL} = B \tag{2.134}$$

that obeys

$$QA_{IL} + A_{IL}\Psi_{IL} + \Psi_{IL}A_{IL} = K + Bc\left(1-K\right) + c\left(1-K\right)B = K + 1 - K = 1 \tag{2.135}$$

and confirm Sen's second conjecture. In [1] it was shown that $\Psi_{IL}$ is actually gauge related to the simple tachyon vacuum in a slightly modified form:

$$\Psi'_{STV} = \left(c + cKBc\right) \frac{1}{1+K} \tag{2.136}$$

Indeed, using the gauge parameter

$$\Lambda = 1 + cBK, \qquad \Lambda^{-1} = \frac{1}{1 + cBK} = 1 - cBK\frac{1}{1 + K} \qquad (2.137)$$

one can check straightforwardly that

$$\Psi'_{STV} = \Lambda\Psi_{IL}\Lambda^{-1} + \Lambda Q\Lambda^{-1}. \qquad (2.138)$$

In section 7.4.2 we give an attempt to regularize $\Psi_{IL}$ using stubs, such that the on-shell action is well-defined.

## 2.7 Feynman graphs and amplitudes

In this last section on OSFT we want to discuss the geometry of amplitudes and the individual Feynman diagrams. As it was stated in the introduction, we expect that every Feynman diagram corresponds to a punctured Riemann surface, such that within an amplitude, the sum of all Feynman diagrams precisely covers the moduli space of the Riemann surfaces in question. This was proven by Barton Zwiebach in [86] to all loop orders. We want to understand now how to obtain those Riemann surfaces and, as an example, write down the tree-level four point amplitude.

The first object we need to define is the open string propagator $P$. It is by definition the inverse of the kinetic operator, hence in our case we expect the relation

$$\{Q, P\} = -1 \qquad (2.139)$$

to hold (The minus sign is just a convention.). Moreover, we require $P$ to be BPZ-self-conjugate. In any gauge theory, we must specify a gauge to invert the kinetic operator and one of the most commonly used gauges in OSFT is *Siegel gauge*,

$$b_0\Psi = 0. \qquad (2.140)$$

It can be motivated for instance by looking at the gauge field component of $\Psi$ at the massless level and notice that Siegel gauge precisely eliminates the unphysical, auxiliary field. The validity of Siegel gauge on the full Hilbert space was analyzed in detail in [12]. From (2.30) we can guess the following expression for the propagator:

$$P = -\frac{b_0}{L_0} = -b_0 \int_0^\infty dt\, e^{-tL_0}. \qquad (2.141)$$

In the second step we defined $\frac{1}{L_0}$ via its Schwinger parametrization[10]: This is not only a convenient way to make sense of the expression but it also has a nice geometric interpretation. Since $L_0$ is the Hamiltonian of the CFT, $e^{-tL_0}$ generates time evolution, it propagates a string by a strip of length $t$. $P$ therefore inserts a superposition of strips of arbitrary lengths into the diagram and $t$ becomes a modulus of the Riemann surfaces obtained. This is in agreement with the expectations: An $n$-point tree diagram contains $n-3$ propagators since we have only cubic vertices. Similarly, the corresponding Riemann surface is an $n$-punctured disk which has $n-3$ moduli parameters.

---

[10]Strictly speaking the Schwinger parametrization is only valid for states with positive $L_0$-eigenvalue. Hence for the tachyon and the massless level this definition is problematic and can lead to divergences, see e. g. [42].

To write down the tree-level four point amplitude we need to sum over the $s$-channel and $t$-channel contributions. There is no $u$-channel because the boundary operators are cyclically ordered and the $u$-channel would break this ordering. The full expression reads

$$\mathcal{A}_4 = -\int_0^\infty dt \left(\left\langle \Psi_1 * \Psi_2, b_0\, e^{-tL_0}\left(\Psi_3 * \Psi_4\right)\right\rangle + \left\langle \Psi_2 * \Psi_3, b_0\, e^{-tL_0}\left(\Psi_4 * \Psi_1\right)\right\rangle\right), \qquad (2.142)$$

the first term can be visualized as

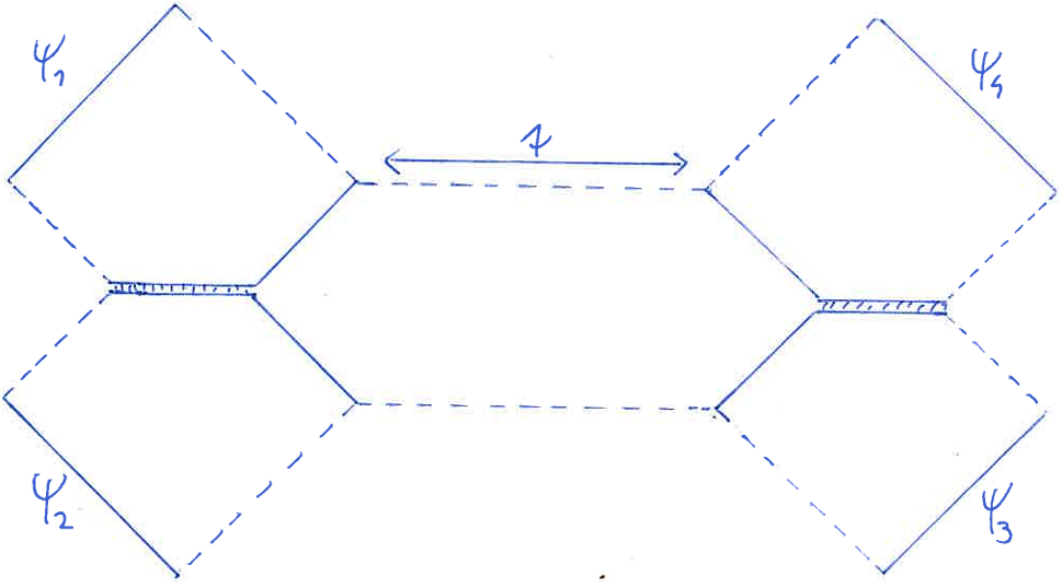

Since we have seen that in the context of analytic solutions, sliver gauge has emerged successful, we can ask if amplitudes and perturbation theory in general can be done in the sliver gauge as well. This is a very subtle topic and has been partially addressed in [43, 42], but there are still open questions. The first task is to derive the propagator: A naive choice might be

$$P \stackrel{?}{=} -\frac{\mathcal{B}_0}{\mathcal{L}_0}. \qquad (2.143)$$

This does obey (2.139) but it is not BPZ-self-conjugate and hence not applicable. A solution was given in [42] by making the propagator ghost number dependent: If we define $\mathcal{P}_\pm$ to be the projector on even (odd) ghost number, then their result is

$$P = -\frac{\mathcal{B}_0}{\mathcal{L}_0} Q \frac{\mathcal{B}_0^*}{\mathcal{L}_0^*} \mathcal{P}_+ - \frac{\mathcal{B}_0^*}{\mathcal{L}_0^*} Q \frac{\mathcal{B}_0}{\mathcal{L}_0} \mathcal{P}_-. \qquad (2.144)$$

This is manifestly BPZ-even and equation (2.139) becomes

$$\begin{aligned}\{Q,P\} &= -Q\frac{\mathcal{B}_0}{\mathcal{L}_0}Q\frac{\mathcal{B}_0^*}{\mathcal{L}_0^*}\mathcal{P}_+ - Q\frac{\mathcal{B}_0^*}{\mathcal{L}_0^*}Q\frac{\mathcal{B}_0}{\mathcal{L}_0}\mathcal{P}_- - \frac{\mathcal{B}_0}{\mathcal{L}_0}Q\frac{\mathcal{B}_0^*}{\mathcal{L}_0^*}\mathcal{P}_+Q - \frac{\mathcal{B}_0^*}{\mathcal{L}_0^*}Q\frac{\mathcal{B}_0}{\mathcal{L}_0}\mathcal{P}_-Q \\ &= -Q\frac{\mathcal{B}_0^*}{\mathcal{L}_0^*}\mathcal{P}_+ - Q\frac{\mathcal{B}_0}{\mathcal{L}_0}\mathcal{P}_- - \frac{\mathcal{B}_0}{\mathcal{L}_0}Q\frac{\mathcal{B}_0^*}{\mathcal{L}_0^*}Q\mathcal{P}_- - \frac{\mathcal{B}_0^*}{\mathcal{L}_0^*}Q\frac{\mathcal{B}_0}{\mathcal{L}_0}Q\mathcal{P}_+ \\ &= -Q\frac{\mathcal{B}_0^*}{\mathcal{L}_0^*}\mathcal{P}_+ - Q\frac{\mathcal{B}_0}{\mathcal{L}_0}\mathcal{P}_- - \frac{\mathcal{B}_0}{\mathcal{L}_0}Q\mathcal{P}_- - \frac{\mathcal{B}_0^*}{\mathcal{L}_0^*}Q\mathcal{P}_+ \\ &= -\mathcal{P}_+ - \mathcal{P}_- = -1 \end{aligned} \quad (2.145)$$

so we see why the ghost number dependence is necessary. A crucial point is that $P$ now contains two Schwinger integrals instead of one which seems to be in conflict with the dimensionality of the moduli space. It is expected though that the presence of $Q$ cancels this infinite overcounting, evidence for that has been given in [42]. In section 6.3.4 we will verify the covering of the moduli space explicitly by computing the tree-level four point amplitude which reads

$$\begin{aligned}\mathcal{A}_4 = -\int_0^\infty dt \int_0^\infty ds \Big( &\left\langle \Psi_1 * \Psi_2, e^{-t\mathcal{L}_0}\mathcal{B}_0 Q\mathcal{B}_0^* e^{-s\mathcal{L}_0^*}(\Psi_3 * \Psi_4)\right\rangle \\ + &\left\langle \Psi_2 * \Psi_3, e^{-t\mathcal{L}_0}\mathcal{B}_0 Q\mathcal{B}_0^* e^{-s\mathcal{L}_0^*}(\Psi_4 * \Psi_1)\right\rangle \Big). \end{aligned} \quad (2.146)$$

Before moving on to closed string field theory we should mention a few words on quantization: So far we have viewed OSFT as a purely classical theory, however, in the end of the day we expect SFT being a fully fledged quantum field theory. The reason why we leave the topic of quantization for the next part is that pure OSFT is not consistent as a quantum theory. As soon as there are loop diagrams present, one would have to allow for the emissions of closed string states, which are not part of the Hilbert space. As discussed in section 7.3, one can formally eliminate those closed string emissions by demanding a *quantum consistency condition*. This only works on a formal level though, for having a physically meaningful theory, one has to consider the full open-closed SFT [52, 88].

# 3 Closed string field theory

Closed string field theory (CSFT) differs in many aspects from OSFT as its algebraic and geometric structure are far more involved [19, 87]. Whileas in OSFT it is pretty easy to understand the action and the challenges come with all the operator technology developed for analytic solutions, for CSFT it is already difficult to write down a consistent action. In this part we will introduce the main concepts which are necessary for that and also comment on outlook and open problems.

## 3.1 The closed string background

As before we start by describing the background CFT of closed strings and focus particularly on the case of a 26-dimensional spacetime with no background fields. Many expressions will carry over almost unchanged from section 2.1 with the main difference that now we have both a holomorphic and an antiholomorphic sector. The most important new ingredient is the *level-matching condition* discussed in the end of this section.

The mode expansion for the 26 scalar fields reads

$$X^\mu(z, \bar{z}) = x^\mu - i\frac{\alpha'}{2} p^\mu \ln |z|^2 + i\sqrt{\frac{\alpha'}{2}} \sum_{n \neq 0} \frac{1}{n} \left( \frac{\alpha_n^\mu}{z^n} + \frac{\bar{\alpha}_n^\mu}{\bar{z}^n} \right) \tag{3.1}$$

and its (anti-)holomorphic derivative is again a conformal primary of dimension (1,0) ((0,1))

$$\partial X^\mu(z) = -i\sqrt{\frac{\alpha'}{2}} \sum_n \frac{\alpha_n^\mu}{z^{n+1}}, \qquad \bar{\partial} X^\mu(z) = -i\sqrt{\frac{\alpha'}{2}} \sum_n \frac{\bar{\alpha}_n^\mu}{\bar{z}^{n+1}}, \tag{3.2}$$

the only difference is the normalization of the zero mode

$$\alpha_0^\mu = \bar{\alpha}_0^\mu = \sqrt{\frac{\alpha'}{2}} p^\mu. \tag{3.3}$$

The $b, c$-ghost system is unchanged with the expansions given by

$$b(z) = \sum_n \frac{b_n}{z^{n+2}}, \qquad c(z) = \sum_n \frac{c_n}{z^{n-1}} \tag{3.4}$$

and similarly for the antiholomorphic sector. We do not impose any doubling trick hence the two sectors are totally independent. It is convenient to define the new zero mode combinations

$$c_0^\pm = \frac{1}{2}(c_0 \pm \bar{c}_0), \qquad b_0^\pm = b_0 \pm \bar{b}_0. \tag{3.5}$$

The commutation relations

$$[\alpha_n^\mu, \alpha_m^\nu] = n\eta^{\mu\nu}\delta_{n+m}, \qquad \{b_n, c_m\} = \delta_{n+m} \tag{3.6}$$

are supplemented by a second copy for the antiholomorphic sector as well as

$$\{b_0^+, c_0^+\} = \{b_0^-, c_0^-\} = 1. \tag{3.7}$$

All other combinations vanish, in particular, modes of different sectors (anti-)commute.

The construction of the Fock space works completely analogously to the open string case after taking into account both sectors. We will also frequently use the Schrödinger representation, where a state is defined as a surface conformally equivalent to the unit disc prepared with arbitrary bulk operator insertions. The definition of the BPZ-conjugate differs by a sign in the sense that we have

$$\Phi^* = \langle 0 | I \circ \Phi(0) \qquad \text{with} \qquad I(z) = \frac{1}{z}. \tag{3.8}$$

As a result, general BPZ-conjugated Fock states are related via

$$\Phi = \alpha^{\mu_1}_{-n_1}...\alpha^{\mu_r}_{-n_r}\overline{\alpha}^{\mu_1}_{-n'_1}...\overline{\alpha}^{\mu_{r'}}_{-n'_{r'}}b_{-m_1}...b_{-m_s}\overline{b}_{-m'_1}...\overline{b}_{-m'_{s'}}c_{-p_1}...c_{-p_t}\overline{c}_{-p'_1}...\overline{c}_{-p'_{t'}}|0,k\rangle,$$

$$\Phi^* = (-1)^{r+t+r'+t'}\langle 0,-k|\,\alpha^{\mu_1}_{n_1}...\alpha^{\mu_r}_{n_r}\overline{\alpha}^{\mu_1}_{n'_1}...\overline{\alpha}^{\mu_{r'}}_{n'_{r'}}b_{m_1}...b_{m_s}\overline{b}_{m_1}...\overline{b}_{m_{s'}}c_{p_1}...c_{p_t}\overline{c}_{p_1}...\overline{c}_{p_{t'}}$$

$$\text{for} \qquad n_i, n'_i \geq 1, \quad m_i, m'_i \geq 2, \quad p_i, p'_i \geq -1. \tag{3.9}$$

The BPZ-product induced by this definition still obeys (2.17) and the basic overlap of the vacua will be normalized as

$$\langle 0,k|\,c_{-1}\overline{c}_{-1}c_0^+ c_0^- c_1\overline{c}_1\,|0,k'\rangle = (2\pi)^{26}\,\delta^{(26)}\,(k-k'). \tag{3.10}$$

Note that now we need a total ghost number of six for having a non-vanishing correlation function.

The energy-momentum tensor has now two independent components denoted $T(z)$, $\overline{T}(\overline{z})$ and its modes $L_n$, $\overline{L}_n$ obey two copies of the Virasoro algebra. Similarly to the ghost fields we define

$$L_0^\pm = L_0 \pm \overline{L}_0. \tag{3.11}$$

To construct the BRST charge we need to include both sectors and get

$$Q = \oint \frac{dz}{2\pi i} j_B(z) + \oint \frac{d\overline{z}}{2\pi i} \overline{j}_B(\overline{z})$$
$$= \oint \frac{dz}{2\pi i}\left(cT^{(m)}(z) + :bc\partial c:(z) + \frac{3}{2}\partial^2 c(z)\right) + \oint \frac{d\overline{z}}{2\pi i}\left(\overline{c}\overline{T}^{(m)}(\overline{z}) + :\overline{b}\overline{c}\overline{\partial}\overline{c}:(\overline{z}) + \frac{3}{2}\overline{\partial}^2\overline{c}(\overline{z})\right). \tag{3.12}$$

$Q$ is still BPZ-odd and fulfills

$$\{Q, \overline{b}(z)\} = \overline{L}(z) \tag{3.13}$$

in addition to (2.30).

A main difference to the open case is that we demand two (related) additional constraints on the string field, namely

$$L_0^- \Phi = 0, \qquad b_0^- \Phi = 0. \tag{3.14}$$

The first one is called the *level matching condition* and arises from rotational invariance of the string. It follows from the second condition if the state is BRST-closed. The space of physical fields is then also not just the BRST-cohomology at ghost number two but the *semirelative cohomology*, defined in the subspace satisfying (3.14). One consequence of that is that the *ghost dilaton*

$$D = (c_1 c_{-1} - \overline{c}_1 \overline{c}_{-1})|0\rangle \tag{3.15}$$

is part of the physical spectrum: In principle it is BRST-exact,

$$D = Q c_0^- |0\rangle, \tag{3.16}$$

but $c_0^- |0\rangle$ does not obey the second condition in (3.14), hence $D$ is physical. The ghost dilaton is important because it can provide an obstruction to solve the CSFT equations of motion perturbatively, see [53, 67].

## 3.2 The classical CSFT action

In this section we will write down the classical action of CSFT ([87]) and start first of all with the kinetic term. Inspired by OSFT we could guess

$$S_{kin} \stackrel{?}{\propto} \frac{1}{2} \langle \Phi, Q\Phi \rangle, \tag{3.17}$$

but this does not work because a classical closed string field has ghost number two, so we only get a total ghost number of five within the two-point function. We need an additional $c$-ghost insertion and the correct choice turns out to be

$$S_{kin} \propto \frac{1}{2} \langle \Phi, c_0^- Q\Phi \rangle. \tag{3.18}$$

It can be understood by varying $S_{kin}$ to obtain

$$\begin{aligned}\delta S_{kin} &\propto \frac{1}{2} \langle \delta\Phi, c_0^- Q\Phi \rangle + \frac{1}{2} \langle \Phi, c_0^- Q\delta\Phi \rangle \\ &= \frac{1}{2} \langle c_0^- Q\Phi, \delta\Phi \rangle - \frac{1}{2} \langle Qc_0^- \Phi, \delta\Phi \rangle\end{aligned} \tag{3.19}$$

The second term can be modified using the subsidiary conditions (3.14), yielding

$$\begin{aligned}\langle Qc_0^- \Phi, \delta\Phi \rangle &= \langle \left(b_0^- c_0^- + c_0^- b_0^-\right) Qc_0^- \Phi, \delta\Phi \rangle = \langle c_0^- Qc_0^- \Phi, b_0^- \delta\Phi \rangle + \langle c_0^- b_0^- Qc_0^- \Phi, \delta\Phi \rangle \\ &= \langle c_0^- L_0^- c_0^- \Phi, \delta\Phi \rangle - \langle c_0^- Qb_0^- c_0^- \Phi, \delta\Phi \rangle = - \langle c_0^- Q\Phi, \delta\Phi \rangle\end{aligned} \tag{3.20}$$

In the end we get the equations of motion

$$c_0^- Q\Phi = 0 \tag{3.21}$$

which together with $b_0^- Q\Phi = 0$ imply the expected

$$Q\Phi = 0. \tag{3.22}$$

One can see from this calculation that the additional insertion has to be compatible with (3.14), so $c_0^-$ is the simplest choice.

Let us now turn to the interactions: First, as it was argued in [85], we expect closed string interactions to be totally (graded) symmetric. Imagine gluing together two strings along some interval and integrate over those degrees of freedom to produce a new closed string. After imposing rotational invariance (which is equivalent to level-matching), there is no way to determine which string came first. On the other hand, consider three strings joined as in the figure:

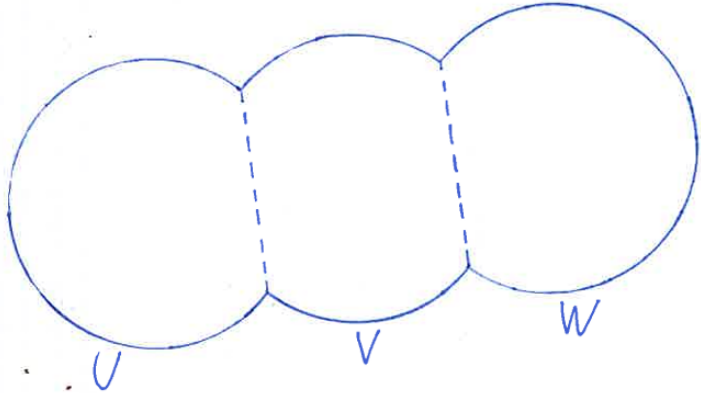

The product $U \cdot (W \cdot V)$ is non-zero whileas $(U \cdot W) \cdot V$ is zero since $U$ and $V$ are not connected. Hence, we have to expect that the string product giving rise to the three-vertex will be non-associative.

We can proceed naively by trying to generalize the Witten product to the closed string: In fact it is straightforward to extend the maps (2.64) to the whole Riemann sphere without imposing the doubling trick and then just symmetrize over the two inputs of the product. However, the output will typically fail to obey (3.14) and therefore not be within our state space. We can define the operator

$$\mathcal{P} = b_0^- \delta \left( L_0^- \right) \tag{3.23}$$

which just projects on the subspace defined by (3.14) and write[11]

$$\Phi_1 \cdot \Phi_2 = \mathcal{P} \left( \Phi_1 * \Phi_2 + \Phi_2 * \Phi_1 \right). \tag{3.24}$$

This is indeed a viable definition and gives rise to a well-defined three-vertex, however, as anticipated, it is non-associative. We conclude that it will be necessary to introduce higher vertices in the spirit of section 2.3 to ensure classical gauge invariance. In the end, the classical action will take the form

$$S_{cl} = \frac{1}{2} \left\langle \Phi, c_0^- Q \Phi \right\rangle + \sum_{n=2}^{\infty} \frac{\kappa^{n-1}}{(n+1)!} \left\langle \Phi, c_0^- l_n \left( \Phi, \Phi, ..., \Phi \right) \right\rangle \tag{3.25}$$

where $\kappa$ is the closed string coupling constant and the $l_n$ are $n$-linear, totally symmetric products of ghost number $3 - 2n$. For deriving the classical equations of motion we need to demand that the $l_n$ are cyclic with respect to the BPZ-product and that their output obeys (3.14), see the calculation (3.20). We then get

$$Q\Phi + \frac{\kappa^{n-1}}{n!} l_n \left( \Phi, \Phi, ..., \Phi \right) = 0. \tag{3.26}$$

Within the next section we will discuss in detail the algebraic properties satisfied by the $l_n$.

---

[11]With a slight abuse of notation we use $*$ for the Witten product in open and closed SFT.

## 3.3 Algebraic structure

In section 2.3 we have defined the notion of an $A_\infty$-algebra as a collection of multilinear maps giving rise to gauge invariant equations of motion. To describe CSFT we need to incorporate the total graded symmetry into this concept. Let us first define the *symmetrized tensor product* of $n$ vectors as

$$v_1 \odot v_2 \odot ... \odot v_n = \sum_\sigma \frac{1}{n!} (-1)^{\epsilon(\sigma)} v_{\sigma(1)} \otimes v_{\sigma(2)} \otimes ... \otimes v_{\sigma(n)} \tag{3.27}$$

where the sum runs over all permutations of $n$ elements and $\epsilon$ is the Grassmann sign which is picked up.[12] Similarly, an $n$-linear map is called totally symmetric if it fulfills

$$l_n \left( v_1 \odot v_2 \odot ... \odot v_n \right) = (-1)^{\epsilon(\sigma)} l_n \left( v_{\sigma(1)} \otimes v_{\sigma(2)} \otimes ... \otimes v_{\sigma(n)} \right) \tag{3.28}$$

for all permutations $\sigma$. Those definitions naturally induce the *symmetrized tensor algebra STV* over a graded vector space $V$ as

$$V^{\odot 0} + V^{\odot 1} + V^{\odot 2} + ... = \sum_{n=0}^\infty V^{\odot n}. \tag{3.29}$$

We can now proceed as in section 2.3 and define the projections $\pi_k$ analogously; the definition of a coderivation however simplifies to

$$\mathbf{l_n} = \sum_{m=n}^\infty \frac{1}{(m-n)!} l_n \odot \mathbb{1}^{\otimes m-n} \tag{3.30}$$

for any totally symmetric $n$-linear map $l_n$. If a coderivation $\mathbf{l}$ over $STV$ squares to zero, its components $l_n = \pi_1 \mathbf{l} \pi_n$ form a so-called $L_\infty$-*algebra*. Let us check what this means for the first few components:

$$l_1 \left( l_1 \left( \Phi \right) \right) = 0$$
$$l_1 l_2 \left( \Phi_1, \Phi_2 \right) + l_2 \left( l_1 \Phi_1, \Phi_2 \right) + (-1)^{|\Phi_1||\Phi_2|} l_2 \left( l_1 \Phi_2, \Phi_1 \right) = 0$$
$$l_2 \left( l_2 \left( \Phi_1, \Phi_2 \right), \Phi_3 \right) + (-1)^{|\Phi_1|(|\Phi_2|+|\Phi_3|)} l_2 \left( l_2 \left( \Phi_2, \Phi_3 \right), \Phi_1 \right) +$$
$$(-1)^{|\Phi_3|(|\Phi_1|+|\Phi_2|)} l_2 \left( l_2 \left( \Phi_3, \Phi_1 \right), \Phi_2 \right) + l_1 l_3 \left( \Phi_1, \Phi_2, \Phi_3 \right) + l_3 \left( l_1 \Phi_1, \Phi_2, \Phi_3 \right) +$$
$$(-1)^{|\Phi_1|(|\Phi_2|+|\Phi_3|)} l_3 \left( l_1 \Phi_2, \Phi_3, \Phi_1 \right) + (-1)^{|\Phi_3|(|\Phi_1|+|\Phi_2|)} l_3 \left( l_1 \Phi_3, \Phi_1, \Phi_2 \right) = 0 \tag{3.31}$$

The first two equations are familiar as the nilpotency of $l_1$ and the Leibniz rule with respect to $l_2$. In the case $l_{n\geq 3} = 0$, the last equation is just the super-Jacobi identity for $l_2$. This means that if we have only odd elements such that $l_2$ is antisymmetric, we get a differential Lie algebra. An $L_\infty$-algebra is therefore a generalization of a Lie algebra in the same way as an $A_\infty$-algebra is a generalization of an associative algebra. Moreover, given any $A_\infty$-algebra we can construct an $L_\infty$-algebra by symmetrizing all the products. Those results suggest that an $L_\infty$-algebra is a good candidate to describe classical CSFT, with $l_1 \equiv Q$.

---

[12]In the case of CSFT we can set the degree equal to the ghost number $|\Phi| = gh(\Phi)$ such that for classical string fields of ghost number two, no signs arise.

To write down the action we also need a symplectic form $\omega$ which we will define, inspired by the kinetic term (3.18), as

$$\omega(\Phi_1, \Phi_2) = (-1)^{|\Phi_1|} \langle \Phi_1, c_0^- \Phi_2 \rangle \tag{3.32}$$

with

$$\omega(\Phi_1, \Phi_2) = -(-1)^{|\Phi_1||\Phi_2|} \omega(\Phi_2, \Phi_1). \tag{3.33}$$

Now the products $l_n$ are called *cyclic* if they obey

$$\omega(\Phi_1, l_n(\Phi_2, ..., \Phi_{n+1})) = -(-1)^{|\Phi_1|} \omega(l_n(\Phi_1, ..., \Phi_n), \Phi_{n+1}), \tag{3.34}$$

reminiscent of equation (2.57).

In the $L_\infty$-case it is convenient to replace the short-cut notation (2.51) by

$$e^\Phi \equiv \mathbb{1} + \Phi + \frac{1}{2}\Phi^{\odot 2} + ... = \sum_{n=0}^{\infty} \frac{1}{n!} \Phi^{\odot n}. \tag{3.35}$$

Then the $L_\infty$-action

$$S(\Phi) = \sum_{n=1}^{\infty} \frac{1}{(n+1)!} \omega\left(\Phi, l_n\left(\Phi^{\odot n}\right)\right) \tag{3.36}$$

can be written in the spirit of (2.58) as

$$S(\Phi) = \int_0^1 dt\, \omega\left(\pi_1 \partial_t e^{\Phi(t)}, \pi_1 \mathbf{l}\left(e^{\Phi(t)}\right)\right) \tag{3.37}$$

where $\Phi(t)$ is again some interpolation between 0 and $\Phi$. The equations of motion are just the Maurer-Cartan equation of the $L_\infty$-algebra and read[13]

$$\pi_1 \mathbf{l}\left(e^\Phi\right) = \sum_{n=1}^{\infty} \frac{1}{n!} l_n\left(\Phi^{\odot n}\right) = 0. \tag{3.38}$$

Classical gauge invariance can now be proven analogously to the $A_\infty$-case: Define

$$\delta \Phi = \sum_{n=1}^{\infty} \frac{1}{(n-1)!} l_n\left(\Lambda, \Phi^{\odot n-1}\right) = \pi_1 \{\mathbf{\Lambda}, \mathbf{l}\} e^\Phi, \tag{3.39}$$

for some ghost number one gauge parameter $\Lambda$, then a calculation according to (2.61) yields the desired result.

---

[13]In the coalgebra notation, the string coupling $\kappa$ is typically set to one.

## 3.4 Quantization and the BV-formalism

So far we have only discussed the classical action with its classical equations of motion. Second-quantizing the theory is actually non-trivial and will result in additional terms in the action. It is done using the so-called *Batalin-Vilkovisky (BV) formalism,* a quite sophisticated method to quantize theories where the BRST-formalism is not applicable anymore. Typically, in theories described via $A_\infty$- or $L_\infty$-algebras, the algebra of gauge transformations only closes *on-shell* (This can be seen by just calculating the commutator of two transformations (2.60).) This implies, if we could define a BRST-operator $\mathfrak{Q}$ for the second-quantized theory[14], it would square to zero only on-shell. Therefore, we would not be able to define the cohomology properly and we would not be able to define observables. In the BV-approach, this problem is circumvented, which makes it the right candidate to quantize CSFT. In the following, we will review the main ideas behind the BV-formalism and apply it to the special case of CSFT. For a more detailed and mathematical treatment, see the excellent reviews [28, 36, 73].

The first step is to extend our field space: So far we have considered only classical string fields of ghost number two and gauge parameters of ghost number one. It turns out that we will also need to introduce "gauge for gauges" in the sense that any meaningful gauge condition has gauge freedom itself. As a result, we must include gauge parameters of all ghost numbers smaller than one, which has been iteratively shown for the open string case in [42]. The BV-formalism now instructs us to double the field content and introduce for any present basis state field $\Phi^i$ an *antifield* $\Phi_i^+$ of opposite statistics. Speaking more mathematically, we just take the degree-shifted cotangent bundle of our space of fields. In CSFT this is particularly simple because we can define the antifields to be ghost number $n$ fields as those with ghost number $5-n$. Then the total field space just includes all string fields with arbitrary ghost number.

Being a cotangent bundle, our total field space comes with a natural symplectic structure given by

$$\omega = (-1)^{|\phi^i|} \delta\Phi^i \wedge \delta\Phi_i^+. \tag{3.40}$$

It induces a second-order differential operator called the *BV-Laplacian*

$$\Delta = \frac{\partial_r}{\partial\Phi^i} \frac{\partial_l}{\partial\Phi_i^+} \tag{3.41}$$

where $\partial_{r/l}$ are the derivatives from the right and left, respectively. Due to the different grading of field and antifield we have

$$\Delta^2 = 0. \tag{3.42}$$

If $F$ and $G$ are two arbitrary functionals on the field space, we can calculate

$$\Delta(F \cdot G) = \Delta F \cdot G + (-1)^{|F|} F \cdot \Delta G + (-1)^{|F|} \{F, G\}_{BV} \tag{3.43}$$

with $\{\cdot, \cdot\}_{BV}$ being the Poisson bracket with respect to $\omega$ called the *BV-antibracket*

$$\{F, G\}_{BV} = \frac{\partial_r F}{\partial\Phi^i} \frac{\partial_l G}{\partial\Phi_i^+} - (-1)^{(|F|+1)(|G|+1)} \frac{\partial_r G}{\partial\Phi^i} \frac{\partial_l F}{\partial\Phi_i^+}. \tag{3.44}$$

---
[14] $\mathfrak{Q}$ should not be confused with the BRST-charge $Q$ of the background CFT.

The antibracket therefore measures the failure of $\Delta$ obeying the Leibniz rule for pointwise mulitplication of functionals in field space.

The next step is to extend the action in a proper way: First, it needs to be defined on the full, extended field space and second, we expect quantum corrections. The main relation that has to be satisfied is

$$\Delta \left( e^{-\frac{1}{\hbar} S_{qu}} \right) = 0, \tag{3.45}$$

i. e. the path integral measure has to be $\Delta$-closed. This condition actually has a geometrical origin (see [73]): In an ordinary path integral

$$\int \mathcal{D}\Phi \, \mathcal{O}_1 ... \mathcal{O}_n e^{-\frac{1}{\hbar} S_{qu}} \tag{3.46}$$

the integration would run only over the fields, not the antifields. If (3.45) is obeyed however, the integration can be done over *any* Lagrangian subspace[15] of the full field space, as long as it is homologous to the one with all antifields set to zero. This powerful result implies that we can choose a more convenient subspace where the integral is well-behaved and can be computed perturbatively. If we use (3.43), we can see that (3.45) is equivalent to

$$\{S_{qu}, S_{qu}\} = -2\hbar \Delta S_{qu}. \tag{3.47}$$

This equation is often called the *quantum master equation* and can be seen as a condition that all the gauge symmetries are correctly lifted to the full quantum theory. To solve it, one typically expands $S_{qu}$ in a power series in $\hbar$

$$S_{qu} = S_0 + \hbar S_1 + \hbar^2 S_2 + ... \tag{3.48}$$

and gets

$$\{S_0, S_0\} = 0 \tag{3.49}$$

$$\{S_0, S_1\} = -\Delta S_0 \tag{3.50}$$

$$\{S_0, S_2\} + \frac{1}{2} \{S_1, S_1\} = -\Delta S_1 \tag{3.51}$$

$$\vdots$$

The first equation is the *classical master equation* and just tells us how to extend the definition of $S_{cl}$ to the total field space. One typically demands the boundary condition

$$S_0 \mid_{\Phi^+ = 0} = S_{cl} \tag{3.52}$$

such that we recover our original action when setting all the antifields to zero. Plugging (3.37) into (3.49) yields the condition

$$\mathbf{l}^2 = 0 \tag{3.53}$$

---

[15] A Lagrangian subspace is a maximal subspace on which the induced symplectic form is vanishing.

that we already assume. Hence there is a very simple solution to (3.49) by just keeping $S_{cl}$ as it is and allow for all ghost numbers. The next equations are not so trivial to solve actually: Since $\Delta S_0 \neq 0$ we need non-zero quantum corrections $S_k$ to all orders. If we make the ansatz

$$S_{k,n+1}(\Phi) = \frac{1}{(n+1)!} \omega\left(\Phi, l_n^k\left(\Phi^{\odot n}\right)\right) \tag{3.54}$$

for the $n+1$-vertex within $S_k$[16], the whole action can be written as

$$S_{qu} = \sum_{k,n=0}^{\infty} \frac{\hbar^k \kappa^{n+2k-2}}{(n+1)!} \omega\left(\Phi, l_n^k\left(\Phi^{\odot n}\right)\right). \tag{3.55}$$

Then the master equation translates into

$$\sum_{k_1+k_2=k} \sum_{n_1+n_2=n_3+1} \sum_{\{r,s\}} \sigma(r,s) \, l_{n_1}^{k_1}\left(\Phi_{r_1}, ..., \Phi_{r_{n_1-1}}, l_{n_2}^{k_2}\left(\Phi_{s_1}, ..., \Phi_{s_{n_2}}\right)\right)$$
$$+ \frac{1}{2} \sum_m (-1)^{|\Phi^m|} l_{n_3+2}^{k-1}\left(\Phi^m, b_0^- \Phi_m, \Phi_1, ..., \Phi_{n_3}\right) = 0. \tag{3.56}$$

The sum over $\{r,s\}$ just means one has to sum over all inequivalent splittings of $\Phi_1...\Phi_{n_3}$ into the two groups entering the first term. $\sigma(r,s)$ is the Grassmann sign which is picked up by this rearrangement. Moreover, the sum in the second term runs over a complete (infinite) set of basis states that obey the subsidiary consitions (3.14). The state $\Phi_m$ is the conjugate of $\Phi^m$ defined via the condition

$$\langle \Phi^m, \Phi_n \rangle = \delta_n^m \tag{3.57}$$

whereas $b_0^- \Phi_n$ can be interpreted as conjugate with respect to the symplectic form $\omega$. The combined object

$$U \equiv \sum_m (-1)^{|\Phi^m|} \Phi^m \otimes b_0^- \Phi_m \tag{3.58}$$

which appears in (3.56) is called the *Poisson bivector* and arises as the dual to the BV-Laplacian $\Delta$.

A set of multilinear products $l_n^k$ which fulfill this condition are said to form a *quantum $L_\infty$-algebra.* The restriction to $l_n^0$ is just the condition for an ordinary $L_\infty$-algebra. Equation (3.56) is sometimes called the *main identity* of CSFT and guarantees quantum consistency of the theory. There is one interesting point that we can already deduce: In QFT, an expansion in $\hbar$ is typically related to an expansion in loops. This means, the additional vertices we got are expected to have *internal loops,* i. e. loops which do not arise from Feynman diagrams. Since we are dealing with Riemann surfaces, the loop expansion is equivalent to a genus expansion. In the next section we will address this geometric interpretation in more detail and give a more visualizable picture of the vertices.

---

[16]At higher $k$, we also might have "zero-vertices", i. e. vacuum bubbles, where this ansatz is not applicable anymore. We will ignore this sublety for now.

## 3.5 Geometric vertices

We already know that the Feynman diagrams generated by the action (3.55) should cover the whole moduli space of Riemann surfaces of the type in question. For an $n$-point function with $k$ loops that would be the space of all $n$-punctured Riemann surfaces with genus $g = k$. Its dimension is given by

$$\dim \mathcal{M}_{g,n} = 6g - 6 + 2n \tag{3.59}$$

for $g \geq 2$, $g = 1$ and $n \geq 1$ as well as $g = 0$ and $n \geq 3$. The two-punctured sphere corresponds to the kinetic term which was already defined earlier. The other cases which are not included in the formula, namely the tree-level and one-loop vacuum bubble, corresponding to sphere and torus diagrams respectively, and the tree-level tadpole are in general not necessary to construct the action.

Since we have vertices of arbitrary order in the action, we know they must include some integration over moduli space themselves. One might be tempted to just define the geometric part of the vertices $\mathcal{V}_{g,n}$ (for simplicity just called vertices in this section) as suitable subspaces of $\mathcal{M}_{g,n}$. This is not enough however: As discussed earlier, the incoming states are in general *off-shell*. In the Schrödinger representation we can view them as surfaces conformally equivalent to a unit disc prepared with some operator insertions. If we want to glue those to our vertex surface, we must specify how this gluing is done, hence we must specify non-overlapping *local coordinates* around every puncture. Let us therefore consider the fibre bundle $\mathcal{P}_{g,n}$ with base space $\mathcal{M}_{g,n}$ and the fibre being the (infinite dimensional) space of local coordinate discs. This space seems at first a promising arena to define our vertices, however, there is a topological obstruction: It does in general not admit global sections. A simple argument to illustrate that is the following: Consider $\mathcal{P}_{0,4}$, the space of coordinate discs on a four-punctured sphere. Three of the punctures can be fixed to arbitrary values using conformal invariance whileas the position of the fourth puncture spans $\mathcal{M}_{0,4}$. Now construct the vector field on the sphere that points from the puncture (with local coordinate $z_4 = 0$) to the point with local coordinate $z_4 = 1$. This vector field must not vanish anywhere, otherwise the coordinate disc would be degenerate. Hence we have a continuous, nowhere vanishing vector field on a sphere, which is in contradiction to the hairy ball theorem.[17] The way out is to identify any two points in $\mathcal{P}_{g,n}$ if they differ by pure phase rotations of the coordinate discs. The resulting bundle $\hat{\mathcal{P}}_{g,n}$ avoids this problem and, as we will see later, still allows for a well-defined gluing procedure for the external states. As a conclusion, we define the vertices $\mathcal{V}_{g,n}$ as sections of $\hat{\mathcal{P}}_{g,n}$ over suitable subsets of $\mathcal{M}_{g,n}$. We already know that the products $l_n^k$ will be totally symmetric, so we also demand that the vertices are symmetrized over all the punctures. To get the algebraic expression $\omega\left(\Phi, l_n^k\left(\Phi^{\odot n}\right)\right)$, we need to integrate some $6g - 6 + 2n$-dimensional form to be defined within section 3.6 over $\mathcal{V}_{g,n}$.

Let us discuss the propagator now: As in OSFT we first must choose a gauge and the most convenient one is again Siegel gauge given by the condition

$$b_0^+ \Phi = 0. \tag{3.60}$$

---

[17]This argument is not totally conclusive because it ignores the fact that punctures can collide. To make it a hundred percent precise, a more detailed analysis is necessary, see [19].

Moreover, the propagator should contain the projector $\mathcal{P}$ such that it does not propagate unphysical states. By the same reasoning as for the open string we get

$$P = \frac{b_0^+}{L_0^+} b_0^- \delta\left(L_0^-\right) = \frac{1}{2\pi} b_0^+ b_0^- \int_0^\infty dt \int_0^{2\pi} ds\, e^{-tL_0^+ + isL_0^-}. \tag{3.61}$$

We have now two moduli integrals which again have a nice geometric interpretation: A propagating closed string forms a cylinder, where $t$ parametrizes its length in similarity to the open string case and $s$ parametrizes its *twist angle*. The dimension counting still works out: If we consider again the tree-level 4-point amplitude, we see that the Feynman contribution with two 3-vertices and one propagator has two real moduli (from the propagator), same as the elementary 4-vertex being a four-punctured sphere. This argument can be generalized to $n$-point diagrams of arbitrary genus.

To build a Feynman diagram out of the vertices and propagators we have available, we need to know how the gluing is precisely done. Let us assume the punctures we want to connect (it does not matter if they belong to the same Riemann surface or not) have local coordinates $z_1$ and $z_2$. We then demand a gluing condition of the form

$$z_1 z_2 = q \tag{3.62}$$

with some complex gluing parameter $q$ of modulus smaller than one. We can encode this condition within a coordinate transformation of, lets say, $z_1$ and write

$$\tilde{z}_1 z_2 = 1 \qquad \text{with} \qquad \tilde{z}_1 = \frac{z_1}{q}. \tag{3.63}$$

This rescaling induces a conformal transformation of the state inserted at the first puncture given by

$$\Phi_1 \to q^{L_0} \bar{q}^{\bar{L}_0} \Phi_1. \tag{3.64}$$

To include all possible surfaces that can be formed, we need to integrate over $q$. By splitting it into its modulus and phase

$$|q| \equiv e^{-t}, \qquad \text{Arg}\, q \equiv s \tag{3.65}$$

we can actually reproduce the integral part of the propagator

$$\Phi_1 \to \int_0^\infty dt \int_0^{2\pi} ds\, e^{-tL_0^+ + isL_0^-} \Phi_1. \tag{3.66}$$

In particular, by integrating over the argument of $q$ we implemented the arbitrariness of the phase of the local coordinates, giving rise to the level matching condition $\delta\left(L_0^-\right)$.

We are now in the position to discuss the conditions on the vertices coming from the requirement that the $n$-point functions span the whole moduli space. If we look at the tree-level 4-point function once again, we need to add the $s$-, $t$- and $u$-channel of two three-vertices connected by a propagator and the elementary four-vertex. If we demand that there is no "gap" left in the moduli space, then the boundary of the Feynman region should coincide with the boundary of the elementary vertex, with opposite orientation. The boundary of the Feynman region corresponds to $t = 0$, namely when the propagator collapses to zero length and in the gluing condition we have $\mid q \mid = 1$.[18] If the subspace of surfaces within the Feynman region is denoted by $R_{0,4}(t,s)$ then we demand

$$\partial \mathcal{V}_{0,4} = -\partial R_{0,4}(t,s) = -R_{0,4}(0,s). \tag{3.67}$$

The same idea should of course hold for all $\mathcal{V}_{g,n}$ and is contained in the formula

$$\partial \mathcal{V}_{g,n} = -\frac{1}{2} \sum_{g_1+g_2=g} \sum_{n_1+n_2=n+2} \{\mathcal{V}_{g_1,n_1}, \mathcal{V}_{g_2,n_2}\} - \frac{1}{2}\widetilde{\Delta}\mathcal{V}_{g-1,n+2}. \tag{3.68}$$

The $\{\cdot,\cdot\}$-product in the first term is defined as follows: We take each surface in $\mathcal{V}_{g_1,n_1}$ and connect it to each surface in $\mathcal{V}_{g_2,n_2}$ with a collapsed propagator, i. e. glue any two punctures with a gluing parameter $\mid q \mid = 1$. The operator $\widetilde{\Delta}$ does the same thing, with the only difference that both punctures belong to the same surface contained in $\mathcal{V}_{g-1,n+2}$. Equation (3.68) means that the subspaces of $\widehat{\mathcal{P}}_{g,n}$ on both sides have to agree in their moduli parameters and in the local coordinates around the punctures. If it is fulfilled, the Feynman diagrams indeed cover the full moduli space. Let us see what we get for the first few choices of $g$ and $n$:

$$\partial \mathcal{V}_{0,3} = 0 \tag{3.69}$$

$$\partial \mathcal{V}_{0,4} = -\frac{1}{2}\{\mathcal{V}_{0,3},\mathcal{V}_{0,3}\} \tag{3.70}$$

$$\partial \mathcal{V}_{0,5} = -\{\mathcal{V}_{0,4},\mathcal{V}_{0,3}\} \tag{3.71}$$

$$\partial \mathcal{V}_{1,1} = -\frac{1}{2}\widetilde{\Delta}\mathcal{V}_{0,3} \tag{3.72}$$

$$\vdots$$

The first condition is automatically fulfilled since $\mathcal{M}_{0,3}$ is zero-dimensional. The second equation is the equivalent of (3.67), formulated using the $\{\cdot,\cdot\}$-product. All the higher ones are also just the consistency equations one would expect by drawing the respective Feynman diagrams.

One should notice the similarity between equations (3.68) and (3.56): If one extracts the terms involving $l_1^0 = Q$ in (3.56), then their structure is basically the same: One just has to associate $Q$ to $\partial$, the composition of the $l_n^k$ to the $\{\cdot,\cdot\}$-product and the insertion of a sum over basis states with $\widetilde{\Delta}$. For this reason, (3.68) is also called the *geometrical master equation* and $\widetilde{\Delta}$ and $\{\cdot,\cdot\}$ the *geometrical BV-Laplacian* and *antibracket,* respectively.

To find an explicit solution to equation (3.68) is still one of the major open problems in modern CSFT. The two main approaches to the problem are *minimal area metrics* and *hyperbolic vertices.* Although some existence theorems have been proven, we are still lacking an explicit description of all vertices. A promising direction might be [32], where a recursion relation for hyperbolic vertices was derived. More progress on this issue could immensely increase the versatility of string field theory and allow for calculations relevant also for string phenomenology for instance.

---

[18]The other boundary for $t \to \infty$ gives rise to a nodal surface contained within the *Deligne-Mumford compactification* of the moduli space. Although those surfaces can become relevant in some context (see e. g. [75]), we will ignore this issue here.

## 3.6 Algebraic vertices

In this last section we finally want to give an explicit expression for the whole algebraic vertices and hence also for the higher products $l_n^k$ entering the action. For that we need to define the $6g - 6 + 2n$-dimensional forms $\Omega^{g,n}$ which can be integrated over the $\mathcal{V}_{g,n}$. Our strategy will be as follows: Let us assume we have found a solution to (3.68), then the explicit form of the $\Omega^{g,n}$ should be such that the main identity (3.56) is automatically implied. First we recall some properties of the $\Omega^{g,n}$ that we expect to hold:

1. $\Omega^{g,n}$ should depend on the $n$ input states $\Phi_i$ such that we can write $\Omega^{g,n} \equiv \Omega^{g,n}_{\Phi_i}$

2. $\Omega^{g,n}$ can (and will) also depend on the surface $\Sigma$ representing the base point in $\mathcal{M}_{g,n}$. In similarity to the Schrödinger representation, we can think of $\Sigma$ with its coordinate disks removed giving rise to a state $|\Sigma\rangle$ in $\mathcal{H}^{\otimes n}$: By specifying boundary conditions on the $n$ coordinate curves corresponding to field configurations at time $t = 0$, we can carry out the path integral over $\Sigma$ to obtain a number. $|\Sigma\rangle$ therefore provides a functional from $n$ constant-time configurations of strings into complex numbers, which is the definition of a quantum state within $\mathcal{H}^{\otimes n}$. Its BPZ-conjugate $\langle \Sigma |$ will enter the definition of $\Omega^{g,n}$.

3. $\Omega^{g,n}$ needs to have the right ghost number that leads to non-vanishing world-sheet correlation functions. In general, on an $n$-loop Riemann surface we need a total ghost number of $6 - 6g$. A classical closed string field has ghost number two, so to compensate that we need precisely $6g - 6 + 2n$ $b$-ghost insertions within $\Omega^{g,n}$.

4. From the previous section we know that we have to relate the boundary operator $\partial$ to the BRST-operator $Q$ to be able to deduce the main identity. By Stokes' theorem

$$\int_{\partial M} \omega = \int_M d\omega \tag{3.73}$$

we can convert $\partial$ into an exterior derivative. We now demand the equation

$$\Omega^{g,n}_{\Sigma Q_i(\Phi_i)} = d\Omega^{g,n(-1)}_{\Phi_i} \tag{3.74}$$

to hold. $\Omega^{g,n(x)}_{\Phi_i}$ is a form of degree $6g - 6 + 2n - x$ defined in a similar way to $\Omega^{g,n}_{\Phi_i}$. The subscript $\Sigma Q_i(\Phi_i)$ just means that $Q$ should act by Leibniz rule on all string fields $\Phi_i$.

Before we are able to give the full definition of $\Omega^{g,n}_{\Phi_i}$ we need to introduce the concept of the *Schiffer variation*. Let us assume we want to describe an infinitesimal deformation of a surface in $\hat{\mathcal{P}}_{g,n}$, or, in other words, parametrize the tangent space $T\hat{\mathcal{P}}_{g,n}$. It has been shown that this can be done using local vector fields $v_i(z_i)$ defined in an annular region around the coordinate curves. We deform the coordinate curve as

$$z_i \to z_i' = z_i + \epsilon v_i \tag{3.75}$$

and glue the interior of the deformed curves back into $\Sigma$ with the original coordinate disks removed. By this procedure we can change the local coordinates, the positions of the punctures as well as the moduli of $\Sigma$. The full proof of this statement is quite elaborate and relies on the Weierstrass gap theorem, see [87]. We just conclude that the tangent vectors $\hat{V}_k$ of $\hat{\mathcal{P}}_{g,n}$ and hence the objects on which our forms $\Omega^{g,n}_{\Phi_i}$ should act on, are given by $n$-tuples of vector fields $\mathbf{v} = (v_1(z_1), v_2(z_2), ..., v_n(z_n))$[19].

Let us see how to incorporate this idea into the operator formalism: We know that a coordinate transformation of the form (3.75) is generated by the energy-momentum tensor as

$$\delta \Phi_i = \epsilon \left( \oint_{C(0)} \frac{dz_i}{2\pi i} v_i(z_i) T(z_i) + \oint_{C(0)} \frac{d\bar{z}_i}{2\pi i} \bar{v}_i(\bar{z}_i) \overline{T}(\bar{z}_i) \right) \Phi_i \tag{3.76}$$

It is convenient to define the operator

$$\mathbf{T}(\mathbf{v}) = \sum_{i=1}^{n} \left( \oint_{C(0)} \frac{dz_i}{2\pi i} v_i(z_i) T(z_i) + \oint_{C(0)} \frac{d\bar{z}_i}{2\pi i} \bar{v}_i(\bar{z}_i) \overline{T}(\bar{z}_i) \right) \tag{3.77}$$

which combines the action on all the external states. We can let it act to the left on the state $\langle \Sigma |$ and get

$$\delta_{\hat{V}} \langle \Sigma | = - \langle \Sigma | \mathbf{T}(\mathbf{v}) \tag{3.78}$$

where $\mathbf{v}$ provides a Schiffer representation of the tangent vector $\hat{V}$. The operators of the form (3.77) obey the algebra

$$[\mathbf{T}(\mathbf{v_1}), \mathbf{T}(\mathbf{v_2})] = \mathbf{T}([\mathbf{v_2}, \mathbf{v_1}]) \tag{3.79}$$

which can be derived from the OPE (2.24). $[\mathbf{v_2}, \mathbf{v_1}]$ just denotes the array of Lie brackets of the (one-complex-dimensional) individual vectors $v_{1i}$, $v_{2i}$ at every puncture.

For later purposes we also define the $b$-ghost version of (3.77) as

$$\mathbf{b}(\mathbf{v}) = \sum_{i=1}^{n} \left( \oint_{C(0)} \frac{dz_i}{2\pi i} v_i(z_i) b(z_i) + \oint_{C(0)} \frac{d\bar{z}_i}{2\pi i} \bar{v}_i(\bar{z}_i) \bar{b}(\bar{z}_i) \right). \tag{3.80}$$

It obeys the relations

$$[\mathbf{T}(\mathbf{v_1}), \mathbf{b}(\mathbf{v_2})] = \mathbf{b}([\mathbf{v_2}, \mathbf{v_1}]) \tag{3.81}$$

and

$$\left\{ \sum_{i=1}^{n} Q_i, \mathbf{b}(\mathbf{v}) \right\} = \mathbf{T}(\mathbf{v}) \tag{3.82}$$

which can be easily verified.

We are now in the position to give the explicit expressions of the basic forms $\Omega^{g,n}_{\Phi_i}$: Their action on a set of $6g - 6 + 2n$ tangent vectors $\hat{V}^{(k)}$ represented by Schiffer fields $\mathbf{v_k}$ is proposed to be

$$\Omega^{g,n}_{\Phi_i}\left(\hat{V}^{(1)}, ..., \hat{V}^{(6g-6+2n)}\right) = N_{g,n} \langle \Sigma | \mathbf{b}(\mathbf{v_1}) ... \mathbf{b}(\mathbf{v_{6g-6+2n}}) | \Phi_1 \rangle ... | \Phi_n \rangle. \tag{3.83}$$

---

[19]In this section we use bold faced letters for $n$-tuples rather than coalgebra objects, in accordance with the notation in [87].

A few comments are in order: First, the normalization constant $N_{g,n}$ is necessary and will be determined later. Second, we have to check if the expression is actually well-defined since it should not depend on the arbitrary choice of phase for every local coordinate. This means, neither a pure rotation of the coordinate curves on $\Sigma$, nor an infinitesimal change of Schiffer vectors

$$v_i(z_i) \to v_i(z_i) + i\epsilon_i z_i \tag{3.84}$$

should affect the result. In fact, from (3.80) we see that this change would result in

$$\mathbf{b}(\mathbf{v}) \to \mathbf{b}(\mathbf{v}) + i\sum_{i=1}^{n} \epsilon_i b_{0(i)}^-. \tag{3.85}$$

Since all string fields $\Phi_i$ in our Hilbert space are expected to be annihilated by $b_0^-$, the additional term gives no contribution. Similarly, a rotation of the coordinate curve is generated by $L_0^-$, which also annihilates the string fields and therefore does not change the result.[20] The moral of the story here is that the level matching conditions (3.14) are not just a physical constraint on the spectrum, but a mathematical necessity for well-defined string interactions. Because of the topological obstruction explained in section 3.5, we need our states to be insensitive to phase rotations.

Let us now turn to equation (3.74): The left hand side can be simplified by commuting through the BRST-operators until they hit $\langle \Sigma |$. From that we get commutator terms

$$\Omega^{g,n}_{\sum Q_i(\Phi_i)}\left(\hat{V}^{(1)}, ..., \hat{V}^{(k)}\right) = N_{g,n}\sum_{l=1}^{k}(-1)^{k-l}\langle\Sigma|\,\mathbf{b}(\mathbf{v_1}),...,\widehat{\mathbf{b}(\mathbf{v_l})}\mathbf{T}(\mathbf{v_l}),...,\mathbf{b}(\mathbf{v_k})\,|\Phi_1\rangle...|\Phi_n\rangle, \tag{3.86}$$

where $\hat{}$ just means that the factor is omitted. The whole sum $Q_1 + ... + Q_n$ annihilates $\langle\Sigma|$ by contour deformation since $\Sigma$ with the coordinate disks removed does not contain any operator insertions or singularities. The above line is therefore the exact result. Let us now also commute through the $\mathbf{T}(\mathbf{v_l})$ factors using the algebra (3.81): We get

$$\Omega^{g,n}_{\sum Q_i(\Phi_i)}\left(\hat{V}^{(1)}, ..., \hat{V}^{(k)}\right)$$
$$= N_{g,n}\sum_{l}(-1)^{k-l}\langle\Sigma|\,\mathbf{T}(\mathbf{v_l})\,\mathbf{b}(\mathbf{v_1}),...,\widehat{\mathbf{b}(\mathbf{v_l})},...,\mathbf{b}(\mathbf{v_k})\,|\Phi_1\rangle...|\Phi_n\rangle$$
$$+ N_{g,n}\sum_{l<j}(-1)^{k+l+j}\langle\Sigma|\,\mathbf{b}([\mathbf{v_l},\mathbf{v_j}])\,\mathbf{b}(\mathbf{v_1}),...,\widehat{\mathbf{b}(\mathbf{l})},...,\widehat{\mathbf{b}(\mathbf{v_j})},...,\mathbf{b}(\mathbf{v_k})\,|\Phi_1\rangle...|\Phi_n\rangle. \tag{3.87}$$

Now let us rewrite the right hand side: From the general definition of the exterior derivative we have

$$d\Omega^{g,n}_{\Phi_i}\left(\hat{V}^{(1)}, ..., \hat{V}^{(k)}\right) = \sum_{l}(-1)^{l+1}\hat{V}^{(l)}\Omega^{g,n}_{\Phi_m}\left(\hat{V}^{(1)}, ..., \widehat{\hat{V}^{(l)}}, ..., \hat{V}^{(k)}\right)$$
$$+ \sum_{l<j}(-1)^{l+j}\Omega^{g,n}_{\Phi_m}\left(\left[\hat{V}^{(l)}, \hat{V}^{(j)}\right], \hat{V}^{(1)}, ..., \widehat{\hat{V}^{(l)}}, ..., \widehat{\hat{V}^{(j)}}, ..., \hat{V}^{(k)}\right). \tag{3.88}$$

---

[20]The situation is actually a bit more subtle since the tangent vectors $\hat{V}^{(k)}$ also change with the base point. The complete derivation can be found in [87].

In the first term, the operator $\hat{V}^{(l)}$ is acting on the surface state $\langle\Sigma|$ as well as on the $b$-ghost insertions, since they are defined on the tangent space $T_p\hat{\mathcal{P}}_{g,n}$. The action on $\langle\Sigma|$ we have calculated in (3.78), it yields

$$N_{g,n}\sum_l (-1)^l \langle\Sigma|\, \mathbf{T}(\mathbf{v_l})\, \mathbf{b}(\mathbf{v_1}),...,\widehat{\mathbf{b}(\mathbf{v_l})},...,\mathbf{b}(\mathbf{v_k})\, |\Phi_1\rangle...|\Phi_n\rangle \tag{3.89}$$

which already equals the first line in (3.87) up to a sign. The action of $\hat{V}^{(l)}$ on the $\mathbf{b}(\mathbf{v_l})$ should therefore combine with the second line in (3.88) to yield the second line in (3.87). We show now that this is indeed the case: First, let us write the action of a Schiffer vector on a local coordinate as

$$z_i' = z_i + \epsilon v_i(z_i) \equiv (1 + \epsilon\delta_{v_i})\, z_i. \tag{3.90}$$

For the Schiffer representations of the tangent vectors we now have

$$\left[\delta_{\mathbf{v}(\hat{V}^{(1)})}, \delta_{\mathbf{v}(\hat{V}^{(2)})}\right] = \delta_{\left[\mathbf{v}(\hat{V}^{(2)}), \mathbf{v}(\hat{V}^{(1)})\right]} = \delta_{\mathbf{v}\left[\hat{V}^{(2)}, \hat{V}^{(1)}\right]}, \tag{3.91}$$

where the first equation follows from simple algebra whereas the second just means that we expect them to furnish a representation of the Lie algebra of tangent vector fields. By expanding, we can derive the following non-trivial relation:

$$\mathbf{v_2} - \mathbf{v_2}\,|_{\Sigma_1} - (\mathbf{v_1} - \mathbf{v_1}\,|_{\Sigma_2}) + \epsilon\mathbf{v}\left[\hat{V}^{(1)}, \hat{V}^{(2)}\right] = \epsilon\,[\mathbf{v_1}, \mathbf{v_2}]. \tag{3.92}$$

Here we abbreviated $\mathbf{v}\left(\hat{V}^{(1/2)}\right)$ with $\mathbf{v_{1/2}}$ and denoted by $\Sigma_{1/2}$ the surface obtained after deformation of $\Sigma$ by $\hat{V}^{(1/2)}$ using (3.78). The difference $\mathbf{v_{2/1}} - \mathbf{v_{2/1}}\,|_{\Sigma_{1/2}}$ is therefore of order $\epsilon$. We could in principle compute $\mathbf{v_{2/1}}\,|_{\Sigma_{1/2}}$ via a Taylor expansion, however, it is not even necessary: The first four terms in the line above exactly correspond to what we get from the action of $\hat{V}^{(i)}$ on the insertions $\mathbf{b}(\mathbf{v_j})$! Combining all remaining terms from (3.88) we find

$$N_{g,n}\sum_{l<j}(-1)^{l+j}\lim_{\epsilon\to 0}\frac{1}{\epsilon}\langle\Sigma|\left(\mathbf{b}(\mathbf{v_j} - \mathbf{v_j}\,|_{\Sigma_l}) - \mathbf{b}(\mathbf{v_l} - \mathbf{v_l}\,|_{\Sigma_j}) + \epsilon\mathbf{b}\left(\mathbf{v}\left(\left[\hat{V}^{(l)}, \hat{V}^{(j)}\right]\right)\right)\right)$$

$$\mathbf{b}(\mathbf{v_1}),...,\widehat{\mathbf{b}(\mathbf{v_l})},...,\widehat{\mathbf{b}(\mathbf{v_j})},...,\mathbf{b}(\mathbf{v_k})\,|\Phi_1\rangle...|\Phi_n\rangle$$

$$=N_{g,n}\sum_{l<j}(-1)^{k+l+j}\langle\Sigma|\,\mathbf{b}([\mathbf{v_l}, \mathbf{v_j}])\,\mathbf{b}(\mathbf{v_1}),...,\widehat{\mathbf{b}(\mathbf{v_l})},...,\widehat{\mathbf{b}(\mathbf{v_j})},...,\mathbf{b}(\mathbf{v_k})\,|\Phi_1\rangle...|\Phi_n\rangle \tag{3.93}$$

which is up to a sign just the second line of (3.87). We have therefore established the desired identity

$$\Omega^{g,n(x)}_{\Sigma Q_i(\Phi_i)} = (-1)^k\, d\Omega^{g,n(x-1)}_{\Phi_i}. \tag{3.94}$$

The difference between the form degree $k$ and the number $x$ is just the real (even) dimension of the moduli space, so for the top form with $x = 0$, the sign factor disappears.

The final definition of the algebraic vertices we got reads

$$\omega\left(\Phi_1, l^g_{n-1}(\Phi_2, ..., \Phi_n)\right) = \int_{\mathcal{V}_{g,n}} \Omega^{g,n}_{\Phi_i}. \tag{3.95}$$

We can extract the higher products $l_n^g$ by using the Poisson bivector defined in (3.56) and using the non-degeneracy of the symplectic form $\omega$. The result is

$$l_n^g(\Phi_1, ..., \Phi_n) = \sum_m (-1)^{|\Phi^m|} b_0^- \Phi_m \int_{\mathcal{V}_{g,n+1}} \Omega^{g,n+1}_{\Phi^m \Phi_1, ..., \Phi_n}. \tag{3.96}$$

The last point to show is that with this definition, the main identity (3.56) is indeed automatically implied. The full rigorous derivation is quite elaborate [87] so we will just sketch the proof. We start with the terms in (3.56) containing $l_1^0 = Q$, i. e. the failure of the BRST-operator to be a derivation of the higher products. It yields

$$Q l_n^g(\Phi_1, ..., \Phi_n) + \sum_i (-1)^{|\Phi_1| + \cdots + |\Phi_{i-1}|} l_n^g(\Phi_1, ..., Q\Phi_i, ..., \Phi_n)$$
$$= \sum_m (-1)^{|\Phi^m|} \left( Q b_0^- \Phi_m \int_{\mathcal{V}_{g,n+1}} \Omega^{g,n+1}_{\Phi^m \Phi_1, ..., \Phi_n} + b_0^- \Phi_m \int_{\mathcal{V}_{g,n+1}} \Omega^{g,n+1}_{\Phi^m (\sum Q_i)(\Phi_1, ..., \Phi_n)} \right). \tag{3.97}$$

The Poisson bivector fulfills the useful identity

$$\sum_m (-1)^{|\Phi^m|} Q\Phi^m \otimes b_0^- \Phi_m + \sum_m (-1)^{|\Phi^m|} \Phi^m \otimes Q b_0^- \Phi_m = 0 \tag{3.98}$$

which is essentially a consequence of $Q$ being BPZ-odd and the level-matching condition (3.14). Using that we can write the expression above simply as

$$\sum_m (-1)^{|\Phi^m|} b_0^- \Phi_m \int_{\mathcal{V}_{g,n+1}} \Omega^{g,n+1}_{(\sum Q_i)(\Phi^m \Phi_1, ..., \Phi_n)}. \tag{3.99}$$

Now comes the key identity (3.74) into play: After applying it together with Stokes theorem the above expression becomes

$$\sum_m (-1)^{|\Phi^m|} b_0^- \Phi_m \int_{\mathcal{V}_{g,n+1}} d\Omega^{g,n+1(-1)}_{\Phi^m \Phi_1, ..., \Phi_n} = \sum_m (-1)^{|\Phi^m|} b_0^- \Phi_m \int_{\partial \mathcal{V}_{g,n+1}} \Omega^{g,n+1(-1)}_{\Phi^m \Phi_1, ..., \Phi_n}. \tag{3.100}$$

The integral runs now over the boundary of $\mathcal{V}_{g,n+1}$ which is constrained by the geometrical BV-equation: It contains Feynman diagrams with precisely one propagator which is collapsed to $t = 0$. This propagator can either connect two elementary vertices or form a loop on one vertex. Let us first consider the tree-diagram case: We will assume that the integral factorises in the following way:

$$\int_{R_{g,n}^{tree}} \Omega^{g,n+1(-1)}_{\Phi^m \Phi_1, ..., \Phi_n} = \sum_p (-1)^{|\Phi^p|} \sum_{g_1+g_2=g} \sum_{n_1+n_2=n} \sum_{\{r,s\}} \sigma(r,s)$$
$$\int_{\mathcal{V}_{g_1,n_1+2}} \Omega^{g_1,n_1+2}_{\Phi^m \Phi_{r_1}, ..., \Phi_{r_{n_1}} b_0^- \Phi_p} \int_{\mathcal{V}_{g_2,n_2+1}} \Omega^{g_2,n_2+1}_{\Phi^p \Phi_{s_1}, ..., \Phi_{s_{n_2}}}. \tag{3.101}$$

Again, similar to (3.56), the sum over $\{r, s\}$ runs over all different splittings of the $n$ string fields into two groups and $\sigma(r, s)$ is the Grassmann sign which is picked up. The meaning of this equation is actually quite intuitive: It says that the collapsed propagator can be represented as an insertion of a complete set of basis states. For a rigorous proof one needs to make sure that the arbitrary phases at the punctures where $\Phi^p$ and $\Phi_p$ are inserted do not affect the result. This can be shown using the level-matching condition (3.14). Moreover, the equation above constrains the normalization constant $N_{g,n}$: From the integration over the sewing angle $s$ we get a factor of $2\pi i$, leading to the equation

$$2\pi i N_{g,n+1} = N_{g_1,n_1+2} N_{g_2,n_2+1}. \tag{3.102}$$

In [48] it was shown that the solution which leads to the correct scattering amplitudes is given by

$$N_{g,n} = (2\pi i)^{-(3g-3+n)}. \tag{3.103}$$

We now use linearity of the fundamental forms $\Omega^{g,n}_{\Phi_i}$ and absorb the last integral of (3.101), which is just a complex number, in the state $b_0^- \Phi_p$, which can be written as

$$\sum_p (-1)^{|\Phi^p|} b_0^- \Phi_p \int_{\mathcal{V}_{g_2,n_2+1}} \Omega^{g_2,n_2+1}_{\Phi^p \Phi_{s_1},...,\Phi_{s_{n_2}}} = l^{g_2}_{n_2}\left(\Phi_{s_1},...,\Phi_{s_{n_2}}\right). \tag{3.104}$$

Similarly, the first integral can be converted into a higher product such that the tree level contribution of (3.100) becomes

$$-\sum_{g_1+g_2=g}\sum_{n_1+n_2=n}\sum_{\{r,s\}} \sigma(r,s)\, l^{g_2}_{n_1+1}\left(\Phi_{r_1},...,\Phi_{r_{n_1}}, l^{g_2}_{n_2}\left(\Phi_{s_1},...,\Phi_{s_{n_2}}\right)\right). \tag{3.105}$$

For the loop contribution we assume the analogous factorization equation

$$\int_{R^{loop}_{g,n}} \Omega^{g,n+1(-1)}_{\Phi^m \Phi_1,...,\Phi_n} = \frac{1}{2}\sum_p (-1)^{|\Phi^p|} \int_{\mathcal{V}_{g-1,n+3}} \Omega^{g-1,n+3}_{\Phi^m b_0^- \Phi_p \Phi^p \Phi_1,...,\Phi_n}. \tag{3.106}$$

The proof is basically the same as for (3.101) and can be found in [87]. Again, we convert the integral into a higher product and write the loop contribution as

$$-\frac{1}{2}\sum_p (-1)^{|\Phi^p|} l^{g-1}_{n+2}\left(b_0^- \Phi_p, \Phi^p, \Phi_1, ..., \Phi_n\right). \tag{3.107}$$

After putting all terms together, we have shown that

$$Ql^g_n(\Phi_1,...,\Phi_n) + \sum_i (-1)^{|\Phi_1|+\cdots+|\Phi_{i-1}|} l^g_{n_1}(\Phi_1,...,Q\Phi_i,...,\Phi_n)$$
$$+ \sum_{\substack{g_1+g_2=g \\ (g_{1/2},n_{1/2})\neq(0,1)}} \sum_{n_1+n_2=n} \sum_{\{r,s\}} \sigma(r,s)\, l^{g_1}_{n_1+1}\left(\Phi_{r_1},...,\Phi_{r_{n_1}}, l^{g_2}_{n_2}\left(\Phi_{s_1},...,\Phi_{s_{n_2}}\right)\right)$$
$$+ \frac{1}{2}\sum_p (-1)^{|\Phi^p|} l^{g-1}_{n+2}\left(b_0^- \Phi_p, \Phi^p, \Phi_1,...,\Phi_n\right) = 0 \tag{3.108}$$

which is exactly the main identity (3.56).

There is one last point left to be addressed: The choice of vertices which solve (3.68) is by far not unique. We can change the local coordinates or the slicing of the moduli space and still obtain a consistent theory. Does this mean that there exist infinitely many inequivalent closed string field theories? Actually this is not the case: In [34] it has been shown that a change of vertices is equivalent to a non-linear field redefinition of the string field,

$$\Phi \to F\left[\Phi\right].$$

The freedom of field redefinitions of this form is actually present in any field theory. In typical QFTs it is just barely used, because usually there exists a certain canonical field for which the action takes the simplest form. For CSFT, we do not know this canonical form yet (if it exists), so in constructing the theory we find a whole family of actions related by field redefinitions. The explicit results of [34] will become important as an inspiration for deforming OSFT using stubs, see section 5.3.3.

# 4 Other string field theories

In this section we want to give a brief overview of other string field theories which have been constructed so far, without providing a detailed discussion.

The first one to mention is bosonic *open/closed SFT* which is just the combination of Witten theory and Zwiebach theory which were described above. In [88], the relevant geometrical master equation is developed whereas [52] discusses the algebraic structure within the coalgebra formalism. The main challenge lies in systematically constructing the new vertices which involve both open and closed insertions and make the theory rather involved.

An additional important family of SFTs are the supersymmetric ones, i. e. those whose background is a supersymmetric CFT. Here, the most problematic new feature is the existence of a new quantum number, the *picture number.* Similar to the ghost number, it has to take a certain value within a correlation function to lead to a non-vanishing result. This implies that so-called *picture changing operators* (PCOs) must be inserted in the products which give rise to the interactions. Typically, they mess up the algebraic properties needed for gauge invariance and therefore complicate the theory.

In the open case, the first consistent theory was found by Berkovits [3] in form of a non-polynomial action of WZW-type. A few years later, in a more bottom-up approach, the *Munich construction* [23] was found, based on an $A_\infty$-algebra. In [17, 18] it was shown that the two theories are equivalent and related by a field redefinition. A more exotic version called *democratic superstring field theory* [47] was found by Kroyter: In this ansatz, the string field contains all possible picture numbers, circumventing the problem with PCO insertions. The action then takes a simple cubic form, although the price to pay is an extended Hilbert space.

Closed superstring field theories have been obtained in the RNS-formalism for type II and heterotic theories, see [8, 74]. As can be anticipated, their explicit construction is complicated, mostly be-cause of the PCO-insertions. Nevertheless, they recently got more attention due to the connection to string phenomenology.

Last but not least, there is the SFT which was established first in history by Kaku and Kikkawa called bosonic *light-cone SFT [41]*. It is fundamentally different from other SFTs because it is based on the light cone gauge and has therefore no BRST-operator and no gauge invariance. The action is surprisingly simple, it is quartic in the open case and cubic in the closed case. A serious drawback is that it cannot describe the zero-momentum sector and can also not be consistently quantized because momentum-integration is ill-defined. Still there has been some recent work on it, for example it has been shown how open lightcone SFT and Witten theory can be matched into each other [26], see also [4].

# Part III

# Selected papers

## 5 Open string field theory with stubs

**Abstract**

There are various reasons why adding stubs to the vertices of open string field theory (OSFT) is interesting: Not only the stubs can tame certain singularities and make the theory more well-behaved, but also the new theory shares a lot of similarities with closed string field theory, which helps to improve our understanding of its structure and possible solutions. In this paper we explore two natural ways of implementing stubs into the framework of OSFT, resulting in an $A_\infty$-algebra giving rise to infinitely many vertices. We find two distinct consistent actions, both generated by a field redefinition, interestingly sharing the same equations of motion. In the last section we illustrate their relationship and physical meaning by applying our construction to nearly marginal solutions.

### 5.1 Introduction and motivation

It is a well-known fact that the algebraic and geometric structures of open and closed string field theory are fundamentally different. The OSFT action (Witten action) consists of a standard kinetic term and a single cubic interaction and reads [85]

$$S(\Psi) = \frac{1}{2}\omega(\Psi, Q\Psi) + \frac{1}{3}\omega(\Psi, \Psi * \Psi). \tag{5.1}$$

The algebraic ingredients are a nilpotent differential given by the BRST-operator, the star product and a symplectic form $\omega$, which all together form a cyclic differential graded algebra. In the following, we will use the coalgebra notation of [84] with $m_2(\Psi_1, \Psi_2) := (-)^{d(\Psi_1)} \Psi_1 * \Psi_2$ and the shifted degree given by $d(\Psi) = gh(\Psi) + 1$. Now the defining algebraic properties are

$$Q^2 = 0, \tag{5.2}$$

$$Qm_2(\Psi_1, \Psi_2) = -m_2(Q\Psi_1, \Psi_2) - (-)^{d(\Psi_1)} m_2(\Psi_1, Q\Psi_2), \tag{5.3}$$

$$m_2(\Psi_1, m_2(\Psi_2, \Psi_3)) = -(-)^{d(\Psi_1)} m_2(m_2(\Psi_1, \Psi_2), \Psi_3), \tag{5.4}$$

$$\omega(\Psi_1, m_2(\Psi_2, \Psi_3)) = -(-)^{d(\Psi_1)} \omega(m_2(\Psi_1, \Psi_2), \Psi_3), \tag{5.5}$$

$$\omega(\Psi_1, Q\Psi_2) = -(-)^{d(\Psi_1)} \omega(Q\Psi_1, \Psi_2). \tag{5.6}$$

In contrast, the classical closed string field theory (CSFT) action [87] contains infinitely many vertices and reads

$$S(\Psi) = \frac{1}{2}\omega'(\Psi, Q\Psi) + \sum_{n=3}^{\infty} \frac{\kappa^{n-2}}{n!} \{\Psi^n\}, \tag{5.7}$$

where the vertices
$$\{\Psi^n\} = \int_{\mathcal{V}_{0,n}} \Omega^{(0)0,n}_{\Psi^n} \tag{5.8}$$
are given by integrating certain differential forms over vertex regions $\mathcal{V}_{0,n}$ in the moduli space of $n$-punctured Riemann surfaces. At quantum level there are also vertices of higher genus to consider, i. e. intrinsic loops already present within the elementary vertices. However, they will not be of interest here for now.

The vertices also can be decomposed into a symplectic form $\omega'$ [21] and higher products $l_n$:
$$\{\Psi^{n+1}\} = \omega'\left(\Psi, l_n\left(\Psi, \Psi, ..., \Psi\right)\right). \tag{5.9}$$
In contrast to the open string star product, the $l_n$ are all totally symmetric. The objects $Q =: l_1$, all higher $l_n$ and $\omega$ give rise to a cyclic $L_\infty$-algebra:
$$l_n l_1\left(\Psi_1, ..., \Psi_n\right) + l_{n-1} l_2\left(\Psi_1, ..., \Psi_n\right) + ... + l_1 l_n\left(\Psi_1, ..., \Psi_n\right) = 0, \quad \forall n \tag{5.10}$$
$$\omega'\left(\Psi_1, l_n\left(\Psi_2, ..., \Psi_{n+1}\right)\right) = -(-)^{d(\Psi_1)} \omega'\left(l_n\left(\Psi_1, ..., \Psi_n\right), \Psi_{n+1}\right). \quad \forall n \tag{5.11}$$

The multiplication of the multilinear products in the first line is defined as
$$l_k l_l\left(\Psi_1, ..., \Psi_{k+l-1}\right) = \sum_\sigma \frac{(-1)^{\epsilon(\sigma)}}{l!\,(k-1)!} l_k\left(l_l\left(\Psi_{\sigma(1)}, ..., \Psi_{\sigma(l)}\right), \Psi_{\sigma(l+1)}, ..., \Psi_{\sigma(k+l-1)}\right), \tag{5.12}$$
where the sum runs over all permutations of $k + l - 1$ elements and the factor $(-1)^{\epsilon(\sigma)}$ is just the sign picked up when permuting the $\Psi$s.

The different algebraic structures have a geometric origin: For a unitary theory it is necessary that the Feynman string diagrams computed from the action cover the full moduli space of $n$-punctured Riemann surfaces exactly once. For the open string where one needs to consider surfaces with boundaries it has been shown by Zwiebach [86] that this is indeed the case for the action (5.1) with only one cubic interaction. In contrast, for the closed string, i. e. surfaces without boundaries, such a simple cubic representation of the action seems not to be possible, see [77].

The simplicity of the OSFT action allowed for the discovery of analytic classical solutions, most importantly the open string tachyon vacuum [70]. In classical CSFT, no such analytic solutions have been found yet [19, 21]. To make some effort into this direction, we want to propose the following strategy: If CSFT cannot be simplified easily, maybe we can make OSFT "more complicated" in the sense that its algebraic and geometric structure resembles that of CSFT? There are two motivations for that: First, it could help to make the very abstract language of CSFT more intuitive and tractable. Second, if we can translate the known analytic solutions to the new, deformed OSFT, it could give an idea how CSFT solutions might look like. In total there are three steps to do:

---

[21] The symplectic form $\omega'$ differs from the $\omega$ contained in the Witten action by an additional insertion of $c_0^-$.

1. Find a consistent deformation of the Witten OSFT such that its mathematical structure resembles CSFT

2. Find analytic solutions of this deformed theory

3. Make an educated guess for analytic solutions of CSFT

The first two tasks will be worked out in this paper while the third one is left for the future.

The deformation we will consider has many additional interesting aspects: First, OSFT exhibits some issues with singularities or ill-defined quantities: For example, there exist "identity-based" solutions [80] (for a recent discussion see [20]) which solve the equations of motion but do not have a well defined action. There are hints that these solutions could be better behaved in the deformed theory. Moreover, the deformed products $M_2, M_3, \ldots$ defined in the next section could allow for a natural definition of a Banach type $A_\infty$-algebra over the space of string fields.

The paper is organized as follows: In the second section the deformation using stubs [88] is discussed in detail, resulting in an explicit definition of the higher products of the $A_\infty$-algebra. The necessary mathematical ingredients, namely homological perturbation theory and homotopy transfer, are introduced as well. As a side result, we give a consistent method of applying homotopy transfer to general homotopy equivalences which are not deformation retracts. Section three deals with analytical solutions and the action of the stubbed theory. Surprisingly, we find two consistent actions with the same equations of motion, both generated by a field redefinition. The two field redefinitions, one coming from homological perturbation theory and one motivated by closed string field theory, are derived explicitly. The last section is devoted to the physical interpretation of our results. We apply the whole construction to a suitable class of solutions and compute the action up to first non-trivial order in the field expansion.

## 5.2 Deforming the Witten action using stubs

We have seen that the algebraic structure of CSFT is that of an $L_\infty$-algebra, so naively we should try to find an $L_\infty$-based deformation of OSFT. However, this would imply a commutative product of open strings, which would hence be fundamentally different from the Witten product. We still want the deformation be equivalent to Witten theory, so the best we can do is to aim for an $A_\infty$-algebra with products $M_n$ obeying

$$M_n M_1 (\Psi_1, \ldots, \Psi_n) + M_{n-1} M_2 (\Psi_1, \ldots, \Psi_n) + \ldots + M_1 M_n (\Psi_1, \ldots, \Psi_n) = 0 \quad \forall n \quad (5.13)$$

and the multiplication given by

$$\begin{aligned} M_k M_l (\Psi_1, \ldots, \Psi_{k+l-1}) &= M_k (M_l (\Psi_1, \ldots, \Psi_l), \Psi_{l+1}, \ldots, \Psi_{k+l-1}) + \\ &+ (-)^{d(\Psi_1)} M_k (\Psi_1, M_l (\Psi_2, \ldots, \Psi_{l+1}), \Psi_{l+2}, \ldots, \Psi_{k+l-1}) + \cdots + \\ &+ (-)^{d(\Psi_1)+\ldots+d(\Psi_{k-1})} M_k (\Psi_1, \ldots, \Psi_{k-1}, M_l (\Psi_k, \ldots, \Psi_{k+l-1})). \end{aligned} \quad (5.14)$$

From now on we will use the tensor coalgebra notation of [84] where the $A_\infty$-relations take the simple form

$$\sum_{k=1}^{n} \mathbf{M_k M_{n+1-k}} = 0. \tag{5.15}$$

Geometrically, to mimic the situation of CSFT, we would like to have a three-vertex that does *not* give rise to a full moduli space cover, such that higher vertices are necessary. One modification that achieves both of these requirements is to attach *stubs* on the three vertex, which means, small pieces of propagating strings on each input. In this way, there will appear some regions of moduli space that are not covered by Feynman diagrams and hence have to be taken over by elementary vertices. The higher products which will give rise to those vertices will indeed form an $A_\infty$-algebra.

The Hamiltonian of our CFT which generates time evolution is $L_0$, so the operator which inserts a strip of length $\lambda$ into a string diagram is $e^{-\lambda L_0}$. The natural modification of the star product which attaches stubs symmetrically on all three inputs is then given by [81]:

$$m_2(\cdot,\cdot) \to M_2(\cdot,\cdot) = e^{-\lambda L_0} m_2\left(e^{-\lambda L_0}\cdot, e^{-\lambda L_0}\cdot\right). \tag{5.16}$$

$M_2$ is cyclic with respect to the simplectic form $\omega$ and $Q$-invariant:

$$\omega\left(\Psi_1, M_2(\Psi_2, \Psi_3)\right) = -(-)^{d(\Psi_1)} \omega\left(M_2(\Psi_1, \Psi_2), \Psi_3\right), \tag{5.17}$$

$$QM_2(\Psi_1, \Psi_2) = -M_2(Q\Psi_1, \Psi_2) - (-)^{d(\Psi_1)} M_2(\Psi_1, Q\Psi_2). \tag{5.18}$$

Those relations follow from $[Q, L_0] = 0$ and the fact that $L_0$ is BPZ-even. It is easy to see that $M_2$ is not associative as expected for an $A_\infty$-algebra:

$$e^{-\lambda L_0} m_2\left(e^{-\lambda L_0}\Psi_1, e^{-2\lambda L_0} m_2\left(e^{-\lambda L_0}\Psi_2, e^{-\lambda L_0}\Psi_3\right)\right)$$
$$+ e^{-\lambda L_0} m_2\left(e^{-2\lambda L_0} m_2\left(e^{-\lambda L_0}\Psi_1, e^{-\lambda L_0}\Psi_2\right), e^{-\lambda L_0}\Psi_3\right) \neq 0. \tag{5.19}$$

The task is now to explicitly determine all the higher products and prove that they satisfy all relevant properties. This will be done in two ways, first in a formal algebraic way using homological perturbation theory and second in a more heuristic way using tree diagrams.

### 5.2.1 Elements of homological perturbation theory

The starting point of homological perturbation theory (HPT) is a *chain homotopy equivalence*. Let $V, W$ be two chain complexes, i. e. graded vector spaces with a nilpotent differential of degree one denoted by $d_V$ ($d_W$). Furthermore, we are given two chain maps $p : V \to W$ and $i : W \to V$ of degree zero as well as a homotopy map $h : V \to V$ of degree minus one obeying the relations

$$pd_V = d_W p \tag{5.20}$$
$$id_W = d_V i \tag{5.21}$$
$$ip - 1 = hd_V + d_V h. \tag{5.22}$$

This ensures that $i$ and $p$ as well as the combination $ip$ leave cohomology classes invariant.

Let us assume that the differential $d_V$ is perturbed by some $\delta$ of degree one such that $(d_V + \delta)^2 = 0$ again. The *homological perturbation lemma* (HPL) now states that one can construct a new homotopy equivalence using the perturbed differential with the other maps given by

$$D_V = d_V + \delta \tag{5.23}$$
$$D_W = d_W + p\delta\,(1 - h\delta)^{-1}\,i \tag{5.24}$$
$$I = (1 - h\delta)^{-1}\,i \tag{5.25}$$
$$P = p\,(1 - \delta h)^{-1} \tag{5.26}$$
$$H = (1 - h\delta)^{-1}\,h. \tag{5.27}$$

The expression $(1 - h\delta)^{-1}$ requires some explanation: Usually it should be understood as a geometric series

$$(1 - h\delta)^{-1} = 1 + h\delta + h\delta h\delta + h\delta h\delta h\delta + ... \quad . \tag{5.28}$$

We demand for the validity of the HPL that this series either converges or terminates, which will be the case in our examples. The proof of the lemma including all relevant relations is straightforward but tedious and can be found in [7, 84].

An important case in literature is that of a *deformation retract* (DR) where $W$ is isomorphic to a subspace of $V$, see e.g. [15]. The maps $i$ and $p$ are then given by the canonical inclusion and projection, respectively, hence it follows trivially that $pi = 1$. If additionally the following *annihilation conditions*

$$hh = 0, \quad hi = 0, \quad ph = 0 \tag{5.29}$$

are satisfied, it is called a *special deformation retract* (SDR). It can be shown that any DR can be turned into an SDR by a redefinition of the maps [7]. If the perturbation lemma is applied to an DR, the perturbed data will in general not form a DR again. However, in the case of an SDR, all relations are preserved after the perturbation, so $PI = 1$ and $I$, $P$ and $H$ also fulfill the annihilation conditions.

### 5.2.2 Transferring algebraic structure: SDR-case

Our main purpose for using the HPL will be transferring some algebraic structure from one side of the homotopy equivalence to the other. Lets consider an SDR with an associative multiplication defined on $V$. It is a natural question to ask if there exists some "induced product" defined on $W$: For example, for $a, b \in W$ the product $ia \cdot ib$ in $V$ need not be in $W$ anymore. The only thing that could be done is to project it to $W$, i. e. define $a \cdot b\,|_W = p\,(ia \cdot ib)$, but this product will in general not be associative anymore. What happens is that an associative algebra on one side of the equivalence becomes an $A_\infty$-algebra after transferring to the other side. A formal way to construct all the higher products is given by the so-called *tensor trick:*

One defines a new SDR over the tensor coalgebra[22] of $V$ ($W$) with the tensorial versions of the maps given by

$$\mathbf{d_{V(W)}} = \sum_{n=1}^{\infty}\sum_{k=0}^{n-1} 1^{\otimes k} \otimes d_{V(W)} \otimes 1^{\otimes n-k-1} \tag{5.30}$$

$$\mathbf{i} = \sum_{n=1}^{\infty} i^{\otimes n} \tag{5.31}$$

$$\mathbf{p} = \sum_{n=1}^{\infty} p^{\otimes n} \tag{5.32}$$

$$\mathbf{h} = \sum_{n=1}^{\infty}\sum_{k=0}^{n-1} 1^{\otimes k} \otimes h \otimes (ip)^{\otimes n-k-1}. \tag{5.33}$$

$\mathbf{d_{V(W)}}$ is the standard tensor coderivation associated to $d_{V(W)}$, see (5.139), $\mathbf{i}$ and $\mathbf{p}$ are cohomomorphisms while $\mathbf{h}$ is defined somewhat asymmetrically which will turn out to be crucial for the formalism to work. One can check directly that those definitions indeed give rise to an SDR again. The product on $V$, denoted by $m_2$, can now be treated as a perturbation $\delta$ of the coderivative $\mathbf{d_V}$,

$$\mathbf{D_V} = \mathbf{d_V} + \mathbf{m_2}, \tag{5.34}$$

where $\mathbf{m_2}$ is the coderivation associated to $m_2$. $\mathbf{D_V}^2 = 0$ will be fulfilled as a consequence of $m_2$ being associative and $d_V$ obeying the Leibniz rule. According to the HPL we will be given a new, perturbed complex with the maps

$$\mathbf{D_W} = \mathbf{d_W} + \mathbf{p m_2} \left(1 - \mathbf{h m_2}\right)^{-1} \mathbf{i} \tag{5.35}$$

$$\boldsymbol{I} = \left(1 - \mathbf{h m_2}\right)^{-1} \mathbf{i} \tag{5.36}$$

$$\boldsymbol{P} = \mathbf{p} \left(1 - \mathbf{m_2 h}\right)^{-1} \tag{5.37}$$

$$\boldsymbol{H} = \mathbf{h} \left(1 - \mathbf{h m_2}\right)^{-1}, \tag{5.38}$$

where $\mathbf{D_W}$ squares to zero and therefore defines an $A_\infty$-algebra on $W$. The higher products can be obtained by expanding $\mathbf{D_W}$, i. e. projecting onto $n$ inputs and one output:

$$M_n = \pi_1 \mathbf{D_W} \pi_n. \tag{5.39}$$

For example we get

$$M_2\left(\cdot, \cdot\right) = p m_2 \left(i\cdot, i\cdot\right) \tag{5.40}$$

as already anticipated above.

---

[22]See the appendix for a short summary on tensor coalgebras.

### 5.2.3 Transferring algebraic structure: Non-SDR-case

We want to apply those concepts now to the problem of attaching stubs in OSFT. The space $V$ should be given by the space of string fields $\mathcal{H}_{BCFT}$ with its grading $d(\Psi) = gh(\Psi) + 1$ and the differential $Q$. By comparing (5.16) with (5.40) it seems natural to define

$$i = p = e^{-\lambda L_0}. \tag{5.41}$$

However, we see immediately that this definition does not give rise to a DR: $i$ and $p$ are not an inclusion and projection anymore and $pi \neq 1$. Still, in principle the HPL holds for arbitrary homotopy equivalences, so there is a chance to succeed anyway.

Our choice is so far completely symmetric, so lets define $W = V = \mathcal{H}_{BCFT}$, $d_W = d_V = Q$ and $h_W = h_V = h$ where we have to solve

$$e^{-2\lambda L_0} - 1 = hQ + Qh \tag{5.42}$$

The simplest (although not unique) solution for $h$ is

$$h = \frac{e^{-2\lambda L_0} - 1}{L_0} b_0. \tag{5.43}$$

It is important to stress that $h$ is well-behaved also for $L_0 = 0$ and does not have any pole.

One could now try to proceed with the tensor trick as above and eventually compute the map $\mathbf{D_W}$, but this runs into problems: Although $\mathbf{D_W}^2 = 0$ still, as guaranteed by the HPL, $\mathbf{D_W}$ is not a coderivation anymore! This means, it does not obey the co-Leibniz rule (5.137). As explained in the Appendix of [44], the condition $pi = 1$ as well as the annihilation relations are necessary (and sufficient) for the tensor trick to work. So there has to be some modification to account for that and it will actually turn out to be surprisingly simple: We can just take the expression we get for $\mathbf{D_W}$ and while expanding to calculate the higher products, *pretend* that all SDR-relations (5.29) are satisfied. More precisely, we define an operator $P_{SDR}$ acting on the space of maps from $T\mathcal{H} \to T\mathcal{H}$ which projects on maps in which all SDR-relations hold. This means, every time that $pi$ occurs, it will be replaced by 1 and every time $hh$, $hi$ or $ph$ occurs, the term will be discarded:

$$\begin{aligned} P_{SDR}(....hh....) &= 0, & P_{SDR}(....hi....) &= 0, \\ P_{SDR}(....ph....) &= 0, & P_{SDR}(....pi....) &= P_{SDR}(....1....) \end{aligned} \tag{5.44}$$

Now our new higher products will be given as

$$M_n = P_{SDR} \pi_1 \mathbf{D_W} \pi_n. \tag{5.45}$$

Their associated coderivations can be added together to form a total map called $\mathbf{M}$,

$$\mathbf{M} = \sum_{n=1}^{\infty} \sum_{m=n}^{\infty} \sum_{k=0}^{m-n} 1^{\otimes k} \otimes M_n \otimes 1^{\otimes m-k-n}. \tag{5.46}$$

It is a coderivation by construction and moreover it squares to zero because in the proof of the HPL, where it is shown that $D_W^2 = 0$ (Eq. 1.2.16 of [84]), the SDR-relations are never used. Since we know that $\mathbf{D_W}^2 = 0$ independently of the SDR-relations, also $\mathbf{M}^2 = 0$ and $\mathbf{M}$ defines the desired $A_\infty$-algebra.

As an example, lets explicitly calculate $M_3$:

$$\begin{aligned}
P_{SDR}\pi_1 \mathbf{D_W}\pi_3 &= P_{SDR}\pi_1 \mathbf{pm_2 hm_2 i}\pi_3 = P_{SDR}\left(pm_2\left(1\otimes h + h\otimes ip\right)\left(1\otimes m_2 + m_2\otimes 1\right)i^{\otimes 3}\right)\\
&= P_{SDR}\Big(pm_2\left(i\cdot,hm_2\left(i\cdot,i\cdot\right)\right) - pm_2\left(m_2\left(i\cdot,i\cdot\right),hi\cdot\right)\\
&\quad + pm_2\left(hi\cdot,ipm_2\left(i\cdot,i\cdot\right)\right) + pm_2\left(hm_2\left(i\cdot,i\cdot\right),ipi\cdot\right)\Big)\\
&= pm_2\left(i\cdot,hm_2\left(i\cdot,i\cdot\right)\right) + pm_2\left(hm_2\left(i\cdot,i\cdot\right),i\cdot\right). \tag{5.47}
\end{aligned}$$

The second and third term in the second line contained an $hi$ and were deleted whereas in the last term $ipi$ was replaced by $i$.

This whole procedure including the proof of the statement may seem quite handwavy, however, there exists a more formal and precise way of arriving at the same result using operad theory [83]. A combinatorial proof using tree diagrams will be given in the next section.

### 5.2.4 Higher products using tree diagrams

The proposal is that $M_n$ is equal to the sum of all distinct, rooted, full binary trees with $n$ leaves such that every leaf represents one input and the root is the output. With every leaf there is one factor of $i$ associated, with every node the product $m_2$, with every internal line $h$ and with the root $p$. In [49] it is argued that this is true for SDRs, since we construct the products in the same way as for an SDR, we conclude that the proposal also holds for our non-SDR case.

In the tree language, the $A_\infty$-relations can be proven directly: The commutator of $\mathbf{Q}$ with an $n$-leaved tree gives a sum of $n-2$ terms, in which one of the $n-2$ internal lines $h$ is replaced by $1-ip$. The 1-terms actually cancel away because of the associativity of $m_2$: The propagator associated with unity connects two nodes $m_2$ which leads to an expression of the form $m_2\left(m_2\left(A,B\right),C\right)$ where $A,B,C$ are three subtrees. In the sum there always exists a second tree with another propagator turned into unity giving rise to the expression $m_2\left(A,m_2\left(B,C\right)\right)$. These two trees cancel away such that only the -$ip$-factors remain in total. Now the other terms occuring in the relation

$$-[\mathbf{Q},\mathbf{M_n}] = \mathbf{M_2 M_{n-1}} + \mathbf{M_3 M_{n-2}} + ... + \mathbf{M_{n-1} M_2} \tag{5.48}$$

can be interpreted as follows: If we project on one output, $\pi_1 \mathbf{M_k M_{n+1-k}}$ gives a sum of trees where one of the terms in $M_{n+1-k}$ is connected with its root to one of the $k$ leaves of one of the trees in $M_k$ and this is done in all possible combinations. The result is a sum of trees with in total $n$ leaves, where one of the internal lines does not contain $h$ but $ip$; the $i$ from the leaf of the left tree and the $p$ from the root of the right tree. But that is exactly the same sum of terms we have on the l.h.s., indeed it is easy to see that each tree occuring on the r.h.s. must also occur on the l.h.s. and vice versa.

An interesting crosscheck of the $\pi_1$-projection of (5.48) can be done by comparing the total number of trees on both sides: The number of full binary trees with $n+1$ leaves is given by the Catalan number

$$C_n = \frac{1}{n+1}\binom{2n}{n}. \tag{5.49}$$

This means that on the l.h.s. there are $C_{n-1}$ trees in $M_n$ and $n-2$ internal lines that can be changed, hence a total of $C_{n-1}(n-2)$ trees. On the r.h.s. we have $2 \cdot C_1 C_{n-2} + 3 \cdot C_2 C_{n-3} + ... + (n-1) C_{n-2} C_1$ trees in total, leading to the equation

$$\sum_{k=2}^{n-1} (n+1-k) C_{n-k} C_{k-1} = (n-2) C_{n-1}. \tag{5.50}$$

The Catalan numbers fulfill the following useful recursive relations:

$$C_{n+1} = \frac{2(2n+1)}{n+2} C_n, \qquad C_{n+1} = \sum_{k=0}^{n} C_{n-k} C_k. \tag{5.51}$$

Using them one can proceed by induction: Assuming equation (5.50) is valid for some $n$, then

$$\sum_{k=2}^{n} (n+2-k) C_{n+1-k} C_{k-1}$$
$$= \sum_{k=2}^{n} (n+2-k) \frac{2(2n-2k+1)}{n-k+2} C_{n-k} C_{k-1}$$
$$= 4 \sum_{k=2}^{n-1} (n+1-k) C_{n-k} C_{k-1} - 2 \sum_{k=2}^{n-1} C_{n-k} C_{k-1} + 2 C_0 C_{n-1}$$
$$= (4n-8) C_{n-1} - 2 \sum_{k=1}^{n-2} C_{n-1-k} C_k + 2 C_{n-1}$$
$$= (4n-6) C_{n-1} - 2 (C_n - C_0 C_{n-1} - C_{n-1} C_0)$$
$$= (4n-2) \frac{n+1}{2(2n-1)} C_n - 2 C_n$$
$$= (n-1) C_n \tag{5.52}$$

as it should be to complete the induction. This shows that the number of terms in the equation (5.48) is the same on both sides.

### 5.2.5 Proof of cyclicity to all orders

Using the tree language it is possible to prove that all higher products $M_n$ are cyclic with respect to the BPZ-product. We have to show

$$\omega(\Psi_1, M_n(\Psi_2, ..., \Psi_{n+1})) = -(-)^{d(\Psi_1)} \omega(M_n(\Psi_1, ..., \Psi_n), \Psi_{n+1}), \tag{5.53}$$

hence we start with a sum of trees on the l.h.s. and use the BPZ-properties of $m_2$ and $h$ as well as $p = i^\dagger = i$ to rewrite it as the sum of trees on the r.h.s.. The explicit steps are:

1. Take the $p$ from the root of the tree and write it to the left side of the product where it can be interpreted as $i$, acting on $\Psi_1$.

2. Take the $m_2$ from the root of the tree and use cyclicity of $m_2$ to apply it on the first two arguments inside of $\omega$. This gives a sign factor of $-(-)^{d(\Psi_1)}$. In general one will be left with two subtrees then with $n+1$ leaves in total.

3. Take the $h$ from the right subtree and use that it is BPZ-even to apply it on the left subtree. This gives an additional sign factor of $(-)^{d(left\ subtree)}$.

4. Take the $m_2$ from the root of the right subtree and apply it on the first two arguments inside of $\omega$. It gives a sign factor of $-(-)^{d(left\ subtree)+1}$ (the $+1$ comes from the $h$ that was shifted in step 3) which cancels the sign factor of step 3. Again, one is left with two subtrees.

5. Repeat steps 3 and 4 until the right subtree only consist of $i$ acting on one input. The total sign factor remains $-(-)^{d(\Psi_1)}$.

6. Remove the $i$ acting on $\Psi_{n+1}$ and let it act as a $p$ on the left subtree.

Now the left subtree fulfills all requirements to be an element of $M_n$ and since we also have the right sign factor, we have obtained a term contained in the r.h.s. of (5.53). The manipulations are all uniquely invertible so we can conclude that the map between the trees is one-to-one and all terms we need are constructed exactly once. As a result, Eq. (5.53) holds and all higher products are cyclic. Moreover, as already suggested by the name, Eq. (5.53) together with the antisymmetry of $\omega$ implies invariance of the vertices under cyclic permutations.

### 5.2.6 Geometric picture

We have now shown that the higher products fulfill all the algebraic requirements but we do not know anything yet about the geometric picture, if they indeed give rise to a full single cover of the moduli space. To answer this question, the tree description of the products turns out to be very useful. Lets consider an arbitrary string tree diagram using the stubbed three vertex: As long as the external states are on-shell, the stubs make no difference because the external legs consist of a semiinfinite strip anyway. On the internal lines instead, the stubs make a difference because all internal strips with a length smaller than $2\lambda$ do not appear. We can conclude that the additional elemantary vertices we need should consist of all tree diagrams with all internal strips having a length smaller than $2\lambda$. The Siegel-gauge string propagator in the Schwinger parametrization is given by

$$\int_0^\infty dt\, e^{-tL_0} b_0 = \frac{b_0}{L_0}. \tag{5.54}$$

The integral over $t$ can be thought of an integral over strips of propagating strings of all different lengths. Following this logic, the propagators in our new vertices should be given as

$$\int_0^{2\lambda} dt\, e^{-tL_0} b_0 = -\frac{e^{-2\lambda L_0} - 1}{L_0} b_0 = -h \tag{5.55}$$

and hence be equal to minus the homotopy![23] We have constructed the higher products by drawing all binary tree diagrams with Witten vertices, $h$ as propagators and $e^{-\lambda L_0}$ on the leaves and the root. Those are in one-to-one correspondence with all the Feynman tree diagrams that should make up the new elementary vertices. This shows that our higher products $M_n$ indeed define a set of vertices which gives rise to a full cover of the moduli space.

## 5.3 Analytic solutions and action(s)

### 5.3.1 Projection cohomomorphism from the HPL

Using the definition
$$\mathbf{m} = \mathbf{Q} + \mathbf{m_2} \tag{5.56}$$
the equations of motion of the original Witten theory can be written in coalgebra language as

$$\mathbf{m} \frac{1}{1 - \Psi} = 0, \tag{5.57}$$

i. e. solutions are Maurer-Cartan elements of the $A_2$-algebra defining the theory. In the same spirit, to find solutions of the deformed theory, we have to solve the equation

$$\mathbf{M} \frac{1}{1 - \Psi} = 0. \tag{5.58}$$

In fact, the homological perturbation lemma already gave us an operator which maps solutions of the Witten theory to solutions of the stubbed theory. This can be seen as follows: The perturbed projection $\mathbf{P}$ is a chain map and hence obeys

$$\mathbf{Pm} = \mathbf{MP}. \tag{5.59}$$

Now lets assume $\Psi^*$ is a solution of the Witten theory, then

$$(\Psi^*)' = \pi_1 \mathbf{P} \frac{1}{1 - \Psi^*} \tag{5.60}$$

obeys

$$\mathbf{M} \frac{1}{1 - (\Psi^*)'} = \mathbf{M} \frac{1}{1 - \pi_1 \mathbf{P} \frac{1}{1-\Psi^*}} = \mathbf{MP} \frac{1}{1 - \Psi^*} = \mathbf{Pm} \frac{1}{1 - \Psi^*} = 0, \tag{5.61}$$

---

[23]It is interesting to notice at this point that the simple but not unique choice of $h$ corresponds to choosing Siegel gauge for the propagator. The minus sign is just a convention and can be absorbed in the definition of $h$.

where Eq. (5.147) was used. It remains to determine the cohomorphism **P** for the case where the homotopy equivalence is not an SDR. Again, as for the higher products above, the simplest way is to take the expression from the HPL

$$\mathbf{P}_{HPL} = \mathbf{p}\left(\mathbf{1} - \mathbf{m_2 h}\right)^{-1} \tag{5.62}$$

and apply the operator $P_{SDR}$ on its components to get

$$P_n = P_{SDR}\pi_1 \mathbf{P}_{HPL}\pi_n. \tag{5.63}$$

The resulting maps can then be packaged into a cohomorphism **P** again. More explicitly, we get for the first few orders

$$\begin{aligned}
P_1 &= p \\
P_2 &= pm_2\left(\cdot, h\cdot\right) + pm_2\left(h\cdot, ip\cdot\right) \\
P_3 &= pm_2\left(\cdot, hm_2\left(\cdot, h\cdot\right)\right) + pm_2\left(\cdot, hm_2\left(h\cdot, ip\cdot\right)\right) + pm_2\left(h\cdot, hm_2\left(ip\cdot, ip\cdot\right)\right) + pm_2\left(hm_2\left(\cdot, h\cdot\right), ip\cdot\right) \\
&\quad + pm_2\left(hm_2\left(h\cdot, ip\cdot\right), ip\cdot\right) + pm_2\left(h\cdot, ipm_2\left(\cdot, h\cdot\right)\right) + pm_2\left(h\cdot, ipm_2\left(h\cdot, ip\cdot\right)\right) \\
\ldots & \quad .
\end{aligned} \tag{5.64}$$

One can now check the equation $\pi_1 \mathbf{Pm} = \pi_1 \mathbf{MP}$ order by order:

$$\begin{aligned}
\pi_1 \mathbf{Pm}\pi_1 &= pQ = Qp = \pi_1 \mathbf{MP}\pi_1 \\
\pi_1 \mathbf{Pm}\pi_2 &= P_2\left(Q\cdot, \cdot\right) + P_2\left(\cdot, Q\cdot\right) + pm_2 \\
&= -pm_2\left(Q\cdot, h\cdot\right) + pm_2\left(hQ\cdot, ip\cdot\right) + pm_2\left(\cdot, hQ\cdot\right) + pm_2\left(h\cdot, ipQ\cdot\right) + pm_2 \\
&= -pm_2\left(Q\cdot, h\cdot\right) - pm_2\left(Qh\cdot, ip\cdot\right) + pm_2\left(\left(ip-1\right)\cdot, ip\cdot\right) \\
&\quad - pm_2\left(\cdot, Qh\cdot\right) + pm_2\left(\cdot, \left(ip-1\right)\cdot\right) + pm_2\left(h\cdot, Qip\cdot\right) + pm_2 \\
&= Qpm_2\left(\cdot, h\cdot\right) + Qpm_2\left(h\cdot, ip\cdot\right) + pm_2\left(ip\cdot, ip\cdot\right) \\
&= QP_2 + M_2\left(p\cdot, p\cdot\right) = \pi_1 \mathbf{MP}\pi_2. \tag{5.65}
\end{aligned}$$

For order three the calculation is already very tedious but it also turns out to work. The important point is that in the manipulations that are necessary, the SDR-relations were never used. This implies, since we know that (5.64) works for SDRs, it also works in our case and (5.64) is a valid definition. Now we have a cohomorphism by construction that obeys the chain map relation (5.59) such that we can construct analytic solutions of the deformed theory.[24]

### 5.3.2 The action

The on-shell action is one of the most important observables in OSFT, for example for the tachyon vacuum its value is equal to minus the energy of the D-brane which has decayed. Since the stubbed theory should be physically equivalent to the original Witten theory, we expect that the values for the on-shell action we get in the two theories coincide. The Witten action can be written in coalgebra notation [84] as

$$S(\Psi) = \int_0^1 dt\, \omega\left(\pi_1 \boldsymbol{\partial}_t \frac{1}{1-\Psi(t)}, \pi_1 \mathbf{m} \frac{1}{1-\Psi(t)}\right), \tag{5.66}$$

where $\Psi(t)$ is any smooth interpolation between $\Psi(0) = 0$ and $\Psi(1) = \Psi$ and $\boldsymbol{\partial}_t$ the coderivation associated to $\partial_t$. Similarly, the stubbed action reads

$$S'(\Psi) = \int_0^1 dt\, \omega\left(\pi_1 \boldsymbol{\partial}_t \frac{1}{1-\Psi(t)}, \pi_1 \mathbf{M} \frac{1}{1-\Psi(t)}\right). \tag{5.67}$$

---

[24]It is actually a non-trivial counting problem to determine the number of terms of $P_n$. It grows faster than the Catalan numbers because at order three we have seven terms and at order four already 33.

We would expect now a relation

$$S'\left((\Psi^*)'\right) \stackrel{?}{=} S(\Psi^*) \tag{5.68}$$

with $\Psi^*$ a MC-element of $\mathbf{m}$. However, this relation turns out to be non-obvious: Instead

$$\begin{aligned}
S(\Psi) &= \int_0^1 dt\, \omega\left(\pi_1 \boldsymbol{\partial}_t \frac{1}{1-\Psi(t)}, \pi_1 \mathbf{m} \frac{1}{1-\Psi(t)}\right) \\
&= \int_0^1 dt\, \omega\left(\pi_1 \boldsymbol{\partial}_t \mathbf{P}^{-1} \mathbf{P} \frac{1}{1-\Psi(t)}, \pi_1 \mathbf{P}^{-1} \mathbf{M} \mathbf{P} \frac{1}{1-\Psi(t)}\right) \\
&= \int_0^1 dt\, \omega\left(\pi_1 \boldsymbol{\partial}_t \mathbf{P}^{-1} \frac{1}{1-\pi_1 \mathbf{P} \frac{1}{1-\Psi(t)}}, \pi_1 \mathbf{P}^{-1} \mathbf{M} \frac{1}{1-\pi_1 \mathbf{P} \frac{1}{1-\Psi(t)}}\right) \\
&= \int_0^1 dt\, \omega\left(\pi_1 \mathbf{P}^{-1} \boldsymbol{\partial}_t \frac{1}{1-(\Psi)'(t)}, \pi_1 \mathbf{P}^{-1} \mathbf{M} \frac{1}{1-(\Psi)'(t)}\right) =: \tilde{S}(\Psi'), \tag{5.69}
\end{aligned}$$

where the last line differs from $S'(\Psi')$ by the insertions of $\mathbf{P}^{-1}$ on both inputs of $\omega$.

The invertibility of $\mathbf{P}$ is actually a delicate question: In general, a cohomomorphism is invertible iff its linear component, i. e. $P_1 = p$ is invertible. Now $p = e^{-\lambda L_0}$ inserts a strip of length $\lambda$, so one would expect the inverse $e^{\lambda L_0}$ to remove a strip of length $\lambda$ from the world sheet, which is not always possible. On the other hand, $e^{\lambda L_0}$ makes sense on a string field expanded in eigenstates of $L_0$ as long as the eigenvalues are finite. From now on, we shall assume that this is the case and $e^{\lambda L_0}$ is well-defined on all string fields in question.

Explicitly, the inversion of $\mathbf{P}$ works as follows: $\mathbf{P}\mathbf{P}^{-1} = \mathbf{P}^{-1}\mathbf{P}$ should be equal to the identity cohomomorphism, which is identity in its linear component and zero in all higher components. One can now solve the components $P_n^{-1} = \pi_1 \mathbf{P}^{-1} \pi_n$ order by order:

$$\begin{aligned}
P_1^{-1} &= p^{-1} \\
P_2^{-1} &= -m_2\left(p^{-1}\cdot, hp^{-1}\cdot\right) - m_2\left(hp^{-1}\cdot, i\cdot\right) \\
P_3^{-1} &= -m_2\left(hp^{-1}\cdot, hm_2(i\cdot, i\cdot)\right) + m_2\left(m_2\left(p^{-1}\cdot, hp^{-1}\cdot\right), hp^{-1}\cdot\right) + m_2\left(m_2\left(hp^{-1}\cdot, i\cdot\right), hp^{-1}\cdot\right) \\
&\cdots \quad . \tag{5.70}
\end{aligned}$$

One can now write out in more detail:

$$\begin{aligned}
\tilde{S}(\Psi) = &\frac{1}{2}\omega\left(p^{-1}\Psi, p^{-1}Q\Psi\right) \\
&+ \frac{1}{3}\omega\left(p^{-1}\Psi, p^{-1}M_2(\Psi,\Psi)\right) + \frac{1}{3}\omega\left(p^{-1}\Psi, P_2^{-1}((Q\Psi,\Psi)+(\Psi,Q\Psi))\right) \\
&+ \frac{1}{3}\omega\left(P_2^{-1}(\Psi,\Psi), p^{-1}Q\Psi\right) + \frac{1}{4}\omega\left(p^{-1}\Psi, p^{-1}M_3(\Psi,\Psi,\Psi)\right) \\
&+ \frac{1}{4}\omega\left(p^{-1}\Psi, P_2^{-1}((M_2(\Psi,\Psi),\Psi)+(\Psi,M_2(\Psi,\Psi)))\right) \\
&+ \frac{1}{4}\omega\left(p^{-1}\Psi, P_3^{-1}((Q\Psi,\Psi,\Psi)+(\Psi,Q\Psi,\Psi)+(\Psi,\Psi,Q\Psi))\right) \\
&+ \frac{1}{4}\omega\left(P_2^{-1}(\Psi,\Psi), p^{-1}M_2(\Psi,\Psi)\right) + \frac{1}{4}\omega\left(P_2^{-1}(\Psi,\Psi), P_2^{-1}((Q\Psi,\Psi)+(\Psi,Q\Psi))\right) \\
&+ \frac{1}{4}\omega\left(P_3^{-1}(\Psi,\Psi,\Psi), p^{-1}Q\Psi\right) + \mathcal{O}\left(\Psi^{\otimes 5}\right).
\end{aligned} \quad (5.71)$$

It seems that the cohomomorphism $\mathbf{P}$ does not define the field redefinition we were looking for, instead it relates the Witten theory to a theory defined by $\tilde{S}(\Psi)$. The equations of motion derived from $\tilde{S}$ are the same as for $S'$, namely $\mathbf{M}\frac{1}{1-\Psi} = 0$, but it is not at all obvious that the two actions agree even on-shell.

The reason is that the HPL does not know anything about the symplectic form $\omega$: To get an invariant action with $S'(\Psi') = S(\Psi)$, not only $\mathbf{m}$ has to transform accordingly, but also $\omega$ would have to go to

$$\omega'(\Psi_1, \Psi_2) = \omega\left(\pi_1 \mathbf{P}^{-1}\frac{1}{1-\Psi_1}, \pi_1 \mathbf{P}^{-1}\frac{1}{1-\Psi_2}\right), \quad (5.72)$$

otherwise $\mathbf{P}$ would fail to be cyclic. As it can be seen from (5.69), $\tilde{S}(\Psi)$ is just a fancy rewriting of the Witten action and therefore defines an equivalent theory. But since we would like to keep the original $\omega$, the field redefinition induced by $\mathbf{P}$ from the HPL and giving rise to $\tilde{S}(\Psi)$ is not exactly what we want. However, since it shares the same equations of motion as $S'(\Psi)$, it might provide a new family of gauge-invariant observables for solutions of $S'(\Psi)$, parametrized by $\lambda$. In the last section we will check this explicitly on a special class of solutions. For now the next task is to derive the originally desired field redefinition which relates the actions $S(\Psi)$ and $S'(\Psi)$. [25]

### 5.3.3 Elements from closed string field theory

In [34], Zwiebach and Hata have shown how to relate slightly different, consistent sets of vertices in CSFT via an infinitesimal field redefinition. Our strategy is now to apply their method to our problem in OSFT and integrate the result to the finite case. This will not only provide us the field redefinition we are looking for, but also give some insight into the rather abstract formalism of CSFT. First, it is useful to collect some basic information about the structure of CSFT.

---
[25]It would also be an interesting direction to examine if there exists some kind of "dual" HPL that directly yields this correct field redefinition.

As already explained in the introduction, the vertices are given by integrating *basic differential forms* $\Omega^{g,n}_{\Psi_1\Psi_2...\Psi_n}$ defined by

$$\Omega^{g,n}_{\Psi_1\Psi_2...\Psi_n}\left(\hat{V}_1, \hat{V}_2, ..., \hat{V}_{6g-6+2n}\right) = (2\pi i)^{-(3g-3+n)} \langle\Sigma|\, \mathbf{b}(\mathbf{v_1})...\mathbf{b}(\mathbf{v_{6g-6+2n}}) |\Psi_1\rangle ... |\Psi_n\rangle. \quad (5.73)$$

They are living in the tangent space of the fibre bundle $\hat{\mathcal{P}}_{g,n}$ over the moduli space $\mathcal{M}_{g,n}$, with the fiber being the space of local coordinates around the punctures modulo phase rotations. The dimension of this bundle is infinite, but the degree of $\Omega^{g,n}_{\Psi_1\Psi_2...\Psi_n}$ is just the real dimension of the base space $\mathcal{M}_{g,n}$. It takes as arguments tangent vectors $\hat{V}_i \in T\hat{\mathcal{P}}_{g,n}$, which represent deformations of the world sheet Riemann surface $\Sigma$, either by changing the moduli or the local coordinates. The $\mathbf{v_i}$ are Schiffer vectors on $\Sigma$, supported around the punctures, which generate those deformations. This means that the local coordinate around the $n$th puncture transforms as

$$z^{(n)} \to z^{(n)} + \epsilon v^{(n)}\left(z^{(n)}\right) \quad (5.74)$$

for some small $\epsilon$. The $b$-ghost insertions are then defined as

$$\mathbf{b}(\mathbf{v}) = \sum_{i=1}^{n} \left(\oint \frac{dz_i}{2\pi i} b(z_i) v^{(i)}(z_i) + \oint \frac{d\bar{z}_i}{2\pi i} \bar{b}(\bar{z}_i) \bar{v}^{(i)}(\bar{z}_i)\right). \quad (5.75)$$

We will be only interested in the classical action without the loop vertices, so the genus $g$ shall be zero from now on. The basic forms can now be integrated over sections of $\hat{\mathcal{P}}_{0,n}$ defining the vertices $\mathcal{V}_{0,n}$.

The quantization procedure makes use of the Batalin-Vilkovisky formalism; although we are not interested in quantum effects, the BV-antibracket is used in constructing the symmetry generator. It is defined as

$$\{A, B\} = \frac{\partial_r A}{\partial \Psi^i}\frac{\partial_l B}{\partial \Psi^*_i} - \frac{\partial_r A}{\partial \Psi^*_i}\frac{\partial_l B}{\partial \Psi^i}, \quad (5.76)$$

where the $\Psi^*_i$ are antifields of opposite parity associated to each basis element of $\mathcal{H}$. The BV-master action takes the same form as (5.7) with the only difference that the $\Psi$ are not restricted in ghost number and run over fields as well as antifields.

### 5.3.4 Constructing the field redefinition

We are looking for a non-linear field redefinition of the form

$$\Psi' = F(\Psi) = \sum_{n=1}^{\infty} F_n\left(\Psi^{\otimes n}\right) = \pi_1 \mathbf{F} \frac{1}{1-\Psi} \quad (5.77)$$

that relates the Witten action to the stubbed $A_\infty$-action in the form (5.67). To be consistent with the results of Zwiebach and Hata [34] we demand

$$S(\Psi') = S'(\Psi). \quad (5.78)$$

Since the kinetic term is identical, we immediately find

$$F_1(\Psi) = \Psi. \tag{5.79}$$

In [34] it is shown that under an infinitesimal field redefinition of the form

$$\Psi \to \Psi + t\{\Psi, e\} + \mathcal{O}(t^2), \tag{5.80}$$

the classical action transforms as

$$S(\Psi) \to S(\Psi) + t\{S, e\} + \mathcal{O}(t^2), \tag{5.81}$$

where $\{\}$ denotes the BV-antibracket. In the paper it is now argued that for any small change of vertices, the change of the action indeed takes this form and the generator $e$ is constructed explicitly:

$$e(u_0) = -\sum_n \kappa^{n-2} \frac{1}{n!} \int_{\mathcal{V}_{0,n}(u_0)} \Omega^{(0)0,n}_{\mathbf{b}(\mathbf{u})\Psi^{\otimes n}}. \tag{5.82}$$

Here we assume that there exists some family of consistent vertex sets $\mathcal{V}_{0,n}(u)$ parametrized by some real number $u$ and everything is evaluated at the point $u_0$. The vector $\mathbf{u}$ is some Schiffer vector which generates a deformation of the $\mathcal{V}_{0,n}(u_0)$ in the direction of $u$, i. e. it generates diffeomorphisms which push $\mathcal{V}_{0,n}(u_0)$ into $\mathcal{V}_{0,n}(u_0 + \delta u)$. For the case of varying the stub length, this Schiffer vector takes a particular simple form: First, lets notice that the stub length $\lambda$ for closed strings is defined as the geodesic distance from the location $|z|=1$ of the local coordinate to the begin of the semiinfinite cylinder associated with the puncture. This implies that $\lambda$ can be changed by just rescaling the coordinate: Sending $z$ to $z' = z + \epsilon z$, the location of $|z'|=1$ corresponds to $|z|=1-\epsilon$, such that $\lambda$ is increased by $\epsilon$. By comparing with Eq. (5.74) we read off

$$u^{(i)} = z^{(i)}. \tag{5.83}$$

The $b$-ghost insertion is then given by

$$\mathbf{b}(\mathbf{u}) = \sum_{i=1}^n \left(\oint \frac{dz_i}{2\pi i} z_i b(z_i) + \oint \frac{d\bar{z}_i}{2\pi i} \bar{z}_i \bar{b}(\bar{z}_i)\right) = \sum_{i=1}^n b_0^{(i)} + \bar{b}_0^{(i)} = \sum_{i=1}^n b_0^{+(i)}. \tag{5.84}$$

We want to use the above expression for $e$ in the context of changing the stub length for open strings, hence a few modifications and simplifications are necessary: First, the combinatorial factor $n!$ originates from total symmetrization of the vertices and is not necessary for open strings. Second, the insertions of $b_0^+$ should get replaced simply by $b_0$ since there is no antiholomorphic sector. Moreover, the string coupling $\kappa$ will be set to one. Now the generator simplifies to

$$e(\lambda) = -\sum_n \int_{\mathcal{V}(\lambda)_{0,n}} \Omega^{0,n}_{\mathbf{b_0}\Psi^{\otimes n}}. \tag{5.85}$$

If we make the ansatz
$$\Psi' = \Psi + \delta\lambda \sum_{n=2}^{\infty} f_n\left(\Psi^{\otimes n}\right) \tag{5.86}$$
as the infinitesimal version of (5.77), then the $f_n$ are determined as
$$f_n\left(\Psi^{\otimes n}\right) = \left\{\Psi, e^{(n)}\right\}, \tag{5.87}$$
where $\delta\lambda$ plays the role of $t$ in (5.80).

To find $f_2(\Psi, \Psi)$ we need to consider $e(\lambda)$ for $n = 3$: The vertex $\mathcal{V}(\lambda)_{0,3}$ is zero dimensional, so there is no integral and the surface state $\langle\Sigma|$ is just the Witten vertex with stubs of length $\lambda$,
$$\langle V_3(\lambda)| = \omega\left(\cdot, M_2\left(\cdot, \cdot\right)\right) = \omega\left(e^{-\lambda L_0}\cdot, e^{-\lambda L_0}\cdot *e^{-\lambda L_0}\cdot\right). \tag{5.88}$$
Inserting into (5.85) yields
$$e^{(3)}(\lambda, \Psi) = -\left(\omega\left(b_0\Psi, M_2(\Psi, \Psi)\right) + \omega\left(\Psi, M_2(b_0\Psi, \Psi)\right) + \omega\left(\Psi, M_2(\Psi, b_0\Psi)\right)\right). \tag{5.89}$$
The BV-bracket with $\Psi$ can be straightforwardly evaluated; after carefully checking the signs the result is

$$f_2(\Psi, \Psi) = -b_0 M_2(\Psi, \Psi) + M_2(b_0\Psi, \Psi) + M_2(\Psi, b_0\Psi). \tag{5.90}$$

We expect now the relation
$$S'\left(\Psi + \delta\lambda f_2(\Psi, \Psi), \lambda\right) = S'(\Psi, \lambda + \delta\lambda) \tag{5.91}$$
to hold up to order 3 in $\Psi$; by directly inserting we can compute explicitly
$$S'\left(\Psi + \delta\lambda f_2(\Psi, \Psi), \lambda\right) = \frac{1}{2}\omega(\Psi, Q\Psi) + \frac{1}{3}\omega(\Psi, M_2(\Psi, \Psi)) + \delta\lambda\, \omega(\Psi, Qf_2(\Psi, \Psi)) + \mathcal{O}\left(\Psi^{\otimes 4}\right). \tag{5.92}$$

The last and most interesting term yields
$$\left(-\omega(\Psi, Qb_0 M_2(\Psi, \Psi)) + \omega(\Psi, QM_2(b_0\Psi, \Psi)) + \omega(\Psi, QM_2(\Psi, b_0\Psi))\right)\delta\lambda$$
$$= \left(-\omega(\Psi, L_0 M_2(\Psi, \Psi)) + \omega(\Psi, b_0 QM_2(\Psi, \Psi)) - \omega(Q\Psi, M_2(b_0\Psi, \Psi)) - \omega(Q\Psi, M_2(\Psi, b_0\Psi))\right)\delta\lambda$$
$$= -\omega(\Psi, L_0 M_2(\Psi, \Psi))\,\delta\lambda, \tag{5.93}$$
where the last three terms in the second line cancel after applying the Leibniz rule and cyclicity. On the other hand,
$$S'(\Psi, \lambda + \delta\lambda) = S'(\Psi, \lambda) + \delta\lambda \frac{d}{d\lambda} S'(\Psi, \lambda)$$
$$= \frac{1}{2}\omega(\Psi, Q\Psi) + \frac{1}{3}\omega(\Psi, M_2(\Psi, \Psi))$$
$$- \frac{1}{3}\delta\lambda\, \omega(L_0\Psi, M_2(\Psi, \Psi)) - \frac{1}{3}\delta\lambda\, \omega(\Psi, M_2(L_0\Psi, \Psi))$$
$$- \frac{1}{3}\delta\lambda\, \omega(\Psi, M_2(\Psi, L_0\Psi)) + \mathcal{O}\left(\Psi^{\otimes 4}\right)$$
$$= \frac{1}{2}\omega(\Psi, Q\Psi) + \frac{1}{3}\omega(\Psi, M_2(\Psi, \Psi)) - \delta\lambda\, \omega(L_0\Psi, M_2(\Psi, \Psi)) + \mathcal{O}\left(\Psi^{\otimes 4}\right), \tag{5.94}$$

which is the same expression up to order $\Psi^{\otimes 3}$ (Again, cyclicity was used in the last line.).

The explicit form of $f_2(\Psi, \Psi)$ suggests the following general structure: We can guess the ansatz

$$f_n\left(\Psi^{\otimes n}\right) = -b_0 M_n\left(\Psi^{\otimes n}\right) + M_n\left(\mathbf{b_0}\left(\Psi^{\otimes n}\right)\right), \tag{5.95}$$

where $\mathbf{b_0}$ again denotes the coderivation associated to $b_0$. At first sight, Eq.(5.95) looks a bit strange now from the coalgebra perspective because it is a commutator of two odd objects, so one would more naturally expect an anticommutator. However, the first term in (5.95) stems from the application of $b_0$ on the first argument of the symplectic form $\omega$, hence the sign contains implicit information about $\omega$. From the discussion about $\mathbf{P}$ from the HPL we could anticipate that $\omega$ has to enter the calculation at some point. The more natural looking expression $\pi_1[\mathbf{b_0}, \mathbf{M_n}]$ would have been independent of $\omega$.

One can prove now that the ansatz (5.95) is indeed correct by directly inserting into the action.

### 5.3.5 Proof of the ansatz for the infinitesimal field redefinition

If we focus solely on terms of order $n+1$ in $\Psi$ we get

$$S'^{(n+1)}(\Psi') = \omega\left(\Psi, Q f_n\left(\Psi^{\otimes n}\right)\right)\delta\lambda + \sum_{k=2}^{n-1} \omega\left(\Psi, M_k\left(f_{n+1-k}\left(\Psi^{\otimes n+1-k}\right), \Psi^{\otimes k-1}\right)\right)\delta\lambda$$
$$+ \frac{1}{n+1}\omega\left(\Psi, M_n\left(\Psi^{\otimes n}\right)\right). \tag{5.96}$$

The last term is the contribution from the original $S(\Psi)$, so the first two terms denoted by $\delta S^{(n+1)}$ should yield the infinitesimal variation

$$\delta S'^{(n+1)} \stackrel{?}{=} \delta\lambda \frac{d}{d\lambda}\frac{1}{n+1}\omega\left(\Psi, M_n\left(\Psi^{\otimes n}\right)\right). \tag{5.97}$$

Inserting the ansatz and performing some straight forward manipulations gives

$$\frac{\delta S'^{(n+1)}}{\delta\lambda} = \omega\left(\Psi, Q\left(-b_0 M_n + M_n \mathbf{b_0}\right)\left(\Psi^{\otimes n}\right)\right)$$
$$+ \sum_{k=2}^{n-1} \omega\left(\Psi, M_k\left(\left(-b_0 M_{n+1-k} + M_{n+1-k}\mathbf{b_0}\right)\left(\Psi^{\otimes n+1-k}\right), \Psi^{\otimes k-1}\right)\right)$$
$$= -\omega\left(\Psi, L_0 M_n\left(\Psi^{\otimes n}\right)\right) + \omega\left(\Psi, b_0[\mathbf{Q}, \mathbf{M_n}]\left(\Psi^{\otimes n}\right)\right) - \omega\left(\Psi, b_0 M_n \mathbf{Q}\left(\Psi^{\otimes n}\right)\right)$$
$$+ \omega\left(\Psi, Q M_n \mathbf{b_0}\left(\Psi^{\otimes n}\right)\right) + \sum_{k=2}^{n-1}\left(-\omega\left(\Psi, M_k\left(b_0 M_{n+1-k}\left(\Psi^{\otimes n+1-k}\right), \Psi^{\otimes k-1}\right)\right)\right.$$
$$\left. + \omega\left(\Psi, M_k\left(M_{n+1-k}\mathbf{b_0}\left(\Psi^{\otimes n+1-k}\right), \Psi^{\otimes k-1}\right)\right)\right). \tag{5.98}$$

The last term of the third line and the first of the fourth actually cancel each other:

$$-\omega\left(\Psi, b_0 M_n \mathbf{Q}\left(\Psi^{\otimes n}\right)\right) + \omega\left(\Psi, Q M_n \mathbf{b_0}\left(\Psi^{\otimes n}\right)\right)$$
$$= -\omega\left(b_0 \Psi, M_n \mathbf{Q}\left(\Psi^{\otimes n}\right)\right) - \omega\left(Q\Psi, M_n \mathbf{b_0}\left(\Psi^{\otimes n}\right)\right)$$
$$= -\omega\left(b_0\Psi, M_n\left(Q\Psi, \Psi, ..., \Psi\right)\right) - \omega\left(b_0\Psi, M_n\left(\Psi, Q\Psi, ..., \Psi\right)\right) - ... - \omega\left(b_0\Psi, M_n\left(\Psi, \Psi, ..., Q\Psi\right)\right)$$
$$- \omega\left(Q\Psi, M_n\left(b_0\Psi, \Psi, ..., \Psi\right)\right) - \omega\left(Q\Psi, M_n\left(\Psi, b_0\Psi, ..., \Psi\right)\right) - ... - \omega\left(Q\Psi, M_n\left(\Psi, \Psi, ..., b_0\Psi\right)\right).$$
(5.99)

Because of cyclicity of the $(n+1)$-vertex the last two lines contain the same terms, just differing by a sign which comes from commuting $Q$ with $b_0$. Therefore they add to zero. The second term in the second line of (5.98) can be further manipulated using the $A_\infty$-relations:

$$\frac{\delta S'^{(n+1)}}{\delta\lambda} = -\omega\left(\Psi, L_0 M_n\left(\Psi^{\otimes n}\right)\right) - \sum_{k=2}^{n-1} \omega\left(\Psi, M_k\left(b_0 M_{n+1-k}\left(\Psi^{\otimes n+1-k}\right), \Psi^{\otimes k-1}\right)\right)$$
$$+ \sum_{k=2}^{n-1} \left(\omega\left(\Psi, M_k\left(M_{n+1-k}\left(\mathbf{b_0}\left(\Psi^{\otimes n+1-k}\right)\right), \Psi^{\otimes k-1}\right)\right) - \omega\left(\Psi, b_0\mathbf{M_k M_{n+1-k}}\left(\Psi^{\otimes n}\right)\right)\right).$$
(5.100)

Now again, the terms in the last line cancel after using cyclicity:

$$\omega\left(\Psi, b_0\mathbf{M_k M_{n+1-k}}\left(\Psi^{\otimes n}\right)\right) = \omega\left(b_0\Psi, M_k\mathbf{M_{n+1-k}}\left(\Psi^{\otimes n}\right)\right)$$
$$= \omega\left(M_k\left(b_0\Psi, M_{n+1-k}\left(\Psi^{\otimes n+1-k}\right), \Psi^{\otimes k-2}\right), \Psi\right)$$
$$+ \omega\left(M_k\left(b_0\Psi, \Psi, M_{n+1-k}\left(\Psi^{\otimes n+1-k}\right), \Psi^{\otimes k-3}\right), \Psi\right)$$
$$+ ... + \omega\left(M_k\left(b_0\Psi, \Psi^{\otimes k-2}, M_{n+1-k}\left(\Psi^{\otimes n+1-k}\right)\right), \Psi\right)$$
$$+ \omega\left(M_k\left(b_0\Psi, \Psi^{\otimes k-1}\right), M_{n+1-k}\left(\Psi^{\otimes n+1-k}\right)\right).$$
(5.101)

All terms except for the last one can be further manipulated using the antisymmetry of $\omega$ and cyclicity of $M_k$. For example,

$$\omega\left(M_k\left(b_0\Psi, M_{n+1-k}\left(\Psi^{\otimes n+1-k}\right), \Psi^{\otimes k-2}\right), \Psi\right)$$
$$= -\omega\left(\Psi, M_k\left(b_0\Psi, M_{n+1-k}\left(\Psi^{\otimes n+1-k}\right), \Psi^{\otimes k-2}\right)\right)$$
$$= \omega\left(M_k\left(\Psi, b_0\Psi, M_{n+1-k}\left(\Psi^{\otimes n+1-k}\right), \Psi^{\otimes k-3}\right), \Psi\right)$$
$$= ... = \omega\left(M_k\left(\Psi^{\otimes k-1}, b_0\Psi\right), M_{n+1-k}\left(\Psi^{\otimes n+1-k}\right)\right),$$
(5.102)

so in all of the terms the $M_{n+1-k}$ can be moved to the outermost right. After all, the terms can be summed up as

$$\omega\left(M_k\left(\Psi^{\otimes k-1}, b_0\Psi\right), M_{n+1-k}\left(\Psi^{\otimes n+1-k}\right)\right) + \omega\left(M_k\left(\Psi^{\otimes k-2}, b_0\Psi, \Psi\right), M_{n+1-k}\left(\Psi^{\otimes n+1-k}\right)\right)$$
$$+ ... + \omega\left(M_k\left(b_0\Psi, \Psi^{\otimes k-1}\right), M_{n+1-k}\left(\Psi^{\otimes n+1-k}\right)\right)$$
$$= \omega\left(M_k\left(\mathbf{b_0}\left(\Psi^{\otimes k}\right)\right), M_{n+1-k}\left(\Psi^{\otimes n+1-k}\right)\right)$$
$$= \omega\left(\Psi, M_{n+1-k}\left(M_k\left(\mathbf{b_0}\left(\Psi^{\otimes k}\right)\right), \Psi^{\otimes n-k}\right)\right), \tag{5.103}$$

which is after the summation over $k$ identical to the first term in the second line of (5.100), just with opposite sign. So we arrive at the expression

$$\frac{\delta S'^{(n+1)}}{\delta \lambda} = -\omega\left(\Psi, L_0 M_n\left(\Psi^{\otimes n}\right)\right) - \sum_{k=2}^{n-1} \omega\left(\Psi, M_k\left(b_0 M_{n+1-k}\left(\Psi^{\otimes n+1-k}\right), \Psi^{\otimes k-1}\right)\right), \tag{5.104}$$

which should now be compared to the result of formula (5.97).

The derivative with respect to $\lambda$ can act on the stubs as well as on the homotopy $h$. The action on $e^{-\lambda L_0}$ inserts a factor of $-L_0$ on every input string field of the $(n+1)$-vertex. Since the vertices are cyclically symmetric, we get $n+1$ identical terms, which cancels the prefactor $\frac{1}{n+1}$. The result is

$$\frac{d}{d\lambda} \frac{1}{n+1} \omega\left(\Psi, M_n\left(\Psi^{\otimes n}\right)\right) \supset -\omega\left(L_0\Psi, M_n\left(\Psi^{\otimes n}\right)\right), \tag{5.105}$$

which is equal to the first term of (5.104). To compute the action on $h$, the tree representation turns out to be useful again: First of all,

$$\frac{d}{d\lambda} h = -2b_0 e^{-2\lambda L_0}, \tag{5.106}$$

hence we get a sum of all possible tree diagrams with one propagator replaced by $-2b_0 e^{-2\lambda L_0}$. We can cut through the diagram along this replaced propagator and think of the factor $e^{-2\lambda L_0}$ as arising from two $e^{-\lambda L_0}$-stubs from the leaf and the root of the two subtrees. Both subtrees are now part of a higher product $M_k$ for some $k$, $2 \leq k \leq n-1$. So the whole expression can be written as a combination of two higher products $M_k$, $M_{n+1-k}$ with a factor $-2b_0$ inserted:

$$\frac{d}{d\lambda} \frac{1}{n+1} \omega\left(\Psi, M_n\left(\Psi^{\otimes n}\right)\right) \supset -\frac{1}{n+1} \sum_{k=2}^{n-1} \omega\left(\Psi, M_k\left(2b_0 M_{n+1-k}\left(\Psi^{\otimes n+1-k}\right), \Psi^{\otimes k-1}\right)\right) +$$
$$\omega\left(\Psi, M_k\left(\Psi, 2b_0 M_{n+1-k}\left(\Psi^{\otimes n+1-k}\right), \Psi^{\otimes k-2}\right)\right) + ... +$$
$$\omega\left(\Psi, M_k\left(\Psi^{\otimes k-1}, 2b_0 M_{n+1-k}\left(\Psi^{\otimes n+1-k}\right)\right)\right). \tag{5.107}$$

Because of cyclicity of the $k+1$-vertex, the different lines contain the same terms so we have

$$\frac{d}{d\lambda} \frac{1}{n+1} \omega\left(\Psi, M_n\left(\Psi^{\otimes n}\right)\right) \supset -\frac{1}{n+1} \sum_{k=2}^{n-1} 2k \cdot \omega\left(\Psi, M_k\left(b_0 M_{n+1-k}\left(\Psi^{\otimes n+1-k}\right), \Psi^{\otimes k-1}\right)\right). \tag{5.108}$$

The last bracket can be further manipulated:

$$\omega\left(\Psi, M_k\left(b_0 M_{n+1-k}\left(\Psi^{\otimes n+1-k}\right), \Psi^{\otimes k-1}\right)\right) = -\omega\left(M_k\left(b_0 M_{n+1-k}\left(\Psi^{\otimes n+1-k}\right), \Psi^{\otimes k-1}\right), \Psi\right)$$
$$= \omega\left(b_0 M_{n+1-k}\left(\Psi^{\otimes n+1-k}\right), M_k\left(\Psi^{\otimes k}\right)\right) = -\omega\left(M_{n+1-k}\left(\Psi^{\otimes n+1-k}\right), b_0 M_k\left(\Psi^{\otimes k}\right)\right)$$
$$= \omega\left(\Psi, M_{n+1-k}\left(\Psi^{\otimes n-k}, b_0 M_k\left(\Psi^{\otimes k}\right)\right)\right) = \omega\left(\Psi, M_{n+1-k}\left(b_0 M_k\left(\Psi^{\otimes k}\right), \Psi^{\otimes n-k}\right)\right). \quad (5.109)$$

In the last step cyclicity of the $(n+1-k)$-vertex was used again. We see that in the sum of (5.108) the $k$th term and the $(n+1-k)$th term are identical so the sum can be rewritten as

$$-\frac{1}{n+1}\sum_{k=2}^{n-1}(n+1)\cdot\omega\left(\Psi, M_k\left(b_0 M_{n+1-k}\left(\Psi^{\otimes n+1-k}\right), \Psi^{\otimes k-1}\right)\right)$$
$$= \sum_{k=2}^{n-1}\omega\left(\Psi, M_k\left(b_0 M_{n+1-k}\left(\Psi^{\otimes n+1-k}\right), \Psi^{\otimes k-1}\right)\right), \quad (5.110)$$

which is precisely the second term in (5.104).

This completes the proof that the ansatz

$$f_n\left(\Psi^{\otimes n}\right) = -b_0 M_n\left(\Psi^{\otimes n}\right) + M_n\left(\mathbf{b_0}\left(\Psi^{\otimes n}\right)\right) \quad (5.111)$$

indeed yields the correct infinitesimal field redefinition.

### 5.3.6 Finite field redefinition

So far we have only been concerned with infinitesimal variations of $\lambda$, now we want to generalize the results to finite changes. We know

$$\Psi_{\lambda+\delta\lambda} = \Psi_\lambda + \delta\lambda f_2^\lambda\left(\Psi_\lambda, \Psi_\lambda\right) + \delta\lambda f_3^\lambda\left(\Psi_\lambda, \Psi_\lambda, \Psi_\lambda\right) + ... = \Psi_\lambda + \delta\lambda\frac{d}{d\lambda}\Psi_\lambda, \quad (5.112)$$

where we have written the superscript $\lambda$ to indicate that the $f_n$ also depend on $\lambda$ explicitly. This equation can be integrated to

$$\Psi_\lambda = \Psi_0 + \int_0^\lambda dt f_2^t\left(\Psi_t, \Psi_t\right) + \int_0^\lambda dt f_3^t\left(\Psi_t, \Psi_t, \Psi_t\right) + ... \quad . \quad (5.113)$$

Inserting $\Psi_\lambda$ back we get a perturbative expansion in the original solution $\Psi_0$:

$$\Psi_\lambda = \Psi_0 + \int_0^\lambda dt\ f_2^t\left(\Psi_0, \Psi_0\right) + \int_0^\lambda dt\ f_3^t\left(\Psi_0, \Psi_0, \Psi_0\right) +$$
$$\int_0^\lambda dt\ f_2^t\left(\int_0^t ds\ f_2^s\left(\Psi_0, \Psi_0\right), \Psi_0\right) + \int_0^\lambda dt\ f_2^t\left(\Psi_0, \int_0^t ds\ f_2^s\left(\Psi_0, \Psi_0\right)\right) + \mathcal{O}\left(\Psi_0^{\otimes 4}\right). \quad (5.114)$$

This formula provides an algorithm to find the associated $A_\infty$-solution to each known solution of the Witten OSFT.

## 5.4 Physical interpretation

To summarize, we found two distinct field redefinitions

$$\tilde{\Psi} = \sum_{n=1}^{\infty} P_n\left(\Psi^{\otimes n}\right), \qquad \Psi' = \sum_{n=1}^{\infty} F_n\left(\Psi^{\otimes n}\right), \tag{5.115}$$

which generate two different actions $\tilde{S}$ and $S'$ via

$$\tilde{S}\left(\tilde{\Psi}\right) = S\left(\Psi\right), \qquad S'\left(\Psi\right) = S\left(\Psi'\right). \tag{5.116}$$

However, both actions share the same equations of motion, namely the Maurer-Cartan equation of the stubbed $A_\infty$-algebra

$$\pi_1 \mathbf{M} \frac{1}{1-\Psi} = 0. \tag{5.117}$$

It remains to examine what the physical meaning of those two actions is, most importantly, if they yield the same on-shell value for a given solution $\Psi^*$.

In [58, 68], a special class of solutions containing a nearly marginal vertex operator is introduced which serves as a useful playground to analyze this question.

### 5.4.1 Nearly marginal solutions

Consider a matter conformal primary field $V$ with weight $h$ smaller but very close to one. The string field

$$\Psi_1 = \mu \cdot cV(0)\left|0\right\rangle, \tag{5.118}$$

with some real coupling constant $\mu$ obeys the Siegel gauge condition

$$b_0 \Psi_1 = 0 \tag{5.119}$$

and will serve as a starting point to find the full solution $\Psi = \sum \Psi_n$ as a perturbative series in the expansion parameter $y = 1 - h$. One can solve for the string coupling $\mu$ using the Witten equations of motion,

$$Q\Psi + \Psi * \Psi = 0, \tag{5.120}$$

to obtain [68]

$$\mu = \frac{y}{C_{VVV}} + \mathcal{O}\left(y^3\right), \tag{5.121}$$

where $C_{VVV}$ denotes the three-point function constant of $V$. We can deduce that $\Psi_1 = \mathcal{O}(y)$ and from the perturbative algorithm for the full solution one can also show that in general $\Psi_n = \mathcal{O}(y^n)$.

The on-shell action in Witten theory can be written compactly as

$$S(\Psi) = -\frac{1}{6}\left\langle\Psi, Q\Psi\right\rangle. \tag{5.122}$$

From
$$Q\Psi_1 = \mu y \cdot c\partial c V(0)|0\rangle \tag{5.123}$$

we see that the action will be of leading order $y^3$ and given by

$$S(\Psi) = -\frac{1}{6}\frac{y^3}{C_{VVV}^2}\langle cV, c\partial cV\rangle = \frac{1}{6}\frac{y^3}{C_{VVV}^2} + \mathcal{O}(y^4), \tag{5.124}$$

if we assume that $V$ is conveniently normalized, $\langle V(z_1), V(z_2)\rangle = z_{12}^{-2h}$.

### 5.4.2 The $A_\infty$-action

The first important observation is that the cohomomorphism **P** simplifies significantly for string fields in Siegel gauge: Since $h$ is proportional to $b_0$, it annihilates $\Psi$ and (5.64) collapses to

$$\tilde{\Psi} = \sum_{n=1}^{\infty} P_n\left(\Psi^{\otimes n}\right) = p\Psi. \tag{5.125}$$

We already know that $\tilde{S}(\tilde{\Psi})$ yields the original value $S(\Psi)$, so now we want to check the expression $S'(\tilde{\Psi})$: $\Psi_1$ is an $L_0$-eigenstate so we straightforwardly get

$$p\Psi_1 = e^{-\lambda L_0}\Psi_1 = e^{\lambda y}\Psi_1. \tag{5.126}$$

For cubic order in $y$ we just have to insert this into the kinetic term (5.122) and get

$$S'(\tilde{\Psi})\Big|_{y^3} = -\frac{1}{6}\langle p\Psi_1, Qp\Psi_1\rangle = \frac{1}{6}\frac{y^3}{C_{VVV}^2}e^{2\lambda y}\Big|_{y^3} = \frac{1}{6}\frac{y^3}{C_{VVV}^2}, \tag{5.127}$$

which agrees with the result above up to terms of $\mathcal{O}(y^4)$.

For a more non-trivial check we can collect the terms of quartic order in $y$: In the action we have to consider the first three terms

$$S'(\tilde{\Psi})\Big|_{y^4} = \left(-\frac{1}{2}\langle\tilde{\Psi}, Q\tilde{\Psi}\rangle - \frac{1}{3}\langle\tilde{\Psi}, M_2(\tilde{\Psi}, \tilde{\Psi})\rangle - \frac{1}{4}\langle\tilde{\Psi}, M_3(\tilde{\Psi}, \tilde{\Psi}, \tilde{\Psi})\rangle\right)\Big|_{y^4}. \tag{5.128}$$

The equations of motion however tell us that

$$\langle\tilde{\Psi}, Q\tilde{\Psi}\rangle + \langle\tilde{\Psi}, M_2(\tilde{\Psi}, \tilde{\Psi})\rangle + \langle\tilde{\Psi}, M_3(\tilde{\Psi}, \tilde{\Psi}, \tilde{\Psi})\rangle = \mathcal{O}(y^5), \tag{5.129}$$

so the expression simplifies to

$$S'(\tilde{\Psi})\Big|_{y^4} = \left(-\frac{1}{4}\langle\tilde{\Psi}, Q\tilde{\Psi}\rangle - \frac{1}{12}\langle\tilde{\Psi}, M_2(\tilde{\Psi}, \tilde{\Psi})\rangle\right)\Big|_{y^4}. \tag{5.130}$$

Plugging in $\tilde{\Psi} = e^{\lambda y}\Psi_1$ and isolating $y^4$-terms yields

$$S'\left(\tilde{\Psi}\right)|_{y^4} = \left(-\frac{1}{4}\left\langle e^{\lambda y}\Psi_1, Qe^{\lambda y}\Psi_1\right\rangle - \frac{1}{12}\left\langle e^{\lambda y}\Psi_1, e^{-\lambda L_0}\left(e^{-\lambda L_0}e^{\lambda y}\Psi_1 * e^{-\lambda L_0}e^{\lambda y}\Psi_1\right)\right\rangle\right)|_{y^4}$$

$$= \left(-\frac{1}{4}e^{2\lambda y}\left\langle \Psi_1, Q\Psi_1\right\rangle - \frac{1}{12}e^{6\lambda y}\left\langle \Psi_1, (\Psi_1 * \Psi_1)\right\rangle\right)|_{y^4}$$

$$= -\frac{1}{2}\frac{\lambda y^4}{C_{VVV}^2}\left\langle cV, c\partial cV\right\rangle - \frac{1}{2}\frac{\lambda y^4}{C_{VVV}^3}\left\langle cV, (cV * cV)\right\rangle. \qquad (5.131)$$

The correlation functions can be calculated by using standard CFT methods, see e.g. [68, 64]; the result is

$$\langle cV, c\partial cV\rangle = -1,$$

$$\langle cV, (cV * cV)\rangle = C_{VVV}\left(\frac{3\sqrt{3}}{4}\right)^{3y} = C_{VVV}\left(1 + 3y \cdot \ln\left(\frac{3\sqrt{3}}{4}\right)\right) + \mathcal{O}\left(y^2\right). \qquad (5.132)$$

We see by inserting into (5.131) that $S'\left(\tilde{\Psi}\right)|_{y^4}$ indeed vanishes and the value of the on-shell action is the same as the original $S(\Psi)$ to order $y^4$. In principle, terms containing $\Psi_2 \propto y^2$ also contribute at this order. However, since any appearance of $\lambda$ automatically comes with a factor $y$, there are no terms of order $y^4$ containing $\lambda$ as well as $\Psi_2$. All $\Psi_2$-contributions are just the ones already present in $S(\Psi)$ and were studied in detail in [68].

We see that to the first two leading orders, the actions $S'\left(\tilde{\Psi}\right)$ and $\tilde{S}\left(\tilde{\Psi}\right) = S(\Psi)$ give the same on-shell result. Since $S'\left(\tilde{\Psi}\right) = S\left(\tilde{\Psi}'\right)$, where

$$\tilde{\Psi}' = \sum_{n=1}^{\infty} F_n\left(\tilde{\Psi}^{\otimes n}\right) = \pi_1 \mathbf{FP}\frac{1}{1-\Psi}, \qquad (5.133)$$

this suggests that the combination $\mathbf{FP}$ gives rise to a gauge transformation rather than a physically distinct solution. However, a full proof of this statement will be left for future publications.

## 5.5 Conclusion and outlook

We succeeded in providing an explicit consistent description of OSFT with stubs and interestingly found two possible actions with the same equations of motion. The field redefinitions used to convert solutions of Witten OSFT to the new theory are given in explicit form. We hope that the analysis of solutions to the stubbed equations of motion can teach us more general properties of solutions to Maurer-Cartan equations for $A_\infty$- or $L_\infty$-algebras, in particular about the solutions of closed string field theory. One way to proceed would be to transform the whole construction to the sliver frame, where many analytic solutions of OSFT are formulated.

Another possible future direction is to examine wether the stubbed theory is "more well-behaved" in the sense that some typical singularities and ambiguities, for example connected to identity-like solutions, are ameliorated.

**Acknowledgements**

We thank Ted Erler, Jakub Vošmera, Branislav Jurčo, Igor Khavkine and Martin Markl for useful discussions. Our work has been funded by the Grant Agency of Czech Republic under the grant EXPRO 20-25775X.

# Appendix

**Tensor coalgebras**

The tensor coalgebra $TV$ associated to a (graded) vector space $V$ is defined as the Fock space

$$V^{\otimes 0} + V^{\otimes 1} + V^{\otimes 2} + ... \tag{5.134}$$

together with the comultiplication $\Delta : TV \to TV \otimes' TV$ given by

$$\Delta (v_1 \otimes ... \otimes v_n) = \sum_{k=0}^{n} (v_1 \otimes ... \otimes v_k) \otimes' (v_{k+1} \otimes ... \otimes v_n) \tag{5.135}$$

on homogeneous elements and extended by linearity. Here the $v_i$ are elements of $V$ and $\otimes'$ denotes the tensor product arising from a comultiplication, in contrast to the usual $\otimes$. We define the projection operator $\pi_n : TV \to TV$ to project any element on its $n$th tensor power component,

$$\pi_n TV = V^{\otimes n}. \tag{5.136}$$

A linear map $\mathbf{d} : TV \to TV$ is called a coderivation if it satisfies the co-Leibniz rule:

$$\Delta \mathbf{d} = (\mathbf{d} \otimes' \mathbf{1} + \mathbf{1} \otimes' \mathbf{d}) \Delta. \tag{5.137}$$

Linear combinations of coderivations are again coderivations as well as their graded commutator

$$[\mathbf{d}_1, \mathbf{d}_2] = \mathbf{d}_1 \mathbf{d}_2 - (-1)^{deg(\mathbf{d}_1)deg(\mathbf{d}_2)} \mathbf{d}_2 \mathbf{d}_1. \tag{5.138}$$

The product $\mathbf{d}_1 \mathbf{d}_2$ is in general not a coderivation. For any $m$-linear map $d_m : V^{\otimes m} \to V$ one can construct an associated coderivation by the formula

$$\mathbf{d} = \sum_{n=m}^{\infty} \sum_{k=0}^{n-m} 1^{\otimes k} \otimes d_m \otimes 1^{\otimes n-k-m}. \tag{5.139}$$

The co-Leibniz rule guarantees that any coderivation is a sum of terms of this form for different $m$. The individual $m$-products can be recovered as

$$d_m = \pi_1 \mathbf{d} \pi_m. \tag{5.140}$$

If an odd coderivation $\mathbf{d}$ obeys
$$\mathbf{d}^2 = 0 \tag{5.141}$$
then its components $d_m$ form an $A_\infty$-algebra.

A linear map $\mathbf{f}$ is called a cohomomorphism if it fulfills
$$\Delta \mathbf{f} = (\mathbf{f} \otimes' \mathbf{f}) \Delta. \tag{5.142}$$

Linear combinations and products of cohomomorphisms are again cohomomorphisms. Given a family of $m$-products $f_m$ one can construct a unique cohomomorphism via
$$\mathbf{f} = \sum_{j=1}^{\infty} \sum_{k=1}^{\infty} \sum_{m_1+\ldots+m_j=k} f_{m_1} \otimes \ldots \otimes f_{m_j}. \tag{5.143}$$

Again, the individual products can be recovered from $\mathbf{f}$ as
$$f_m = \pi_1 \mathbf{f} \pi_m. \tag{5.144}$$

Of special importance are elements of $TV$ of the form
$$1 + v + v \otimes v + v \otimes v \otimes v + \ldots =: \frac{1}{1-v} \tag{5.145}$$
for some $v \in V$. They fulfill the following useful properties:
$$\pi_1 \mathbf{f} \frac{1}{1-v} = \sum_{m=1}^{\infty} f_m\left(v^{\otimes m}\right), \tag{5.146}$$
$$\mathbf{f} \frac{1}{1-v} = \frac{1}{1 - \pi_1 \mathbf{f} \frac{1}{1-v}} \tag{5.147}$$
for any cohomomorphism $\mathbf{f}$.

A bilinear map $\langle \omega |: TV \times TV \to \mathbb{C}$ is called a symplectic form if it satisfies
$$\langle \omega | \, v_1 \otimes v_2 =: \omega(v_1, v_2) = -(-1)^{deg(v_1)deg(v_2)} \omega(v_2, v_1). \tag{5.148}$$

A multilinear product $m_k$ is called cyclic with respect to $\omega$ if it fulfills
$$\omega(v_1, m_k(v_2, \ldots, v_{k+1})) = -(-1)^{deg(v_1)deg(m_k)} \omega(m_k(v_1, \ldots, v_k), v_{k+1}). \tag{5.149}$$

A coderivation $\mathbf{d}$ is cyclic if all of its components $d_m = \pi_1 \mathbf{d} \pi_m$ are cyclic or equivalently
$$\langle \omega | \, \pi_2 \mathbf{d} = 0. \tag{5.150}$$

Given two symplectic forms $\langle \omega |$, $\langle \omega' |$, a cohomomorphism $\mathbf{f}$ is cyclic if
$$\langle \omega' | \, \pi_2 \mathbf{f} = \langle \omega | \, \pi_2. \tag{5.151}$$

# 6 More on stubs in open string field theory

**Abstract**

We continue our analysis of open string field theory based on $A_\infty$-algebras obtained from Witten's theory by attaching stubs to the elementary vertex. Classical solutions of the new theory can be obtained from known analytic solutions in Witten's theory by applying a cohomomorphism. In a previous work two such cohomomorphisms were found, one non-cyclic, obtained from the homological perturbation lemma and another one by geometric methods. Here we show that to first order in the stub length the two resulting maps are related by a combination of a gauge transformation and a term vanishing on-shell. We also extend our construction to more general gauges and explicitly calculate the first few orders of the new $A_\infty$-algebra solutions in the sliver frame.

## 6.1 Introduction and motivation

Open string field theory (OSFT) as introduced in [85] is described by the action

$$S(\Psi) = \frac{1}{2}\omega(\Psi, Q\Psi) + \frac{1}{3}\omega(\Psi, m_2(\Psi, \Psi)) \quad (6.1)$$

where $Q$ is the BRST-operator, $m_2$ is the Witten star product and $\omega$ is the BPZ-product. Due to the simplicity of the action, a lot of analytical methods have been developed [20] which enabled the discovery of classical analytical solutions, most importantly the tachyon vacuum [70]. Despite those successes, it is as well of interest to analyze modifications of the theory. For once, we would like to have a formulation of OSFT which structure is closer to that of closed string field theory (CSFT): In this way we can not only gain more insight into the more complicated CSFT, but also get a step closer to explicitly formulating a combined open-closed SFT [88]. Moreover, we expect certain singular behaviour of the Witten theory, for instance concerning identity based solutions [1, 20], to be ameliorated in a modified version.

In [71] a specific modification of OSFT is discussed where stubs are attached to the Witten three-vertex[26], i. e.

$$m_2(\cdot, \cdot) \to M_2(\cdot, \cdot) = e^{-\lambda L_0} m_2\left(e^{-\lambda L_0}\cdot, e^{-\lambda L_0}\cdot\right). \quad (6.2)$$

This new product is not associative, which makes it necessary to introduce infinitely many higher products $M_n$ for $n \geq 3$ to ensure gauge invariance of the action. Those higher products are however only unique up to a gauge choice. The whole set $Q$, $M_2$, $M_3$, ... then forms a cyclic $A_\infty$-algebra. Explicitly, the higher products were constructed using a slightly generalized version of homotopy transfer resulting in the following definition: $M_n$ is equal to the sum of all distinct, rooted, full planar binary trees with $n$ leaves such that every leaf represents one input and the root is the output. With every leaf there is one factor of $i = e^{-\lambda L_0}$ associated, with every node the product $m_2$, with every internal line the homotopy $h$, which in Siegel gauge takes the form $h = \frac{e^{-2\lambda L_0}-1}{L_0}b_0$ and with the root $p = e^{-\lambda L_0}$. So for instance,

$$M_3(\Psi_1, \Psi_2, \Psi_3) = e^{-\lambda L_0} m_2 \left( e^{-\lambda L_0}\Psi_1, \frac{e^{-2\lambda L_0}-1}{L_0} b_0 m_2 \left( e^{-\lambda L_0}\Psi_2, e^{-\lambda L_0}\Psi_3 \right) \right)$$
$$+ e^{-\lambda L_0} m_2 \left( \frac{e^{-2\lambda L_0}-1}{L_0} b_0 m_2 \left( e^{-\lambda L_0}\Psi_1, e^{-\lambda L_0}\Psi_2 \right), e^{-\lambda L_0}\Psi_3 \right). \quad (6.3)$$

---

[26]Stubs in open and closed string field theories have been introduced by Zwiebach [87, 88] and further studied in [54, 81]. Toy models for stubs were considered in [5]. More recent works which deal with stubs include [13, 22].

The higher vertices have a nice geometric interpretation: They consist of all the string Feynman diagrams where the propagator is replaced by an integral over strips of the form

$$h = -\int_0^{2\lambda} dt\, e^{-tL_0} b_0 = \frac{e^{-2\lambda L_0}-1}{L_0} b_0. \quad (6.4)$$

Comparing with the standard Schwinger representation of the propagator

$$-\int_0^\infty dt\, e^{-tL_0} b_0 = -\frac{b_0}{L_0} \quad (6.5)$$

one sees that the vertices cover exactly those Riemann surfaces which are missed by the ordinary Feynman diagrams after the inclusion of stubs. This ensures that all Feynman diagrams including the new higher elementary vertices generate a full single cover of the moduli space of bordered Riemann surfaces.

Now the main interest in this stubbed theory lies in studying its classical solutions and how to obtain them from solutions of the standard Witten theory. Defining

$$\mathbf{m} = \mathbf{Q} + \mathbf{m_2}, \qquad \mathbf{M} = \mathbf{Q} + \mathbf{M_2} + \mathbf{M_3} + ... \quad (6.6)$$

as the coderivations encoding the DGA of Witten theory and the $A_\infty$-algebra of the stubbed theory, respectively, the equations of motion can be written concisely as

$$\mathbf{m}\frac{1}{1-\Psi} = 0, \qquad \mathbf{M}\frac{1}{1-\Psi'} = 0. \quad (6.7)$$

In [71], two cohomomorphisms $\mathbf{P}$ and $\mathbf{F}$ were introduced and discussed in detail, both having the property of mapping solutions of one theory to solutions of the other. $\mathbf{P}$ is a non-linear map derived via the homological perturbation lemma and explicitly given up to quadratic order by

$$\pi_1 \mathbf{P}\frac{1}{1-\Psi} = e^{-\lambda L_0}\Psi + e^{-\lambda L_0} m_2 \left( \Psi, \frac{e^{-2\lambda L_0}-1}{L_0} b_0 \Psi \right)$$
$$+ e^{-\lambda L_0} m_2 \left( \frac{e^{-2\lambda L_0}-1}{L_0} b_0 \Psi, e^{-2\lambda L_0}\Psi \right) + \mathcal{O}\left(\Psi^3\right). \quad (6.8)$$

Now $\mathbf{P}$ obeys the chain map relation

$$\mathbf{MP} = \mathbf{Pm}, \quad (6.9)$$

i. e. **P** intertwines between the two algebras. As a result, it maps solutions of Witten theory to solutions of the stubbed theory,

$$\mathbf{M}\frac{1}{1-\pi_1\mathbf{P}\frac{1}{1-\Psi}} = 0 \text{ if } \mathbf{m}\frac{1}{1-\Psi} = 0, \tag{6.10}$$

where the relation (6.181) was used. In contrast, **F** was derived by geometrical methods and reads

$$\pi_1\mathbf{F}\frac{1}{1-\Psi'} = \Psi' + \int_0^\lambda dt \ (e^{-tL_0}m_2\left(e^{-tL_0}b_0\Psi', e^{-tL_0}\Psi'\right) + e^{-tL_0}m_2\left(e^{-tL_0}\Psi', e^{-tL_0}b_0\Psi'\right)$$
$$- e^{-tL_0}b_0m_2\left(e^{-tL_0}\Psi', e^{-tL_0}\Psi'\right)) + \mathcal{O}\left(\Psi'^3\right). \tag{6.11}$$

It fulfills the opposite intertwining relation

$$\mathbf{FM} = \mathbf{mF} \tag{6.12}$$

which gives rise to

$$\mathbf{m}\frac{1}{1-\pi_1\mathbf{F}\frac{1}{1-\Psi'}} = 0 \text{ if } \mathbf{M}\frac{1}{1-\Psi'} = 0 \tag{6.13}$$

hence it maps solutions of the stubbed theory to solutions of the cubic theory. [27]

A fundamental difference between the two occurs by examining the action. Since the Witten action can be written as [84] [28]

$$S(\Psi) = \int_0^1 dt \, \omega\left(\pi_1\partial_t\frac{1}{1-\Psi(t)}, \pi_1\mathbf{m}\frac{1}{1-\Psi(t)}\right) \tag{6.14}$$

we would expect the stubbed action to read

$$S'(\Psi) = \int_0^1 dt \, \omega\left(\pi_1\partial_t\frac{1}{1-\Psi(t)}, \pi_1\mathbf{M}\frac{1}{1-\Psi(t)}\right) \tag{6.15}$$

and indeed, $S'$ is generated by the cohomomorphism **F** via the relation

$$S'(\Psi) = S\left(\pi_1\mathbf{F}\frac{1}{1-\Psi}\right). \tag{6.16}$$

On the contrary we get

$$S\left(\pi_1\mathbf{P}^{-1}\frac{1}{1-\Psi}\right) = \int_0^1 dt \, \omega\left(\pi_1\mathbf{P}^{-1}\partial_t\frac{1}{1-\Psi(t)}, \pi_1\mathbf{P}^{-1}\mathbf{M}\frac{1}{1-\Psi(t)}\right) =: \tilde{S}(\Psi), \tag{6.17}$$

---

[27]The fact that they are naturally defined in the opposite direction stems from the manifestly different derivation, see [71].
[28]Here, $\Psi(t)$ is a smooth interpolation with the properties $\Psi(0) = 0$ and $\Psi(1) = \Psi$.

where $\tilde{S}$ leads to the same equations of motion as $S'$ but takes a significantly different form. The reason is that $\mathbf{P}$ (and so also $\mathbf{P^{-1}}$) is not a cyclic cohomomorphism with respect to $\omega$, hence it does not obey

$$\omega\left(\mathbf{P}\cdot,\mathbf{P}\cdot\right)=\omega\left(\cdot,\cdot\right). \tag{6.18}$$

We can also form the combined transformation $\mathbf{T} =: \mathbf{FP}$ which obeys

$$\mathbf{Tm} = \mathbf{mT} \tag{6.19}$$

and is therefore a symmetry of the Witten equations of motion. However, it manifestly changes the action such that it cannot be a symmetry of the full quantum theory, it rather behaves like some generalized symmetry. Those rather surprising facts raise some interesting questions:

1. How are the two actions physically related? And connected to that

2. what is the physical meaning of the combined transformation $\mathbf{T}$?

3. Is there a more general family of actions which leads to the expected equations of motion?

Moreover, most of the known analytic solutions are formulated in the sliver frame, so to study those explicitly in the stubbed theory we have to ask:

1. Can the whole construction of the higher products and the cohomomorphisms be generalized to the sliver frame?

2. How will explicit solutions of the stubbed theory look like?

3. Can we infer some general structure of Maurer-Cartan elements of $A_\infty$-algebras?

Those questions shall be addressed and answered within this work.

## 6.2 Systematic analysis of the intertwining cohomomorphism

The purpose of this section is to gain a better understanding of the cohomomorphisms $\mathbf{P}$ and $\mathbf{F}$. Therefore we find a general strategy how to obtain intertwining cohomomorphisms and then analyze $\mathbf{P}$ and $\mathbf{F}$ from this perspective. We are looking for a non-linear field redefinition[29] of the form

$$\Psi \to \Psi' = A_1\left(\Psi\right) + A_2\left(\Psi,\Psi\right) + A_3\left(\Psi,\Psi,\Psi\right) + ... \tag{6.20}$$

---

[29]Field redefinitions have been studied using cohomomorphisms already in the context of open superstring field theory with the goal of relating the Berkovits theory to the Munich construction, see [17, ?, 18]). In our work we use them to find the field redefinition between the stubbed theory and Witten theory.

with $A_n$ being a collection of multi-linear maps which ensures that $\Psi'$ is a classical solution of the stubbed theory as long as $\Psi$ is a solution of Witten theory, i. e.

$$\mathbf{M}\frac{1}{1-\Psi'} = 0 \quad \text{if} \quad \mathbf{m}\frac{1}{1-\Psi} = 0. \tag{6.21}$$

Since in the tensor algebra formalism finite transformations are encoded in cohomomorphisms[30], it is natural to package the $A_n$ into a cohomorphism $\mathbf{A}$ in the standard way s. t.

$$\Psi' = \pi_1 \mathbf{A} \frac{1}{1-\Psi}. \tag{6.22}$$

Now applying (6.181)

$$\mathbf{M}\frac{1}{1-\Psi'} = \mathbf{M}\frac{1}{1-\pi_1\mathbf{A}\frac{1}{1-\Psi}} = \mathbf{M}\mathbf{A}\frac{1}{1-\Psi} \tag{6.23}$$

naively suggests to solve for $\mathbf{MA} = \mathbf{m}$, but that is in general not possible since a coderivation composed with a cohomomorphism does not yield a coderivation. However, the combination $\mathbf{A}^{-1}\mathbf{MA}$ is always a coderivation and solving

$$\mathbf{A}^{-1}\mathbf{MA} = \mathbf{m} \quad \text{or} \quad \mathbf{MA} = \mathbf{Am} \tag{6.24}$$

still implies (6.21), hence this is the fundamental relation we will try to solve. In the context of homotopy transfer it appeared as the chain map relation and simply states that $\mathbf{A}$ acts as an intertwiner between the two coderivations which define the algebras. It is easy to see that the set of all intertwiners forms a vector space.

### 6.2.1 Infinitesimal treatment

The object $\mathbf{M}(\lambda)$ is in fact a continuous family of coderivations parametrized by the stub length $\lambda \in [0, \infty)$, obeying $\mathbf{M}(0) = \mathbf{m}$. Similarly, $\mathbf{A}(\lambda_1, \lambda_2)$ is a continuous family with $\mathbf{A}(\lambda_1, \lambda_2) = \mathbf{1}$ for $\lambda_1 = \lambda_2$, hence we can write (6.24) more generally as

$$\mathbf{M}(\lambda_2)\mathbf{A}(\lambda_1, \lambda_2) = \mathbf{A}(\lambda_1, \lambda_2)\mathbf{M}(\lambda_1). \tag{6.25}$$

Suppose we want to know the infinitesimal cohomomorphism which takes us from any fixed $\lambda$ to $\lambda + \delta\lambda$: It will take the form $\mathbf{A}(\lambda, \lambda + \delta\lambda) = \mathbf{1} + \delta\lambda\mathbf{a}(\lambda) + \mathcal{O}(\delta\lambda^2)$ with $\mathbf{a}(\lambda)$ some coderivation. Plugging into (6.25) straightforwardly yields

$$[\mathbf{a}(\lambda), \mathbf{M}(\lambda)] = \frac{d}{d\lambda}\mathbf{M}(\lambda). \tag{6.26}$$

Since this is an equation of coderivations, it is sufficient to examine the projection to one output. Choosing $\lambda = 0$, (6.26) acting on $n$ inputs becomes

---
[30]For a short summary of the tensor algebra formalism see Appendix A.

$$
\begin{align}
n = 1: \quad & \pi_1 \left[ \mathbf{a_1}(0), \mathbf{Q} \right] \pi_1 = \mathbf{0} \tag{6.27} \\
n = 2: \quad & \pi_1([\mathbf{a_1}(0), \mathbf{m_2}] + [\mathbf{a_2}(0), \mathbf{Q}])\pi_2 = \pi_1(-\mathbf{L_0 m_2} - \mathbf{m_2 L_0})\pi_2 \tag{6.28} \\
n = 3: \quad & \pi_1([\mathbf{a_2}(0), \mathbf{m_2}] + [\mathbf{a_3}(0), \mathbf{Q}])\pi_3 = -2\pi_1 \mathbf{m_2}(\mathbf{b_0 m_2})\pi_3 \tag{6.29} \\
n \geq 4: \quad & \pi_1 \left( [\mathbf{a_{n-1}}(0), \mathbf{m_2}] + [\mathbf{a_n}(0), \mathbf{Q}] \right) \pi_n = \mathbf{0} \tag{6.30}
\end{align}
$$

where $\mathbf{L_0}$ and $\mathbf{b_0}$ are the coderivations associated to $L_0$ and $b_0$, see (6.173). At $\lambda = 0$, $\frac{d}{d\lambda} M_n(\lambda)$ vanishes for $n \geq 4$ since it contains $n - 2$ factors of $h$, which are of order $\lambda$. It is straightforward to write down those equations for any finite $\lambda$.

### 6.2.2 Finite transformation

To find the finite intertwiner $\mathbf{A}(\lambda)$, we first need to solve (6.26) for all $\lambda$, which results in

$$
[\mathbf{a_1}(\lambda), \mathbf{M}_n(\lambda)] + [\mathbf{a_2}(\lambda), \mathbf{M}_{n-1}(\lambda)] + ... + [\mathbf{a_n}(\lambda), \mathbf{Q}] = \frac{d}{d\lambda} \mathbf{M}_n(\lambda). \tag{6.31}
$$

for all $n$. This equation has a structural resemblance of the parallel transport equation in gauge theories [60]. If we have some matter field $\psi(x^\mu)$ defined along some curve $\gamma$ in spacetime parametrized by $t$ and some gauge connection $B_\mu$, then we say that $\psi$ is parallel transported along $\gamma$ if it fulfills

$$
\frac{d\psi(\gamma(t))}{dt} = -\frac{dx^\mu}{dt} B_\mu \psi(\gamma(t)). \tag{6.32}
$$

The solution is given by the path-ordered exponential

$$
\psi(\gamma(t)) = \mathcal{P} e^{-\int_0^t ds\, B_\mu \frac{dx^\mu}{ds}} \psi(\gamma(0)). \tag{6.33}
$$

In our case the "connection" $\mathbf{a}(\lambda)$ acts on $\mathbf{M}$ via a commutator. This implies that after integration the path-ordered exponential has to act in the adjoint way, i. e.

$$
\mathbf{M}(\lambda) = \mathcal{P} e^{\int_0^\lambda dt\, \mathbf{a}(t)} \mathbf{m} \mathcal{P} e^{-\int_0^\lambda dt\, \mathbf{a}(t)}, \tag{6.34}
$$

hence we can identify the finite cohomomorphism as

$$
\mathbf{A}(\lambda) = \mathcal{P} e^{\int_0^\lambda dt\, \mathbf{a}(t)}. \tag{6.35}
$$

By expanding the exponential we can write the new solution[31] as

$$\pi_1 \mathbf{A} \frac{1}{1-\Psi} = \Psi + \int_0^\lambda dt\, a_1^t \Psi + \int_0^\lambda dt\, a_2^t (\Psi, \Psi) + \int_0^\lambda dt\, a_3^t (\Psi, \Psi, \Psi) + \int_0^\lambda dt\, a_1^t \left(\int_0^t ds\, a_1^s \Psi\right)$$
$$+ \int_0^\lambda dt\, a_1^t \left(\int_0^t ds\, a_2^s (\Psi, \Psi)\right) + \int_0^\lambda dt\, a_2^t \left(\left(\int_0^t ds\, a_1^s \Psi\right), \Psi\right)$$
$$+ \int_0^\lambda dt\, a_2^t \left(\Psi, \left(\int_0^t ds\, a_1^s \Psi\right)\right) + \int_0^\lambda dt\, a_3^t \left(\left(\int_0^t ds\, a_1^s \Psi\right), \Psi, \Psi\right)$$
$$+ \int_0^\lambda dt\, a_3^t \left(\Psi, \left(\int_0^t ds\, a_1^s \Psi\right), \Psi\right) + \int_0^\lambda dt\, a_3^t \left(\Psi, \Psi, \left(\int_0^t ds\, a_1^s \Psi\right)\right)$$
$$+ \int_0^\lambda dt\, a_1^t \left(\int_0^t ds\, a_3^s (\Psi, \Psi, \Psi)\right) + \int_0^\lambda dt\, a_2^t \left(\int_0^t ds\, a_2^s (\Psi, \Psi), \Psi\right)$$
$$+ \int_0^\lambda dt\, a_2^t \left(\Psi, \int_0^t ds\, a_2^s (\Psi, \Psi)\right) + \mathcal{O}\left(\Psi^{\otimes 4}\right) + \mathcal{O}\left(\lambda^3\right). \tag{6.36}$$

To lighten the notation we have denoted the $t$-dependence of $a(t)$ by a corresponding superscript. In general it is non-trivial to solve the integrals explicitly, the only cohomomorphism we have available in integrated form is $\mathbf{P}$, which was constructed in a manifestly finite way by the homological perturbation lemma.

### 6.2.3 The cyclic cohomomorphism F

It is instructive to check the linearized equations explicitly for the two known cohomomorphisms $\mathbf{F}$ and $\mathbf{P}$. $\mathbf{F}$ was already derived in the infinitesimal form (see [71]) which is given by

$$f_n(\lambda) = \pi \left(-\mathbf{b_0} \mathbf{M_n}(\lambda) + \mathbf{M_n}(\lambda) \mathbf{b_0}\right) \pi_n \tag{6.37}$$

for $n \geq 2$ and $f_1 = 0$ (here $\mathbf{b_0}$ is the coderivation associated to $b_0$). To be consistent with the previous analysis we have to consider $\mathbf{F}^{-1}$ instead of $\mathbf{F}$ though, but on the infinitesimal level this will result only in an overall sign change. By expanding around $\lambda = 0$ we see that only $f_2$ is non-vanishing and the relevant equations become[32]

$$\pi_1 \left[\mathbf{Q}, \mathbf{f_2}(0)\right] \pi_2 = \pi_1 \left(-[\mathbf{Q}, \mathbf{b_0 m_2}] + [\mathbf{Q}, \mathbf{m_2 b_0}]\right) \pi_2 = \pi_1 (-\mathbf{L_0 m_2} - \mathbf{m_2 L_0}) \pi_2, \tag{6.38}$$

$$\pi_1 \left[\mathbf{m_2}, \mathbf{f_2}(0)\right] \pi_3 = \pi_1 \left(-\mathbf{m_2}(\mathbf{b_0 m_2}) + \mathbf{b_0 m_2 m_2} + \mathbf{m_2}(\mathbf{m_2 b_0}) - \mathbf{m_2 b_0 m_2}\right) \pi_3$$
$$= -2\pi_1 \mathbf{m_2}(\mathbf{b_0 m_2}) \pi_3. \tag{6.39}$$

---

[31] We thank Jakub Vošmera for useful discussions and suggestions on that.
[32] In the following $[\cdot, \cdot]$ will always denote a commutator and $\{\cdot, \cdot\}$ will always denote an anticommutator, regardless of the Grassmannality of the entries.

The first equation follows in a simple way from $\{Q, b_0\} = L_0$ and the Leibniz rule $\{\mathbf{Q}, \mathbf{m_2}\} = 0$, whereas the second one uses associativity $\{\mathbf{m_2}, \mathbf{m_2}\} = 0$ and follows after expanding all the coderivations. [33] The finite form of $\mathbf{F}$ is then given according to (6.36) by

$$\pi_1 \mathbf{F} \frac{1}{1-\Psi'} = \Psi' + \int_0^\lambda dt \left( e^{-tL_0} m_2 \left( e^{-tL_0} b_0 \Psi', e^{-tL_0} \Psi' \right) + e^{-tL_0} m_2 \left( e^{-tL_0} \Psi', e^{-tL_0} b_0 \Psi' \right) \right.$$
$$\left. - e^{-tL_0} b_0 m_2 \left( e^{-tL_0} \Psi', e^{-tL_0} \Psi' \right) \right) + \mathcal{O}\left(\Psi'^3\right). \tag{6.40}$$

### 6.2.4 The non-cyclic cohomomorphism P

In [71], $\mathbf{P}$ was derived from the homological perturbation lemma as

$$\mathbf{P} = \mathbf{p}\left(1 - \mathbf{m_2 h}\right)^{-1} \tag{6.41}$$

and given explicitly as a finite transformation which reads to the first few orders

$$P_1 = p$$
$$P_2 = pm_2(\cdot, h\cdot) + pm_2(h\cdot, ip\cdot)$$
$$P_3 = pm_2(\cdot, hm_2(\cdot, h\cdot)) + pm_2(\cdot, hm_2(h\cdot, ip\cdot)) + pm_2(h\cdot, hm_2(ip\cdot, ip\cdot)) + pm_2(hm_2(\cdot, h\cdot), ip\cdot)$$
$$+ pm_2(hm_2(h\cdot, ip\cdot), ip\cdot) + pm_2(h\cdot, ipm_2(\cdot, h\cdot)) + pm_2(h\cdot, ipm_2(h\cdot, ip\cdot)) \ldots \quad . \tag{6.42}$$

Here, the individual maps are given by

$$i = p = e^{-\lambda L_0}, \qquad h = \frac{e^{-2\lambda L_0} - 1}{L_0} b_0. \tag{6.43}$$

As it was argued and proven in [71, 13], $\mathbf{P}$ obeys the chain map relation $\mathbf{MP} = \mathbf{Pm}$ provided one assumes the side conditions $h^2 = hi = ph = 0$ as well as $pi = 1$. In practice this means that in the expansion of $\mathbf{P}$ every $pi$ appearing should be replaced by unity and every term containing one of the side conditions should be set to zero.

We are now interested in the infinitesimal form of $\mathbf{P}$, i. e. taking the "path-ordered logarithm". Expanding around $\lambda = 0$ as $\mathbf{P} = \mathbf{1} + \delta\lambda \mathbf{g}$ yields

$$g_1 = -L_0$$
$$g_2 = -2\pi_1 \mathbf{m_2 b_0} \pi_2$$
$$g_{n \geq 3} = 0 \tag{6.44}$$

and the relevant equations become

$$\pi_1[\mathbf{g_1}, \mathbf{Q}]\pi_1 = -\pi_1[\mathbf{L_0}, \mathbf{Q}]\pi_1 = 0 \tag{6.45}$$

---
[33]When two coderivations appear in parentheses it means that the first one always has to act on the output of the second one, see Appendix.

$$\pi_1 \left([\mathbf{g_1}, \mathbf{m_2}] + [\mathbf{g_2}, \mathbf{Q}]\right) \pi_2 = -\pi_1 \left([\mathbf{L_0}, \mathbf{m_2}] + 2[\mathbf{m_2 b_0}, \mathbf{Q}]\right) \pi_2 = \pi_1 \{\mathbf{L_0}, \mathbf{m_2}\} \pi_2 \qquad (6.46)$$

$$\pi_1 \left([\mathbf{g_2}, \mathbf{m_2}] + [\mathbf{g_3}, \mathbf{Q}]\right) \pi_3 = -2\pi_1 \left[(\mathbf{m_2 b_0}), \mathbf{m_2}\right] \pi_3 = -2\pi_1 \mathbf{m_2} \left(\mathbf{b_0 m_2}\right) \pi_3. \qquad (6.47)$$

Again, all of them can be checked straightforwardly by using the well-known commutation relations of the operators that occur. However, we will now use the structure of $\mathbf{f}$ and $\mathbf{g}$ to determine a more general family of solutions of (6.26). It is worth pointing out that although $\mathbf{F}$ and $\mathbf{P}$ look quite similar when expanded around $\lambda = 0$, their finite versions are fundamentally different: While we have $\mathbf{P}$ available explicitly, $\mathbf{F}$ is only known as a path-ordered exponential. In fact, both expressions are given as expansions in the tensor algebra but to calculate the action of $\mathbf{F}$ we need an extra expansion in the number of integrals.

### 6.2.5 The symmetry map T

At this point it is actually an interesting task to examine the combined map $\mathbf{T} = \mathbf{FP}$ in more detail. We know from $\mathbf{MP} = \mathbf{Pm}$ and $\mathbf{MF^{-1}} = \mathbf{F^{-1} m}$ that

$$\mathbf{Tm} = \mathbf{mT} \qquad (6.48)$$

holds, hence $\mathbf{T}$ commutes with the equations of motion of the Witten theory. This is the generic condition for a symmetry of the equations of motion, however, $\mathbf{T}$ does not preserve the action, which manifestly changes (see (6.17)). Hence it cannot generate a gauge symmetry but rather some kind of generalized symmetry that we will work out now. It is useful to linearize $\mathbf{T}$ around $\lambda = 0$: If $\mathbf{T} = \mathbf{1} + \delta\lambda \mathbf{t}$ then $\mathbf{t}$ is given by

$$\begin{aligned} t_1 &= g_1 - f_1 = -L_0 \\ t_2 &= g_2 - f_2 = -\pi_1 \{\mathbf{b_0}, \mathbf{m_2}\} \pi_2 \\ t_{n \geq 3} &= 0 \end{aligned} \qquad (6.49)$$

and it induces the transformation

$$\Psi \to \Psi - \delta\lambda \left(L_0 \Psi + b_0 m_2 (\Psi, \Psi) + m_2 (b_0 \Psi, \Psi) + m_2 (\Psi, b_0 \Psi)\right). \qquad (6.50)$$

Let us define a gauge parameter $\Lambda = b_0 \Psi$, then (6.50) can be rewritten as

$$\Psi \to \Psi - \delta\lambda \left(Q\Lambda + m_2 (\Lambda, \Psi) + m_2 (\Psi, \Lambda) + b_0 \left(Q\Psi + m_2 (\Psi, \Psi)\right)\right). \qquad (6.51)$$

We see that the transformation we get is a combination of an infinitesimal gauge transformation and a term being proportional to the equations of motion. While the gauge transformation part was expected in a symmetry of the equations of motion, the other part is more interesting and more unconventional. It is this part which is responsible for the change of the off-shell action, however it is also clear that the value of the on-shell action is preserved. This was already conjectured in [71], since this value has physical significance and we did not expect to find a physically distinct solution by applying $\mathbf{T}$.

Extending our analysis to higher orders in $\lambda$ quickly becomes cumbersome and we leave this problem for the future. In principle it is possible that the flow of $\mathbf{T}(\lambda)$ leaves the gauge orbit for a finite $\lambda$, although we do not expect this for physical reasons, see [71].

### 6.2.6 More general solution for the intertwiner

Given $\mathbf{M}(\lambda)$, the most general solution of (6.24) is actually hard to describe explicitly. However, we will now construct a more general family of intertwiners that includes $\mathbf{F}^{-1}$, which was found already. The equations (6.31) are linear in $\mathbf{a}$ with an inhomogeneity so once we found a solution we can add an arbitrary solution of the homogenous equation

$$[\mathbf{a}_1(\lambda), \mathbf{M}_n(\lambda)] + [\mathbf{a}_2(\lambda), \mathbf{M}_{n-1}(\lambda)] + ... + [\mathbf{a}_n(\lambda), \mathbf{Q}] = 0. \tag{6.52}$$

This is the infinitesimal form of

$$\mathbf{AM} = \mathbf{MA}, \tag{6.53}$$

the equation for the symmetry discussed in section 6.2.5. For $n=1$, it reduces to

$$[a_1, Q] = 0 \tag{6.54}$$

which is solved by various operators. The simplest example are arbitrary linear combinations of arbitrary products of Virasoro operators. Another example would be $a_1(\cdot) = m_2(\Phi, \cdot)$ where $\Phi$ is any $Q$-closed string field of ghost number zero. For simplicity we will focus just on linear combinations of Virasoros. Motivated by the form of (6.49) we make the ansatz

$$a_1 = \sum_k v_k L_k, \qquad a_{n\geq 2} = \sum_k v_k \pi_1 \{\mathbf{b_k}, \mathbf{M_n}\} \pi_n. \tag{6.55}$$

This can be directly inserted into (6.52) acting on $n$ elements:

$$\sum_k v_k \pi_1 \left( [\mathbf{L_k}, \mathbf{M_n}] + \sum_{i=2}^{n-1} [(\mathbf{b_k M_i}), \mathbf{M_{n+1-i}}] + [(\mathbf{M_i b_k}), \mathbf{M_{n+1-i}}] + [(\mathbf{b_k M_n}), \mathbf{Q}] + [(\mathbf{M_n b_k}), \mathbf{Q}] \right) \pi_n$$

$$= \sum_k v_k \pi_1 \left( \mathbf{L_k M_n} - \mathbf{M_n L_k} + \sum_{i=2}^{n-1} \mathbf{b_k M_i M_{n+1-i}} - \mathbf{M_{n+1-i}}(\mathbf{b_k M_i}) + \mathbf{M_i b_k M_{n+1-i}} \right) \pi_n$$

$$- \sum_k v_k \pi_1 \left( \mathbf{M_{n+1-i}}(\mathbf{M_i b_k}) + \mathbf{b_k M_n Q} - \mathbf{Q b_k M_n} + \mathbf{M_n b_k Q} - \mathbf{Q M_n b_k} \right) \pi_n$$

$$= \sum_k v_k \pi_1 \left( \mathbf{b_k Q M_n} - \mathbf{M_n Q b_k} + \sum_{i=2}^{n-1} \mathbf{b_k M_i M_{n+1-i}} - \mathbf{M_{n+1-i}}(\mathbf{b_k M_i}) \right) \pi_n$$

$$+ \sum_k v_k \pi_1 \left( \mathbf{M_i b_k M_{n+1-i}} - \mathbf{M_{n+1-i}}(\mathbf{M_i b_k}) + \mathbf{b_k M_n Q} - \mathbf{Q M_n b_k} \right) \pi_n. \tag{6.56}$$

Using the $A_\infty$-relation

$$\{\mathbf{M_n}, \mathbf{Q}\} + \sum_{i=2}^{n-1} \mathbf{M_i M_{n+1-i}} = 0 \tag{6.57}$$

this expression can be shown to vanish, which proves our ansatz to be correct.

We can also show that our solution for **a** is again a combination of a gauge transformation and field redefinition proportional to the equations of motion: If we define analogously to section 6.2.5

$$\Lambda = \sum_k v_k b_k \Psi \tag{6.58}$$

then we have

$$\pi_1 \mathbf{a} \frac{1}{1-\Psi} = Q\Lambda + \sum_{n=2}^{\infty} M_n\left(\Lambda, \Psi^{\otimes n-1}\right) + M_n\left(\Psi, \Lambda, \Psi^{\otimes n-2}\right) + ... + M_n\left(\Psi^{\otimes n-1}, \Lambda\right)$$
$$+ \sum_k v_k b_k \left(Q\Psi + \sum_{n=2}^{\infty} M_n\left(\Psi^{\otimes n}\right)\right). \tag{6.59}$$

To sum up, the family

$$a_1 = \sum_k v_k L_k, \qquad a_{n \geq 2} = \sum_k \pi_1\left((v_k + \delta_{0k}) \mathbf{b_k M_n} + (v_k - \delta_{0k}) \mathbf{M_n b_k}\right) \pi_n \tag{6.60}$$

provides an infinitesimal intertwiner for all possible $v_k$ where $v_k = 0$ corresponds to $\mathbf{F}^{-1}$.

### 6.2.7 Cyclicity and invariance of the action

In this section we shall analyze under which conditions the transformation **A** generates the expected $A_\infty$-action $S'$ (6.15). For that, **A** not only needs to be an intertwiner but also be compatible with the symplectic form $\omega$, i. e. the last equation in

$$S(\Psi) = \int_0^1 dt\, \omega\left(\pi_1 \partial_t \frac{1}{1-\Psi(t)}, \pi_1 \mathbf{m} \frac{1}{1-\Psi(t)}\right)$$
$$= \int_0^1 dt\, \omega\left(\pi_1 \partial_t \left(\mathbf{A}^{-1} \mathbf{A} \frac{1}{1-\Psi(t)}\right), \pi_1 \mathbf{A}^{-1} \mathbf{M A} \frac{1}{1-\Psi(t)}\right)$$
$$= \int_0^1 dt\, \omega\left(\pi_1 \partial_t \mathbf{A}^{-1} \frac{1}{1-\pi_1 \mathbf{A} \frac{1}{1-\Psi(t)}}, \pi_1 \mathbf{A}^{-1} \mathbf{M} \frac{1}{1-\pi_1 \mathbf{A} \frac{1}{1-\Psi(t)}}\right)$$
$$= \int_0^1 dt\, \omega\left(\pi_1 \mathbf{A}^{-1} \partial_t \frac{1}{1-\Psi'(t)}, \pi_1 \mathbf{A}^{-1} \mathbf{M} \frac{1}{1-\Psi'(t)}\right)$$
$$= \int_0^1 dt\, \omega\left(\pi_1 \partial_t \frac{1}{1-\Psi'(t)}, \pi_1 \mathbf{M} \frac{1}{1-\Psi'(t)}\right) = S'(\Psi) \tag{6.61}$$

needs to be true. Note that $\partial_t$ and **A** commute since **A** does not depend on $t$.

Cyclicity of a cohomomorphism is actually a delicate question: In [84] it is stated as the condition

$$\omega\left(\mathbf{A}\cdot, \mathbf{A}\cdot\right) = \omega\left(\cdot, \cdot\right) \tag{6.62}$$

but one has to be precise on what type of elements it is supposed to act. It is quite clear that the relation is too restrictive to act on arbitrary elements: One would get

$$\omega\left(A_1(\Psi_1) + A_2(\Psi_1, \Psi_2) + A_3(\Psi_1, \Psi_2, \Psi_3) + \cdots, A_1(\phi_1) + A_2(\phi_1, \phi_2) + A_3(\phi_1, \phi_2, \phi_3) + \cdots\right)$$
$$= \omega(\Psi_1, \phi_1) \tag{6.63}$$

which would imply (at least in the case where $A_1$ is invertible, which is equivalent to $\mathbf{A}$ being invertible) that every output of $A_{n\geq 2}$ is orthogonal to any possible $\Psi$. Since $\omega$ is non-degenerate, we would conclude that the $A_{n\geq 2}$ all have to be identically zero, which is not what we want. Even if we only allow group-like inputs, i. e. elements of the form $\frac{1}{1-\Psi}$, the same argument shows that the two inputs have to be identical. By looking at (6.14) however we see that we need a generalization of that by allowing coderivations to act on the group-like inputs. The equation

$$\omega\Big(\pi_1\left(\mathbf{A_1 d_1}(\Psi) + \mathbf{A_2 d_1}\left(\Psi^{\otimes 2}\right) + \mathbf{A_1 d_2}\left(\Psi^{\otimes 2}\right) + \cdots\right),$$
$$\pi_1\left(\mathbf{A_1 d'_1}(\Psi) + \mathbf{A_2 d'_1}\left(\Psi^{\otimes 2}\right) + \mathbf{A_1 d'_2}\left(\Psi^{\otimes 2}\right) + \cdots\right)\Big)$$
$$= \omega\left(\pi_1 \mathbf{d}\frac{1}{1-\Psi}, \pi_1 \mathbf{d'}\frac{1}{1-\Psi}\right) \tag{6.64}$$

actually makes sense also for non-trivial $A_{n\geq 2}$ because we can have

$$\omega\left(A_1 \pi_1 \mathbf{d}\frac{1}{1-\Psi}, A_1 \pi_1 \mathbf{d'}\frac{1}{1-\Psi}\right) = \omega\left(\pi_1 \mathbf{d}\frac{1}{1-\Psi}, \pi_1 \mathbf{d'}\frac{1}{1-\Psi}\right) \tag{6.65}$$

while the higher terms of a given order in $\Psi$ cancel each other, even if $\mathbf{d} \neq \mathbf{d'}$. For an infinitesimal $\mathbf{A} = \mathbf{1} + \epsilon \mathbf{a} + \mathcal{O}(\epsilon^2)$ it boils down to the condition that $\mathbf{a}$ is cyclic coderivation. To sum up, we define a cohomomorphism $\mathbf{A}$ to be cyclic with respect to $\omega$ if

$$\omega\left(\pi_1 \mathbf{A d}\frac{1}{1-\Psi}, \pi_1 \mathbf{A d'}\frac{1}{1-\Psi}\right) = \omega\left(\pi_1 \mathbf{d}\frac{1}{1-\Psi}, \pi_1 \mathbf{d'}\frac{1}{1-\Psi}\right) \tag{6.66}$$

for arbitrary $\mathbf{d}$, $\mathbf{d'}$ and $\Psi$.

With this definition we can immediately analyze the cyclicity properties of $\mathbf{F}^{-1}$ and $\mathbf{P}$: With the infinitesimal $\mathbf{f}$ given by (6.37) we get

$$\begin{aligned}\omega(\Psi_1, f_n(\Psi_2, ..., \Psi_{n+1})) &= -\omega(\Psi_1, b_0 M_n(\Psi_2, ..., \Psi_{n+1})) + \omega(\Psi_1, M_n(\mathbf{b_0}(\Psi_2, ..., \Psi_{n+1}))) \\ &= -\omega(b_0 \Psi_1, M_n(\Psi_2, ..., \Psi_{n+1})) - \omega(M_n(\Psi_1, \mathbf{b_0}(\Psi_2, ..., \Psi_n)), \Psi_{n+1}) \\ &\quad - \omega(M_n(\Psi_1, ..., \Psi_n), b_0 \Psi_{n+1}) \\ &= -\omega(M_n(b_0 \Psi_1, \Psi_2, ..., \Psi_n), \Psi_{n+1}) - \omega(M_n(\Psi_1, \mathbf{b_0}(\Psi_2, ..., \Psi_n)), \Psi_{n+1}) \\ &\quad + \omega(b_0 M_n(\Psi_1, ..., \Psi_n), \Psi_{n+1}) \\ &= -\omega(M_n \mathbf{b_0}(\Psi_1, \Psi_2, ..., \Psi_n), \Psi_{n+1}) + \omega(b_0 M_n(\Psi_1, ..., \Psi_n), \Psi_{n+1}) \\ &= -\omega(f_n(\Psi_1, ..., \Psi_n) \Psi_{n+1}), \tag{6.67}\end{aligned}$$

hence **f** is cyclic and so is the finite version

$$\mathbf{F^{-1}} = \mathcal{P}e^{\int_0^\lambda dt\, \mathbf{f}(t)}. \tag{6.68}$$

In contrast, for **P** we already see at first order that $f_1' = -L_0$ is not cyclic because the sign does not match:

$$-\omega\left(\Psi_1, L_0\Psi_2\right) = -\omega\left(L_0\Psi_1, \Psi_2\right). \tag{6.69}$$

This implies that $P_1 = e^{-\lambda L_0}$ is not "unitary" with respect to the BPZ-product as it would be required for a cyclic cohomomorphism.

From the results of section 2.6 we also deduce that $\mathbf{F^{-1}}$ is not unique as a cyclic intertwiner: Taking a BPZ-odd choice for $a_1$ in (6.60), i. e. demanding $v_k = -\left(-1\right)^k v_{-k}$ leads to a cyclic coderivation and in turn to a cyclic cohomomorphism by a similar argument as in (6.67).

## 6.3 Generalized stubs

We now want to go a step further and allow for more general stub operators, especially non-BPZ-even ones. The motivation behind that is that we want to apply our construction to explicit analytic solutions of OSFT. While it is in principle straightforward to do that, we face a technical problem: The most important solutions, like for instance the tachyon vacuum ([70, 27]), are formulated in the sliver frame in terms of the $KBc$-algebra. The action of $i$, $p$ and $h$ would take us outside the $KBc$-algebra and is therefore impractical for actual calculations. It would be much more natural to use the sliver frame analogue of the stub operator, i. e. replace $e^{-\lambda L_0}$ by $e^{-\lambda \mathcal{L}_0}$. We will first discuss general aspects of non-BPZ-even stub operators and provide a careful treatment of the operator $e^{-\lambda \mathcal{L}_0}$ in section 6.3.4.

### 6.3.1 Algebraic aspects

Let us consider a generalized stub operator of the form

$$e^{-\lambda \sum v_k L_k} =: e^{-\lambda L} \tag{6.70}$$

with some real coefficients $v_k$. An important example is given by the family

$$L = L_t = L_0 + 2\sum_{k=1}^{\infty} \frac{(-1)^{k+1}}{4k^2 - 1} e^{-2tk} L_{2k} \tag{6.71}$$

which interpolates between the Siegel gauge and sliver gauge stub: For $t = 0$ we get $\mathcal{L}_0$, whereas in the limit of $t \to \infty$ we recover $L_0$. The most important new algebraic aspect is that $e^{-\lambda L}$ is not BPZ-even since in general $L^* \neq L$. Hence, the naive choice $p = i = e^{-\lambda L}$ would not result in cyclic products, we need to define

$$p = e^{-\lambda L^*} \tag{6.72}$$

instead. While this small change seems innocuous at first sight, it also affects the Hodge-Kodaira relation and therefore our possible choices of $h$, which we use to construct the higher vertices.

To motivate our general construction of $h$ let us first consider the special case of $L = \mathcal{L}_0$, postponing the discussion of potential geometrical subtleties to section 6.3.4. The right-hand side of

$$hQ + Qh = ip - 1 \tag{6.73}$$

evaluates to

$$e^{-\lambda \mathcal{L}_0} e^{-\lambda \mathcal{L}_0^*} - 1 = e^{(e^{-\lambda} - 1)(\mathcal{L}_0 + \mathcal{L}_0^*)} - 1 \tag{6.74}$$

using the algebraic relations [70]

$$x^{\mathcal{L}_0} y^{\mathcal{L}_0^*} = \left(\frac{1}{1 + \frac{x}{y} - x}\right)^{\mathcal{L}_0^*} \left(\frac{1}{1 + \frac{y}{x} - y}\right)^{\mathcal{L}_0}, \qquad x^{\mathcal{L}_0^*} x^{\mathcal{L}_0} = e^{\left(1 - \frac{1}{x}\right)(\mathcal{L}_0 + \mathcal{L}_0^*)}. \tag{6.75}$$

A natural choice for $h$ would now be

$$h_{\hat{\mathcal{B}}_0} = \frac{e^{(e^{-\lambda} - 1)(\mathcal{L}_0 + \mathcal{L}_0^*)} - 1}{(\mathcal{L}_0 + \mathcal{L}_0^*)} (\mathcal{B}_0 + \mathcal{B}_0^*) = -\int_0^{1 - e^{-\lambda}} dt\, (\mathcal{B}_0 + \mathcal{B}_0^*)\, e^{-t(\mathcal{L}_0 + \mathcal{L}_0^*)}. \tag{6.76}$$

This expression is manifestly non-singular: If $\mathcal{L}_0 + \mathcal{L}_0^*$ yields zero on some state (which would for example formally the case for the sliver state) then no pole is produced. It corresponds to the propagator in $\hat{\mathcal{B}}_0$-gauge

$$-\int_0^\infty dt\, (\mathcal{B}_0 + \mathcal{B}_0^*)\, e^{-t(\mathcal{L}_0 + \mathcal{L}_0^*)} = -\frac{\mathcal{B}_0 + \mathcal{B}_0^*}{\mathcal{L}_0 + \mathcal{L}_0^*} \equiv -\frac{\hat{\mathcal{B}}_0}{\hat{\mathcal{L}}_0}. \tag{6.77}$$

From an algebraic perspective, this propagator is quite convenient and leads to a simple set of vertices.

The solution for $h_{\hat{\mathcal{B}}_0}$ relied heavily on the special algebraic properties of $\mathcal{L}_0$ and $\mathcal{L}_0^*$. To find a solution for a generic $L$, most importantly $L_t$ defined above, we can take

$$h_{\hat{B}} = -\int_0^\lambda dt\, e^{-tL} (B + B^*)\, e^{-tL^*} \tag{6.78}$$

with $B$ defined as

$$B = \sum_k v_k B_k. \tag{6.79}$$

For $L = \mathcal{L}_0$ and $B = \mathcal{B}_0$ it reduces to (6.76). Plugging into the Hodge-Kodaira relation and using $\{Q, B\} = L$ we get

$$Qh_{\hat{B}} + h_{\hat{B}} Q = -\int_0^\lambda dt\, e^{-tL} (L + L^*)\, e^{-tL^*} = \int_0^\lambda dt\, \frac{d}{dt} \left(e^{-tL} e^{-tL^*}\right) = e^{-\lambda L} e^{-\lambda L^*} - 1 = ip - 1 \tag{6.80}$$

as desired. Hence we succeeded to find a well-defined solution for the homotopy for any generalized stub $e^{-\lambda L}$.

### 6.3.2 Geometric aspects

As we discussed in [71] already, we have to ensure that the Feynman diagrams constructed out of the vertices and the propagator provide a full single cover of the moduli space of bordered punctured Riemann surfaces. This implies that the higher elementary vertices must include precisely those surfaces which are missed after attaching the stubs. What changes compared to [71] is that $e^{-\lambda L}$ is not the time evolution operator in radial quantization anymore and induces a non-trivial distortion of the worldsheet surface [42].

We have seen above that every choice of stub operator $e^{-\lambda L}$ is naturally associated to a gauge condition given by
$$B\Psi = 0. \tag{6.81}$$
However, the homotopy $h_{\hat{B}}$ we constructed in (6.78) is part of the propagator in $\hat{B}$-gauge, i. e. where the gauge condition
$$(B + B^*)\Psi = 0 \tag{6.82}$$
is imposed. We could ask now if there is also a choice for $h$ that corresponds to $B$-gauge directly. After all, analytic solutions have been found in $\mathcal{B}_0$-gauge, not $\hat{\mathcal{B}}_0$-gauge, where they become singular. Hence an $h$ that corresponds to sliver gauge (i. e. $\mathcal{B}_0\Psi = 0$) would seem more natural. To answer that, we will use some geometric input from computing amplitudes in general linear $B$-gauges.

In [42] a condition on $B = \sum_k v_k B_k$ was given that ensures that all tree level amplitudes can be computed unambiguously. In terms of the vector field $v(\xi) = \sum_k v_k \xi^{k+1}$ this condition reads
$$\operatorname{Re}\left(\bar{\xi} v(\xi)\right) > 0 \quad \text{for} \quad |\xi| = 1. \tag{6.83}$$
It is for example obeyed for the family
$$B_t =: e^{tL_0} \mathcal{B}_0 e^{-tL_0} = b_0 + 2\sum_{k=1}^{\infty} \frac{(-1)^{k+1}}{4k^2 - 1} e^{-2tk} b_{2k} \tag{6.84}$$
associated to the interpolation (6.71) as long as $t$ is strictly greater than zero. This means that for sliver gauge the condition is marginally violated, see section 6.3.4. For this section we will assume that (6.83) holds for our choice of $v_k$. The propagator in a general $B$-gauge was derived in [42] to be
$$-\frac{B^*}{L^*} Q \frac{B}{L} \quad \text{on odd ghost number states,} \quad -\frac{B}{L} Q \frac{B^*}{L^*} \quad \text{on even ghost number states,} \tag{6.85}$$
so it contains two Schwinger parameters instead of one. The result is an infinite overcounting of the moduli space, every surface is now additionally integrated over from zero to infinity. The reason why the theory is still unitary and produces the right values for amplitudes is the presence of $Q$ in the propagator, which cancels the overcounting.

To derive the correct form of $h_B$ in $B$-gauge let us analyze the on-shell four-amplitude in Witten theory as well as in the stubbed theory. In Witten theory there is no elementary 4-vertex and the whole amplitude is given by the Feynman region:

$$\mathcal{A}_4 = -\omega\left(\Psi_1, m_2\left(\Psi_2, \frac{B}{L}Q\frac{B^*}{L^*}m_2\left(\Psi_3, \Psi_4\right)\right)\right) + \text{perm.} \tag{6.86}$$

Here, perm. stands for the t-channel contribution obtained by a cyclic permutation where $\frac{B}{L}Q\frac{B^*}{L^*}m_2$ acts on $\Psi_2$ and $\Psi_3$. In the stubbed theory we have to sum the Feynman region and the vertex region:

$$\mathcal{A}_4 = -\omega\left(\Psi_1, M_2\left(\Psi_2, \frac{B}{L}Q\frac{B^*}{L^*}M_2\left(\Psi_3, \Psi_4\right)\right)\right) + \text{perm.} + \omega\left(\Psi_1, M_3\left(\Psi_2, \Psi_3, \Psi_4\right)\right), \tag{6.87}$$

where

$$M_3\left(\cdot, \cdot, \cdot\right) = e^{-\lambda L^*} m_2\left(e^{-\lambda L}\cdot, h_B m_2\left(e^{-\lambda L}\cdot, e^{-\lambda L}\cdot\right)\right) + e^{-\lambda L^*} m_2\left(h_B m_2\left(e^{-\lambda L}\cdot, e^{-\lambda L}\cdot\right), e^{-\lambda L}\cdot\right). \tag{6.88}$$

If we only focus on the s-channel (including the part of $M_3$ that "extends" to the s-channel) the expression becomes

$$\mathcal{A}_{4s} = -\omega\left(\Psi_1, e^{-\lambda L^*} m_2\left(e^{-\lambda L}\Psi_2, e^{-\lambda L}\frac{B}{L}Q\frac{B^*}{L^*}e^{-\lambda L^*} m_2\left(e^{-\lambda L}\Psi_3, e^{-\lambda L}\Psi_4\right)\right)\right)$$
$$+ \omega\left(e^{-\lambda L}\Psi_1, m_2\left(e^{-\lambda L}\Psi_2, h_B m_2\left(e^{-\lambda L}\Psi_3, e^{-\lambda L}\Psi_4\right)\right)\right) \tag{6.89}$$

Since we take the external states to be on-shell, the stub operators acting directly on $\Psi_i$ do not matter and we get

$$\mathcal{A}_{4s} = -\omega\left(\Psi_1, m_2\left(\Psi_2, e^{-\lambda L}\frac{B}{L}Q\frac{B^*}{L^*}e^{-\lambda L^*} m_2\left(\Psi_3, \Psi_4\right)\right)\right) + \omega\left(\Psi_1, m_2\left(\Psi_2, h_B m_2\left(\Psi_3, \Psi_4\right)\right)\right). \tag{6.90}$$

Comparing with (6.86) would give us $h_B$ when acting on a ghost number two state as

$$e^{-\lambda L}\frac{B}{L}Q\frac{B^*}{L^*}e^{-\lambda L^*} - \frac{B}{L}Q\frac{B^*}{L^*} \tag{6.91}$$

up to $Q$-exact terms. The sliver gauge propagator was dependent on the ghost number of the input; if we demand that this property should also hold for the homotopy we arrive at

$$h_B = \left(e^{-\lambda L}\frac{B}{L}Q\frac{B^*}{L^*}e^{-\lambda L^*} - \frac{B}{L}Q\frac{B^*}{L^*}\right)P_+ + \left(e^{-\lambda L}\frac{B^*}{L^*}Q\frac{B}{L}e^{-\lambda L^*} - \frac{B^*}{L^*}Q\frac{B}{L}\right)P_-, \tag{6.92}$$

where $P_+$ ($P_-$) is the projector on states of even (odd) ghost number. Now we can verify that $h_B$ also obeys the Hodge-Kodaira relation (6.73). Note that the dependence on the ghost number is crucial for that to work. We point out that this construction of $h_B$ was purely governed by the consistency of the on-shell amplitudes. We see that it is natural for $h_B$ to be in the same gauge as the propagator which is used to compute amplitudes. An interesting point is that (6.92) can contain poles if $L$ or $L^*$ give zero on some state, in contrast to $h_{\hat{\mathcal{B}}_0}$ (6.76). This is particularly important in the sliver frame limit and will be discussed in section 6.3.4.

To sum up, we are now able to construct the higher products in exactly the same way as in [71] but with $i = e^{-\lambda L}$, $p = e^{-\lambda L^*}$ and $h$ equal to (6.92) or (6.78), depending on the purposes.

### 6.3.3 Intertwining cohomomorphisms

Let us see now what changes regarding the intertwiners if we are working with generalized stubs. Actually the construction of **P** is very simple: One can just use the modified homotopy transfer formula of [71] and replace $i$, $p$ and $h$ by their generalized counterparts defined above. The only necessary algebraic ingredients were the chain map relations $Qi = iQ$, $Qp = pQ$ as well as (6.73), which we have shown to be true. Hence we conclude that **P** constructed this way obeys the perturbed chain map relation

$$\mathbf{Pm} = \mathbf{MP} \tag{6.93}$$

as desired.

To find a cyclic intertwiner that also preserves the form of the action is a bit more involved: Motivated by the explicit form of $\mathbf{f}(\lambda)$ in the standard frame we propose the following ansatz:

$$f_1 = 0 \quad \text{and} \quad f_n(\lambda) = \pi_1 \left( \mathbf{X}^* \mathbf{M_n}(\lambda) - \mathbf{M_n}(\lambda) \mathbf{X} \right) \pi_n \tag{6.94}$$

Cyclicity then follows automatically by (6.67) with $X$ replacing $b_0$. Looking at equations (6.38), (6.39) in combination with (6.26) we can deduce the following conditions on $X$:

$$\{Q, X\} = L, \tag{6.95}$$

$$X + X^* = -\frac{d}{d\lambda} h \mid_{\lambda=0} \tag{6.96}$$

In $\hat{B}$-gauge it is actually simple because

$$-\frac{d}{d\lambda} h_{\hat{B}} \mid_{\lambda=0} = \hat{B} \tag{6.97}$$

and the natural choice $X = B$ provides a solution.

For $B$- gauge the situation is slightly more complicated, we get

$$-\frac{d}{d\lambda} h_B \mid_{\lambda=0} = \left( BQ \frac{B^*}{L^*} + \frac{B}{L} QB^* \right) P_+ + \left( L \frac{B^*}{L^*} Q \frac{B}{L} + \frac{B^*}{L^*} Q \frac{B}{L} L^* \right) P_-. \tag{6.98}$$

To solve both conditions we can define

$$X = BQ \frac{B^*}{L^*} P_+ + L \frac{B^*}{L^*} Q \frac{B}{L} P_-. \tag{6.99}$$

To prove that our ansatz is indeed correct we need to insert into (6.31):

$$\pi_1 \left( [\mathbf{X}^* \mathbf{M_n}, \mathbf{Q}] - [\mathbf{M_n} \mathbf{X}, \mathbf{Q}] + \sum_{i=2}^{n-1} [(\mathbf{X}^* \mathbf{M_{n+1-i}}), \mathbf{M_i}] - [\mathbf{M_{n+1-i}} \mathbf{X}, \mathbf{M_i}] \right) \pi_n = \frac{d}{d\lambda} M_n \tag{6.100}$$

The r. h. s. consists of two parts: One where the derivative acts on the stubs and one where it acts on the homotopy. The action on the stubs just brings down a factor of $-L$ or $-L^*$ whereas the second part consists of all tree diagrams with one internal line replaced by $\frac{d}{d\lambda} h_B(\lambda)$. The replaced line divides the tree into two subtrees with $i$ and $n+1-i$ leaves, respectively. The sum of those subtrees form the products $M_i$ and $M_{n+1-i}$ again such that we can write

$$\frac{d}{d\lambda} M_n = \pi_1 \left( -\mathbf{L}^* \mathbf{M_n} - \mathbf{M_n} \mathbf{L} - \sum_{i=2}^{n-1} \mathbf{M_{n+1-i}}, ((\mathbf{X} + \mathbf{X}^*) \mathbf{M_i}) \right) \pi_n. \tag{6.101}$$

Indeed, the leaf of $M_{n+1-i}$ and the root of $M_i$ combine to give

$$e^{-\lambda L} (-X - X^*) e^{-\lambda L^*} = e^{-\lambda L} \frac{d}{d\lambda} h_B \mid_{\lambda=0} e^{-\lambda L^*} = \frac{d}{d\lambda} h_B(\lambda) \tag{6.102}$$

as desired. We can now manipulate the l. h. s. using the $A_\infty$-relation

$$\{\mathbf{M_n}, \mathbf{Q}\} + \sum_{i=2}^{n-1} \mathbf{M_{n+1-i}} \mathbf{M_i} = 0 \tag{6.103}$$

and get

$$\pi_1 \Big( \mathbf{X}^* \{\mathbf{M_n}, \mathbf{Q}\} - \{\mathbf{Q}, \mathbf{X}^*\} \mathbf{M_n} - \mathbf{M_n} \{\mathbf{X}, \mathbf{Q}\} + \{\mathbf{M_n}, \mathbf{Q}\} \mathbf{X}$$
$$+ \sum_{i=2}^{n-1} \mathbf{X}^* \mathbf{M_{n+1-i}} \mathbf{M_i} - \mathbf{M_i} (\mathbf{X}^* \mathbf{M_{n+1-i}}) - \mathbf{M_{n+1-i}} \mathbf{X} \mathbf{M_i} + \mathbf{M_i} (\mathbf{M_{n+1-i}} \mathbf{X}) \Big) \pi_n$$
$$= \pi_1 \left( -\mathbf{L}^* \mathbf{M_n} - \mathbf{M_n} \mathbf{L} - \sum_{i=2}^{n-1} \mathbf{M_{n+1-i}}, ((\mathbf{X} + \mathbf{X}^*) \mathbf{M_i}) \right) \pi_n \tag{6.104}$$

and hence exactly the r. h. s. The finite cohomomorphism can now be computed again as the path-ordered exponential:[34]

$$\mathbf{F} = \mathcal{P} e^{\int_0^\lambda dt\, \mathbf{f}(t)}. \tag{6.105}$$

### 6.3.4 The sliver frame limit

As already stated above, the sliver gauge condition marginally violates the regularity condition (6.83) which has the following geometric reason: The stub operator $e^{-\lambda L_t}$, which is also used in the Schwinger representation of the propagator, does not attach just a rectangular strip, but a more general surface to the world sheet. For $t = 0$ this distortion of the strip becomes singular in the sense that the string midpoint is pushed to an infinite distance. This means that every stub $e^{-\lambda \mathcal{L}_0}$ independently of $\lambda$ covers an infinitely long region on the Riemann surface. The same problem also concerns the operator $e^{-\lambda(\mathcal{L}_0 + \mathcal{L}_0^*)}$. This raises the question of potential singularities and it also makes it less obvious which region of moduli space will be covered by the higher vertices.

---

[34]Observe that $\mathbf{F}$ is now defined for simplicity in the same direction as $\mathbf{P}$, so it corresponds to $\mathbf{F}^{-1}$ in [71].

Moreover, there is also a caveat related to the Schwinger representation: Since naively one would expect

$$\frac{1}{L_t} = \lim_{\Lambda \to \infty} \int_0^\Lambda dt\, e^{-tL_t} = \lim_{\Lambda \to \infty} \frac{1}{L_t}\left(1 - e^{-\Lambda L_t}\right), \quad (6.106)$$

we must ensure that the second term gives zero contribution, i. e. it must produce a surface on the boundary of the moduli space. It has been shown that this is the case for $t$ strictly bigger than zero but not for $t = 0$ [42].

There are indeed singularities showing up in the sliver frame limit: If we look at the homotopy $h_{\mathcal{B}_0}$ (6.92) it takes the form

$$h_{\mathcal{B}_0} = \left(e^{-\lambda \mathcal{L}_0}\frac{\mathcal{B}_0}{\mathcal{L}_0}Q\frac{\mathcal{B}_0^*}{\mathcal{L}_0^*}e^{-\lambda \mathcal{L}_0^*} - \frac{\mathcal{B}_0}{\mathcal{L}_0}Q\frac{\mathcal{B}_0^*}{\mathcal{L}_0^*}\right)P_+ + \left(e^{-\lambda \mathcal{L}_0}\frac{\mathcal{B}_0^*}{\mathcal{L}_0^*}Q\frac{\mathcal{B}_0}{\mathcal{L}_0}e^{-\lambda \mathcal{L}_0^*} - \frac{\mathcal{B}_0^*}{\mathcal{L}_0^*}Q\frac{\mathcal{B}_0}{\mathcal{L}_0}\right)P_-. \quad (6.107)$$

Within the $KBc$-algebra there exist a couple of states on which this expression is singular: While at ghost number zero there are no singularities showing up, at ghost number one we have $e^{-\frac{K}{2}}cKe^{-\frac{K}{2}}$, $e^{-\frac{K}{2}}Kce^{-\frac{K}{2}}$, $e^{-\frac{K}{2}}cKBce^{-\frac{K}{2}}$ and $e^{-\frac{K}{2}}ce^{-\frac{K}{2}}$ and at ghost number two $e^{-\frac{K}{2}}cKcKe^{-\frac{K}{2}}$ and $e^{-\frac{K}{2}}KcKce^{-\frac{K}{2}}$ (and linear combinations thereof). The appearance of the zero-momentum tachyon in this list might seem surprising since it has $\mathcal{L}_0$-eigenvalue minus one, but the operator $e^{-\lambda \mathcal{L}_0^*}$ creates a level zero state out of it, see section 6.4.1. One should notice that for the ghost number two states, it is not $\frac{1}{\mathcal{L}_0^*}$ but $\frac{1}{\mathcal{L}_0}$ on the left side of the expression which creates the singularity.

We can increase the range of definition of $h_{\mathcal{B}_0}$ to all $Q$-closed states, which includes $e^{-\frac{K}{2}}cKBce^{-\frac{K}{2}}$, $e^{-\frac{K}{2}}cKcKe^{-\frac{K}{2}}$ and $e^{-\frac{K}{2}}KcKce^{-\frac{K}{2}}$ by the following trick: We rewrite the sliver gauge propagator as

$$\frac{\mathcal{B}_0^*}{\mathcal{L}_0^*}Q\frac{\mathcal{B}_0}{\mathcal{L}_0}P_- \cong \left(1 - Q\frac{\mathcal{B}_0^*}{\mathcal{L}_0^*}\right)\frac{\hat{\mathcal{B}}_0}{\hat{\mathcal{L}}_0}\left(1 - \frac{\mathcal{B}_0}{\mathcal{L}_0}Q\right)P_- \quad (6.108)$$

and likewise for even ghost numbers. This expression is manifestly BPZ-even and produces well-defined results on all $Q$-closed states while being equivalent to the original version on all non-problematic states[35]. Now the only true and non-curable singularities essentially occur at the states $e^{-\frac{K}{2}}cKe^{-\frac{K}{2}}$, $e^{-\frac{K}{2}}Kce^{-\frac{K}{2}}$ and $e^{-\frac{K}{2}}ce^{-\frac{K}{2}}$.

They are however not a consequence of adding stubs, they stem from the fact that the sliver gauge propagator itself has poles on those states. The difference is that now a part of these pole contributions is moved to the internal vertices and hence also creates singularities in the equations of motion. Our strategy for making sense of the stubbed theory in the sliver frame and for computing analytic solutions will be to use the interpolating stub $e^{-\lambda L_t}$ and understand it in the limit $t \to 0$. This means for the geometric interpretation, $\mathcal{L}_0$ ($\mathcal{B}_0$) should be replaced by $L_t$ ($B_t$) in every expression while in the end we let $t$ go to zero. From an algebraic point of view, this is unproblematic for the $h_{\hat{\mathcal{B}}_0}$-vertices, where no poles appear. The $h_{\mathcal{B}_0}$-vertices that were motivated from the calculations of amplitudes are also fine as long as they are restricted to on-shell states because typical representatives of the cohomology do not include the above-mentioned problematic $KBc$-states. It would be interesting if the range of definition of $h_{\mathcal{B}_0}$ can be extended to the full Hilbert space by including suitable projectors and treat the problematic states separately. We will leave this problem for future work and in this paper just analyze the singularities that appear case by case. Moreover, so far it has not been proven in general that all amplitudes can be defined consistently in sliver gauge and a full proof of this statement lies beyond the scope of this paper as well, some useful references include [43, 66].

---

[35] A nice side effect is that in this way we can construct an interpolation between $h_{\mathcal{B}_0}$ and $h_{\hat{\mathcal{B}}_0}$: If we define $h_{int}(\alpha) = \left(1 - \alpha Q\frac{\mathcal{B}_0^*}{\mathcal{L}_0^*}\right)\frac{\hat{\mathcal{B}}_0}{\hat{\mathcal{L}}_0}\left(1 - \alpha \frac{\mathcal{B}_0}{\mathcal{L}_0}Q\right)P_- + \left(1 - \alpha Q\frac{\mathcal{B}_0}{\mathcal{L}_0}\right)\frac{\hat{\mathcal{B}}_0}{\hat{\mathcal{L}}_0}\left(1 - \alpha\frac{\mathcal{B}_0^*}{\mathcal{L}_0^*}Q\right)P_+$ then $h_{int}(0) = h_{\hat{\mathcal{B}}_0}$ and $h_{int}(1) = h_{\mathcal{B}_0}$.

To proceed, we now want to give an argument why the inclusion of higher vertices is necessary from a geometrical viewpoint, directly in the sliver frame. Let us once again consider the s-channel contribution to the on-shell four-point amplitude:

$$\mathcal{A}_{4s} = -\omega\left(\Psi_1, m_2\left(\Psi_2, e^{-\lambda\mathcal{L}_0}\frac{\mathcal{B}_0}{\mathcal{L}_0}Q\frac{\mathcal{B}_0^*}{\mathcal{L}_0^*}e^{-\lambda\mathcal{L}_0^*}m_2\left(\Psi_3, \Psi_4\right)\right)\right). \tag{6.109}$$

Using Schwinger parameters we can write that as

$$\mathcal{A}_{4s} = \int_0^\infty dt \int_0^\infty ds\,\omega\left(m_2\left(\Psi_1, \Psi_2\right), e^{-(\lambda+t)\mathcal{L}_0}\mathcal{B}_0 Q \mathcal{B}_0^* e^{-(\lambda+s)\mathcal{L}_0^*}m_2\left(\Psi_3, \Psi_4\right)\right). \tag{6.110}$$

Now we use the fact that $Q$ annihilates the on-shell states $\Psi_3$ and $\Psi_4$ and get

$$\mathcal{A}_{4s} = \int_0^\infty dt \int_0^\infty ds\,\omega\left(m_2\left(\Psi_1, \Psi_2\right), e^{-(\lambda+t)\mathcal{L}_0}\mathcal{B}_0 \mathcal{L}_0^* e^{-(\lambda+s)\mathcal{L}_0^*}m_2\left(\Psi_3, \Psi_4\right)\right). \tag{6.111}$$

We see that the $\mathcal{L}_0^*$-insertion can be written as a derivative which localizes the $s$-integral:

$$\mathcal{A}_{4s} = \lim_{\Lambda\to\infty}\int_0^\infty dt\,\omega\left(m_2\left(\Psi_1, \Psi_2\right), e^{-(\lambda+t)\mathcal{L}_0}\mathcal{B}_0\left(e^{-\lambda\mathcal{L}_0^*} - e^{-(\lambda+\Lambda)\mathcal{L}_0^*}\right)m_2\left(\Psi_3, \Psi_4\right)\right). \tag{6.112}$$

As mentioned already, the cut-off term containing $e^{-\Lambda\mathcal{L}_0^*}$ is not guaranteed to yield a vanishing contribution but we will first of all focus on the first term. The next step is to use the state-operator correspondence and write the star product in the form outlined in [70, 69]:

$$\Psi_1(0)\left|0\right\rangle * \Psi_2(0)\left|0\right\rangle = \left(\frac{8}{9}\right)^{h_1+h_2} e^{\ln\frac{2}{3}\mathcal{L}_0^*}\Psi_1\left(\frac{1}{\sqrt{3}}\right)\Psi_2\left(-\frac{1}{\sqrt{3}}\right)\left|0\right\rangle, \tag{6.113}$$

where the $h_i$ are the conformal weights of the primary fields $\Psi_i$. Since we take our external states to be on-shell, we can omit the prefactor and get for the first term

$$\mathcal{A}_{4s}^{(1)} = \int_0^\infty dt\,\langle 0|\,\Psi_2\left(\sqrt{3}\right)\Psi_1\left(-\sqrt{3}\right)e^{-\left(\lambda+t-\ln\frac{2}{3}\right)\mathcal{L}_0}\mathcal{B}_0 e^{-\left(\lambda-\ln\frac{2}{3}\right)\mathcal{L}_0^*}\Psi_3\left(\frac{1}{\sqrt{3}}\right)\Psi_4\left(-\frac{1}{\sqrt{3}}\right)\left|0\right\rangle. \tag{6.114}$$

To commute the two exponentials we can use the formula (6.75) and arrive at

$$\mathcal{A}_{4s}^{(1)} = \int_0^\infty dt \, \langle 0| \Psi_2\left(\sqrt{3}\right) \Psi_1\left(-\sqrt{3}\right) \mathcal{B}_0 \tag{6.115}$$

$$\left(\frac{1}{1+e^{-t}-\frac{2}{3}e^{-\lambda-t}}\right)^{\mathcal{L}_0^*} \left(\frac{1}{1+e^t-\frac{2}{3}e^{-\lambda}}\right)^{\mathcal{L}_0} \Psi_3\left(\frac{1}{\sqrt{3}}\right) \Psi_4\left(-\frac{1}{\sqrt{3}}\right) |0\rangle. \tag{6.116}$$

The operator $x^{\mathcal{L}_0}$ is the scaling operator in the sliver frame and acts on the upper-half-plane coordinates as $z \to \tan(x \arctan z)$. Similarly, the $\mathcal{L}_0^*$-exponential can be made acting to the left where it transforms the coordinates as $z \to \cot\left(x \arctan \frac{1}{z}\right)$. The scaling of the operators can be omitted again and we get

$$\mathcal{A}_{4s}^{(1)} = \int_0^\infty dt \, \langle 0| \Psi_2\left(\cot\left(\frac{\frac{\pi}{6}}{1+e^{-t}-\frac{2}{3}e^{-\lambda-t}}\right)\right) \Psi_1\left(\cot\left(-\frac{\frac{\pi}{6}}{1+e^{-t}-\frac{2}{3}e^{-\lambda-t}}\right)\right)$$

$$\mathcal{B}_0 \Psi_3\left(\tan\left(\frac{\frac{\pi}{6}}{1+e^t-\frac{2}{3}e^{-\lambda}}\right)\right) \Psi_4\left(\tan\left(-\frac{\frac{\pi}{6}}{1+e^t-\frac{2}{3}e^{-\lambda}}\right)\right) |0\rangle. \tag{6.117}$$

This is our final expression for the s-channel contribution in terms of a four-point function dependent on one real modulus $t$. This four-point function is some function of the cross-ratio of the insertion points given by

$$c_s = \frac{(z_1-z_2)(z_3-z_4)}{(z_1-z_3)(z_2-z_4)} = 4\frac{\cot\left(\frac{\frac{\pi}{6}}{1+e^{-t}-\frac{2}{3}e^{-\lambda-t}}\right)\tan\left(\frac{\frac{\pi}{6}}{1+e^t-\frac{2}{3}e^{-\lambda}}\right)}{\left(\cot\left(\frac{\frac{\pi}{6}}{1+e^{-t}-\frac{2}{3}e^{-\lambda-t}}\right)+\tan\left(\frac{\frac{\pi}{6}}{1+e^t-\frac{2}{3}e^{-\lambda}}\right)\right)^2}. \tag{6.118}$$

$c_s$ is a useful parameter of the moduli space of four-punctured disks so by analyzing its range we can see which portion of the moduli space is covered [61].

Let us first consider Witten theory with $\lambda = 0$: $c_s(t)$ is now a monotonically decreasing function with $c_s(0) = \frac{1}{2}$ and $\lim_{t\to\infty} c_s(t) = 0$. This is an expected result: We consider just one specific ordering of the operators here and choosing the standard locations 0, 1 and $\infty$ for three of them, $c_s$ is just given by the second location $z_2$ and should therefore lie between 0 and 1. Hence the portion of the moduli space we expect to be covered is the unit interval and the s-channel covers half of it. The t-channel contribution can be simply found by a cyclic permutation $z_i \to z_{i+1}$ modulo 4 and we get

$$c_t = \frac{(z_2-z_3)(z_4-z_1)}{(z_2-z_4)(z_3-z_1)} = 1 - c_s \tag{6.119}$$

so indeed the other half of the unit interval is covered.

Now let us see what happens if we add stubs: For $\lambda > 0$, $c_s(t)$ is still a monotonically decreasing function with $\lim_{t\to\infty} c_s(t) = 0$ but $c_s(0) < \frac{1}{2}$. This means that the interval $(c_s(0), 1 - c_s(0))$ is not covered by the Feynman diagrams and adding higher vertices is necessary also from a geometrical point of view. We can see that more explictly by setting $t$ to zero in (6.118) to get

$$c_s(t=0) = \sin^2\left(\frac{e^\lambda}{1-3e^\lambda}\frac{\pi}{2}\right) \tag{6.120}$$

This function is monotically decreasing, which means the uncovered region gets bigger as the stub length is increased. An interesting point is that in the limit of infinitely long stubs we get $c_s(t=0) = \frac{1}{4}$, hence the Feynman region covers precisely half of the moduli space.

Finally we want to analyze the cut-off term of the Schwinger parametrization given by

$$\mathcal{A}^{(2)}_{4s} = -\lim_{\Lambda \to \infty} \int_0^\infty dt\, \omega\left(m_2(\Psi_1, \Psi_2), e^{-(\lambda+t)\mathcal{L}_0} \mathcal{B}_0 e^{-(\lambda+\Lambda)\mathcal{L}_0^*} m_2(\Psi_3, \Psi_4)\right). \tag{6.121}$$

Going through the same steps as before we can calculate the cross-ratio and get

$$c_s = 4 \frac{\cot\left(\frac{\frac{\pi}{6}}{1+e^{-t+\Lambda}-\frac{2}{3}e^{-\lambda-t}}\right) \tan\left(\frac{\frac{\pi}{6}}{1+e^{t-\Lambda}-\frac{2}{3}e^{-\lambda-\Lambda}}\right)}{\left(\cot\left(\frac{\frac{\pi}{6}}{1+e^{-t+\Lambda}-\frac{2}{3}e^{-\lambda-t}}\right) + \tan\left(\frac{\frac{\pi}{6}}{1+e^{t-\Lambda}-\frac{2}{3}e^{-\lambda-\Lambda}}\right)\right)^2}. \tag{6.122}$$

In the limit of $\Lambda \to \infty$ we have $c_s \to 0$, hence the cut-off term indeed yields a contribution only at the boundary of the moduli space, as expected.

## 6.4 The tachyon vacuum in the stubbed theory

In this section we want to apply the cohomomorphisms we found on the most important classical solution of Witten theory, namely the tachyon vacuum. It is explicitly given by [70, 20]

$$\Psi_{TV} = e^{-\frac{K}{2}} c \frac{KB}{1-e^{-K}} c e^{-\frac{K}{2}} \tag{6.123}$$

with the elements of the $KBc$-algebra defined for instance in [20, 63]. It obeys the sliver gauge condition

$$\mathcal{B}_0 \Psi_{TV} = 0, \tag{6.124}$$

hence we want to use the associated stub operator $e^{-\lambda \mathcal{L}_0}$.

### 6.4.1 Action of P

We now want to apply **P** onto $\Psi_{TV}$ since it is the only cohomomorphism we have available in closed form. The linear term is just $e^{-\lambda \mathcal{L}_0^*} \Psi_{TV}$, which can be computed by expanding $\Psi_{TV}$ in formal eigenstates of $\mathcal{L}_0^*$ (see Appendix B). The resulting general formula is

$$e^{-\lambda \mathcal{L}_0^*}\left(e^{-\alpha K} f(K, B, c) e^{-\alpha K}\right) = e^{-\left((\alpha+\frac{1}{2})e^\lambda - \frac{1}{2}\right)K} f\left(e^\lambda K, e^\lambda B, e^{-\lambda} c\right) e^{-\left((\alpha+\frac{1}{2})e^\lambda - \frac{1}{2}\right)K} \tag{6.125}$$

from which

$$e^{-\lambda \mathcal{L}_0^*} \Psi_{TV} = e^{-K\left(e^\lambda - \frac{1}{2}\right)} c \frac{KB}{1-e^{-e^\lambda K}} c e^{-K\left(e^\lambda - \frac{1}{2}\right)} \tag{6.126}$$

follows.

To calculate the quadratic terms we first need to compute $h\Psi_{TV}$. Here a major difference to the standard frame occurs because suppose we have a solution in Siegel gauge and apply

$$h = \frac{e^{-2\lambda L_0} - 1}{L_0} b_0, \qquad (6.127)$$

the result vanishes and hence the new solution is just given by the linear term only. It is of course tempting to use our result for $h$ in sliver gauge and apply it to $\Psi_{TV}$ which results in

$$h_{\mathcal{B}_0}\Psi_{TV} = \left( e^{-\lambda\mathcal{L}_0} \frac{\mathcal{B}_0^*}{\mathcal{L}_0^*} Q \frac{\mathcal{B}_0}{\mathcal{L}_0} e^{-\lambda\mathcal{L}_0^*} - \frac{\mathcal{B}_0^*}{\mathcal{L}_0^*} Q \frac{\mathcal{B}_0}{\mathcal{L}_0} \right) \Psi_{TV}. \qquad (6.128)$$

However, as we have seen in the last section, the equations of motion using $h_{\mathcal{B}_0}$ exhibit singularities on certain states including the zero-momentum tachyon $e^{-\frac{K}{2}} c e^{-\frac{K}{2}}$, so it is not clear if we will end up with a well-defined solution. Actually, $h_{\mathcal{B}_0}\Psi_{TV}$ is ill-defined as can be shown as follows: We will focus on the first term here, the second term is dealt with in section 6.4.2.

The action of $\mathcal{B}_0$ can be determined by the formula

$$\mathcal{B}_0 \left( e^{-\frac{K}{2}} \Psi e^{-\frac{K}{2}} \right) = e^{-\frac{K}{2}} B^- \Psi e^{-\frac{K}{2}} \qquad (6.129)$$

where $B^- = \frac{1}{2}(\mathcal{B}_0 - \mathcal{B}_0^*)$ acts as a star algebra derivative and obeys

$$B^- K = B, \qquad B^- B = 0, \qquad B^- c = 0, \qquad (6.130)$$

the result is

$$\mathcal{B}_0 e^{-\lambda\mathcal{L}_0^*} \Psi_{TV} = \left(1 - e^\lambda\right) e^{-K\left(e^\lambda - \frac{1}{2}\right)} \left[\frac{KB}{1 - e^{-e^\lambda K}}, c\right] e^{-K\left(e^\lambda - \frac{1}{2}\right)}. \qquad (6.131)$$

To apply $\frac{1}{\mathcal{L}_0}$ it is convenient to use the Schwinger representation

$$\frac{1}{\mathcal{L}_0} = \int_0^\infty dt \, e^{-t\mathcal{L}_0} \qquad (6.132)$$

and use an expansion in eigenstates of $\mathcal{L}_0$ (see Appendix B). The calculation is then analogous to (6.126) and results in

$$\frac{\mathcal{B}_0}{\mathcal{L}_0} e^{-\lambda\mathcal{L}_0^*} \Psi_{TV} = \int_0^\infty dt \left(e^{-t} - e^{\lambda-t}\right) e^{-K\left(e^{\lambda-t} - e^{-t} + \frac{1}{2}\right)} \left[\frac{KB}{1 - e^{-e^{\lambda-t}K}}, c\right] e^{-K\left(e^{\lambda-t} - e^{-t} + \frac{1}{2}\right)} \qquad (6.133)$$

This Schwinger integral is actually divergent: If we expand the integrand for large $t$ we get

$$\int_0^\infty dt \left(e^{-\lambda} - 1\right) e^{-\frac{K}{2}} [B, c] e^{-\frac{K}{2}} \qquad (6.134)$$

which is an infinite integral over an expression independent of $t$. The problem can also be seen from (6.131) already: If we look at the $\mathcal{L}_0$-level expansion, we see that the expression contains the term

$$\left(e^{-\lambda}-1\right)e^{-\frac{K}{2}}[B,c]\,e^{-\frac{K}{2}} \tag{6.135}$$

which has $\mathcal{L}_0$-eigenvalue zero and applying $\frac{1}{\mathcal{L}_0}$ is ill-defined. It is straightforward to show that the situation does not improve by applying all the other operators in (6.128), especially since the divergence is not $Q$-closed: By isolating

$$(h_{\mathcal{B}_0}\Psi_{TV})_{div} = e^{-\lambda\mathcal{L}_0}\frac{\mathcal{B}_0^*}{\mathcal{L}_0^*}Q\frac{1}{0}\left(e^{-\lambda}-1\right)e^{-\frac{K}{2}}[B,c]\,e^{-\frac{K}{2}} \tag{6.136}$$

and directly applying $Q$ we get

$$Q\,(h_{\mathcal{B}_0}\Psi_{TV})_{div} = e^{-\lambda\mathcal{L}_0}\frac{1}{0}\left(e^{-\lambda}-1\right)e^{-\frac{K}{2}}Q\,[B,c]\,e^{-\frac{K}{2}} = \frac{2}{0}\left(e^{-\lambda}-1\right)e^{-\frac{K}{2}}cKBce^{-\frac{K}{2}} \neq 0. \tag{6.137}$$

To get rid of the singularity we will instead use the vertices constructed from $h_{\hat{\mathcal{B}}_0}$ in $\hat{\mathcal{B}}_0$-gauge (6.76) given by

$$h_{\hat{\mathcal{B}}_0} = \frac{e^{(e^{-\lambda}-1)\hat{\mathcal{L}}_0}-1}{\hat{\mathcal{L}}_0}\hat{\mathcal{B}}_0 = -\int_0^{1-e^{-\lambda}}dt\,e^{-t\hat{\mathcal{L}}_0}\hat{\mathcal{B}}_0. \tag{6.138}$$

First we use the formula

$$\hat{\mathcal{B}}_0\Psi = B\Psi + (-)^{\mathrm{gh}(\Psi)}\Psi B \tag{6.139}$$

to write

$$\hat{\mathcal{B}}_0\Psi_{TV} = e^{-\frac{K}{2}}\left[\frac{KB}{1-e^{-K}},c\right]e^{-\frac{K}{2}}. \tag{6.140}$$

Then we exponentiate the well-known relation

$$\hat{\mathcal{L}}_0\Psi = K\Psi + \Psi K \tag{6.141}$$

to get

$$e^{-t\hat{\mathcal{L}}_0}\Psi = e^{-tK}\Psi e^{-tK}, \tag{6.142}$$

now the total action of $h$ becomes

$$h_{\hat{\mathcal{B}}_0}\Psi_{TV} = -\int_0^{1-e^{-\lambda}}dt\,e^{-K\left(t+\frac{1}{2}\right)}\left[\frac{KB}{1-e^{-K}},c\right]e^{-K\left(t+\frac{1}{2}\right)}. \tag{6.143}$$

As a second intermediate result we compute $ip\Psi_{TV}$ using (6.74) and (6.142):

$$e^{-\lambda\mathcal{L}_0}e^{-\lambda\mathcal{L}_0^*}\Psi_{TV} = e^{-K\left(\frac{3}{2}-e^{-\lambda}\right)}c\frac{KB}{1-e^{-K}}ce^{-K\left(\frac{3}{2}-e^{-\lambda}\right)}. \tag{6.144}$$

Combining the results and applying the projection operator $e^{-\lambda \mathcal{L}_0^*}$ once more yields the full quadratic part of the solution according to (6.42):

$$pm_2\left(\Psi_{TV}, h_{\hat{\mathcal{B}}_0}\Psi_{TV}\right) + pm_2\left(h_{\hat{\mathcal{B}}_0}\Psi_{TV}, ip\Psi_{TV}\right)$$

$$= \int_0^{1-e^{-\lambda}} dt\, e^{\lambda}(-e^{-K\left(e^{\lambda}-\frac{1}{2}\right)}c\frac{K^2 B}{\left(1-e^{-e^{\lambda}K}\right)^2}e^{-K\left(e^{\lambda}t+e^{\lambda}\right)}ce^{-K\left(e^{\lambda}+e^{\lambda}t-\frac{1}{2}\right)} \quad (6.145)$$

$$+ e^{-K\left(e^{\lambda}-\frac{1}{2}\right)}c\frac{KB}{1-e^{-e^{\lambda}K}}c\frac{K}{1-e^{-e^{\lambda}K}}e^{-K\left(2e^{\lambda}t+2e^{\lambda}-\frac{1}{2}\right)}$$

$$- e^{-K\left(e^{\lambda}-\frac{1}{2}\right)}c\frac{KB}{1-e^{-e^{\lambda}K}}e^{-K\left(e^{\lambda}t+e^{\lambda}\right)}c\frac{K}{1-e^{-e^{\lambda}K}}e^{-K\left(e^{\lambda}+e^{\lambda}t-\frac{1}{2}\right)}$$

$$+ e^{-K\left(e^{\lambda}+e^{\lambda}t-\frac{1}{2}\right)}\frac{K}{1-e^{-e^{\lambda}K}}e^{-K\left(e^{\lambda}t+2e^{\lambda}-1\right)}c\frac{KB}{1-e^{-e^{\lambda}K}}e^{-K\left(2e^{\lambda}-\frac{3}{2}\right)}$$

$$- e^{-K\left(e^{\lambda}+e^{\lambda}t-\frac{1}{2}\right)}\frac{K}{1-e^{-e^{\lambda}K}}ce^{-K\left(e^{\lambda}t+2e^{\lambda}-1\right)}\frac{KB}{1-e^{-e^{\lambda}K}}ce^{-K\left(2e^{\lambda}-\frac{3}{2}\right)}$$

$$- e^{-K\left(e^{\lambda}+e^{\lambda}t-\frac{1}{2}\right)}c\frac{K^2 B}{\left(1-e^{-e^{\lambda}K}\right)^2}e^{-K\left(e^{\lambda}t+2e^{\lambda}-1\right)}ce^{-K\left(2e^{\lambda}-\frac{3}{2}\right)}). \quad (6.146)$$

This expression is well-defined since (6.76) has no poles. An interesting point is that it fails to be twist symmetric, so we conclude that the application of **P** in general breaks twist symmetry.

### 6.4.2 A simpler stubbed theory?

Since the last result is quite complicated already at quadratic order in $\Psi_{TV}$, one might ask if there exists a stubbed theory with simpler solutions. For example one may try the following: Let us replace $\mathcal{L}_0$ by $\mathcal{L}_0^*$ in every formula, i. e. switch the operators $i$ and $p$ such that the new product becomes

$$M_2(\cdot, \cdot) = e^{-\lambda \mathcal{L}_0}\left(e^{-\lambda \mathcal{L}_0^*}\cdot, e^{-\lambda \mathcal{L}_0^*}\cdot\right) \quad (6.147)$$

This choice is less natural than the original one since in the three vertex only $\mathcal{L}_0^*$s would appear instead of the scaling operator $\mathcal{L}_0$. However, the homotopy in sliver gauge would become

$$h'_{\mathcal{B}_0} = \left(e^{-\lambda \mathcal{L}_0^*}\frac{\mathcal{B}_0}{\mathcal{L}_0}Q\frac{\mathcal{B}_0^*}{\mathcal{L}_0^*}e^{-\lambda \mathcal{L}_0} - \frac{\mathcal{B}_0}{\mathcal{L}_0}Q\frac{\mathcal{B}_0^*}{\mathcal{L}_0^*}\right)P_+ + \left(e^{-\lambda \mathcal{L}_0^*}\frac{\mathcal{B}_0^*}{\mathcal{L}_0^*}Q\frac{\mathcal{B}_0}{\mathcal{L}_0}e^{-\lambda \mathcal{L}_0} - \frac{\mathcal{B}_0^*}{\mathcal{L}_0^*}Q\frac{\mathcal{B}_0}{\mathcal{L}_0}\right)P_- \quad (6.148)$$

so one would naively assume

$$h'_{\mathcal{B}_0}\Psi_{TV} = 0 \quad (6.149)$$

because of the gauge condition. However, there is an important caveat here: $\Psi_{TV}$ contains the ghost number one state $-\frac{1}{2}e^{-\frac{K}{2}}cKBce^{-\frac{K}{2}}$ which is annihilated by $\mathcal{B}_0$ and $\mathcal{L}_0$, so the result is ambiguous. To make sense of it it is useful to write

$$e^{-\frac{K}{2}}cKBce^{-\frac{K}{2}} = Q\left(e^{-\frac{K}{2}}Bce^{-\frac{K}{2}}\right) \quad (6.150)$$

and use (6.108) such that we arrive at

$$h'_{\mathcal{B}_0} \Psi_{TV} = \frac{1}{2}\left(1 - e^{-\lambda \mathcal{L}_0^*}\right) \frac{\mathcal{B}_0^*}{\mathcal{L}_0^*} e^{-\frac{K}{2}} cKBc e^{-\frac{K}{2}}. \tag{6.151}$$

This result is also compatible with the Hodge-Kodaira relation (6.73). Using the same techniques as above we compute

$$h'_{\mathcal{B}_0} \Psi_{TV} = \frac{1}{2} \int_0^\lambda dt\, e^t e^{-K\left(e^t - \frac{1}{2}\right)} [KB, c] e^{-K\left(e^t - \frac{1}{2}\right)} \tag{6.152}$$

and obtain to quadratic order in $\Psi_{TV}$

$$\pi_1 \mathbf{P} \frac{1}{1-\Psi_{TV}} = e^{-\frac{K}{2}} c \frac{KB}{1-e^{-e^{-\lambda}K}} c e^{-\frac{K}{2}}$$
$$+ \frac{1}{2}\int_0^\lambda dt\, e^{t-\lambda}\Big(e^{-\frac{K}{2}} c \frac{KB}{1-e^{-K}} e^{-e^{t-\lambda}K} \{c, K\} e^{-K\left(e^{t-\lambda} - e^{-\lambda} + \frac{1}{2}\right)}$$
$$- e^{-\frac{K}{2}} c \frac{KB}{1-e^{-K}} cK e^{-K\left(2e^{t-\lambda} - e^{-\lambda} + \frac{1}{2}\right)}$$
$$+ e^{-K\left(e^{t-\lambda} - e^{-\lambda} + \frac{1}{2}\right)} \{c, K\} e^{-K\left(e^{t-\lambda} + e^{-\lambda} - e^{-2\lambda}\right)} \frac{KB}{1-e^{-K}} c e^{-K\left(\frac{1}{2} + e^{-\lambda} - e^{-2\lambda}\right)}$$
$$- e^{-\left(2e^{t-\lambda} + \frac{1}{2} - e^{-2\lambda}\right)} Kc \frac{KB}{1-e^{-K}} c e^{-K\left(\frac{1}{2} + e^{-\lambda} - e^{-2\lambda}\right)}\Big) + \cdots \tag{6.153}$$

This result is indeed a bit simpler than (6.146) but also here no obvious simplification is visible.

### 6.4.3 Action of F

We can also ask what happens if we use the cyclic cohomomorphism (6.94) where the operator $X$ was defined in (6.99) and apply it on $\Psi_{TV}$:

$$\pi_1 \mathbf{F} \frac{1}{1-\Psi_{TV}} = \pi_1 \mathcal{P}\exp\left(\int_0^\lambda dt \left(\sum_{n=2}^\infty \pi_1 \left(\mathbf{X}^* \mathbf{M_n}(\lambda) - \mathbf{M_n}(\lambda)\mathbf{X}\right)\pi_n\right)\right) \frac{1}{1-\Psi_{TV}}. \tag{6.154}$$

At this stage the result is not very satisfying, but we can analyze it for an infinitesimal $\lambda \ll 1$: The sum collapses then to the $n=2$ term and yields

$$\pi_1 \mathbf{F} \frac{1}{1-\Psi_{TV}} = \Psi_{TV} + \lambda \left(X^* m_2\left(\Psi_{TV}^{\otimes 2}\right) - m_2\left(X\Psi_{TV}, \Psi_{TV}\right) - m_2\left(\Psi_{TV}, X\Psi_{TV}\right)\right) + \mathcal{O}\left(\lambda^2\right). \tag{6.155}$$

Let us again first try to use $h_{\mathcal{B}_0}$ for the vertices: The first becomes explicitly

$$X^* m_2\left(\Psi_{TV}, \Psi_{TV}\right) = \frac{\mathcal{B}_0}{\mathcal{L}_0} Q \mathcal{B}_0^* m_2\left(\Psi_{TV}, \Psi_{TV}\right). \tag{6.156}$$

Using the equations motion for $\Psi_{TV}$ we can write this as

$$-\frac{\mathcal{B}_0}{\mathcal{L}_0}Q\mathcal{B}_0^*Q\Psi_{TV} = -\frac{\mathcal{B}_0}{\mathcal{L}_0}Q\mathcal{L}_0^*\Psi_{TV}. \tag{6.157}$$

The action of $\mathcal{L}_0^*$ is straightforward to calculate using (6.190), the result is

$$\mathcal{L}_0^*\Psi_{TV} = \left\{K, e^{-\frac{K}{2}}c\frac{KB}{1-e^{-K}}ce^{-\frac{K}{2}}\right\} - e^{-\frac{K}{2}}c\frac{e^{-K}K^2B}{(1-e^{-K})^2}ce^{-\frac{K}{2}} \tag{6.158}$$

Applying $Q$ yields

$$\begin{aligned}Q\mathcal{L}_0^*\Psi_{TV} &= \left\{K, e^{-\frac{K}{2}}cKc\frac{KB}{1-e^{-K}}ce^{-\frac{K}{2}}\right\} + \left\{K, e^{-\frac{K}{2}}c\frac{KB}{1-e^{-K}}cKce^{-\frac{K}{2}}\right\} \\ &\quad - \left\{K, e^{-\frac{K}{2}}c\frac{K^2}{1-e^{-K}}ce^{-\frac{K}{2}}\right\} - e^{-\frac{K}{2}}cKc\frac{e^{-K}K^2B}{(1-e^{-K})^2}ce^{-\frac{K}{2}} \\ &\quad + e^{-\frac{K}{2}}c\frac{e^{-K}K^3}{(1-e^{-K})^2}ce^{-\frac{K}{2}} - e^{-\frac{K}{2}}c\frac{e^{-K}K^2B}{(1-e^{-K})^2}cKce^{-\frac{K}{2}}.\end{aligned} \tag{6.159}$$

To check if we can apply $\frac{\mathcal{B}_0}{\mathcal{L}_0}$ we have to isolate the level zero contribution in the $\mathcal{L}_0$-expansion. It is given by

$$Q\mathcal{L}_0^*\Psi_{TV}\big|_{\mathcal{L}_0=0} = e^{-\frac{K}{2}}cKcKe^{-\frac{K}{2}} + e^{-\frac{K}{2}}KcKce^{-\frac{K}{2}} \tag{6.160}$$

This state is non-vanishing and not annihilated by $\mathcal{B}_0$, so the application of $\frac{\mathcal{B}_0}{\mathcal{L}_0}$ leads again to an ill-defined result as anticipated.

Hence, to get something well-defined, we should use $h_{\hat{\mathcal{B}}_0}$ which leads to

$$X\Psi_{TV} = \mathcal{B}_0\Psi_{TV} = 0 \tag{6.161}$$

because of the gauge condition, so we arrive at

$$\pi_1\mathbf{F}\frac{1}{1-\Psi_{TV}} = \Psi_{TV} + \lambda\mathcal{B}_0^*m_2\left(\Psi_{TV}^{\otimes 2}\right) + \mathcal{O}\left(\lambda^2\right). \tag{6.162}$$

This can be calculated explicitly using the same methods as in the previous section:

$$\begin{aligned}\mathcal{B}_0^*m_2\left(\Psi_{TV}^{\otimes 2}\right) &= \mathcal{B}_0^*\left(e^{-\frac{K}{2}}c\frac{KB}{1-e^{-K}}ce^{-K}c\frac{KB}{1-e^{-K}}ce^{-\frac{K}{2}}\right) \\ &= e^{-\frac{K}{2}}\frac{KB}{1-e^{-K}}ce^{-K}c\frac{KB}{1-e^{-K}}ce^{-\frac{K}{2}} + e^{-\frac{K}{2}}c\frac{KB}{1-e^{-K}}ce^{-K}c\frac{KB}{1-e^{-K}}e^{-\frac{K}{2}} \\ &\quad + e^{-\frac{K}{2}}c\frac{KB}{1-e^{-K}}e^{-K}c\frac{KB}{1-e^{-K}}ce^{-\frac{K}{2}} \\ &= e^{-\frac{K}{2}}\left\{\frac{KB}{1-e^{-K}}c\frac{K}{e^K-1},c\right\}e^{-\frac{K}{2}} + e^{-\frac{K}{2}}c\frac{K^2B}{e^K+e^{-K}-2}ce^{-\frac{K}{2}} \\ &\quad - e^{-\frac{K}{2}}\left\{\frac{KB}{e^K-1}c\frac{K}{1-e^{-K}},c\right\}e^{-\frac{K}{2}}.\end{aligned} \tag{6.163}$$

In the end we get

$$\pi_1 \mathbf{F} \frac{1}{1 - \Psi_{TV}} = e^{-\frac{K}{2}} c \frac{KB}{1 - e^{-K}} c e^{-\frac{K}{2}}$$
$$+ \lambda \left( e^{-\frac{K}{2}} \left\{ \frac{KB}{1 - e^{-K}} c \frac{K}{e^K - 1}, c \right\} e^{-\frac{K}{2}} + e^{-\frac{K}{2}} c \frac{K^2 B}{e^K + e^{-K} - 2} c e^{-\frac{K}{2}} \right.$$
$$\left. - e^{-\frac{K}{2}} \left\{ \frac{KB}{e^K - 1} c \frac{K}{1 - e^{-K}}, c \right\} e^{-\frac{K}{2}} \right)$$
$$+ \mathcal{O}\left(\lambda^2\right). \tag{6.164}$$

In contrast to our result for $\pi_1 \mathbf{P} \frac{1}{1 - \Psi_{TV}}$, this expression is twist symmetric and from the discussion in section 6.2.5 we also deduce that it is gauge equivalent to (6.146) to first order in $\lambda$. Indeed, comparing the infinitesimal actions of $\mathbf{P}$ and $\mathbf{F}$

$$\pi_1 \mathbf{P} \frac{1}{1 - \Psi_{TV}} = \Psi_{TV} - \delta\lambda \mathcal{L}_0^* \Psi_{TV} - \delta\lambda \left( m_2 \left( X^* \Psi_{TV}, \Psi_{TV} \right) + m_2 \left( \Psi_{TV}, X^* \Psi_{TV} \right) \right) \tag{6.165}$$

$$\pi_1 \mathbf{F} \frac{1}{1 - \Psi_{TV}} = \Psi_{TV} + \delta\lambda X^* m_2 \left( \Psi_{TV}, \Psi_{TV} \right) \tag{6.166}$$

one sees that with the definition $\Lambda =: X^* \Psi_{TV}$ the difference can be written as

$$\Delta \Psi_{TV} = \delta\lambda \left( Q\Lambda + m_2 \left( \Lambda, \Psi_{TV} \right) + m_2 \left( \Psi_{TV}, \Lambda \right) + X^* \left( Q \Psi_{TV} + m_2 \left( \Psi_{TV}, \Psi_{TV} \right) \right) \right). \tag{6.167}$$

Since $\Psi_{TV}$ obeys the original Witten's equation of motion, $\Delta \Psi_{TV}$ is just a gauge transformation.

## 6.5 Conclusion and outlook

In this paper we addressed and answered some pending questions concerning stubs in open string field theory and moreover generalized the whole framework to more general gauges. We systematically constructed a family of cohomomorphisms which map the Witten theory to the stubbed theory, including the two maps $\mathbf{F}^{-1}$ and $\mathbf{P}$ already found in [71]. To first order in $\lambda$, the difference between two such maps always consists of a gauge transformation and a part proportional to the equations of motion, according to the conjecture of [71] that the on-shell value of the action is independent of the cohomomorphism. When moving to a different coordinate frame we encountered some algebraic constraints on the homotopy operator to get consistent and well-defined results. We have shown that there is a solution for $h_{\mathcal{B}_0}$ under the assumption that perturbation theory in the sliver frame is geometrically consistent. While for on-shell amplitudes this operator $h_{\mathcal{B}_0}$ is well-dedfined, it exhibits singular behaviour on some off-shell states in the Hilbert space. We concluded therefore that for the calculation of classical solutions we should rather pick the $\hat{\mathcal{B}}_0$-gauge for the homotopy. This we could explicitly verify by computing the tachyon vacuum in the stubbed theory up to second order. In a separate calculation we also argued from a geometrical viewpoint why the inclusion of higher vertices in the sliver frame is necessary.

At this time it is not clear yet if those explicit expressions can indeed lead us to more general properties of $A_\infty$-solutions or even solutions of CSFT but it is certainly a possible direction for further research. Especially the relation to [54], where the closed string cohomology was identified in a purely open string setup, is for sure worth to investigate. Another promising path to follow is the connection to [13], where stubs in OSFT are represented through an auxiliary field. It would be interesting if this formalism can be generalized to the sliver frame as well and what the auxiliary field would look like for specific classical solutions. In this context one could also make the role of twist symmetry more precise and analyze under which conditions it can be preserved. Finally, it would be worth to explicitly explore if the stubbed theory solves issues with singularities, for instance concerning identity-based solutions like $\Psi = c\,(1 - K)$ [1, 20].

### Acknowledgements

We thank Ted Erler, Jakub Vošmera, Martin Markl, Harold Erbin and Atakan Firat for useful discussions. Our work has been funded by the Grant Agency of Czech Republic under the grant EXPRO 20-25775X.

### Tensor coalgebras

The tensor coalgebra $TV$ associated to a (graded) vector space $V$ is defined as the Fock space

$$V^{\otimes 0} + V^{\otimes 1} + V^{\otimes 2} + \ldots \tag{6.168}$$

together with the comultiplication $\Delta : TV \to TV \otimes' TV$ given by

$$\Delta\left(v_1 \otimes \ldots \otimes v_n\right) = \sum_{k=0}^{n} \left(v_1 \otimes \ldots \otimes v_k\right) \otimes' \left(v_{k+1} \otimes \ldots \otimes v_n\right) \tag{6.169}$$

on homogeneous elements and extended by linearity. Here the $v_i$ are elements of $V$ and $\otimes'$ denotes the tensor product arising from a comultiplication, in contrast to the usual $\otimes$. We define the projection operator $\pi_n : TV \to TV$ to project any element on its $n$th tensor power component,

$$\pi_n TV = V^{\otimes n}. \tag{6.170}$$

A linear map $\mathbf{d} : TV \to TV$ is called a coderivation if it satisfies the co-Leibniz rule:

$$\Delta \mathbf{d} = \left(\mathbf{d} \otimes' \mathbf{1} + \mathbf{1} \otimes' \mathbf{d}\right) \Delta. \tag{6.171}$$

Linear combinations of coderivations are again coderivations as well as their graded commutator

$$[\mathbf{d}_1, \mathbf{d}_2] = \mathbf{d}_1 \mathbf{d}_2 - (-1)^{deg(\mathbf{d}_1)deg(\mathbf{d}_2)}\, \mathbf{d}_2 \mathbf{d}_1. \tag{6.172}$$

The product $\mathbf{d_1 d_2}$ is associative but in general not a coderivation. For any $m$-linear map $d_m : V^{\otimes m} \to V$ one can construct an associated coderivation by the formula

$$\mathbf{d_m} = \sum_{n=m}^{\infty} \sum_{k=0}^{n-m} 1^{\otimes k} \otimes d_m \otimes 1^{\otimes n-k-m}. \tag{6.173}$$

The co-Leibniz rule guarantees that any coderivation is a sum of terms of this form for different $m$. Given two coderivations of this form, we use the notation $(\mathbf{d_1 d_2})$ to denote the subset of terms where $d_{1m}$ acts on the output of $d_{2m'}$. The individual $m$-products can be recovered from a general coderivation as

$$d_m = \pi_1 \mathbf{d} \pi_m. \tag{6.174}$$

If an odd coderivation $\mathbf{d}$ obeys

$$\mathbf{d}^2 = 0 \tag{6.175}$$

then its components $d_m$ can be regarded as products of an $A_\infty$-algebra.

A linear map $\mathbf{f}$ is called a cohomomorphism if it fulfills

$$\Delta \mathbf{f} = (\mathbf{f} \otimes' \mathbf{f}) \Delta. \tag{6.176}$$

Linear combinations and products of cohomomorphisms are again cohomomorphisms. Given a family of $m$-products $f_m$ one can construct a unique cohomomorphism via

$$\mathbf{f} = \sum_{j=1}^{\infty} \sum_{k=1}^{\infty} \sum_{m_1+\ldots+m_j=k} f_{m_1} \otimes \ldots \otimes f_{m_j}. \tag{6.177}$$

Again, the individual products can be recovered from $\mathbf{f}$ as

$$f_m = \pi_1 \mathbf{f} \pi_m. \tag{6.178}$$

Of special importance are elements of $TV$ of the form

$$1 + v + v \otimes v + v \otimes v \otimes v + \ldots =: \frac{1}{1-v} \tag{6.179}$$

for some $v \in V$. They fulfill the following useful properties:

$$\pi_1 \mathbf{f} \frac{1}{1-v} = \sum_{m=1}^{\infty} f_m \left( v^{\otimes m} \right), \tag{6.180}$$

$$\mathbf{f} \frac{1}{1-v} = \frac{1}{1 - \pi_1 \mathbf{f} \frac{1}{1-v}} \tag{6.181}$$

for any cohomomorphism $\mathbf{f}$.

A bilinear map $\langle \omega | : TV \times TV \to \mathbb{C}$ is called a symplectic form if it satisfies

$$\langle \omega | (v_1 \otimes v_2) =: \omega(v_1, v_2) = -(-1)^{deg(v_1) deg(v_2)} \omega(v_2, v_1). \tag{6.182}$$

A multilinear product $m_k$ is called cyclic with respect to $\omega$ if it fulfills

$$\omega(v_1, m_k(v_2, ..., v_{k+1})) = -(-1)^{deg(v_1)deg(m_k)} \omega(m_k(v_1, ..., v_k), v_{k+1}). \tag{6.183}$$

A coderivation $\mathbf{d}$ is cyclic if all of its components $d_m = \pi_1 \mathbf{d} \pi_m$ are cyclic or equivalently

$$\langle \omega | \pi_2 \mathbf{d} = 0. \tag{6.184}$$

Given two symplectic forms $\langle \omega |$, $\langle \omega' |$, a cohomomorphism $\mathbf{f}$ is cyclic [36] if

$$\langle \omega' | \pi_2 \mathbf{f} = \langle \omega | \pi_2. \tag{6.185}$$

### Eigenstates of $\mathcal{L}_0$ and $\mathcal{L}_0^*$

The operator $\mathcal{L}_0$ obeys the familiar relation [20]

$$\mathcal{L}_0 \left( e^{-\frac{K}{2}} X e^{-\frac{K}{2}} \right) = e^{-\frac{K}{2}} \frac{1}{2} \mathcal{L}^- X e^{-\frac{K}{2}} \tag{6.186}$$

where $\mathcal{L}^- = \mathcal{L}_0 - \mathcal{L}_0^*$ and $X$ is typically an element of the $KBc$-algebra. $\mathcal{L}^-$ acts as a derivation of the star algebra and fulfills

$$\frac{1}{2} \mathcal{L}^- K = K, \qquad \frac{1}{2} \mathcal{L}^- B = B, \qquad \frac{1}{2} \mathcal{L}^- c = -c \tag{6.187}$$

so we can deduce that states of the form

$$e^{-\frac{K}{2}} K^m c K^n e^{-\frac{K}{2}} \quad \text{or} \quad e^{-\frac{K}{2}} K^m c K^n B c K^r e^{-\frac{K}{2}} \tag{6.188}$$

are eigenstates of $\mathcal{L}_0$ with eigenvalue $m + n - 1$ or $m + n + r - 1$, respectively.

To find the eigenstates of $\mathcal{L}_0^*$ we notice that equation (6.186) can be alternatively written as

$$\mathcal{L}_0 X = \frac{1}{2} \mathcal{L}^- X + \frac{1}{2} (KX + XK), \tag{6.189}$$

since the anticommutator with $K$ cancels the terms coming from the action of $\frac{1}{2} \mathcal{L}^-$ on the security strips $e^{-\frac{K}{2}}$. Now writing

$$-\mathcal{L}_0^* X = \frac{1}{2} \mathcal{L}^- X - \frac{1}{2} (KX + XK) \tag{6.190}$$

we see that the sign of $K$ in the security strips has to change. Indeed, by a straightforward calculation one can show that

$$e^{\frac{K}{2}} K^m c K^n e^{\frac{K}{2}} \quad \text{or} \quad e^{\frac{K}{2}} K^m c K^n B c K^r e^{\frac{K}{2}} \tag{6.191}$$

---

[36]See section 2.7 for a more careful treatment.

are formal eigenstates of $\mathcal{L}_0^*$ with eigenvalue $-m-n+1$ or $-m-n-r+1$, respectively. They are in fact ill-defined due to the appearance of inverse wedge states. However, we can still use them as a formal device to determine the action of $\mathcal{L}_0^*$ on the string fields in question and get a well-defined final result. So for example the tachyon vacuum can be expanded as

$$\Psi_{TV} = \sum_{m,n,r=0}^{\infty} \frac{(-1)^{m+n+r} B_n}{m!n!r!} \left(e^{\frac{K}{2}} K^m c B K^n c K^r e^{\frac{K}{2}}\right). \tag{6.192}$$

We can derive the two useful formulas

$$e^{-\lambda \mathcal{L}_0}\left(e^{-\alpha K} f(K,B,c) e^{-\alpha K}\right) = e^{-\left((\alpha-\frac{1}{2})e^{-\lambda}+\frac{1}{2}\right)K} f\left(e^{-\lambda}K, e^{-\lambda}B, e^{\lambda}c\right) e^{-\left((\alpha-\frac{1}{2})e^{-\lambda}+\frac{1}{2}\right)K} \tag{6.193}$$

$$e^{-\lambda \mathcal{L}_0^*}\left(e^{-\alpha K} f(K,B,c) e^{-\alpha K}\right) = e^{-\left((\alpha+\frac{1}{2})e^{\lambda}-\frac{1}{2}\right)K} f\left(e^{\lambda}K, e^{\lambda}B, e^{-\lambda}c\right) e^{-\left((\alpha+\frac{1}{2})e^{\lambda}-\frac{1}{2}\right)K} \tag{6.194}$$

As a crosscheck we can verify the algebra

$$[\mathcal{L}_0, \mathcal{L}_0^*] = \mathcal{L}_0 + \mathcal{L}_0^* \tag{6.195}$$

by acting on an arbitrary function of $K$ and using the eigenstates:

$$\mathcal{L}_0^* F(K) = \mathcal{L}_0^* e^{\frac{K}{2}} \sum_n f_n K^n e^{\frac{K}{2}} = -\sum_n n f_n K^n e^K \tag{6.196}$$

if $F(K)$ is expanded as $F(K) = \sum_n f_n K^n e^K$. On the other hand,

$$\mathcal{L}_0 F(K) = \frac{1}{2}\mathcal{L}^- F(K) + K F(K) = K(F(K) + F'(K)) = 2\sum_n f_n K^{n+1} e^K + \sum_n n f_n K^n e^K. \tag{6.197}$$

Proceeding in the same way yields

$$\mathcal{L}_0 \mathcal{L}_0^* F(K) = -\mathcal{L}_0 \sum_n n f_n K^n e^K = -\sum_n n^2 f_n K^n e^K - 2\sum_n n f_n K^{n+1} e^K \tag{6.198}$$

$$\mathcal{L}_0^* \mathcal{L}_0 F(K) = -2\sum_n (n+1) f_n K^{n+1} e^K - \sum_n n^2 f_n K^n e^K. \tag{6.199}$$

Now one can straightforwardly see that

$$[\mathcal{L}_0, \mathcal{L}_0^*] F(K) = 2\sum_n f_n K^{n+1} e^K = \mathcal{L}_0 F(K) + \mathcal{L}_0^* F(K) \tag{6.200}$$

as expected. An alternative way to show that is to use the fact that on functions of $K$, $\mathcal{L}_0$ and $\mathcal{L}_0^*$ can be represented as $K\frac{d}{dK} + K$ and $-K\frac{d}{dK} + K$ respectively, and use

$$\left[K\frac{d}{dK} + K, -K\frac{d}{dK} + K\right] = 2K. \tag{6.201}$$

# 7 Sliver frame stubs in OSFT via auxiliary fields

**Abstract**

In this short paper we want to generalize some recent concepts related to stubs in open string field theory. First, we modify the auxiliary field method by Erbin and Firat [13] to non-BPZ even sliver frame stubs. We then also show that the construction is consistent at the full quantum level without any additional assumptions. Finally, we apply the method explicitly to the tachyon vacuum and the simplest identity-like solution.

## 7.1 Introduction

Deforming Witten's open string field theory with stubs has a couple of interesting motivations and features. First of all, the stubbed OSFT builds a bridge between the structures of open and closed string field theory and will also be of relevance in a combined open-closed SFT. Second, it can help us to understand general properties of classical solutions to $A_\infty$-type theories. Moreover, it is also expected to tame certain singularities present in Witten theory, for example associated to identity-like solutions.

The starting point is the following: Instead of Witten theory defined by the action

$$S(\Psi) = -\frac{1}{2}\omega(\Psi, Q\Psi) - \frac{1}{3}\omega(\Psi, m_2(\Psi, \Psi)) \tag{7.1}$$

we consider a modification of the two-product

$$m_2(\cdot, \cdot) \to M_2(\cdot, \cdot) = e^{-\lambda L_0} m_2\left(e^{-\lambda L_0}\cdot, e^{-\lambda L_0}\cdot\right). \tag{7.2}$$

This new product $M_2$ is not associative anymore, hence to ensure gauge invariance, we need to introduce higher products $M_{n\geq 3}$ to obtain an $A_\infty$-algebra as described in detail in [71, 72]. Erbin and Firat have shown in [13] that it is also possible to obtain the stubbed $A_\infty$-theory from a cubic theory containing an auxiliary string field. By integrating out the auxiliary field in a specific way one can either get the stubbed theory or go back to Witten theory. Their results were then generalized to the full quantum level by Maccaferri et al in [50].

In both works it is a necessary condition for the stub operator to be BPZ-even, which does not include the sliver frame stub $e^{-\lambda \mathcal{L}_0}$. Since most analytical solutions are formulated in the sliver frame, it is of interest to use $e^{-\lambda \mathcal{L}_0}$ since many algebraic manipulations become much simpler. While the higher products and the maps relating solutions were already constructed in [72], in this work we want to focus on the auxiliary field method and the quantum theory. We point out that geometric aspects of sliver frame stubs are actually very subtle and some aspects are adressed in [72], see also [42]. In this paper we focus solely on algebraic aspects.

The paper is organized as follows: Section 2.1 is just a summary of the work by Erbin and Firat which is included for self-consistency and does not contain any new results. In section 2.2 we work out the generalization to the sliver frame and also discuss an alternative way to obtain Witten theory by integrating out the auxiliary field. Section 2.3 is then devoted to cohomomorphisms and how to obtain classical solutions of the stubbed equations of motion. In section 3 we show that our construction also works at the full quantum level without any additional constraints. Finally, in section 4 we apply the formalism to two known analytic solutions, namely the tachyon vacuum and one identity-like solution.

## 7.2 Sliver frame stubs as an auxiliary field

### 7.2.1 Ordinary stubs as an auxiliary field

In this section we want to give an overview of the results by Erbin and Firat [13], who managed to decribe the stubbed $A_\infty$-theory in a cubic way using an additional auxiliary string field. By integrating out the auxiliary field via homotopy transfer one obtains the infinitely many higher products.

The first step is to double the Hilbert space and turn the string field into a two-vector:

$$\Psi \in \mathcal{H} \quad \rightarrow \quad \Phi = \begin{pmatrix} \Psi \\ \Sigma \end{pmatrix} \in \mathcal{H} \oplus \mathcal{H}. \tag{7.3}$$

The proposed action for the combined string field $\Phi$ reads explicitly

$$S = -\frac{1}{2}\omega\left(\Psi, Q\Psi\right) - \frac{1}{2}\omega\left(\Sigma, \frac{Q}{1 - e^{-2\lambda L_0}}\Sigma\right)$$
$$+ \frac{1}{3}\omega\left(\Sigma - e^{-\lambda L_0}\Psi, m_2\left(\Sigma - e^{-\lambda L_0}\Psi, \Sigma - e^{-\lambda L_0}\Psi\right)\right) \tag{7.4}$$

and is invariant under the infinitesimal gauge transformations

$$\delta\Psi = Q\Lambda_1 + e^{-\lambda L_0} m_2\left(\Sigma - e^{-\lambda L_0}\Psi, \Lambda_2 - e^{-\lambda L_0}\Lambda_1\right)$$
$$+ e^{-\lambda L_0} m_2\left(\Lambda_2 - e^{-\lambda L_0}\Lambda_1, \Sigma - e^{-\lambda L_0}\Psi\right), \tag{7.5}$$

$$\delta\Sigma = Q\Lambda_2 + \left(e^{-2\lambda L_0} - 1\right) m_2\left(\Sigma - e^{-\lambda L_0}\Psi, \Lambda_2 - e^{-\lambda L_0}\Lambda_1\right)$$
$$+ \left(e^{-2\lambda L_0} - 1\right) m_2\left(\Lambda_2 - e^{-\lambda L_0}\Lambda_1, \Sigma - e^{-\lambda L_0}\Psi\right). \tag{7.6}$$

Here, $\Lambda_1$ and $\Lambda_2$ are two independent gauge parameters both of ghost number zero. The algebraic ingredients giving rise to this action are a differential

$$V_1\left(\Phi\right) = \begin{pmatrix} Q\Psi \\ Q\Sigma \end{pmatrix}, \tag{7.7}$$

an associative product
$$V_2(\Phi_1, \Phi_2) = \begin{pmatrix} e^{-\lambda L_0} m_2 \left(\Sigma_1 - e^{-\lambda L_0}\Psi_1, \Sigma_2 - e^{-\lambda L_0}\Psi_2\right) \\ \left(e^{-2\lambda L_0} - 1\right) m_2 \left(\Sigma_1 - e^{-\lambda L_0}\Psi_1, \Sigma_2 - e^{-\lambda L_0}\Psi_2\right) \end{pmatrix} \tag{7.8}$$
and a bilinear form
$$\Omega(\Phi_1, \Phi_2) = \omega(\Psi_1, \Psi_2) + \omega\left(\Sigma_1, \frac{1}{1 - e^{-2\lambda L_0}}\Sigma_2\right). \tag{7.9}$$
Together, $V_1$, $V_2$ and $\Omega$ form a cyclic differential graded algebra. We see that $\Omega$ is ill-defined on states in the kernel of $L_0$. To solve that, we just assume that $\Sigma$ can only take values outside of $\ker L_0$, such that our Hilbert space is effectively $\mathcal{H} \oplus (\mathcal{H} \setminus \ker L_0)$. In [50] it has been shown that this action can be obtained by adding a trivial, non-propagating field to Witten theory and then performing an orthogonal rotation amongst the two fields.

To obtain the non-polynomial stubbed action one has to integrate out the auxiliary field $\Sigma$ using homotopy transfer. We therefore define the natural inclusion and projection maps
$$i_S = \begin{pmatrix} 1 \\ 0 \end{pmatrix}, \quad p_S = \begin{pmatrix} 1 & 0 \end{pmatrix} \tag{7.10}$$
and choose for the homotopy
$$h_S = \begin{pmatrix} 0 & 0 \\ 0 & -\frac{b_0}{L_0} \end{pmatrix} \tag{7.11}$$
which fulfills the Hodge-Kodaira relation
$$h_S V_1 + V_1 h_S = i_S p_S - 1. \tag{7.12}$$
Those maps give rise to a special deformation retract which means they obey $p_S i_S = 1$ as well as the side conditions
$$h_S i_S = 0, \quad p_S h_S = 0, \quad h_S h_S = 0. \tag{7.13}$$
In [71] it has been shown that those conditions are actually not necessary to perform homotopy transfer, only the Hodge-Kodaira relation is needed. However, if they are obeyed, we can follow the standard procedure known in the literature without any modifications. Applying the tensor trick (see for instance [84]) and adding $\mathbf{V_2}$ as a perturbation, we get from the homological perturbation lemma the new products as
$$M_n = \pi_1 \mathbf{p_S} \mathbf{V_2} (1 - \mathbf{h_S} \mathbf{V_2})^{-1} \mathbf{i_S} \pi_n, \quad n \geq 2. \tag{7.14}$$
or more explicitly
$$M_2(\cdot, \cdot) = e^{-\lambda L_0} m_2 \left(e^{-\lambda L_0}\cdot, e^{-\lambda L_0}\cdot\right)$$
$$M_3(\cdot, \cdot, \cdot) = e^{-\lambda L_0} m_2 \left(e^{-\lambda L_0}\cdot, \frac{e^{-2\lambda L_0} - 1}{L_0} b_0 m_2 \left(e^{-\lambda L_0}\cdot, e^{-\lambda L_0}\cdot\right)\right)$$
$$+ e^{-\lambda L_0} m_2 \left(\frac{e^{-2\lambda L_0} - 1}{L_0} b_0 m_2 \left(e^{-\lambda L_0}\cdot, e^{-\lambda L_0}\cdot\right), e^{-\lambda L_0}\cdot\right) \tag{7.15}$$
$$\vdots$$

In [13] it has been shown that they are exactly equivalent to the higher products constructed in [71].

If we want to get back Witten theory, we have to project on the subspace of $\mathcal{H} \oplus (\mathcal{H} \setminus \ker L_0)$ given by elements of the form $\begin{pmatrix} \Psi \\ -2\sinh(\lambda L_0)\Psi \end{pmatrix}$. The inclusion, projection and homotopy defined in [13] read

$$i_W = \begin{pmatrix} 1 \\ -2\sinh(\lambda L_0) \end{pmatrix}, \quad p_W = \frac{1}{1+4\sinh^2(\lambda L_0)}\begin{pmatrix} 1 & -2\sinh(\lambda L_0) \\ -2\sinh(\lambda L_0) & 4\sinh^2(\lambda L_0) \end{pmatrix}, \quad (7.16)$$

$$h_W = -\frac{b_0}{L_0}\frac{1}{1+4\sinh^2(\lambda L_0)}\begin{pmatrix} 4\sinh^2(\lambda L_0) & 2\sinh(\lambda L_0) \\ 2\sinh(\lambda L_0) & 1 \end{pmatrix}. \quad (7.17)$$

Again, they form a special deformation retract so homotopy transfer works straightforwardly. Especially, the "magical relation"

$$h_W V_2 = 0 \quad (7.18)$$

holds which can be seen by direct calculation. In fact, by writing out (7.14) in components and using the side conditions, one can see that $h_W V_2$ appears in all terms contained in the products with more then two inputs. This means that all the higher products vanish identically and we are left with only a two-product and, as a result, a cubic theory. If we write down the action with input fields $i_W \Psi$ and the perturbed product we get

$$S = -\frac{1}{2}\Omega\left(i_W\Psi, V_1(i_W\Psi)\right) - \frac{1}{3}\Omega\left(i_W\Psi, p_W V_2(i_W\Psi, i_W\Psi)\right)$$
$$= -\frac{1}{2}\omega\left(e^{\lambda L_0}\Psi, Qe^{\lambda L_0}\Psi\right) - \frac{1}{3}\omega\left(e^{\lambda L_0}\Psi, m_2\left(e^{\lambda L_0}\Psi, e^{\lambda L_0}\Psi\right)\right) \quad (7.19)$$

which is exactly the Witten action up to the field redefinition $\Psi \to e^{\lambda L_0}\Psi$.

We want to point out that the Witten product $m_2$ was not directly obtained from homotopy transfer: First, the $p_W$ in (7.16) is a two-by-two matrix, hence the target space of the homotopy transfer is still $\mathcal{H} \oplus (\mathcal{H} \setminus \ker L_0)$. Only after writing down the action using $\Omega$ and undoing the field redefinition $\Psi \to e^{\lambda L_0}\Psi$ it is possible to read off $m_2$. In the next section we will give a more direct way of obtaining Witten theory.

### 7.2.2 Generalization to sliver frame stubs

Now we turn to the actual task and replace the stub operator by its sliver frame counterpart. The whole construction above was only valid for BPZ-even stub operators, hence if we want to use $e^{-\lambda \mathcal{L}_0}$ we need to generalize it. We propose to use the following product and bilinear form (the differential will be unchanged)[37]:

$$V_2'(\Phi_1, \Phi_2) = \begin{pmatrix} e^{-\lambda \mathcal{L}_0^*} m_2\left(\Sigma_1 - e^{-\lambda \mathcal{L}_0}\Psi_1, \Sigma_2 - e^{-\lambda \mathcal{L}_0}\Psi_2\right) \\ \left(e^{-\lambda \mathcal{L}_0} e^{-\lambda \mathcal{L}_0^*} - 1\right) m_2\left(\Sigma_1 - e^{-\lambda \mathcal{L}_0}\Psi_1, \Sigma_2 - e^{-\lambda \mathcal{L}_0}\Psi_2\right) \end{pmatrix} \quad (7.20)$$

---
[37]From now on, primed objects always refer to the sliver frame stubs.

$$\Omega'(\Phi_1, \Phi_2) = \omega(\Psi_1, \Psi_2) + \omega\left(\Sigma_1, \frac{1}{1 - e^{-\lambda \mathcal{L}_0} e^{-\lambda \mathcal{L}_0^*}} \Sigma_2\right). \tag{7.21}$$

One can show straightforwardly that the new product is again associative and cyclic with respect to $\Omega'$, so all algebraic relations are preserved. The new action reads

$$\begin{aligned} S' = &-\frac{1}{2}\omega(\Psi, Q\Psi) - \frac{1}{2}\omega\left(\Sigma, \frac{Q}{1 - e^{-\lambda \mathcal{L}_0} e^{-\lambda \mathcal{L}_0^*}} \Sigma\right) \\ &+ \frac{1}{3}\omega\left(\Sigma - e^{-\lambda \mathcal{L}_0}\Psi, m_2\left(\Sigma - e^{-\lambda \mathcal{L}_0}\Psi, \Sigma - e^{-\lambda \mathcal{L}_0}\Psi\right)\right) \end{aligned} \tag{7.22}$$

and is invariant under the gauge transformations

$$\begin{aligned} \delta\Psi = &Q\Lambda_1 + e^{-\lambda \mathcal{L}_0^*} m_2\left(\Sigma - e^{-\lambda \mathcal{L}_0}\Psi, \Lambda_2 - e^{-\lambda \mathcal{L}_0}\Lambda_1\right) \\ &+ e^{-\lambda \mathcal{L}_0^*} m_2\left(\Lambda_2 - e^{-\lambda \mathcal{L}_0}\Lambda_1, \Sigma - e^{-\lambda \mathcal{L}_0}\Psi\right), \end{aligned} \tag{7.23}$$

$$\begin{aligned} \delta\Sigma = &Q\Lambda_2 + \left(e^{-\lambda \mathcal{L}_0} e^{-\lambda \mathcal{L}_0^*} - 1\right) m_2\left(\Sigma - e^{-\lambda \mathcal{L}_0}\Psi, \Lambda_2 - e^{-\lambda \mathcal{L}_0}\Lambda_1\right) \\ &+ \left(e^{-\lambda \mathcal{L}_0} e^{-\lambda \mathcal{L}_0^*} - 1\right) m_2\left(\Lambda_2 - e^{-\lambda \mathcal{L}_0}\Lambda_1, \Sigma - e^{-\lambda \mathcal{L}_0}\Psi\right). \end{aligned} \tag{7.24}$$

The operator $O := \frac{1}{1 - e^{-\lambda \mathcal{L}_0} e^{-\lambda \mathcal{L}_0^*}}$ appearing in the kinetic term actually deserves attention. First of all, we can use the formulas of [72, 70] to write

$$e^{-\lambda \mathcal{L}_0} e^{-\lambda \mathcal{L}_0^*} = e^{(e^{-\lambda} - 1)(\mathcal{L}_0 + \mathcal{L}_0^*)} \tag{7.25}$$

so in fact $O$ is a function of $\hat{\mathcal{L}} = \mathcal{L}_0 + \mathcal{L}_0^*$. In general, the exponential of $\hat{\mathcal{L}}$ acts on an arbitrary string field as

$$e^{-t\hat{\mathcal{L}}}\Psi = e^{-tK}\Psi e^{-tK}, \tag{7.26}$$

hence the action of $O$ can be defined via its Laplace transform. One state in the kernel of $\hat{\mathcal{L}}$ for which applying $O$ could be problematic is the sliver state $e^{-\infty K}$. However, this state is of formal nature and singular in many senses so we just choose to exclude it from our Hilbert space. We will see that in the end, after integrating out the auxiliary field, $O$ does not appear anymore.

The homotopy transfer to the stubbed theory is essentially unchanged and again generated by (7.10). For the homotopy, the most compatible choice is

$$h'_S = \begin{pmatrix} 0 & 0 \\ 0 & -\frac{\mathcal{B}_0 + \mathcal{B}_0^*}{\mathcal{L}_0 + \mathcal{L}_0^*} \end{pmatrix} : \tag{7.27}$$

The reason is that in the higher products $h'_S$ always acts on the output of $V'_2$ such that the expression

$$h'_S\left(e^{-\lambda \mathcal{L}_0} e^{-\lambda \mathcal{L}_0^*} - 1\right) = h'_S\left(e^{(e^{-\lambda} - 1)(\mathcal{L}_0 + \mathcal{L}_0^*)} - 1\right) \tag{7.28}$$

appears. Hence, with this choice, $h'_S V'_2$ is manifestly non-singular and well-defined on all states. The new higher products one receives are then explicitly given by

$$M'_2(\cdot, \cdot) = e^{-\lambda \mathcal{L}_0^*} m_2 \left( e^{-\lambda \mathcal{L}_0} \cdot, e^{-\lambda \mathcal{L}_0} \cdot \right),$$

$$M'_3(\cdot, \cdot, \cdot) = e^{-\lambda \mathcal{L}_0^*} m_2 \left( e^{-\lambda \mathcal{L}_0} \cdot, \frac{e^{(e^{-\lambda}-1)(\mathcal{L}_0+\mathcal{L}_0^*)} - 1}{(\mathcal{L}_0 + \mathcal{L}_0^*)} (\mathcal{B}_0 + \mathcal{B}_0^*) m_2 \left( e^{-\lambda \mathcal{L}_0} \cdot, e^{-\lambda \mathcal{L}_0} \cdot \right) \right)$$

$$+ e^{-\lambda \mathcal{L}_0^*} m_2 \left( \frac{e^{(e^{-\lambda}-1)(\mathcal{L}_0+\mathcal{L}_0^*)} - 1}{(\mathcal{L}_0 + \mathcal{L}_0^*)} (\mathcal{B}_0 + \mathcal{B}_0^*) m_2 \left( e^{-\lambda \mathcal{L}_0} \cdot, e^{-\lambda \mathcal{L}_0} \cdot \right), e^{-\lambda \mathcal{L}_0} \cdot \right). \quad (7.29)$$

$\vdots$

They are identical to the products with curly stubs derived in [72]. We also note that the operator $O$ does not appear anymore and all quantities are well-defined.

The transfer to Witten theory is actually a bit more subtle: It is possible to construct maps $\tilde{i}_W$ and $\tilde{p}_W$ which lead directly to Witten theory, without any field redefinition being necessary. Let us first see how this works for the ordinary stubs where we define

$$\tilde{i}_W = \begin{pmatrix} e^{-\lambda L_0} \\ e^{-2\lambda L_0} - 1 \end{pmatrix}, \qquad \tilde{p}_W = \begin{pmatrix} e^{-\lambda L_0} & -1 \end{pmatrix}. \quad (7.30)$$

They do not satisfy the typical properties of an inclusion and a projection, but this is by no means necessary; they only need to commute with the differential and obey the Hodge-Kodaira relation, which can be solved by

$$\tilde{h}_W = \frac{b_0}{L_0} \begin{pmatrix} e^{-2\lambda L_0} - 1 & -e^{-\lambda L_0} \\ e^{-3\lambda L_0} - e^{-\lambda L_0} & -e^{-2\lambda L_0} \end{pmatrix}. \quad (7.31)$$

Now the maps $\tilde{i}_W$, $\tilde{p}_W$ and $\tilde{h}_W$ even satisfy all the SDR-conditions. The tensorial structure of $\tilde{i}$ and $\tilde{p}$ implies that the resulting theory is defined on just one copy of $\mathcal{H}$, so we can directly calculate all the products without computing the action first. Most importantly, the magical relation $\tilde{h}_W V_2 = 0$ is satisfied again and therefore only the differential and the two-product are non-vanishing. They are given by

$$\tilde{p}_W V_1 \tilde{i}_W = Q, \qquad \tilde{p}_W V_2 \tilde{i}_W^{\otimes 2} = m_2 \quad (7.32)$$

hence we indeed end up with Witten theory if we take the bilinear form on the target space to be $\omega$.

For the generalization to the sliver frame stubs we propose the following maps:

$$\tilde{i}'_W = \begin{pmatrix} e^{-\lambda \mathcal{L}_0^*} \\ e^{-\lambda \mathcal{L}_0} e^{-\lambda \mathcal{L}_0^*} - 1 \end{pmatrix}, \qquad \tilde{p}'_W = \begin{pmatrix} e^{-\lambda \mathcal{L}_0} & -1 \end{pmatrix}. \quad (7.33)$$

$$\tilde{h}'_W = \frac{\mathcal{B}_0 + \mathcal{B}_0^*}{\mathcal{L}_0 + \mathcal{L}_0^*} \begin{pmatrix} e^{-\lambda \mathcal{L}_0^*} e^{-\lambda \mathcal{L}_0} - 1 & -e^{-\lambda \mathcal{L}_0^*} \\ e^{-\lambda \mathcal{L}_0} e^{-\lambda \mathcal{L}_0^*} e^{-\lambda \mathcal{L}_0} - e^{-\lambda \mathcal{L}_0} & -e^{-\lambda \mathcal{L}_0} e^{-\lambda \mathcal{L}_0^*} \end{pmatrix}. \quad (7.34)$$

Again, $\tilde{i}'_W$, $\tilde{p}'_W$ and $\tilde{h}'_W$ form an SDR and obey $\tilde{h}'_W V'_2 = 0$, so we can directly transfer to Witten theory:
$$\tilde{p}'_W V'_1 \tilde{i}'_W = Q, \qquad \tilde{p}'_W V'_2 \tilde{i}'^{\otimes 2}_W = m_2. \tag{7.35}$$

### 7.2.3 Cohomomorphisms and classical solutions

We want to study now how to obtain classical solutions of the curly auxiliary field theory, i. e. solve the equations of motion derived from (7.22). In coalgebra notation they read

$$(\mathbf{V'_1} + \mathbf{V'_2}) \frac{1}{1 - \Phi'} \equiv \mathbf{V'} \frac{1}{1 - \Phi'} = 0, \tag{7.36}$$

where the schematic notation

$$\frac{1}{1 - \Phi'} =: \sum_{n=0}^{\infty} \Phi'^{\otimes n} \tag{7.37}$$

is used. Given an analytic solution $\Psi$ of Witten theory, for example the tachyon vacuum, we can ask if there exists a non-linear map $A$ such that $A(\Psi)$ solves the equation above. A sufficient condition is that $\mathbf{A}$ is a cohomomorphism that intertwines between the algebras of Witten theory and the curly auxiliary theory, i. e.

$$\mathbf{V'A} = \mathbf{Am}. \tag{7.38}$$

for $\mathbf{m} \equiv \mathbf{Q} + \mathbf{m_2}$. This relation arises as the chain map relation automatically obeyed by the interacting inclusion $\tilde{\mathbf{I}}'_\mathbf{W}$ coming from the homological perturbation lemma. Using the explicit expression for $\tilde{\mathbf{I}}'_\mathbf{W}$ we can directly observe that

$$\tilde{\mathbf{I}}'_\mathbf{W} = \left(1 - \tilde{\mathbf{h}}'_\mathbf{W} \mathbf{V'_2}\right)^{-1} \tilde{\mathbf{i}}'_\mathbf{W} = \tilde{\mathbf{i}}'_\mathbf{W} \tag{7.39}$$

because of the magical relation $\tilde{\mathbf{h}}'_\mathbf{W} \mathbf{V'_2} = 0$. So using (7.33) we conclude that

$$\Phi = \begin{pmatrix} e^{-\lambda \mathcal{L}_0^*} \Psi \\ e^{-\lambda \mathcal{L}_0} e^{-\lambda \mathcal{L}_0^*} \Psi - \Psi \end{pmatrix} \tag{7.40}$$

is a solution of the curly auxiliary field theory if $\Psi$ is a solution of Witten theory. By direct computation we can see that $\tilde{\mathbf{I}}'_\mathbf{W}$ is also cyclic, i. e. it preserves the action:[38]

$$S = -\frac{1}{2} \Omega \left(\Phi, V'_1(\Phi)\right) - \frac{1}{3} \Omega \left(\Phi, V'_2(\Phi, \Phi)\right) = -\frac{1}{2} \omega \left(\Psi, Q(\Psi)\right) - \frac{1}{3} \omega \left(\Psi, m_2(\Psi, \Psi)\right). \tag{7.41}$$

We can also ask if it is possible to transfer directly from Witten theory to the curly-stubbed theory as it is discussed in [72]. For this we must combine the two homotopy transfers, so first include from Witten theory to the curly auxiliary field theory and then project down to the stubbed theory.

---

[38]Cyclicity of a cohomomorphism is actually a delicate question and was discussed in more detail in [72].

The combined map is hence the product $\mathbf{P'_S}\tilde{\mathbf{I}}'_\mathbf{W}$, where $\mathbf{P'_S} = \mathbf{p'_S}(1 - \mathbf{V'_2}\mathbf{h'_S})^{-1}$ is the interacting projection of the homotopy transfer to the stubbed theory. We can compute explicitly

$$\pi_1 \mathbf{P'_S}\tilde{\mathbf{I}}'_\mathbf{W}\pi_1 = p'_S \tilde{i}'_W = e^{-\lambda \mathcal{L}_0^*}$$

$$\pi_1 \mathbf{P'_S}\tilde{\mathbf{I}}'_\mathbf{W}\pi_2 = p'_S V'_2 h'_S \tilde{i}'^{\otimes 2}_W = e^{-\lambda \mathcal{L}_0^*} m_2\left(\cdot, \frac{\mathcal{B}_0 + \mathcal{B}_0^*}{\mathcal{L}_0 + \mathcal{L}_0^*}\left(e^{-\lambda \mathcal{L}_0}e^{-\lambda \mathcal{L}_0^*} - 1\right)\cdot\right)$$
$$+ e^{-\lambda \mathcal{L}_0^*} m_2\left(\frac{\mathcal{B}_0 + \mathcal{B}_0^*}{\mathcal{L}_0 + \mathcal{L}_0^*}\left(e^{-\lambda \mathcal{L}_0}e^{-\lambda \mathcal{L}_0^*} - 1\right)\cdot, e^{-\lambda \mathcal{L}_0}e^{-\lambda \mathcal{L}_0^*}\cdot\right) \quad (7.42)$$

and observe that $\mathbf{P'_S}\tilde{\mathbf{I}}'_\mathbf{W}$ is exactly equal to the non-cyclic cohomomorphism $\mathbf{P}$ constructed in [72] up to second order. In the appendix we give a full proof that they are equal to all orders.

## 7.3 Curly quantum stubs

In this section we want to generalize the curly stubbed theory to the full quantum level as it was done in [50] for BPZ-even stubs. For that we need to turn the $A_\infty$-algebra into a quantum $A_\infty$-algebra, i. e. the action has to obey the quantum BV master equation

$$\frac{1}{2}(S, S)_{BV} + \Delta S = 0 \quad (7.43)$$

where $(\cdot, \cdot)_{BV}$ is the BV-bracket and $\Delta$ is the BV-Laplacian.

Suppose we are given any action of the form

$$S(\Psi) = \sum_{n=1}^{\infty} \frac{1}{n+1} \omega\left(\Psi, m_n\left(\Psi^{\otimes n}\right)\right) \quad (7.44)$$

with some higher products $m_n$ and a bilinear form $\omega$. If we package the products into a coderivation as $\mathbf{m} = \mathbf{m_1} + \mathbf{m_2} + ...$, then $S$ can be written in the form

$$S(\Psi) = \int_0^1 dt\, \omega\left(\pi_1 \partial_t \frac{1}{1 - \Psi(t)}, \pi_1 \mathbf{m} \frac{1}{1 - \Psi(t)}\right) \quad (7.45)$$

where $\Psi(t)$ is a smooth interpolation between $\Psi(0) = 0$ and $\Psi(1) = \Psi$, see for instance [84]. Using this representation, the BV master equation can be shown to take the following form (see [52]):

$$\int_0^1 dt\, \omega\left(\pi_1 \partial_t \frac{1}{1 - \Psi(t)}, \pi_1 \left(\mathbf{m}^2 + \mathbf{m}\mathbf{U}\right) \frac{1}{1 - \Psi(t)}\right) = 0. \quad (7.46)$$

Here, $\mathbf{U}$ is the tensor coalgebra extension of the Poisson bi-vector

$$U = \frac{(-)^{v^a}}{2} v_a \wedge v^a, \quad (7.47)$$

where the $v_a$ provide a basis of $\mathcal{H}$. The basis vectors are taken to be canonically normalized via $\omega(v^a, v_b) = -\omega(v_b, v^a) = \delta^a_b$, which means that $U$ implicitly depends on $\omega$. $\mathbf{U}$ is then defined by inserting $v_a$ and $v^a$ in all possible combinations of positions in a tensor product. This can be understood from the fact that $U$ is the dual object to the BV-Laplacian $\Delta$: Since $\Delta$ is a second order differential operator, it removes two elements from a tensor product. Similarly, $U$ inserts two elements into a tensor product. $\Delta$ is not a derivation with respect to the tensor product and likewise $\mathbf{U}$ is not a coderivation on the tensor coalgebra and does not obey the co-Leibniz rule.

If we first consider ordinary Witten theory, we immediately notice that $\mathbf{m}^2 = 0$ already. The only non-trivial contribution of the second term is $m_2(U)$, which is however actually ill-defined and non-zero [82]. This resembles the fact that pure open string field theory is incomplete on the quantum level because it does not include the emission of closed strings. The inconsistency is removed in the full quantized open-closed string field theory. Since we are purely interested in the open sector, we will just formally demand

$$m_2(U) =: 0 \tag{7.48}$$

as a quantum consistency condition.

Now let us consider the extended action (7.22): Again we know that $\mathbf{V'}^2 = (\mathbf{V'_1} + \mathbf{V'_2})^2 = 0$ so we only have to examine $V'_2(U')$. First, we choose a basis of $\mathcal{H} \oplus \mathcal{H}$ as

$$\left\{ \begin{pmatrix} v_a \\ 0 \end{pmatrix}, \begin{pmatrix} 0 \\ w_b \end{pmatrix} \right\}, \tag{7.49}$$

where $v_a$ and $w_b$ can in principle be independent. By looking at the symplectic form $\Omega'$ (7.21) we see that the basis vectors of the auxiliary field change non-trivially under dualization. In the end we get

$$U' = \frac{(-)^{v^a}}{2} v_a \wedge v^a + \frac{(-)^{w^a}}{2} w_a \wedge \left(1 - e^{-\lambda \mathcal{L}_0} e^{-\lambda \mathcal{L}_0^*}\right) w^a \tag{7.50}$$

which can be directly inserted into

$$V'_2(U') = \begin{pmatrix} e^{-\lambda \mathcal{L}_0^*} \left( \frac{(-)^{v^a}}{2} m_2 \left( e^{-\lambda \mathcal{L}_0} v_a \wedge e^{-\lambda \mathcal{L}_0} v^a \right) \right. \\ \left. + \frac{(-)^{w^a}}{2} m_2 \left( w_a \wedge \left(1 - e^{-\lambda \mathcal{L}_0} e^{-\lambda \mathcal{L}_0^*}\right) w^a \right) \right) \\ \left( e^{-\lambda \mathcal{L}_0} e^{-\lambda \mathcal{L}_0^*} - 1 \right) \left( \frac{(-)^{v^a}}{2} m_2 \left( e^{-\lambda \mathcal{L}_0} v_a \wedge e^{-\lambda \mathcal{L}_0} v^a \right) \right. \\ \left. + \frac{(-)^{w^a}}{2} m_2 \left( w_a \wedge \left(1 - e^{-\lambda \mathcal{L}_0} e^{-\lambda \mathcal{L}_0^*}\right) w^a \right) \right) \end{pmatrix}. \tag{7.51}$$

We can relate the two bases of $\mathcal{H}$ according to

$$w_a = e^{-\lambda \mathcal{L}_0} v_a \tag{7.52}$$

This is a well-defined basis change; if we take for instance Fock states as basis states, the operator $e^{-\lambda \mathcal{L}_0}$ just produces a number. From the normalization condition $\omega(v^a, v_b) = \delta^a_b$ it follows that

$$v^a = e^{-\lambda \mathcal{L}_0^*} w^a \tag{7.53}$$

and we see that the term in parentheses becomes

$$\frac{(-)^{w^a}}{2} m_2\left(w_a \wedge e^{-\lambda \mathcal{L}_0} e^{-\lambda \mathcal{L}_0^*} w^a\right) + \frac{(-)^{w^a}}{2} m_2\left(w_a \wedge \left(1 - e^{-\lambda \mathcal{L}_0} e^{-\lambda \mathcal{L}_0^*}\right) w^a\right) = \frac{(-)^{w^a}}{2} m_2\left(w_a \wedge w^a\right). \tag{7.54}$$

This is precisely the original quantum consistency condition, so we have shown that quantizing the theory with sliver frame stubs does not require any additional constraints.

Integrating out the auxiliary field works very similarly to the classical case: We start with the SDR between $\mathcal{H} \oplus \mathcal{H}$ and $\mathcal{H}$ given by the natural inclusion and projection maps (7.10) and the homotopy (7.27) and lift it to the tensor coalgebra as usual. The difference is now that we perturb not only by the product $\mathbf{V'_2}$, but by the combination $\mathbf{V'_2} + \mathbf{U'}$. By the homological perturbation lemma we get for the quantum higher products

$$\left(M'_{qu}\right)_n = \pi_1 \mathbf{M'_{qu}} \pi_n = \pi_1 \left(\mathbf{Q} + \mathbf{p_S} \mathbf{V'_2} \frac{1}{1 + \mathbf{h'_S} \mathbf{V'_2} + \mathbf{h'_S} \mathbf{U'}} \mathbf{i_S}\right) \pi_n. \tag{7.55}$$

Note that although $\mathbf{U'}$ is not a coderivation, $\mathbf{M'_{qu}}$ is, since it obeys the co-Leibniz rule[39]. However, in general it does not square to zero because only

$$\left(\mathbf{M'_{qu}} + \mathbf{p_S} \mathbf{U'}\right)^2 = 0 \tag{7.56}$$

has to be true, see [50]. As a result, the quantum products $\left(M'_{qu}\right)_n$ do not form an ordinary $A_\infty$-algebra but a quantum $A_\infty$-algebra, which contains extra higher loop terms. The geometrical interpretation of that is that after introducing stubs, the moduli space of higher genus Riemann surfaces with boundary is not covered anymore by the Feynman diagrams. One needs extra vertices with intrinsic loops in the action, i.e. loops that do not arise from the Feynman rules.

## 7.4 Explicit solutions

One of the main motivations of introducing stubs in OSFT is to study the structure of solutions of $A_\infty$-theories, so in this section we want to apply our results to some known classical solutions.

### 7.4.1 Tachyon vacuum

Let us first consider the tachyon vacuum given by

$$\Psi_{TV} = e^{-\frac{K}{2}} c \frac{KB}{1 - e^{-K}} c e^{-\frac{K}{2}} \tag{7.57}$$

---

[39]GS wants to thank Jojiro Totsuka-Yoshinaka for discussions on that, see also [45]

with the elements of the $KBc$-algebra defined for instance in [20, 63].[40] It obeys the sliver gauge condition

$$\mathcal{B}_0 \Psi_{TV} = 0. \tag{7.58}$$

In order to calculate $\Phi'$ we first need to know the action of $e^{-\lambda \mathcal{L}_0^*}$ on $\Psi_{TV}$. The calculation was done in detail in [72]; what basically happens is that the width of the security strips $e^{-\frac{K}{2}}$ changes and the $K$ in the denominator gets rescaled by $e^\lambda$. The result is

$$e^{-\lambda \mathcal{L}_0^*} \Psi_{TV} = e^{-K\left(e^\lambda - \frac{1}{2}\right)} c \frac{KB}{1 - e^{-e^\lambda K}} c e^{-K\left(e^\lambda - \frac{1}{2}\right)}. \tag{7.59}$$

For the second component we use again the formulas (7.25) and (7.26) to get

$$e^{-\lambda \mathcal{L}_0} e^{-\lambda \mathcal{L}_0^*} \Psi_{TV} = e^{-K\left(\frac{3}{2} - e^{-\lambda}\right)} c \frac{KB}{1 - e^{-K}} c e^{-K\left(\frac{3}{2} - e^{-\lambda}\right)}. \tag{7.60}$$

Now we can write down the full result as

$$\Phi_{TV} = \begin{pmatrix} e^{-K\left(e^\lambda - \frac{1}{2}\right)} c \frac{KB}{1 - e^{-e^\lambda K}} c e^{-K\left(e^\lambda - \frac{1}{2}\right)} \\ e^{-K\left(\frac{3}{2} - e^{-\lambda}\right)} c \frac{KB}{1 - e^{-K}} c e^{-K\left(\frac{3}{2} - e^{-\lambda}\right)} - e^{-\frac{K}{2}} c \frac{KB}{1 - e^{-K}} c e^{-\frac{K}{2}} \end{pmatrix}. \tag{7.61}$$

This result is remarkably simple compared to the expression for the tachyon vacuum in the non-polynomial stubbed theory, cf. formula 4.24 in [72]. The complexity comes in only after integrating out the auxiliary field.

### 7.4.2 Identity-like solutions

Another interesting application of stubs is the possibility of taming certain singularities appearing in Witten theory. Here we want to take a short look at one example, namely the famous identity-like solution

$$\Psi_{IL} = c\left(1 - K\right) \tag{7.62}$$

It obeys the Witten equations of motion as can be checked easily:

$$Q\Psi_{IL} + m_2\left(\Psi_{IL}, \Psi_{IL}\right) = cKc\left(1 - K\right) + c\left(1 - K\right)c\left(1 - K\right) = cKc\left(1 - K\right) - cKc\left(1 - K\right) = 0. \tag{7.63}$$

In [1] it was shown that $\Psi_{IL}$ is actually gauge related to the tachyon vacuum in the form

$$\Psi_{STV} = c\left(1 + K\right) Bc \frac{1}{1 + K} \tag{7.64}$$

called simple tachyon vacuum, which was derived in [27]. Indeed, using the gauge parameter

$$X = 1 + cBK, \qquad X^{-1} = 1 - cBK \frac{1}{1 + K} \tag{7.65}$$

---

[40]Within calculations using $KBc$-elements we will write the Witten product $m_2$ just as a normal product, according to the literature.

one can prove by direct computation that

$$\Psi_{STV} = X\Psi_{IL}X^{-1} + XQX^{-1}. \tag{7.66}$$

The problems start when one tries to compute the action: Since $\Psi_{IL}$ is represented as a worldsheet strip of width zero, one would have to calculate a correlator on a cylinder of vanishing circumference. The result then always depends on the chosen regularization and is hence ambiguous.

In a theory with stubs however, this problem is avoided so we have good reasons to believe that all quantities are well-defined. In [?, 20] a proposal for a criterion is made that "good" solutions should obey: Every function of $K$ appearing in the string field should be bounded for $K \geq 0$. This is clearly not the case for $\Psi_{IL}$. If we calculate the associated $\Phi_{IL}$ according to the formula (7.40) we find

$$\Phi_{IL} = \begin{pmatrix} e^{-\lambda \mathcal{L}_0^*} c(1-K) \\ e^{-\lambda \mathcal{L}_0} e^{-\lambda \mathcal{L}_0^*} c(1-K) - c(1-K) \end{pmatrix} = \begin{pmatrix} e^{-\frac{K}{2}(e^\lambda - 1)} c\left(e^{-\lambda} - K\right) e^{-\frac{K}{2}(e^\lambda - 1)} \\ e^{(e^{-\lambda}-1)K} c(1-K) e^{(e^{-\lambda}-1)K} - c(1-K) \end{pmatrix} \tag{7.67}$$

where we used again the formulas derived in [72]. Although some of the functions of $K$ are now exponentially damped, there is still the $cK$-term in the $\Sigma$-field which is unbounded. If we try to compute the action we see from (7.41) that we just end up with the Witten action and hence run into the same problems as before. However, if we integrate out the auxiliary field and transfer to the non-polynomial theory, the situation changes: In section two we calculated the cohomomorphism we need to apply and have shown that it is actually identical to $\mathbf{P}$ from [72, 71]. To second order in the string field we get

$$\Psi'_{IL} \equiv \pi_1 \mathbf{P} \frac{1}{1 - \Psi_{IL}} = e^{-\lambda \mathcal{L}_0^*} c(1-K) + e^{-\lambda \mathcal{L}_0^*} m_2 \left( c(1-K), \frac{\mathcal{B}_0 + \mathcal{B}_0^*}{\mathcal{L}_0 + \mathcal{L}_0^*} \left( e^{-\lambda \mathcal{L}_0} e^{-\lambda \mathcal{L}_0^*} - 1 \right) c(1-K) \right)$$
$$+ e^{-\lambda \mathcal{L}_0^*} m_2 \left( \frac{\mathcal{B}_0 + \mathcal{B}_0^*}{\mathcal{L}_0 + \mathcal{L}_0^*} \left( e^{-\lambda \mathcal{L}_0} e^{-\lambda \mathcal{L}_0^*} - 1 \right) c(1-K), e^{-\lambda \mathcal{L}_0} e^{-\lambda \mathcal{L}_0^*} c(1-K) \right). \tag{7.68}$$

To simplify the action of $\frac{1}{\mathcal{L}_0 + \mathcal{L}_0^*}$ we use a Schwinger parameter and write

$$\frac{\left(e^{-\lambda \mathcal{L}_0} e^{-\lambda \mathcal{L}_0^*} - 1\right)}{\mathcal{L}_0 + \mathcal{L}_0^*} c(1-K) = \int_0^\infty dt \left( e^{(e^{-\lambda}-1-t)(\mathcal{L}_0 + \mathcal{L}_0^*)} c(1-K) - e^{-t(\mathcal{L}_0 + \mathcal{L}_0^*)} c(1-K) \right)$$
$$= -\int_0^{1-e^{-\lambda}} dt \, e^{-tK} c(1-K) e^{-tK}. \tag{7.69}$$

Together with the well-known formula

$$(\mathcal{B}_0 + \mathcal{B}_0^*) \Psi = B\Psi + (-)^{\text{gh}(\Psi)} \Psi B \tag{7.70}$$

we can write the final result as

$$\Psi'_{IL} = e^{-\frac{K}{2}(e^\lambda-1)} c\left(e^{-\lambda} - K\right) e^{-\frac{K}{2}(e^\lambda-1)}$$

$$- e^{-\frac{K}{2}(e^\lambda-1)} c\left(e^{-\lambda} - K\right) \int_0^{1-e^{-\lambda}} dt\, e^{-e^\lambda tK} (Bc - cB)\left(1 - e^\lambda K\right) e^{-\left((t+\frac{1}{2})e^\lambda - \frac{1}{2}\right)K}$$

$$- \int_0^{1-e^{-\lambda}} dt\, \Big(e^{-\left((t+\frac{1}{2})e^\lambda - \frac{1}{2}\right)K} (Bc - cB)\left(1 - e^\lambda K\right)$$

$$e^{(1-e^\lambda - e^\lambda t)K} c\left(e^{-\lambda} - K\right) e^{-\left((t+\frac{1}{2})e^\lambda - e^{-\lambda} + \frac{1}{2}\right)K}\Big) + \mathcal{O}\left(\Psi^3\right). \tag{7.71}$$

We see that every function of $K$ appearing is now exponentially damped and therefore bounded for positive $K$.[41] Since the higher orders of the cohomomorphism are constructed out of the same ingredients as the first two, we conclude that this result is true to all orders. We could now plug in the result into the stubbed action and get

$$S = -\frac{1}{2}\omega\left(\Psi'_{IL}, Q\Psi'_{IL}\right) - \frac{1}{3}\omega\left(\Psi'_{IL}, M_2\left(\Psi'_{IL}, \Psi'_{IL}\right)\right) - \frac{1}{4}\omega\left(\Psi'_{IL}, M_3\left(\Psi'_{IL}, \Psi'_{IL}, \Psi'_{IL}\right)\right) - \cdots \tag{7.72}$$

with the sliver frame products defined in [72]. This expression does not contain any singular correlators anymore and should hence yield a well-defined result. In [72] it has been argued that the on-shell value of the action is preserved even for the non-cyclic cohomomorphism **P**, so we expect the result to be the total energy of the tachyon vacuum, namely $-\frac{1}{2\pi^2}$. The calculation can in principle be done up to some finite order to check the claim and also study the convergence properties, but this goes beyond the scope of this paper.

## 7.5 Conclusion

In this paper we succeeded to extend the formalism developed by Erbin and Firat to non-BPZ-even stubs, in particular the sliver frame stub. We have also shown, following the route outlined by Maccaferri et al, that the sliver frame stubbed theory is consistent at the full quantum level without any additional assumptions. By applying our construction to the tachyon vacuum, we could see that its auxiliary field form is remarkably simple, in contrast to its expression in the stubbed theory. Moreover, the identity-like solution $c\left(1 - K\right)$ takes a non-singular form in the stubbed theory and gives rise to an action which is computable unambiguously. For future work it would be of interest to calculate this action up to some finite order and study its value and convergence properties. In general, examining the structure of solutions via the auxiliary field method can teach us important algebraic properties which can be relevant also in different context. For example, it might be possible to formulate other theories like open superstring field theory [23] or even closed string field theory [87] in a cubic way using auxiliary fields.

---

[41] The only potential problem could come from the lower integration bound in the second line. However, usually a continuous superposition of states with width going to zero as it appears for instance in the single tachyon vacuum [27] is much better behaved than a single identity-like state.

**Acknowledgements**

GS wants to thank Martin Schnabl for suggesting the topic and collaboration as well as Carlo Maccaferri, Alberto Ruffino, Harold Erbin and Atakan Firat for useful discussions. Our work has been funded by the Grant Agency of Czech Republic under the grant EXPRO 20-25775X.

## Appendix

In this appendix we want to prove the statement of the end of section 2.3, that $\mathbf{P'_S \tilde{I}'_W}$ is equal to the cohomomorphism $\mathbf{P}$ from [72] to all orders. The proof closely follows a related one in the appendix of [13].

The first step is to write the product $V'_2$ as

$$V'_2 = \tilde{i}'_W m_2 \tilde{p}'^{\otimes 2}_W \tag{7.73}$$

using the maps $\tilde{i}'_W$ and $\tilde{p}'_W$ we used earlier to project directly onto Witten theory. The equation can be lifted to the tensor coalgebra in the following way:

$$\mathbf{V'_2} = \mathbf{\Pi} \left( \tilde{\mathbf{i}}'_\mathbf{W} \mathbf{m_2} \tilde{\mathbf{p}}'_\mathbf{W} \right). \tag{7.74}$$

Here, $\Pi$ is a formal object that sets the combination $\tilde{i}'_W \tilde{p}'_W$ equal to unity whenever it appears in any tensor power. We can write the cohomomorphism $\mathbf{P'_S \tilde{I}'_W}$ now as

$$\mathbf{P'_S \tilde{I}'_W} = \mathbf{p'_S} \left( 1 - \mathbf{\Pi} \left( \tilde{\mathbf{i}}'_\mathbf{W} \mathbf{m_2} \tilde{\mathbf{p}}'_\mathbf{W} \right) \mathbf{h'_S} \right)^{-1} \tilde{\mathbf{i}}'_\mathbf{W} \tag{7.75}$$

and observe some relations: First, the combination $\tilde{p}'_W h'_S \tilde{i}'_W$ is equal to

$$\tilde{p}'_W h'_S \tilde{i}'_W = \begin{pmatrix} e^{-\lambda \mathcal{L}_0} & -1 \end{pmatrix} \begin{pmatrix} 0 & 0 \\ 0 & -\frac{\mathcal{B}_0 + \mathcal{B}^*_0}{\mathcal{L}_0 + \mathcal{L}^*_0} \end{pmatrix} \begin{pmatrix} e^{-\lambda \mathcal{L}^*_0} \\ e^{-\lambda \mathcal{L}_0} e^{-\lambda \mathcal{L}^*_0} - 1 \end{pmatrix} = \frac{e^{-\lambda \mathcal{L}_0} e^{-\lambda \mathcal{L}^*_0} - 1}{\mathcal{L}_0 + \mathcal{L}^*_0} (\mathcal{B}_0 + \mathcal{B}^*_0). \tag{7.76}$$

Second, the combination $i_S p_S$ that also appears in the tensor algebra quantity $\mathbf{h'_S}$ combines with $\tilde{\mathbf{i}}'_\mathbf{W}$ and $\tilde{\mathbf{p}}'_\mathbf{W}$ to

$$\tilde{p}'_W i_S p_S \tilde{i}'_W = \begin{pmatrix} e^{-\lambda \mathcal{L}_0} & -1 \end{pmatrix} \begin{pmatrix} 1 & 0 \\ 0 & 0 \end{pmatrix} \begin{pmatrix} e^{-\lambda \mathcal{L}^*_0} \\ e^{-\lambda \mathcal{L}_0} e^{-\lambda \mathcal{L}^*_0} - 1 \end{pmatrix} = e^{-\lambda \mathcal{L}_0} e^{-\lambda \mathcal{L}^*_0}. \tag{7.77}$$

This implies that we can write

$$\tilde{\mathbf{p}}'_\mathbf{W} \mathbf{h'_S} \tilde{\mathbf{i}}'_\mathbf{W} = \sum_{n=1}^{\infty} \sum_{k=0}^{n-1} 1^{\otimes k} \otimes \frac{e^{-\lambda \mathcal{L}_0} e^{-\lambda \mathcal{L}^*_0} - 1}{\mathcal{L}_0 + \mathcal{L}^*_0} (\mathcal{B}_0 + \mathcal{B}^*_0) \otimes \left( e^{-\lambda \mathcal{L}_0} e^{-\lambda \mathcal{L}^*_0} \right)^{\otimes n-k-1}. \tag{7.78}$$

which is precisely equal to the homotopy in $\hat{\mathcal{B}}$-gauge homotopy $\mathbf{h}_{\hat{\mathcal{B}}}$ defined in [72]. Moreover, $p'_S \tilde{i}'_W = e^{-\lambda \mathcal{L}^*_0}$ and hence equal to the projection $p$ used in [72]. We see that if we could ignore the projector $\Pi$ for a moment, we could write

$$\mathbf{P'_S \tilde{I}'_W} \stackrel{?}{=} \mathbf{p} \left( 1 - \mathbf{m_2 h}_{\hat{\mathcal{B}}} \right)^{-1} \tag{7.79}$$

which can be compared to
$$\mathbf{P} = P_{SDR}\left(\mathbf{p}\left(\mathbf{1} - \mathbf{m_2 h_{\hat{\mathcal{B}}}}\right)^{-1}\right). \tag{7.80}$$

Here, $P_{SDR}$ is a formal object that implements the SDR-conditions $pi = 1$, $h_{\hat{\mathcal{B}}}i = ph_{\hat{\mathcal{B}}} = h_{\hat{\mathcal{B}}}h_{\hat{\mathcal{B}}} = 0$ ($h_{\hat{\mathcal{B}}}h_{\hat{\mathcal{B}}} = 0$ is already obeyed), for a more precise definition see [71]. We see that the expressions are equal except for the projectors $\Pi$ and $P_{SDR}$, so we have to find out how to relate them. Let us look at the individual terms explicitly and observe the effect of $\Pi$: As usual it is enough to consider the projection on one output. For one or two inputs we do not get any constraints, so the first non-trivial term without $\Pi$ would be

$$\pi_1 \mathbf{P'_S \tilde{I}'_W} \pi_3 \stackrel{?}{=} p'_S \tilde{i}'_W m_2 \left(\tilde{p}'_W \otimes \tilde{p}'_W\right) \left(\mathbb{1} \otimes h'_S + h'_S \otimes i_S p_S\right) \left(\tilde{i}'_W \otimes \tilde{i}'_W\right) \left(\mathbb{1} \otimes m_2 + m_2 \otimes \mathbb{1}\right)$$
$$\left(\tilde{p}'_W \otimes \tilde{p}'_W \otimes \tilde{p}'_W\right) \left(\mathbb{1} \otimes \mathbb{1} \otimes h'_S + \mathbb{1} \otimes h'_S \otimes i_S p_S + h'_S \otimes i_S p_S \otimes i_S p_S\right) \left(\tilde{i}'_W \otimes \tilde{i}'_W \otimes \tilde{i}'_W\right) \tag{7.81}$$

The terms containing $\tilde{i}'_W \tilde{p}'_W$ are

$$p'_S \tilde{i}'_W m_2 \left(\tilde{p}'_W \otimes \tilde{p}'_W\right) \left(\mathbb{1} \otimes h'_S + h'_S \otimes i_S p_S\right) \left(\tilde{i}'_W \tilde{p}'_W \otimes \tilde{i}'_W m_2 \left(\tilde{p}'_W \otimes \tilde{p}'_W\right) + \tilde{i}'_W m_2 \left(\tilde{p}'_W \otimes \tilde{p}'_W\right) \otimes \tilde{i}'_W \tilde{p}'_W\right)$$
$$\left(\mathbb{1} \otimes \mathbb{1} \otimes h'_S + \mathbb{1} \otimes h'_S \otimes i_S p_S + h'_S \otimes i_S p_S \otimes i_S p_S\right) \left(\tilde{i}'_W \otimes \tilde{i}'_W \otimes \tilde{i}'_W\right) \tag{7.82}$$

and we see that due to $\tilde{p}'_W \tilde{i}'_W = 1$ in many of the terms it does not make any difference if $\tilde{i}'_W \tilde{p}'_W$ is replaced by unity. The problematic terms are

$$p'_S \tilde{i}'_W m_2 \left(\tilde{p}'_W \otimes \tilde{p}'_W\right) \left(h'_S \otimes i_S p_S\right) \left(\tilde{i}'_W \tilde{p}'_W \otimes \tilde{i}'_W m_2 \left(\tilde{p}'_W \otimes \tilde{p}'_W\right)\right) \left(h'_S \otimes i_S p_S \otimes i_S p_S\right) \left(\tilde{i}'_W \otimes \tilde{i}'_W \otimes \tilde{i}'_W\right),$$
$$p'_S \tilde{i}'_W m_2 \left(\tilde{p}'_W \otimes \tilde{p}'_W\right) \left(h'_S \otimes i_S p_S\right) \left(\tilde{i}'_W m_2 \left(\tilde{p}'_W \otimes \tilde{p}'_W\right) \otimes \tilde{i}'_W \tilde{p}'_W\right) \left(\mathbb{1} \otimes \mathbb{1} \otimes h'_S\right) \left(\tilde{i}'_W \otimes \tilde{i}'_W \otimes \tilde{i}'_W\right),$$
$$p'_S \tilde{i}'_W m_2 \left(\tilde{p}'_W \otimes \tilde{p}'_W\right) \left(h'_S \otimes i_S p_S\right) \left(\tilde{i}'_W m_2 \left(\tilde{p}'_W \otimes \tilde{p}'_W\right) \otimes \tilde{i}'_W \tilde{p}'_W\right) \left(\mathbb{1} \otimes h'_S \otimes i_S p_S\right) \left(\tilde{i}'_W \otimes \tilde{i}'_W \otimes \tilde{i}'_W\right),$$
$$p'_S \tilde{i}'_W m_2 \left(\tilde{p}'_W \otimes \tilde{p}'_W\right) \left(h'_S \otimes i_S p_S\right) \left(\tilde{i}'_W m_2 \left(\tilde{p}'_W \otimes \tilde{p}'_W\right) \otimes \tilde{i}'_W \tilde{p}'_W\right) \left(h'_S \otimes i_S p_S \otimes i_S p_S\right) \left(\tilde{i}'_W \otimes \tilde{i}'_W \otimes \tilde{i}'_W\right),$$
$$p'_S \tilde{i}'_W m_2 \left(\tilde{p}'_W \otimes \tilde{p}'_W\right) \left(\mathbb{1} \otimes h'_S\right) \left(\tilde{i}'_W m_2 \left(\tilde{p}'_W \otimes \tilde{p}'_W\right) \otimes \tilde{i}'_W \tilde{p}'_W\right) \left(\mathbb{1} \otimes \mathbb{1} \otimes h'_S\right) \left(\tilde{i}'_W \otimes \tilde{i}'_W \otimes \tilde{i}'_W\right),$$
$$p'_S \tilde{i}'_W m_2 \left(\tilde{p}'_W \otimes \tilde{p}'_W\right) \left(\mathbb{1} \otimes h'_S\right) \left(\tilde{i}'_W m_2 \left(\tilde{p}'_W \otimes \tilde{p}'_W\right) \otimes \tilde{i}'_W \tilde{p}'_W\right) \left(\mathbb{1} \otimes h'_S \otimes i_S p_S\right) \left(\tilde{i}'_W \otimes \tilde{i}'_W \otimes \tilde{i}'_W\right)$$

and

$$p'_S \tilde{i}'_W m_2 \left(\tilde{p}'_W \otimes \tilde{p}'_W\right) \left(\mathbb{1} \otimes h'_S\right) \left(\tilde{i}'_W m_2 \left(\tilde{p}'_W \otimes \tilde{p}'_W\right) \otimes \tilde{i}'_W \tilde{p}'_W\right) \left(h'_S \otimes i_S p_S \otimes i_S p_S\right) \left(\tilde{i}'_W \otimes \tilde{i}'_W \otimes \tilde{i}'_W\right).$$

The first and fifth term vanish because they contain $\left(\tilde{p}'_W h'_S \tilde{i}'_W\right)^2 = h_{\hat{\mathcal{B}}}^2 = 0$. The second term would actually vanish if $\tilde{i}'_W \tilde{p}'_W = \mathbb{1}$ because of $p_S h'_S = 0$. Otherwise we get the combination $\tilde{p}'_W i_S p_S \tilde{i}'_W \tilde{p}'_W h'_S \tilde{i}'_W$ which is equal to $iph_{\hat{\mathcal{B}}}$ in the notation of [72] and hence gets eliminated by $P_{SDR}$. In the third and fourth term we would have $\tilde{p}'_W i_S p_S \tilde{i}'_W \tilde{p}'_W i_S p_S \tilde{i}'_W$ replaced by $\tilde{p}'_W i_S p_S \tilde{i}'_W$ which corresponds to replacing $ipip$ by $ip$ as implemented by $P_{SDR}$. Similarly, in the last two terms the combination $\tilde{p}'_W h'_S \tilde{i}'_W \tilde{p}'_W i_S p_S \tilde{i}'_W = h_{\hat{\mathcal{B}}} ip$ would vanish if $\tilde{i}'_W \tilde{p}'_W = \mathbb{1}$ due to $h'_S i_S = 0$ whileas in the other formula $h_{\hat{\mathcal{B}}} i$ is deleted by $P_{SDR}$. The structure is the same for all orders since no new combinations of operators appear so we can conclude that the projectors $\Pi$ and $P_{SDR}$ have the same effect on all terms that appear. This completes the proof of equivalence between $\mathbf{P'_S \tilde{I}'_W}$ and $\mathbf{P}$.

# 8 A boundary term for open string field theory

**Abstract**

We consider Witten's open string field theory in the presence of a non-trivial boundary of spacetime. For the kinetic term, we derive a Gibbons-Hawking-type contribution that has to be added to the action to guarantee a well-defined variational principle. The derivation is done first in a heuristic way and then confirmed by a path integral based approach using the CFT operator formalism. In the last section we discuss the boundary contributions coming from the cubic vertex, although it is problematic to apply consistent boundary conditions on the string field due to the non-locality of the vertex.

## 8.1 Introduction and motivation

There exist many field theories defined on some spacetime manifold $M$ whose precise definition becomes inconsistent if $M$ has a non-vanishing boundary. It is often perfectly fine to ignore this issue because the boundary is significantly far away and one is only concerned about what happens in the bulk. However, in some situations the boundary behaviour has to be taken account, not only for mathematical consistency but also for physical reasons. The main example is Einstein gravity, where the Einstein-Hilbert action

$$S = \frac{1}{16\pi G_N} \int_M d^4x \sqrt{-g} R \tag{8.1}$$

has to be supplemented with the Gibbons-Hawking term

$$S_{GH} = \frac{1}{8\pi G_N} \int_{\partial M} d^3x \sqrt{|h|} K \tag{8.2}$$

to give rise to a well-defined variational principle. Moreover, the boundary term is necessary to produce the correct values of the ADM-mass and black hole entropy [35]. Since string field theory should contain gravity as one of its subsectors, it is natural to assume that also string field theory should contain a non-trivial boundary term. Additional evidence for why the boundary terms are needed comes from the discrepancy between the results of [21] where it is shown that the on-shell action of closed string field theory vanishes up to possible boundary contributions and [67, 53], from which it follows that the action for at least some classical solutions corresponding to nearly marginal deformations in a linear-dilaton background is non-zero.

Although both arguments concern closed string field theory, for simplicity we want to start with open string field theory and analyze its variational principle in the presence of a boundary in spacetime. The Witten action [85] of open string field theory[42] consists of two terms and reads

$$S(\Psi) = -\frac{1}{2} \langle \Psi, Q\Psi \rangle - \frac{1}{3} \langle \Psi, \Psi * \Psi \rangle \tag{8.3}$$

---

[42]For a recent comprehensive review of the subject see [76].

and we expect to get non-vanishing boundary contributions from both of the terms. While for the kinetic term it will be quite straightforward to obtain them, the cubic term is much more subtle due to its non-local nature. The paper is organized as follows: After reviewing some general facts on surface terms in field theory we will start by computing the kinetic boundary term in a heuristic way using partial integration. In section 8.4 we will reproduce the result in the operator formalism using the fact that the boundary violates conformal invariance. Section 8.5 is of a more speculative character and deals with the cubic boundary term: It is not clear if it even makes sense to have a localized boundary where boundary conditions need to be applied in a non-local theory (see [14, 56] for an analysis of the initial value problem in non-local theories). We will discuss where boundary terms are expected to appear and why their calculation is non-trivial.

## 8.2 General philosophy

Consider an ordinary field theory with a set of fields $\phi^i(x^\mu)$ and an action

$$S[\phi] = \int_M d^D x \sqrt{-g}\, \mathcal{L}\left([\phi], x^\mu\right), \tag{8.4}$$

where $[\phi]$ denotes a dependence on all of the fields and their derivatives and the integral runs over some spacetime manifold $M$ [10]. When we vary $S[\phi]$ to obtain the equations of motion, we want to be left with an expression of the form

$$\int_M d^D x \sqrt{-g} \left(\text{eom}_i [\phi]\, \delta\phi^i\right) \tag{8.5}$$

which allows to deduce

$$\text{eom}_i [\phi] = 0 \qquad \forall i. \tag{8.6}$$

However, to isolate $\delta\phi^i$, it is in general necessary to apply partial integration, which gives rise to surface terms evaluated at $\partial M$:

$$\delta S[\phi] = \int_M d^D x \sqrt{-g} \left(\text{eom}_i [\phi]\, \delta\phi^i\right) + \int_{\partial M} d^{D-1} x \sqrt{|h|}\, X\left([\phi], [\delta\phi], x^\mu\right) \tag{8.7}$$

Here, $h$ denotes the determinant of the induced metric on $\partial M$. For the variational principle to be well-posed, we need (8.6) to be sufficient for making $\delta S[\phi]$ vanish, so we need some additional strategy to eliminate $X\left([\phi], [\delta\phi], x^\mu\right)$. There are two possibilities:

1. Impose boundary conditions at $\partial M$ for the fields $\phi^i$ and/or their derivatives.

2. Add an explicit boundary term of the form $S_B[\phi] = \int_{\partial M} d^{D-1} x \sqrt{|h|}\, \mathcal{L}_B\left([\phi], x^\mu\right)$ that cancels $X$, i. e.

$$\delta S_B[\phi] = -\int_{\partial M} d^{D-1} x \sqrt{|h|}\, X\left([\phi], [\delta\phi], x^\mu\right) \tag{8.8}$$

It is instructive to analyze those two strategies with the help of the simple example of a free real scalar field in flat space: We start with the action

$$S\left[\phi\right] = \int_M d^4x \; \frac{1}{2} \phi \,\Box\, \phi. \tag{8.9}$$

Varying yields straightforwardly

$$\delta S\left[\phi\right] = \int_M d^4x \; \frac{1}{2} \left(\delta\phi \,\Box\, \phi + \phi \,\Box\, \delta\phi\right) = \int_M d^4x \; \frac{1}{2} \left(\delta\phi \,\Box\, \phi - \partial_\mu \phi \partial^\mu \delta\phi\right) + \int_{\partial M} d^{D-1}x \; \frac{1}{2} n_\mu \phi \partial^\mu \delta\phi$$
$$= \int_M d^4x \; \delta\phi \,\Box\, \phi + \int_{\partial M} d^{D-1}x \; \frac{1}{2} n_\mu \left(\phi \partial^\mu \delta\phi - \partial^\mu \phi \, \delta\phi\right), \tag{8.10}$$

where $n_\mu$ denotes the normal vector to the boundary. We can read off directly

$$\text{eom}\left[\phi\right] = \Box\phi, \qquad X\left(\left[\phi\right],\left[\delta\phi\right]\right) = \frac{1}{2} n_\mu \left(\phi \partial^\mu \delta\phi - \partial^\mu \phi \, \delta\phi\right) \tag{8.11}$$

One way to proceed now would be to apply strategy one and impose Neumann boundary conditions for $\phi$, i. e. demand

$$n_\mu \partial^\mu \phi \mid_{\partial M} = 0. \tag{8.12}$$

This would imply $n_\mu \partial^\mu \delta\phi \mid_{\partial M} = 0$ as well and hence kill the full boundary term. However, Neumann boundary conditions allow only for a subspace of the physically interesting solutions, we might be also interested in field configurations which obey the Dirichlet condition $\delta\phi \mid_{\partial M} = 0$. But this does not kill the full boundary term, we are left with

$$X_{rem}\left(\left[\phi\right],\left[\delta\phi\right]\right) = \frac{1}{2} n_\mu \phi \partial^\mu \delta\phi. \tag{8.13}$$

Let us examine the possibilities of strategy two: We would need to be able to write $X\left(\left[\phi\right],\left[\delta\phi\right]\right)$ as $\delta$ of something, or in other words, if we regard the space of fields $\phi\left(x^\mu\right)$ as a Banach manifold with $\delta$ being the exterior derivative, then $X$ must be an exact one-form.[43] A necessary condition for that to be true is that $X$ is closed, $\delta X = 0$. By direct calculation we find that

$$\delta X\left(\left[\phi\right],\left[\delta\phi\right]\right) = \frac{1}{2} n_\mu \left(\delta\phi \wedge \partial^\mu \delta\phi - \partial^\mu \delta\phi \wedge \delta\phi\right) = \delta\phi \wedge n_\mu \partial^\mu \delta\phi \neq 0, \tag{8.14}$$

so strategy two alone is not sufficient to kill the full boundary term. However, after imposing Dirichlet boundary conditions, we find

$$X_{rem}\left(\left[\phi\right],\left[\delta\phi\right]\right) = \frac{1}{2} n_\mu \phi \partial^\mu \delta\phi = \frac{1}{2} n_\mu \partial^\mu \left(\phi \delta\phi\right) = \delta\left(\frac{1}{4} n_\mu \partial^\mu \phi^2\right), \tag{8.15}$$

---

[43]In the case of string field theory, this is subtle because we do not have a precise and consistent definition of the Hilbert space of string fields available yet. With a slight abuse of language, we will still use the terminology of differential forms for expressions using the variational $\delta$.

so adding the term

$$S_B[\phi] = -\frac{1}{4}\int_{\partial M} d^3x\, n_\mu \partial^\mu \phi^2 \tag{8.16}$$

to the action eliminates all boundary contributions. To summarize, there are two requirements that a "good" set of boundary conditions should satisfy: It should be general enough to include all kinds of physically relevant solutions, but at the same time strict enough to make $X_{rem}$ $\delta$-exact.

## 8.3 Kinetic boundary term

We want to apply all of that now to the case of open string field theory and focus first of all on the kinetic term

$$S_{kin}(\Psi) = -\frac{1}{2}\langle \Psi, Q\Psi \rangle. \tag{8.17}$$

To start, we need to make the dependence on the spacetime coordinates and derivatives explicit, i. e. pass from momentum space to position space. A general Fock space state has the form

$$\Psi = \alpha^{\mu_1}_{-n_1}...\alpha^{\mu_r}_{-n_r} b_{-m_1}...b_{-m_s} c_{-p_1}...c_{-p_t} |0,k\rangle, \qquad n_i \geq 1,\ \ m_i \geq 2,\ \ p_i \geq -1 \tag{8.18}$$

while its BPZ-conjugate is given by

$$\Psi^* = (-1)^{\Sigma n_i + \Sigma m_j + \Sigma p_k + r + t} \langle 0,k | \alpha^{\mu_1}_{n_1}...\alpha^{\mu_r}_{n_r} b_{m_1}...b_{m_s} c_{p_1}...c_{p_t} \tag{8.19}$$

The vacua form the basic overlap

$$\langle 0,k| c_{-1} c_0 c_1 |0,k'\rangle = (2\pi)^{26}\,\delta^{(26)}(k+k'). \tag{8.20}$$

with the momentum dependent vacuum defined as

$$|0,k\rangle =:\, e^{ik_\mu \hat{X}^\mu(0,0)}:|0\rangle = e^{ik_\mu \hat{x}^\mu}|0\rangle, \tag{8.21}$$

($|0\rangle$ denotes the $SL(2)$-invariant vacuum). The zero-mode factor of our Hilbert space is just $L^2(M)$, where $M$ is the 26-dimensional spacetime, so the operator $\hat{x}^\mu$ can act by multiplication with $x^\mu$ (We assume that $M$ is just flat space with no background fields). The overlap (8.20) can now be represented as an integral over spacetime:

$$\langle 0,k|c_{-1}c_0c_1|0,k'\rangle = \int d^{26}x\,\langle 0|e^{ik_\mu x^\mu} c_{-1}c_0c_1 e^{ik'_\nu x^\nu}|0\rangle'$$

$$= \int d^{26}x\,e^{i(k_\mu+k'_\mu)x^\mu}\langle 0|c_{-1}c_0c_1|0\rangle' = (2\pi)^{26}\,\delta^{(26)}(k+k') \tag{8.22}$$

Here, $\langle 0|\,|0\rangle'$ denotes the reduced BPZ-product without the $x$-integration, which is written out explicitly. The position dependent form of the string field is a linear superposition of Fock space states (8.18) of the form

$$|\Psi\rangle = \int \frac{d^{26}k}{(2\pi)^{26}} f(k)\,\alpha^{\mu_1}_{-n_1}...\alpha^{\mu_r}_{-n_r} b_{-m_1}...b_{-m_s} c_{-p_1}...c_{-p_t} |0,k\rangle$$

$$= \alpha^{\mu_1}_{-n_1}...\alpha^{\mu_r}_{-n_r} b_{-m_1}...b_{-m_s} c_{-p_1}...c_{-p_t} F(x)|0\rangle \equiv |\Psi(x)\rangle \tag{8.23}$$

where $F(x)$ is the Fourier transform of $f(k)$.[44] The zero mode $p_\mu$ can now be represented as a derivative:
$$\alpha_{0\mu} = \sqrt{2\alpha'} p_\mu = -i\sqrt{2\alpha'} \partial_\mu. \tag{8.24}$$
This is the only place where a spacetime derivative is appearing, so from on now on we can focus solely on terms involving $\alpha_{0\mu}$.

The relation which will be modified is the cyclicity condition of $Q$[45],
$$\langle \Psi_1, Q\Psi_2 \rangle = \langle Q\Psi_1, \Psi_2 \rangle. \tag{8.25}$$
From
$$Q = \sum_n c_{-n} L_n^{(m)} + \text{ghosts} \tag{8.26}$$
we can infer that there will be terms with one derivative proportional to $-i\sqrt{2\alpha'} \sum_{n\neq 0} c_{-n} \alpha_n^\mu \partial_\mu$ and one term with two derivatives given by $-\alpha' c_0 \partial^\mu \partial_\mu$. Examining first the two-derivative term, we see that it can be commuted through until it hits the $x$-dependent function; explicitly we get

$$-\alpha' \int_M d^{26}x \, \langle \Psi_1(x) | c_0 \partial_\mu \partial^\mu | \Psi_2(x) \rangle' = -\alpha' \int_M d^{26}x \, \partial_\mu \left( \langle \Psi_1(x) | c_0 | \partial^\mu \Psi_2(x) \rangle' \right)$$
$$+ \alpha' \int_M d^{26}x \, \langle \partial_\mu \Psi_1(x) | c_0 | \partial^\mu \Psi_2(x) \rangle'$$
$$= -\alpha' \int_{\partial M} d^{25}x \, n_\mu \left( \langle \Psi_1(x) | c_0 | \partial^\mu \Psi_2(x) \rangle' \right) + \alpha' \int_M d^{26}x \, \partial^\mu \left( \langle \partial_\mu \Psi_1(x) | c_0 | \Psi_2(x) \rangle' \right)$$
$$- \alpha' \int_M d^{26}x \, \langle \partial^\mu \partial_\mu \Psi_1(x) | c_0 | \Psi_2(x) \rangle'$$
$$= -\alpha' \int_{\partial M} d^{25}x \, n_\mu \left( \langle \Psi_1(x) | c_0 | \partial^\mu \Psi_2(x) \rangle' \right) + \alpha' \int_{\partial M} d^{25}x \, n_\mu \left( \langle \partial^\mu \Psi_1(x) | c_0 | \Psi_2(x) \rangle' \right)$$
$$- \alpha' \int_M d^{26}x \, \langle \partial^\mu \partial_\mu \Psi_1(x) | c_0 | \Psi_2(x) \rangle' \tag{8.27}$$

and find two boundary terms in analogy with section two. The one-derivative terms yield

$$-i\sqrt{2\alpha'} \sum_{n\neq 0} \int_M d^{26}x \, \langle \Psi_1(x) | c_{-n} \alpha_n^\mu \partial_\mu | \Psi_2(x) \rangle'$$
$$= -i\sqrt{2\alpha'} \sum_{n\neq 0} \int_{\partial M} d^{25}x \, n_\mu \langle \Psi_1(x) | c_{-n} \alpha_n^\mu | \Psi_2(x) \rangle' + i\sqrt{2\alpha'} \sum_{n\neq 0} \int_M d^{26}x \, \langle \partial_\mu \Psi_1(x) | c_{-n} \alpha_n^\mu | \Psi_2(x) \rangle'. \tag{8.28}$$

Substituting $\Psi_1 = \Psi$, $\Psi_2 = \delta\Psi$ all boundary terms together read

$$\int_{\partial M} d^{25}x \, n_\mu \Big( \alpha' \langle \partial^\mu \Psi(x) | c_0 | \delta\Psi(x) \rangle' - \alpha' \langle \Psi(x) | c_0 | \partial^\mu \delta\Psi(x) \rangle'$$
$$- i\sqrt{2\alpha'} \sum_{n\neq 0} \langle \Psi(x) | c_{-n} \alpha_n^\mu | \delta\Psi(x) \rangle' \Big). \tag{8.29}$$

---
[44]In the presence of a boundary, the integral over $k$ might get replaced by a discrete sum.
[45]We assume that $\Psi$ has ghost number one such that Grassmann signs can be omitted.

If we proceed as in the example and apply Dirichlet (Neumann) boundary conditions, the first and last (first and second, respectively) term vanish and the remainder is an exact one-form:

$$-\int_{\partial M} d^{25}x\, n_\mu \left( \alpha' \langle \Psi(x)| c_0 |\partial^\mu \delta\Psi(x)\rangle' + i\sqrt{2\alpha'} \sum_{n\neq 0} \langle \Psi(x)| c_{-n}\alpha_n^\mu |\delta\Psi(x)\rangle' \right). \qquad (8.30)$$

It can be canceled by adding the boundary term

$$S_B = \int_{\partial M} d^{25}x\, n_\mu \left( \alpha' \langle \Psi(x)| c_0 |\partial^\mu \Psi(x)\rangle' + i\sqrt{\frac{\alpha'}{2}} \sum_{n\neq 0} \langle \Psi(x)| c_{-n}\alpha_n^\mu |\Psi(x)\rangle' \right)$$

$$= -\int_{\partial M} d^{25}x\, n_\mu \langle \Psi(x)| (c\partial X^\mu)_0 |\Psi(x)\rangle' \qquad (8.31)$$

to the action. This Gibbons-Hawking-type term for the kinetic part is one of the main results.

## 8.4 Breaking of conformal invariance

The derivation in the last section is probably the simplest and fastest way to compute the boundary term, however, it does not connect naturally to the operator formalism of the CFT. Here we will present an alternative method which can be applied to the cubic term as well.

We have seen that the cyclicity relation of $Q$,

$$\langle \Psi_1, Q\Psi_2 \rangle = \langle Q\Psi_1, \Psi_2 \rangle, \qquad (8.32)$$

is violated in the presence of a boundary in target space. In the CFT formalism this relation can be derived by writing

$$\langle \Psi_1, Q\Psi_2 \rangle = \langle 0| \Psi_1^*(\infty) \oint_{C(0)} \frac{dz}{2\pi i} j_B(z) \Psi_2(0) |0\rangle \qquad (8.33)$$

and deforming the integration contour until it encycles the other insertion at infinity.[46] Since contour deformation is a mathematically well-established technique valid in any two dimensional CFT, this is in sharp contradiction with our result from the last section.

The contradiction is resolved in [46], where the authors show that conformal invariance is broken by the target space boundary. In a nutshell, their argument goes as follows: By using the explicit expression of the worldsheet energy momentum tensor[47]

$$T_{ab} = -\frac{1}{\alpha'} \left( \partial_a X^\mu \partial_b X_\mu - \frac{1}{2}\eta_{ab}\eta^{cd}\partial_c X^\mu \partial_d X_\mu \right) \qquad (8.34)$$

---

[46]Here and in the whole rest of the paper the doubling trick is used.

[47]Here, the worldsheet is a flat strip with coordinates $\sigma$, $\tau$ and a flat metric $\eta_{ab}$. For later purposes we assume that in addition to $0 \leq \sigma \leq \pi$ we have $\tau_{min} \leq \tau \leq \tau_{max}$, such that the worldsheet has a finite area $V_W = \pi(\tau_{max} - \tau_{min})$. In the end of the calculation we can savely take the limit $\tau_{max/min} \to \pm\infty$.

we can write
$$\partial^a T_{ab} = -\frac{1}{\alpha'}\partial^a \partial_a X^\mu \partial_b X_\mu = -2\pi \frac{\delta S_P}{\delta X^\mu}\partial_b X^\mu \tag{8.35}$$
where $S_P$ denotes the Polyakov action. We see that $T_{ab}$ is classically conserved on-shell, as expected. To make sense of the r. h. s. in the quantum theory, it first has to be normal-ordered, i. e. using the standard prescription
$$: X^\mu(\sigma) X^\nu(\sigma') := X^\mu(\sigma) X^\nu(\sigma') + \frac{\alpha'}{2}\eta^{\mu\nu}\ln\left(|\sigma - \sigma'|^2\right) \tag{8.36}$$
we get
$$: \frac{\delta S_P}{\delta X_\mu}\partial_b X_\mu := \frac{\delta S_P}{\delta X^\mu}\partial_b X^\mu - 26\partial_b \delta^{(2)}(0). \tag{8.37}$$
Now, Eq. (8.35) can be written as an operator equation in the path integral as
$$\langle :\partial^a T_{ab}(\sigma): \mathcal{O}(\sigma_1)...\mathcal{O}(\sigma_n)\rangle = -2\pi \int \mathcal{D}X \,:\frac{\delta S_P}{\delta X^\mu}\partial_b X^\mu:(\sigma)\,\mathcal{O}(\sigma_1)...\mathcal{O}(\sigma_n)\,e^{-S_P}$$
$$= 2\pi \int \mathcal{D}X \,\frac{\delta}{\delta X^\mu(\sigma)}\left(\partial_b X^\mu(\sigma)\,\mathcal{O}(\sigma_1)...\mathcal{O}(\sigma_n)\,e^{-S_P}\right). \tag{8.38}$$

We assumed for simplicity that there are no contact terms, $\sigma \neq \sigma_i$. In ordinary cases one would now argue that the path integral of the total functional derivative vanishes, but there may be contributions from the boundary of spacetime: If we expand $X^\mu$ in mutually orthogonal modes with $V_W$ being the volume of the worldsheet,
$$X^\mu = x^\mu + \sum_{n\neq 0} x_n^\mu X_n(\sigma), \quad \int d^2\sigma\, X_n X_m = \delta_{mn}, \quad \frac{\delta}{\delta X^\mu(\sigma)} = V_W^{-1}\frac{\partial}{\partial x^\mu} + \sum_{n\neq 0} X_n(\sigma)\frac{\partial}{\partial x_n^\mu} \tag{8.39}$$
then the boundary in field configuration space corresponds to large values of the coefficients $x^\mu$, $x_n^\mu$. The non-constant modes $X_n$ are exponentially suppressed by $e^{-S_P}$, but this does not hold for the constant mode since $X^\mu$ appears only derivated in $S_P$. So we are only allowed to neglect the non-constant modes in (8.38) and are left with
$$\langle \partial^a T_{ab}(\sigma)\,\mathcal{O}(\sigma_1)...\mathcal{O}(\sigma_n)\rangle = 2\pi V_W^{-1}\int_M d^{26}x\,\frac{\partial}{\partial x^\mu}\langle \partial_b X^\mu(\sigma)\,\mathcal{O}(\sigma_1)...\mathcal{O}(\sigma_n)\rangle'$$
$$= 2\pi V_W^{-1}\int_{\partial M} d^{25}x\, n_\mu\,\langle \partial_b X^\mu(\sigma)\,\mathcal{O}(\sigma_1)...\mathcal{O}(\sigma_n)\rangle'. \tag{8.40}$$

This violation of the conformal Ward identity will give rise to extra terms in the contour deformation: We will focus solely on the spacetime-dependent part and ignore the ghost contribution to get
$$\langle 0|\,\Psi_1^*(\infty)\left(\oint_{C_1}\frac{dz}{2\pi i}\left(cT^{(m)}\right) - \oint_{C_2}\frac{d\bar{z}}{2\pi i}\left(\bar{c}\bar{T}^{(m)}\right)\right)(z,\bar{z})\,\Psi_2(0)\,|0\rangle$$
$$\subset \langle 0|\,\Psi_1^*(\infty)\oint_{C(0)}\frac{dz}{2\pi i}j_B(z)\,\Psi_2(0)\,|0\rangle \tag{8.41}$$

where we made the doubling trick explicit (Here, $C_{1/2}$ denotes the upper/lower semicircle of the contour.). Now we can use the complex divergence theorem

$$\int_R d^2z \left(\partial v^z + \bar{\partial} v^{\bar{z}}\right) = i \oint_{\partial R} \left(v^z d\bar{z} - v^{\bar{z}} dz\right) \tag{8.42}$$

to rewrite the integral: Setting $v^z =: \left(\bar{c}\overline{T}^{(m)}\right)$ supported in the lower and $v^{\bar{z}} =: \left(cT^{(m)}\right)$ supported in the upper half plane yields

$$\langle 0| \Psi_1^* (\infty) \left(\oint_{C_1} \frac{dz}{2\pi i} \left(cT^{(m)}\right) - \oint_{C_2} \frac{d\bar{z}}{2\pi i} \left(\bar{c}\overline{T}^{(m)}\right)\right) (z, \bar{z}) \Psi_2 (0) |0\rangle$$
$$= \frac{1}{2\pi} \langle 0| \Psi_1^* (\infty) \int_{Int(C)} d^2z \left(\bar{\partial} \left(cT^{(m)}\right) + \partial \left(\bar{c}\overline{T}^{(m)}\right)\right) (z, \bar{z}) \Psi_2 (0) |0\rangle$$
$$= \frac{1}{\pi} \langle 0| \Psi_1^* (\infty) \int_{UpInt(C)} d^2z \, \bar{\partial} \left(cT^{(m)}\right) (z, \bar{z}) \Psi_2 (0) |0\rangle. \tag{8.43}$$

In the last step we observed that both terms actually yield the same integral, so we can restrict to the first term intergrated in the upper half plane only. $c(z)$ is still holomorphic so $\bar{\partial} c = 0$, whileas $\bar{\partial} T^{(m)}$ will give rise to a non-trivial boundary contribution. For the cyclicity relation (8.32) we need to deform the contour from an infinitesimal circle around zero to an infinitesimal circle around infinity hence we have to cover the whole complex plane. In the end we get

$$\langle \Psi_1, Q\Psi_2\rangle - \langle Q\Psi_1, \Psi_2\rangle = \frac{1}{\pi} \langle 0| \Psi_1^* (\infty) \int_{UHP} d^2z \, \bar{\partial} \left(cT^{(m)}\right) (z, \bar{z}) \Psi_2 (0) |0\rangle. \tag{8.44}$$

To apply (8.40) we need to transform back to the sigma model coordinates: First, denoting $x = \text{Re } z$, $y = \text{Im } z$ and using the fact that the energy-momentum-tensor is still symmetric and traceless, we get

$$\left(c\bar{\partial}T^{(m)}\right)(z, \bar{z}) = \frac{1}{4}\left(c^x + ic^y\right)\left(\partial^j T_{jx} - i\partial^j T_{jy}\right)(x, y). \tag{8.45}$$

Now using the usual map from the strip to the upper half plane including the Wick rotation $z = e^{i(-\tau+\sigma)}$, the expression can be straightforwardly transformed into

$$\left(c\bar{\partial}T^{(m)}\right)(z, \bar{z}) = \frac{1}{4}\left(c^\sigma - c^\tau\right)\left(\partial^a T_{a\sigma} - \partial^a T_{a\tau}\right)(\sigma, \tau). \tag{8.46}$$

Finally we can insert into (8.40), which yields

$$\langle \Psi_1, Q\Psi_2\rangle - \langle Q\Psi_1, \Psi_2\rangle$$
$$= V_W^{-1} \int_{\partial M} d^{25}x \, n_\mu \langle 0| \Psi_1^* (\tau_{max}) \int d^2\sigma \left(c^\sigma - c^\tau\right)\left(\partial_\sigma X^\mu - \partial_\tau X^\mu\right)(\sigma, \tau) \Psi_2 (\tau_{min}) |0\rangle' \tag{8.47}$$

where $d^2z = 2d^2\sigma$ was used. The integral can now be computed explicitly using the mode expansions

$$(c^\sigma - c^\tau)(\sigma, \tau) \equiv -c^-(\sigma^-) = -\sum_n c_n e^{(n-1)i\sigma^-} \tag{8.48}$$

$$(\partial_\sigma X^\mu - \partial_\tau X^\mu)(\sigma, \tau) \equiv -2(\partial_- X^\mu)(\sigma^-) = 2\sum_{n\neq 0} i\sqrt{\frac{\alpha'}{2}} \alpha_n^\mu e^{(n+1)i\sigma^-} + 2i\alpha' p^\mu e^{i\sigma^-}. \tag{8.49}$$

At this point it is advantageous to keep the zero mode explicit, as we will see in a moment. We get

$$\langle \Psi_1, Q\Psi_2\rangle - \langle Q\Psi_1, \Psi_2\rangle = \int_{\partial M} d^{25}x\, n_\mu \langle \Psi_1(x)| \left(-i\sqrt{2\alpha'} \sum_{n\neq 0} c_{-n}\alpha_n^\mu - 2i\alpha' c_0 p^\mu \right) |\Psi_2(x)\rangle'. \tag{8.50}$$

This is already almost what we would expect from (8.29) but one has to be careful how $p^\mu \propto \partial^\mu$ is acting. The position of the insertion of $(c\partial X^\mu)_0$ was basically arbitrary since we could have as well started the calculation from the other side, i. e. $Q$ acting on $\Psi_1$. Naively the result would be the same because of momentum conservation, but we have a reduced correlation function, where the direction normal to the boundary is not integrated over. It hence makes a difference if $\partial^\mu$ is acting on $\Psi_1$ or $\Psi_2$.

The correct result can actually be derived via the same method: The operator $p^\mu$ can be represented as a contour integral

$$p^\mu \Psi_2(0) = i\frac{1}{\alpha'} \oint_{C(0)} \frac{dz}{2\pi i} \partial X^\mu(z) \Psi_2(0) \tag{8.51}$$

where the doubling trick is used again. If we want $p^\mu$ act on $\Psi_1(\infty)$, the contour has to be deformed around the whole Riemann sphere and the difference can be expressed with the help of (8.42) as

$$-2i\alpha' \int_{\partial M} d^{25}x\, n_\mu \left(\langle \Psi_1(x)| c_0 p^\mu |\Psi_2(x)\rangle' - p^\mu \langle \Psi_1(x)| c_0 |\Psi_2(x)\rangle'\right)$$
$$= \frac{1}{\pi} \int_{\partial M} d^{25}x\, n_\mu \int d^2z\, \langle \Psi_1(x)| c_0 \bar\partial \partial X^\mu(z, \bar z) |\Psi_2(x)\rangle', \tag{8.52}$$

using the same arguments as in (8.43). This expression would normally vanish as it contains the classical equations of motion inside a correlation function. In the presence of a boundary however, we find analogously to (8.38)

$$\langle \partial_+ \partial_- X^\mu(\sigma) \mathcal{O}(\sigma_1)...\mathcal{O}(\sigma_n)\rangle = \frac{\pi}{2}\alpha' \int \mathcal{D}X \frac{\delta S_P}{\delta X^\mu}(\sigma) \mathcal{O}(\sigma_1)...\mathcal{O}(\sigma_n) e^{-S_P}$$
$$= -\frac{\pi}{2}\alpha' \int \mathcal{D}X \frac{\delta}{\delta X^\mu(\sigma)} \left(\mathcal{O}(\sigma_1)...\mathcal{O}(\sigma_n) e^{-S_P}\right)$$
$$= -\frac{\pi}{2}\alpha' V_W^{-1} \int_{\partial M} d^{25}x\, n_\mu \langle \mathcal{O}(\sigma_1)...\mathcal{O}(\sigma_n)\rangle'. \tag{8.53}$$

If we write (8.52) again as a 26-dimensional integral to have a full correlation function we can insert and get

$$\frac{1}{\pi} \int_M d^{26}x\, \partial_\mu \int d^2z\, \langle \Psi_1(x)|\, c_0 \bar{\partial}\partial X^\mu(z,\bar{z}) |\Psi_2(x)\rangle'$$
$$= \frac{1}{\pi}\partial_\mu \int d^2z\, \langle \Psi_1(x)|\, c_0 \bar{\partial}\partial X^\mu(z,\bar{z}) |\Psi_2(x)\rangle$$
$$= -\frac{\alpha'}{2} V_W^{-1} \int_{\partial M} d^{25}x\, n_\mu \partial^\mu \int d^2z\, \langle \Psi_1(x)|\, c_0 |\Psi_2(x)\rangle'$$
$$= -\alpha' \int_{\partial M} d^{25}x\, n_\mu \partial^\mu \left(\langle \Psi_1(x)|\, c_0 |\Psi_2(x)\rangle'\right). \tag{8.54}$$

(In the last line we used $\int d^2z = 2\int d^2\sigma = 2V_W$.) This in general non-zero, so the only chance to get a well-defined result is that the operator $c\partial X^\mu$ shall act *symmetrically* on all of the string fields involved:

$$\int_{\partial M} d^{25}x\, n_\mu \left(\alpha' \partial^\mu \langle \Psi_1(x)|\, c_0 |\Psi_2(x)\rangle' - \alpha' \langle \Psi_1(x)|\, c_0 \partial^\mu |\Psi_2(x)\rangle' \right.$$
$$\left. -i\sqrt{2\alpha'} \sum_{n\neq 0} \langle \Psi_1(x)|\, c_{-n}\alpha_n^\mu |\Psi_2(x)\rangle' \right). \tag{8.55}$$

(The minus sign between the first two terms comes from commuting $c_0$ with $\Psi_1$.) We see that the difference between the first two terms is just (8.54), as expected. If we now set $\Psi_1 = \Psi$ and $\Psi_2 = \delta\Psi$ we precisely reproduce (8.29).

## 8.5 Cubic boundary term

In this section we discuss the cubic vertex

$$S_{int}(\Psi) = -\frac{1}{3} \langle \Psi, \Psi * \Psi \rangle. \tag{8.56}$$

and possible boundary contribution that we expect to get from it. This is actually very subtle since the interaction is non-local, whileas the boundary is localized in spacetime. Traditionally, the vertex is defined as

$$\langle \Psi_1, \Psi_2 * \Psi_3 \rangle =: \langle f_1 \circ \Psi_1(0)\, f_2 \circ \Psi_2(0)\, f_3 \circ \Psi_3(0) \rangle \tag{8.57}$$

with

$$f_j(z) = \tan\left(\frac{2}{3}\left(\arctan z - \pi + \frac{j\pi}{2}\right)\right). \tag{8.58}$$

We will however see that this definition has to be modified in the presence of a boundary. The cyclicity relation

$$\langle f_1 \circ \Psi_1(0)\, f_2 \circ \Psi_2(0)\, f_3 \circ \Psi_3(0) \rangle = \langle f_1 \circ \Psi_3(0)\, f_2 \circ \Psi_1(0)\, f_3 \circ \Psi_2(0) \rangle \tag{8.59}$$

is proven using the map

$$\tilde{f}(z) = \tan\left(\arctan z + \frac{\pi}{3}\right) = \frac{z + \sqrt{3}}{1 - \sqrt{3}z} \tag{8.60}$$

which fulfills $\tilde{f} \circ f_j(z) = f_{j+1}(z)$ modulo three. This transformation is not expected to give rise to any boundary term which can be argued as follows[48]: Consider the unit disc coordinate

$$\xi = \frac{1 + iz}{1 - iz} \tag{8.61}$$

where the functions defining the three-vertex take the form

$$g_1(\xi) = e^{-\frac{2\pi i}{3}} \xi^{\frac{2}{3}}, \qquad g_2(\xi) = \xi^{\frac{2}{3}}, \qquad g_3(\xi) = e^{\frac{2\pi i}{3}} \xi^{\frac{2}{3}}. \tag{8.62}$$

Here, cyclicity is shown via a simple rotation by $\frac{2\pi}{3}$ and rotational symmetry should not be broken by the target space boundary.

To analyze where boundary terms may arise, let us consider an arbitrary tachyon field $t(x) = \int \frac{d^{26}k}{(2\pi)^{26}} T(k) \, c e^{ik^\mu x_\mu} |0\rangle$: One gets

$$S_{int}(t(x)) \propto \int d^{26}x \left(e^{-\alpha' \ln \frac{4}{3\sqrt{3}} \Box} t(x)\right)^3 \tag{8.63}$$

(for a derivation see for instance [62]). It is instructive to expand this expression in powers of $\alpha'$ and check different orders separately. While the first order is very similar to the example in section 8.2, at second order we get up to prefactors

$$\int d^{26}x \left(\frac{1}{2} t^2(x) \Box^2 t(x) + t(x) (\Box t(x))^2\right). \tag{8.64}$$

We can now straightforwardly compute the associated boundary term as outlined in section 8.2 and get

$$X = \int_{\partial M} d^{25}x \, n_\mu \left(\frac{1}{2} t^2 \partial^\mu \Box \delta t - t \partial^\mu t \Box \delta t + 3 t \Box t \partial^\mu \delta t + \partial_\nu t \partial^\nu t \partial^\mu \delta t - 3 t \partial^\mu \Box t \delta t - 4 \partial^\mu t \Box t \delta t\right), \tag{8.65}$$

$$\delta X = \int_{\partial M} d^{25}x \, n_\mu \left(4 t \delta t \wedge \partial^\mu \Box \delta t + 4 \partial^\mu t \delta t \wedge \Box \delta t + 8 \Box t \delta t \wedge \partial^\mu \delta t + 4 t \Box \delta t \wedge \partial^\mu \delta t + 2 \partial_\nu t \partial^\nu \delta t \wedge \partial^\mu \delta t\right). \tag{8.66}$$

Even after applying boundary conditions, $\delta X$ is not zero: For Neumann conditions we are left with the first term and Dirichlet conditions would only eliminate the first three terms. The only possibility we have is to apply either Neumann and Dirichlet conditions simultaneously or put a condition on the third derivative of $\delta t$. Both are physically problematic because important subsets of the solution space are eliminated (see [56] for additional discussions). Naively we would expect that at higher orders in $\alpha'$, higher derivatives of $t(x)$ need to be constrained. At infinite order, we would then end up with boundary conditions on all derivatives of $t(x)$ and hence, in the case where the field $t(x)$ is analytic, it would be fully determined by the boundary conditions alone. This would clearly be an unphysical situation and a reflection of the above-mentioned fact that a sharply located boundary may not be meaningful in a non-local theory. However, we must be careful when summing up all infinitely many orders of $\alpha'$ because it might change the behaviour: In [2], the authors consider non-local kinetic operators in the context of the initial value problem, where one would a priori expect an infinite number of possible initial conditions. Indeed, it is shown that for a subclass of operators, this is not the case and only a finite number of initial conditions are sufficient. Although our problem is fundamentally different, it is possible that upon including all orders, a similar simplification occurs. Another possible route to follow is to consider an asymptotic boundary in the spirit of conformal compactifications. Then boundary conditions would be replaced by fall-off conditions on the string field and the problem might be ameliorated.

---
[48]GS wants to thank Barton Zwiebach for this observation.

We see from this example that as we expected, boundary terms do not arise from cyclic permutations; (8.63) is manifestly cyclic even for three different tachyon fields. The crucial step is to isolate the first string field, i. e. move the differential operator $e^{-\alpha' \ln \frac{4}{3\sqrt{3}} \Box}$ to act on the other two string fields. In fact, equation (8.57) should already contain the boundary contribution:

$$\langle \Psi_1, \Psi_2 * \Psi_3 \rangle =: \langle f_1 \circ \Psi_1(0) f_2 \circ \Psi_2(0) f_3 \circ \Psi_3(0) \rangle - X_{int}(\Psi_1, \Psi_2, \Psi_3) \tag{8.67}$$

Indeed, if one defines

$$S_{int} =: -\frac{1}{3} \langle f_1 \circ \Psi(0) f_2 \circ \Psi(0) f_3 \circ \Psi(0) \rangle \tag{8.68}$$

as the cyclic expression, then varying yields

$$\delta S_{int} = -\langle \delta\Psi, \Psi * \Psi \rangle - X(\delta\Psi, \Psi, \Psi) \tag{8.69}$$

as expected.

Calculating $X$ explicitly for the full string field is troublesome, from the above calculation we can expect that it will take a very complicated form. One might be tempted to apply the function $f_1^{-1}(z)$ to (8.57), however, $f_1(z)$ does not map the upper half plane to itself and is therefore not a symmetry of the 3-vertex. One could use the sliver frame representation of [70] where the vertex reads

$$S_{int} = -\frac{1}{3} \left\langle s \circ \Psi\left(\frac{3\pi}{4}\right) s \circ \Psi\left(\frac{\pi}{4}\right) s \circ \Psi\left(-\frac{\pi}{4}\right) \right\rangle_{C_\pi}. \tag{8.70}$$

$s$ induces the scale transformation $w \to \frac{2}{3}w$ which is generated by the operator

$$U_3 = \left(\frac{2}{3}\right)^{\mathcal{L}_0} \tag{8.71}$$

where $\mathcal{L}_0$ is the zero-mode of the energy momentum tensor in the sliver frame. Writing (8.70) as a two-point function leads to the expression

$$S_{int} = -\frac{1}{3} \left\langle \Psi, U_3^* \left(s \circ \Psi\left(\frac{\pi}{4}\right) s \circ \Psi\left(-\frac{\pi}{4}\right)\right) \right\rangle_{C_\pi}. \tag{8.72}$$

This is exactly the step where the boundary terms come in, where $U_3$ is moved to the other side of the BPZ-product. To use the method of section 8.4 to calculate those boundary terms, we would have to write $U_3$ as a contour integral. This is not easy though, since $U_3$ is just the exponential of a contour integral, hence we cannot proceed straightforwardly.

To sum up, the cubic boundary terms exhibit serious difficulties, on the one hand because of non-locality and on the other hand also because of computational issues. A careful treatment is left for future publications.

## 8.6 Conclusion and outlook

In this paper we analyzed Witten's open string field theory in the case where spacetime contains a boundary. We found non-trivial boundary contributions from the variational principle both for the kinetic as well as for the cubic term: In the former case, the cyclicity relation for the BRST-operator is violated. We calculated a Gibbons-Hawking-like surface term (8.31) that can be added to the action to guarantee a well-defined variational principle if the string field obeys either Neumann or Dirichlet conditions. The derivation was done in two different ways, one rather heuristic way and one using the breaking of conformal invariance in the path-integral formalism, both yield the same result. For the cubic term we saw that boundary terms arise when rewriting the manifestly cyclic form as a BPZ-product. However, it is not straightforward to find suitable boundary or fall-off conditions due to the non-locality of the three-vertex, which is one of the most interesting prospects for further research. As another future direction, it could be enlightening to extend the study to more general backgrounds, i. e. linear dilaton or Liouville CFTs.

**Note added**

Shortly after this work was completed, two other papers on the same topic appeared which were created independently. In [31] the authors compute the kinetic boundary term for the closed string and [51] deals with gauge invariance and boundary modes.

**Acknowledgements**

GS wants to thank Martin Schnabl for suggesting the topic and collaboration as well as Ted Erler, Atakan Firat, Carlo Maccaferri, Riccardo Poletti, Alberto Ruffino, Jaroslav Scheinpflug, Raphaela Wutte and Barton Zwiebach for useful discussions. This work was co-funded by the European Union and supported by the Czech Ministry of Education, Youth and Sports (Project No. FORTE – CZ.02.01.01/00/22_008/0004632).